\pdfoutput=1	
\documentclass[9pt,twocolumn,twoside]{pnas-new}
\setboolean{displaywatermark}{false}

\usepackage{framed}
\usepackage{moreverb}
\usepackage{subcaption}
\usepackage{xspace}
\usepackage{flushend}			%

\usepackage{xfrac}			%

\usepackage{amsfonts}
\usepackage{amssymb}    
\usepackage{amsmath}
\usepackage[thmmarks,amsmath,hyperref]{ntheorem}
\usepackage{graphicx}
\usepackage{times}
\usepackage[scaled]{helvet} 		%
\usepackage{textcomp}
\usepackage{color}
\usepackage{colortbl}
\usepackage{booktabs}
\usepackage{listings}
\usepackage{multicol}
\usepackage{times}
\usepackage{pifont}			%
\usepackage{microtype}

\usepackage{tikz}

\usepackage{libertine}
\usepackage{libertinust1math}
\usepackage[scaled=0.8]{beramono}
\usepackage[T1]{fontenc}

\usepackage{url}
\usepackage{multirow}
\usepackage{array}

\usepackage[normalem]{ulem}

\usepackage[dvipsnames]{xcolor}

\usepackage{siunitx}

\newcommand\code[1]             {{\bf\ttfamily #1}}
\newcommand\codelight[1]             {{\ttfamily #1}}

\def\cameraready{}

\ifdefined\cameraready
	\newcommand\remove[1]	{}
	
\else
	\newcommand\remove[1]	{\textcolor{red}{#1}}
	
\fi

\newcommand\rmtext[1]		{}

\newcommand{\stkout}[1]{\ifmmode\text{\sout{\ensuremath{#1}}}\else\sout{#1}\fi}

\newcommand\adc     {\textsc{\Large adc}}

\newcounter{rownumber}[figure]
\renewcommand{\therownumber}{R\arabic{rownumber}}
\newcommand{\rowcountstep}{\refstepcounter{rownumber}\therownumber}

\makeatletter
\renewcommand{\p@subsection}{\thesection.}
\renewcommand{\p@figure}{\ifnum\pdfstrcmp{\thesection}{SM}=0 SM.F\fi}
\renewcommand{\p@table}{\ifnum\pdfstrcmp{\thesection}{SM}=0 SM.T\fi}
\makeatother 

\definecolor{listinggreen}{rgb}{0,0.6,0}
\definecolor{listinggray}{rgb}{0.5,0.5,0.5}
\definecolor{listingmauve}{rgb}{0.58,0,0.82}
\definecolor{listingkeywordcolor}{rgb}{1.0,0.4,0.0}
\definecolor{listinglightgray}{rgb}{0.9863,0.9863,0.9863}

\definecolor{a}{rgb}{0.9,0.95,0.95}
\definecolor{b}{rgb}{0.99,0.99,0.99}

\lstdefinelanguage{FSharp}
{
	morekeywords	= {
    let,
    type,
    Measure,
	},
	sensitive	= false,
	morecomment	= [l]{\#},
	morecomment	= [s]{(*}{*)},
}

\lstdefinelanguage{Newton}
{
	morekeywords	= {
		signal,
		derivation,
		symbol,
		name,
		invariant,
		constant,
		English,
		sensor,
		name,
		none,
		dot,
		cross,
		derivative,
		integral,
		interface,
		i2c,
		spi,
		analog,
		write,
		read,
		delay,
		range,
		erasuretoken,
		uncertainty,
		accuracy,
		precision,
		Gaussian,
		exponential,
		biexponential,
		to,
		bits,
		dimensionless,
		include
	},
	sensitive	= false,
	morecomment	= [l]{\#},
	morecomment	= [s]{/*}{*/},
}

\lstdefinelanguage{UxHwC}
{
	morekeywords	= {
		static,
		void,
		float,
		int8_t,
		int16_t,
		uint16_t,
		if,
		else,
		for,
		while,
		return,
	},
	morekeywords=[2]{
		UxHwFloatUniformDist,
    },
	sensitive	= false,
	morecomment	= [l]{//},
	morecomment	= [s]{/*}{*/},
	keywordstyle=[2]\color{Bittersweet}\bfseries,
}

\usepackage{hyperref}
\hypersetup{
    colorlinks=false, %
    linktoc=all,     %
    linkcolor=blue,  %
}

\templatetype{pnasresearcharticle} 

\title{Digital Methods to Quantify Sensor Output Uncertainty in Real Time}

\author[a]{Orestis Kaparounakis}
\author[a,b,\textdagger]{Phillip Stanley-Marbell}

\affil[a]{Physical Computation Laboratory, Department of Engineering, University of Cambridge, Cambridge CB3 0FA, UK.}
\affil[b]{Signaloid}

\leadauthor{Kaparounakis}

\authorcontributions{Contributions: Orestis Kaparounakis and Phillip
Stanley-Marbell conceived the original idea. Orestis Kaparounakis picked the
sensor, wrote the work-specific code, performed the measurement experiments, ran
the benchmarking, and wrote this article. Phillip Stanley-Marbell provided
frequent feedback.}
\authordeclaration{The research results presented in the article are part of a commercial activity at Signaloid in which both of the authors have a commercial interest.}
\correspondingauthor{\textsuperscript{\textdagger}To whom correspondence should be addressed. E-mail: {ps751@cam.ac.uk}}

\vspace{-0.1in}

\usepackage{pgfkeys}

    \newcommand{\getvariable}[1]{\pgfkeysvalueof{/#1}}
    \newcommand{\considereduncertain}[1]{\getvariable{ConsideredUncertain#1}}
    \pgfkeys{
    /ConsideredUncertainKVdd/.initial=Yes,
/ConsideredUncertainVdd25/.initial=Yes,
/ConsideredUncertainKvPtat/.initial=Yes,
/ConsideredUncertainKtPtat/.initial=Yes,
/ConsideredUncertainVPtat25/.initial=Yes,
/ConsideredUncertainAlphaPtat/.initial=Yes,
/ConsideredUncertainGainEe/.initial=Yes,
/ConsideredUncertainTgc/.initial=Yes,
/ConsideredUncertainKsTa/.initial=Yes,
/ConsideredUncertainStep/.initial=Yes,
/ConsideredUncertainCt2/.initial=Yes,
/ConsideredUncertainCt3/.initial=Yes,
/ConsideredUncertainKsToScale/.initial=No,
/ConsideredUncertainKsTo0/.initial=Yes,
/ConsideredUncertainKsTo1/.initial=Yes,
/ConsideredUncertainKsTo2/.initial=Yes,
/ConsideredUncertainKsTo3/.initial=Yes,
/ConsideredUncertainAlphaScale/.initial=No,
/ConsideredUncertainOffsetSp0/.initial=Yes,
/ConsideredUncertainOffsetSp1/.initial=Yes,
/ConsideredUncertainAlphaSp0/.initial=Yes,
/ConsideredUncertainAlphaSp1/.initial=Yes,
/ConsideredUncertainCpKta/.initial=Yes,
/ConsideredUncertainKtaScale1/.initial=No,
/ConsideredUncertainCpKv/.initial=Yes,
/ConsideredUncertainKvScale/.initial=No,
/ConsideredUncertainAccRemScale/.initial=No,
/ConsideredUncertainAccColumnScale/.initial=No,
/ConsideredUncertainAccRowScale/.initial=No,
/ConsideredUncertainAlphaRef/.initial=Yes,
/ConsideredUncertainAccRowI0/.initial=Yes,
/ConsideredUncertainAccRowI1/.initial=Yes,
/ConsideredUncertainAccRowI2/.initial=Yes,
/ConsideredUncertainAccRowI3/.initial=Yes,
/ConsideredUncertainAccRowI4/.initial=Yes,
/ConsideredUncertainAccRowI5/.initial=Yes,
/ConsideredUncertainAccRowI6/.initial=Yes,
/ConsideredUncertainAccRowI7/.initial=Yes,
/ConsideredUncertainAccRowI8/.initial=Yes,
/ConsideredUncertainAccRowI9/.initial=Yes,
/ConsideredUncertainAccRowI10/.initial=Yes,
/ConsideredUncertainAccRowI11/.initial=Yes,
/ConsideredUncertainAccRowI12/.initial=Yes,
/ConsideredUncertainAccRowI13/.initial=Yes,
/ConsideredUncertainAccRowI14/.initial=Yes,
/ConsideredUncertainAccRowI15/.initial=Yes,
/ConsideredUncertainAccRowI16/.initial=Yes,
/ConsideredUncertainAccRowI17/.initial=Yes,
/ConsideredUncertainAccRowI18/.initial=Yes,
/ConsideredUncertainAccRowI19/.initial=Yes,
/ConsideredUncertainAccRowI20/.initial=Yes,
/ConsideredUncertainAccRowI21/.initial=Yes,
/ConsideredUncertainAccRowI22/.initial=Yes,
/ConsideredUncertainAccRowI23/.initial=Yes,
/ConsideredUncertainAccColumnI0/.initial=Yes,
/ConsideredUncertainAccColumnI1/.initial=Yes,
/ConsideredUncertainAccColumnI2/.initial=Yes,
/ConsideredUncertainAccColumnI3/.initial=Yes,
/ConsideredUncertainAccColumnI4/.initial=Yes,
/ConsideredUncertainAccColumnI5/.initial=Yes,
/ConsideredUncertainAccColumnI6/.initial=Yes,
/ConsideredUncertainAccColumnI7/.initial=Yes,
/ConsideredUncertainAccColumnI8/.initial=Yes,
/ConsideredUncertainAccColumnI9/.initial=Yes,
/ConsideredUncertainAccColumnI10/.initial=Yes,
/ConsideredUncertainAccColumnI11/.initial=Yes,
/ConsideredUncertainAccColumnI12/.initial=Yes,
/ConsideredUncertainAccColumnI13/.initial=Yes,
/ConsideredUncertainAccColumnI14/.initial=Yes,
/ConsideredUncertainAccColumnI15/.initial=Yes,
/ConsideredUncertainAccColumnI16/.initial=Yes,
/ConsideredUncertainAccColumnI17/.initial=Yes,
/ConsideredUncertainAccColumnI18/.initial=Yes,
/ConsideredUncertainAccColumnI19/.initial=Yes,
/ConsideredUncertainAccColumnI20/.initial=Yes,
/ConsideredUncertainAccColumnI21/.initial=Yes,
/ConsideredUncertainAccColumnI22/.initial=Yes,
/ConsideredUncertainAccColumnI23/.initial=Yes,
/ConsideredUncertainAccColumnI24/.initial=Yes,
/ConsideredUncertainAccColumnI25/.initial=Yes,
/ConsideredUncertainAccColumnI26/.initial=Yes,
/ConsideredUncertainAccColumnI27/.initial=Yes,
/ConsideredUncertainAccColumnI28/.initial=Yes,
/ConsideredUncertainAccColumnI29/.initial=Yes,
/ConsideredUncertainAccColumnI30/.initial=Yes,
/ConsideredUncertainAccColumnI31/.initial=Yes,
/ConsideredUncertainAlphaTemp0/.initial=Yes,
/ConsideredUncertainAlphaTemp1/.initial=Yes,
/ConsideredUncertainAlphaTemp2/.initial=Yes,
/ConsideredUncertainAlphaTemp3/.initial=Yes,
/ConsideredUncertainAlphaTemp4/.initial=Yes,
/ConsideredUncertainAlphaTemp5/.initial=Yes,
/ConsideredUncertainAlphaTemp6/.initial=Yes,
/ConsideredUncertainAlphaTemp7/.initial=Yes,
/ConsideredUncertainAlphaTemp8/.initial=Yes,
/ConsideredUncertainAlphaTemp9/.initial=Yes,
/ConsideredUncertainAlphaTemp10/.initial=Yes,
/ConsideredUncertainAlphaTemp11/.initial=Yes,
/ConsideredUncertainAlphaTemp12/.initial=Yes,
/ConsideredUncertainAlphaTemp13/.initial=Yes,
/ConsideredUncertainAlphaTemp14/.initial=Yes,
/ConsideredUncertainAlphaTemp15/.initial=Yes,
/ConsideredUncertainAlphaTemp16/.initial=Yes,
/ConsideredUncertainAlphaTemp17/.initial=Yes,
/ConsideredUncertainAlphaTemp18/.initial=Yes,
/ConsideredUncertainAlphaTemp19/.initial=Yes,
/ConsideredUncertainAlphaTemp20/.initial=Yes,
/ConsideredUncertainAlphaTemp21/.initial=Yes,
/ConsideredUncertainAlphaTemp22/.initial=Yes,
/ConsideredUncertainAlphaTemp23/.initial=Yes,
/ConsideredUncertainAlphaTemp24/.initial=Yes,
/ConsideredUncertainAlphaTemp25/.initial=Yes,
/ConsideredUncertainAlphaTemp26/.initial=Yes,
/ConsideredUncertainAlphaTemp27/.initial=Yes,
/ConsideredUncertainAlphaTemp28/.initial=Yes,
/ConsideredUncertainAlphaTemp29/.initial=Yes,
/ConsideredUncertainAlphaTemp30/.initial=Yes,
/ConsideredUncertainAlphaTemp31/.initial=Yes,
/ConsideredUncertainAlphaTemp32/.initial=Yes,
/ConsideredUncertainAlphaTemp33/.initial=Yes,
/ConsideredUncertainAlphaTemp34/.initial=Yes,
/ConsideredUncertainAlphaTemp35/.initial=Yes,
/ConsideredUncertainAlphaTemp36/.initial=Yes,
/ConsideredUncertainAlphaTemp37/.initial=Yes,
/ConsideredUncertainAlphaTemp38/.initial=Yes,
/ConsideredUncertainAlphaTemp39/.initial=Yes,
/ConsideredUncertainAlphaTemp40/.initial=Yes,
/ConsideredUncertainAlphaTemp41/.initial=Yes,
/ConsideredUncertainAlphaTemp42/.initial=Yes,
/ConsideredUncertainAlphaTemp43/.initial=Yes,
/ConsideredUncertainAlphaTemp44/.initial=Yes,
/ConsideredUncertainAlphaTemp45/.initial=Yes,
/ConsideredUncertainAlphaTemp46/.initial=Yes,
/ConsideredUncertainAlphaTemp47/.initial=Yes,
/ConsideredUncertainAlphaTemp48/.initial=Yes,
/ConsideredUncertainAlphaTemp49/.initial=Yes,
/ConsideredUncertainAlphaTemp50/.initial=Yes,
/ConsideredUncertainAlphaTemp51/.initial=Yes,
/ConsideredUncertainAlphaTemp52/.initial=Yes,
/ConsideredUncertainAlphaTemp53/.initial=Yes,
/ConsideredUncertainAlphaTemp54/.initial=Yes,
/ConsideredUncertainAlphaTemp55/.initial=Yes,
/ConsideredUncertainAlphaTemp56/.initial=Yes,
/ConsideredUncertainAlphaTemp57/.initial=Yes,
/ConsideredUncertainAlphaTemp58/.initial=Yes,
/ConsideredUncertainAlphaTemp59/.initial=Yes,
/ConsideredUncertainAlphaTemp60/.initial=Yes,
/ConsideredUncertainAlphaTemp61/.initial=Yes,
/ConsideredUncertainAlphaTemp62/.initial=Yes,
/ConsideredUncertainAlphaTemp63/.initial=Yes,
/ConsideredUncertainAlphaTemp64/.initial=Yes,
/ConsideredUncertainAlphaTemp65/.initial=Yes,
/ConsideredUncertainAlphaTemp66/.initial=Yes,
/ConsideredUncertainAlphaTemp67/.initial=Yes,
/ConsideredUncertainAlphaTemp68/.initial=Yes,
/ConsideredUncertainAlphaTemp69/.initial=Yes,
/ConsideredUncertainAlphaTemp70/.initial=Yes,
/ConsideredUncertainAlphaTemp71/.initial=Yes,
/ConsideredUncertainAlphaTemp72/.initial=Yes,
/ConsideredUncertainAlphaTemp73/.initial=Yes,
/ConsideredUncertainAlphaTemp74/.initial=Yes,
/ConsideredUncertainAlphaTemp75/.initial=Yes,
/ConsideredUncertainAlphaTemp76/.initial=Yes,
/ConsideredUncertainAlphaTemp77/.initial=Yes,
/ConsideredUncertainAlphaTemp78/.initial=Yes,
/ConsideredUncertainAlphaTemp79/.initial=Yes,
/ConsideredUncertainAlphaTemp80/.initial=Yes,
/ConsideredUncertainAlphaTemp81/.initial=Yes,
/ConsideredUncertainAlphaTemp82/.initial=Yes,
/ConsideredUncertainAlphaTemp83/.initial=Yes,
/ConsideredUncertainAlphaTemp84/.initial=Yes,
/ConsideredUncertainAlphaTemp85/.initial=Yes,
/ConsideredUncertainAlphaTemp86/.initial=Yes,
/ConsideredUncertainAlphaTemp87/.initial=Yes,
/ConsideredUncertainAlphaTemp88/.initial=Yes,
/ConsideredUncertainAlphaTemp89/.initial=Yes,
/ConsideredUncertainAlphaTemp90/.initial=Yes,
/ConsideredUncertainAlphaTemp91/.initial=Yes,
/ConsideredUncertainAlphaTemp92/.initial=Yes,
/ConsideredUncertainAlphaTemp93/.initial=Yes,
/ConsideredUncertainAlphaTemp94/.initial=Yes,
/ConsideredUncertainAlphaTemp95/.initial=Yes,
/ConsideredUncertainAlphaTemp96/.initial=Yes,
/ConsideredUncertainAlphaTemp97/.initial=Yes,
/ConsideredUncertainAlphaTemp98/.initial=Yes,
/ConsideredUncertainAlphaTemp99/.initial=Yes,
/ConsideredUncertainAlphaTemp100/.initial=Yes,
/ConsideredUncertainAlphaTemp101/.initial=Yes,
/ConsideredUncertainAlphaTemp102/.initial=Yes,
/ConsideredUncertainAlphaTemp103/.initial=Yes,
/ConsideredUncertainAlphaTemp104/.initial=Yes,
/ConsideredUncertainAlphaTemp105/.initial=Yes,
/ConsideredUncertainAlphaTemp106/.initial=Yes,
/ConsideredUncertainAlphaTemp107/.initial=Yes,
/ConsideredUncertainAlphaTemp108/.initial=Yes,
/ConsideredUncertainAlphaTemp109/.initial=Yes,
/ConsideredUncertainAlphaTemp110/.initial=Yes,
/ConsideredUncertainAlphaTemp111/.initial=Yes,
/ConsideredUncertainAlphaTemp112/.initial=Yes,
/ConsideredUncertainAlphaTemp113/.initial=Yes,
/ConsideredUncertainAlphaTemp114/.initial=Yes,
/ConsideredUncertainAlphaTemp115/.initial=Yes,
/ConsideredUncertainAlphaTemp116/.initial=Yes,
/ConsideredUncertainAlphaTemp117/.initial=Yes,
/ConsideredUncertainAlphaTemp118/.initial=Yes,
/ConsideredUncertainAlphaTemp119/.initial=Yes,
/ConsideredUncertainAlphaTemp120/.initial=Yes,
/ConsideredUncertainAlphaTemp121/.initial=Yes,
/ConsideredUncertainAlphaTemp122/.initial=Yes,
/ConsideredUncertainAlphaTemp123/.initial=Yes,
/ConsideredUncertainAlphaTemp124/.initial=Yes,
/ConsideredUncertainAlphaTemp125/.initial=Yes,
/ConsideredUncertainAlphaTemp126/.initial=Yes,
/ConsideredUncertainAlphaTemp127/.initial=Yes,
/ConsideredUncertainAlphaTemp128/.initial=Yes,
/ConsideredUncertainAlphaTemp129/.initial=Yes,
/ConsideredUncertainAlphaTemp130/.initial=Yes,
/ConsideredUncertainAlphaTemp131/.initial=Yes,
/ConsideredUncertainAlphaTemp132/.initial=Yes,
/ConsideredUncertainAlphaTemp133/.initial=Yes,
/ConsideredUncertainAlphaTemp134/.initial=Yes,
/ConsideredUncertainAlphaTemp135/.initial=Yes,
/ConsideredUncertainAlphaTemp136/.initial=Yes,
/ConsideredUncertainAlphaTemp137/.initial=Yes,
/ConsideredUncertainAlphaTemp138/.initial=Yes,
/ConsideredUncertainAlphaTemp139/.initial=Yes,
/ConsideredUncertainAlphaTemp140/.initial=Yes,
/ConsideredUncertainAlphaTemp141/.initial=Yes,
/ConsideredUncertainAlphaTemp142/.initial=Yes,
/ConsideredUncertainAlphaTemp143/.initial=Yes,
/ConsideredUncertainAlphaTemp144/.initial=Yes,
/ConsideredUncertainAlphaTemp145/.initial=Yes,
/ConsideredUncertainAlphaTemp146/.initial=Yes,
/ConsideredUncertainAlphaTemp147/.initial=Yes,
/ConsideredUncertainAlphaTemp148/.initial=Yes,
/ConsideredUncertainAlphaTemp149/.initial=Yes,
/ConsideredUncertainAlphaTemp150/.initial=Yes,
/ConsideredUncertainAlphaTemp151/.initial=Yes,
/ConsideredUncertainAlphaTemp152/.initial=Yes,
/ConsideredUncertainAlphaTemp153/.initial=Yes,
/ConsideredUncertainAlphaTemp154/.initial=Yes,
/ConsideredUncertainAlphaTemp155/.initial=Yes,
/ConsideredUncertainAlphaTemp156/.initial=Yes,
/ConsideredUncertainAlphaTemp157/.initial=Yes,
/ConsideredUncertainAlphaTemp158/.initial=Yes,
/ConsideredUncertainAlphaTemp159/.initial=Yes,
/ConsideredUncertainAlphaTemp160/.initial=Yes,
/ConsideredUncertainAlphaTemp161/.initial=Yes,
/ConsideredUncertainAlphaTemp162/.initial=Yes,
/ConsideredUncertainAlphaTemp163/.initial=Yes,
/ConsideredUncertainAlphaTemp164/.initial=Yes,
/ConsideredUncertainAlphaTemp165/.initial=Yes,
/ConsideredUncertainAlphaTemp166/.initial=Yes,
/ConsideredUncertainAlphaTemp167/.initial=Yes,
/ConsideredUncertainAlphaTemp168/.initial=Yes,
/ConsideredUncertainAlphaTemp169/.initial=Yes,
/ConsideredUncertainAlphaTemp170/.initial=Yes,
/ConsideredUncertainAlphaTemp171/.initial=Yes,
/ConsideredUncertainAlphaTemp172/.initial=Yes,
/ConsideredUncertainAlphaTemp173/.initial=Yes,
/ConsideredUncertainAlphaTemp174/.initial=Yes,
/ConsideredUncertainAlphaTemp175/.initial=Yes,
/ConsideredUncertainAlphaTemp176/.initial=Yes,
/ConsideredUncertainAlphaTemp177/.initial=Yes,
/ConsideredUncertainAlphaTemp178/.initial=Yes,
/ConsideredUncertainAlphaTemp179/.initial=Yes,
/ConsideredUncertainAlphaTemp180/.initial=Yes,
/ConsideredUncertainAlphaTemp181/.initial=Yes,
/ConsideredUncertainAlphaTemp182/.initial=Yes,
/ConsideredUncertainAlphaTemp183/.initial=Yes,
/ConsideredUncertainAlphaTemp184/.initial=Yes,
/ConsideredUncertainAlphaTemp185/.initial=Yes,
/ConsideredUncertainAlphaTemp186/.initial=Yes,
/ConsideredUncertainAlphaTemp187/.initial=Yes,
/ConsideredUncertainAlphaTemp188/.initial=Yes,
/ConsideredUncertainAlphaTemp189/.initial=Yes,
/ConsideredUncertainAlphaTemp190/.initial=Yes,
/ConsideredUncertainAlphaTemp191/.initial=Yes,
/ConsideredUncertainAlphaTemp192/.initial=Yes,
/ConsideredUncertainAlphaTemp193/.initial=Yes,
/ConsideredUncertainAlphaTemp194/.initial=Yes,
/ConsideredUncertainAlphaTemp195/.initial=Yes,
/ConsideredUncertainAlphaTemp196/.initial=Yes,
/ConsideredUncertainAlphaTemp197/.initial=Yes,
/ConsideredUncertainAlphaTemp198/.initial=Yes,
/ConsideredUncertainAlphaTemp199/.initial=Yes,
/ConsideredUncertainAlphaTemp200/.initial=Yes,
/ConsideredUncertainAlphaTemp201/.initial=Yes,
/ConsideredUncertainAlphaTemp202/.initial=Yes,
/ConsideredUncertainAlphaTemp203/.initial=Yes,
/ConsideredUncertainAlphaTemp204/.initial=Yes,
/ConsideredUncertainAlphaTemp205/.initial=Yes,
/ConsideredUncertainAlphaTemp206/.initial=Yes,
/ConsideredUncertainAlphaTemp207/.initial=Yes,
/ConsideredUncertainAlphaTemp208/.initial=Yes,
/ConsideredUncertainAlphaTemp209/.initial=Yes,
/ConsideredUncertainAlphaTemp210/.initial=Yes,
/ConsideredUncertainAlphaTemp211/.initial=Yes,
/ConsideredUncertainAlphaTemp212/.initial=Yes,
/ConsideredUncertainAlphaTemp213/.initial=Yes,
/ConsideredUncertainAlphaTemp214/.initial=Yes,
/ConsideredUncertainAlphaTemp215/.initial=Yes,
/ConsideredUncertainAlphaTemp216/.initial=Yes,
/ConsideredUncertainAlphaTemp217/.initial=Yes,
/ConsideredUncertainAlphaTemp218/.initial=Yes,
/ConsideredUncertainAlphaTemp219/.initial=Yes,
/ConsideredUncertainAlphaTemp220/.initial=Yes,
/ConsideredUncertainAlphaTemp221/.initial=Yes,
/ConsideredUncertainAlphaTemp222/.initial=Yes,
/ConsideredUncertainAlphaTemp223/.initial=Yes,
/ConsideredUncertainAlphaTemp224/.initial=Yes,
/ConsideredUncertainAlphaTemp225/.initial=Yes,
/ConsideredUncertainAlphaTemp226/.initial=Yes,
/ConsideredUncertainAlphaTemp227/.initial=Yes,
/ConsideredUncertainAlphaTemp228/.initial=Yes,
/ConsideredUncertainAlphaTemp229/.initial=Yes,
/ConsideredUncertainAlphaTemp230/.initial=Yes,
/ConsideredUncertainAlphaTemp231/.initial=Yes,
/ConsideredUncertainAlphaTemp232/.initial=Yes,
/ConsideredUncertainAlphaTemp233/.initial=Yes,
/ConsideredUncertainAlphaTemp234/.initial=Yes,
/ConsideredUncertainAlphaTemp235/.initial=Yes,
/ConsideredUncertainAlphaTemp236/.initial=Yes,
/ConsideredUncertainAlphaTemp237/.initial=Yes,
/ConsideredUncertainAlphaTemp238/.initial=Yes,
/ConsideredUncertainAlphaTemp239/.initial=Yes,
/ConsideredUncertainAlphaTemp240/.initial=Yes,
/ConsideredUncertainAlphaTemp241/.initial=Yes,
/ConsideredUncertainAlphaTemp242/.initial=Yes,
/ConsideredUncertainAlphaTemp243/.initial=Yes,
/ConsideredUncertainAlphaTemp244/.initial=Yes,
/ConsideredUncertainAlphaTemp245/.initial=Yes,
/ConsideredUncertainAlphaTemp246/.initial=Yes,
/ConsideredUncertainAlphaTemp247/.initial=Yes,
/ConsideredUncertainAlphaTemp248/.initial=Yes,
/ConsideredUncertainAlphaTemp249/.initial=Yes,
/ConsideredUncertainAlphaTemp250/.initial=Yes,
/ConsideredUncertainAlphaTemp251/.initial=Yes,
/ConsideredUncertainAlphaTemp252/.initial=Yes,
/ConsideredUncertainAlphaTemp253/.initial=Yes,
/ConsideredUncertainAlphaTemp254/.initial=Yes,
/ConsideredUncertainAlphaTemp255/.initial=Yes,
/ConsideredUncertainAlphaTemp256/.initial=Yes,
/ConsideredUncertainAlphaTemp257/.initial=Yes,
/ConsideredUncertainAlphaTemp258/.initial=Yes,
/ConsideredUncertainAlphaTemp259/.initial=Yes,
/ConsideredUncertainAlphaTemp260/.initial=Yes,
/ConsideredUncertainAlphaTemp261/.initial=Yes,
/ConsideredUncertainAlphaTemp262/.initial=Yes,
/ConsideredUncertainAlphaTemp263/.initial=Yes,
/ConsideredUncertainAlphaTemp264/.initial=Yes,
/ConsideredUncertainAlphaTemp265/.initial=Yes,
/ConsideredUncertainAlphaTemp266/.initial=Yes,
/ConsideredUncertainAlphaTemp267/.initial=Yes,
/ConsideredUncertainAlphaTemp268/.initial=Yes,
/ConsideredUncertainAlphaTemp269/.initial=Yes,
/ConsideredUncertainAlphaTemp270/.initial=Yes,
/ConsideredUncertainAlphaTemp271/.initial=Yes,
/ConsideredUncertainAlphaTemp272/.initial=Yes,
/ConsideredUncertainAlphaTemp273/.initial=Yes,
/ConsideredUncertainAlphaTemp274/.initial=Yes,
/ConsideredUncertainAlphaTemp275/.initial=Yes,
/ConsideredUncertainAlphaTemp276/.initial=Yes,
/ConsideredUncertainAlphaTemp277/.initial=Yes,
/ConsideredUncertainAlphaTemp278/.initial=Yes,
/ConsideredUncertainAlphaTemp279/.initial=Yes,
/ConsideredUncertainAlphaTemp280/.initial=Yes,
/ConsideredUncertainAlphaTemp281/.initial=Yes,
/ConsideredUncertainAlphaTemp282/.initial=Yes,
/ConsideredUncertainAlphaTemp283/.initial=Yes,
/ConsideredUncertainAlphaTemp284/.initial=Yes,
/ConsideredUncertainAlphaTemp285/.initial=Yes,
/ConsideredUncertainAlphaTemp286/.initial=Yes,
/ConsideredUncertainAlphaTemp287/.initial=Yes,
/ConsideredUncertainAlphaTemp288/.initial=Yes,
/ConsideredUncertainAlphaTemp289/.initial=Yes,
/ConsideredUncertainAlphaTemp290/.initial=Yes,
/ConsideredUncertainAlphaTemp291/.initial=Yes,
/ConsideredUncertainAlphaTemp292/.initial=Yes,
/ConsideredUncertainAlphaTemp293/.initial=Yes,
/ConsideredUncertainAlphaTemp294/.initial=Yes,
/ConsideredUncertainAlphaTemp295/.initial=Yes,
/ConsideredUncertainAlphaTemp296/.initial=Yes,
/ConsideredUncertainAlphaTemp297/.initial=Yes,
/ConsideredUncertainAlphaTemp298/.initial=Yes,
/ConsideredUncertainAlphaTemp299/.initial=Yes,
/ConsideredUncertainAlphaTemp300/.initial=Yes,
/ConsideredUncertainAlphaTemp301/.initial=Yes,
/ConsideredUncertainAlphaTemp302/.initial=Yes,
/ConsideredUncertainAlphaTemp303/.initial=Yes,
/ConsideredUncertainAlphaTemp304/.initial=Yes,
/ConsideredUncertainAlphaTemp305/.initial=Yes,
/ConsideredUncertainAlphaTemp306/.initial=Yes,
/ConsideredUncertainAlphaTemp307/.initial=Yes,
/ConsideredUncertainAlphaTemp308/.initial=Yes,
/ConsideredUncertainAlphaTemp309/.initial=Yes,
/ConsideredUncertainAlphaTemp310/.initial=Yes,
/ConsideredUncertainAlphaTemp311/.initial=Yes,
/ConsideredUncertainAlphaTemp312/.initial=Yes,
/ConsideredUncertainAlphaTemp313/.initial=Yes,
/ConsideredUncertainAlphaTemp314/.initial=Yes,
/ConsideredUncertainAlphaTemp315/.initial=Yes,
/ConsideredUncertainAlphaTemp316/.initial=Yes,
/ConsideredUncertainAlphaTemp317/.initial=Yes,
/ConsideredUncertainAlphaTemp318/.initial=Yes,
/ConsideredUncertainAlphaTemp319/.initial=Yes,
/ConsideredUncertainAlphaTemp320/.initial=Yes,
/ConsideredUncertainAlphaTemp321/.initial=Yes,
/ConsideredUncertainAlphaTemp322/.initial=Yes,
/ConsideredUncertainAlphaTemp323/.initial=Yes,
/ConsideredUncertainAlphaTemp324/.initial=Yes,
/ConsideredUncertainAlphaTemp325/.initial=Yes,
/ConsideredUncertainAlphaTemp326/.initial=Yes,
/ConsideredUncertainAlphaTemp327/.initial=Yes,
/ConsideredUncertainAlphaTemp328/.initial=Yes,
/ConsideredUncertainAlphaTemp329/.initial=Yes,
/ConsideredUncertainAlphaTemp330/.initial=Yes,
/ConsideredUncertainAlphaTemp331/.initial=Yes,
/ConsideredUncertainAlphaTemp332/.initial=Yes,
/ConsideredUncertainAlphaTemp333/.initial=Yes,
/ConsideredUncertainAlphaTemp334/.initial=Yes,
/ConsideredUncertainAlphaTemp335/.initial=Yes,
/ConsideredUncertainAlphaTemp336/.initial=Yes,
/ConsideredUncertainAlphaTemp337/.initial=Yes,
/ConsideredUncertainAlphaTemp338/.initial=Yes,
/ConsideredUncertainAlphaTemp339/.initial=Yes,
/ConsideredUncertainAlphaTemp340/.initial=Yes,
/ConsideredUncertainAlphaTemp341/.initial=Yes,
/ConsideredUncertainAlphaTemp342/.initial=Yes,
/ConsideredUncertainAlphaTemp343/.initial=Yes,
/ConsideredUncertainAlphaTemp344/.initial=Yes,
/ConsideredUncertainAlphaTemp345/.initial=Yes,
/ConsideredUncertainAlphaTemp346/.initial=Yes,
/ConsideredUncertainAlphaTemp347/.initial=Yes,
/ConsideredUncertainAlphaTemp348/.initial=Yes,
/ConsideredUncertainAlphaTemp349/.initial=Yes,
/ConsideredUncertainAlphaTemp350/.initial=Yes,
/ConsideredUncertainAlphaTemp351/.initial=Yes,
/ConsideredUncertainAlphaTemp352/.initial=Yes,
/ConsideredUncertainAlphaTemp353/.initial=Yes,
/ConsideredUncertainAlphaTemp354/.initial=Yes,
/ConsideredUncertainAlphaTemp355/.initial=Yes,
/ConsideredUncertainAlphaTemp356/.initial=Yes,
/ConsideredUncertainAlphaTemp357/.initial=Yes,
/ConsideredUncertainAlphaTemp358/.initial=Yes,
/ConsideredUncertainAlphaTemp359/.initial=Yes,
/ConsideredUncertainAlphaTemp360/.initial=Yes,
/ConsideredUncertainAlphaTemp361/.initial=Yes,
/ConsideredUncertainAlphaTemp362/.initial=Yes,
/ConsideredUncertainAlphaTemp363/.initial=Yes,
/ConsideredUncertainAlphaTemp364/.initial=Yes,
/ConsideredUncertainAlphaTemp365/.initial=Yes,
/ConsideredUncertainAlphaTemp366/.initial=Yes,
/ConsideredUncertainAlphaTemp367/.initial=Yes,
/ConsideredUncertainAlphaTemp368/.initial=Yes,
/ConsideredUncertainAlphaTemp369/.initial=Yes,
/ConsideredUncertainAlphaTemp370/.initial=Yes,
/ConsideredUncertainAlphaTemp371/.initial=Yes,
/ConsideredUncertainAlphaTemp372/.initial=Yes,
/ConsideredUncertainAlphaTemp373/.initial=Yes,
/ConsideredUncertainAlphaTemp374/.initial=Yes,
/ConsideredUncertainAlphaTemp375/.initial=Yes,
/ConsideredUncertainAlphaTemp376/.initial=Yes,
/ConsideredUncertainAlphaTemp377/.initial=Yes,
/ConsideredUncertainAlphaTemp378/.initial=Yes,
/ConsideredUncertainAlphaTemp379/.initial=Yes,
/ConsideredUncertainAlphaTemp380/.initial=Yes,
/ConsideredUncertainAlphaTemp381/.initial=Yes,
/ConsideredUncertainAlphaTemp382/.initial=Yes,
/ConsideredUncertainAlphaTemp383/.initial=Yes,
/ConsideredUncertainAlphaTemp384/.initial=Yes,
/ConsideredUncertainAlphaTemp385/.initial=Yes,
/ConsideredUncertainAlphaTemp386/.initial=Yes,
/ConsideredUncertainAlphaTemp387/.initial=Yes,
/ConsideredUncertainAlphaTemp388/.initial=Yes,
/ConsideredUncertainAlphaTemp389/.initial=Yes,
/ConsideredUncertainAlphaTemp390/.initial=Yes,
/ConsideredUncertainAlphaTemp391/.initial=Yes,
/ConsideredUncertainAlphaTemp392/.initial=Yes,
/ConsideredUncertainAlphaTemp393/.initial=Yes,
/ConsideredUncertainAlphaTemp394/.initial=Yes,
/ConsideredUncertainAlphaTemp395/.initial=Yes,
/ConsideredUncertainAlphaTemp396/.initial=Yes,
/ConsideredUncertainAlphaTemp397/.initial=Yes,
/ConsideredUncertainAlphaTemp398/.initial=Yes,
/ConsideredUncertainAlphaTemp399/.initial=Yes,
/ConsideredUncertainAlphaTemp400/.initial=Yes,
/ConsideredUncertainAlphaTemp401/.initial=Yes,
/ConsideredUncertainAlphaTemp402/.initial=Yes,
/ConsideredUncertainAlphaTemp403/.initial=Yes,
/ConsideredUncertainAlphaTemp404/.initial=Yes,
/ConsideredUncertainAlphaTemp405/.initial=Yes,
/ConsideredUncertainAlphaTemp406/.initial=Yes,
/ConsideredUncertainAlphaTemp407/.initial=Yes,
/ConsideredUncertainAlphaTemp408/.initial=Yes,
/ConsideredUncertainAlphaTemp409/.initial=Yes,
/ConsideredUncertainAlphaTemp410/.initial=Yes,
/ConsideredUncertainAlphaTemp411/.initial=Yes,
/ConsideredUncertainAlphaTemp412/.initial=Yes,
/ConsideredUncertainAlphaTemp413/.initial=Yes,
/ConsideredUncertainAlphaTemp414/.initial=Yes,
/ConsideredUncertainAlphaTemp415/.initial=Yes,
/ConsideredUncertainAlphaTemp416/.initial=Yes,
/ConsideredUncertainAlphaTemp417/.initial=Yes,
/ConsideredUncertainAlphaTemp418/.initial=Yes,
/ConsideredUncertainAlphaTemp419/.initial=Yes,
/ConsideredUncertainAlphaTemp420/.initial=Yes,
/ConsideredUncertainAlphaTemp421/.initial=Yes,
/ConsideredUncertainAlphaTemp422/.initial=Yes,
/ConsideredUncertainAlphaTemp423/.initial=Yes,
/ConsideredUncertainAlphaTemp424/.initial=Yes,
/ConsideredUncertainAlphaTemp425/.initial=Yes,
/ConsideredUncertainAlphaTemp426/.initial=Yes,
/ConsideredUncertainAlphaTemp427/.initial=Yes,
/ConsideredUncertainAlphaTemp428/.initial=Yes,
/ConsideredUncertainAlphaTemp429/.initial=Yes,
/ConsideredUncertainAlphaTemp430/.initial=Yes,
/ConsideredUncertainAlphaTemp431/.initial=Yes,
/ConsideredUncertainAlphaTemp432/.initial=Yes,
/ConsideredUncertainAlphaTemp433/.initial=Yes,
/ConsideredUncertainAlphaTemp434/.initial=Yes,
/ConsideredUncertainAlphaTemp435/.initial=Yes,
/ConsideredUncertainAlphaTemp436/.initial=Yes,
/ConsideredUncertainAlphaTemp437/.initial=Yes,
/ConsideredUncertainAlphaTemp438/.initial=Yes,
/ConsideredUncertainAlphaTemp439/.initial=Yes,
/ConsideredUncertainAlphaTemp440/.initial=Yes,
/ConsideredUncertainAlphaTemp441/.initial=Yes,
/ConsideredUncertainAlphaTemp442/.initial=Yes,
/ConsideredUncertainAlphaTemp443/.initial=Yes,
/ConsideredUncertainAlphaTemp444/.initial=Yes,
/ConsideredUncertainAlphaTemp445/.initial=Yes,
/ConsideredUncertainAlphaTemp446/.initial=Yes,
/ConsideredUncertainAlphaTemp447/.initial=Yes,
/ConsideredUncertainAlphaTemp448/.initial=Yes,
/ConsideredUncertainAlphaTemp449/.initial=Yes,
/ConsideredUncertainAlphaTemp450/.initial=Yes,
/ConsideredUncertainAlphaTemp451/.initial=Yes,
/ConsideredUncertainAlphaTemp452/.initial=Yes,
/ConsideredUncertainAlphaTemp453/.initial=Yes,
/ConsideredUncertainAlphaTemp454/.initial=Yes,
/ConsideredUncertainAlphaTemp455/.initial=Yes,
/ConsideredUncertainAlphaTemp456/.initial=Yes,
/ConsideredUncertainAlphaTemp457/.initial=Yes,
/ConsideredUncertainAlphaTemp458/.initial=Yes,
/ConsideredUncertainAlphaTemp459/.initial=Yes,
/ConsideredUncertainAlphaTemp460/.initial=Yes,
/ConsideredUncertainAlphaTemp461/.initial=Yes,
/ConsideredUncertainAlphaTemp462/.initial=Yes,
/ConsideredUncertainAlphaTemp463/.initial=Yes,
/ConsideredUncertainAlphaTemp464/.initial=Yes,
/ConsideredUncertainAlphaTemp465/.initial=Yes,
/ConsideredUncertainAlphaTemp466/.initial=Yes,
/ConsideredUncertainAlphaTemp467/.initial=Yes,
/ConsideredUncertainAlphaTemp468/.initial=Yes,
/ConsideredUncertainAlphaTemp469/.initial=Yes,
/ConsideredUncertainAlphaTemp470/.initial=Yes,
/ConsideredUncertainAlphaTemp471/.initial=Yes,
/ConsideredUncertainAlphaTemp472/.initial=Yes,
/ConsideredUncertainAlphaTemp473/.initial=Yes,
/ConsideredUncertainAlphaTemp474/.initial=Yes,
/ConsideredUncertainAlphaTemp475/.initial=Yes,
/ConsideredUncertainAlphaTemp476/.initial=Yes,
/ConsideredUncertainAlphaTemp477/.initial=Yes,
/ConsideredUncertainAlphaTemp478/.initial=Yes,
/ConsideredUncertainAlphaTemp479/.initial=Yes,
/ConsideredUncertainAlphaTemp480/.initial=Yes,
/ConsideredUncertainAlphaTemp481/.initial=Yes,
/ConsideredUncertainAlphaTemp482/.initial=Yes,
/ConsideredUncertainAlphaTemp483/.initial=Yes,
/ConsideredUncertainAlphaTemp484/.initial=Yes,
/ConsideredUncertainAlphaTemp485/.initial=Yes,
/ConsideredUncertainAlphaTemp486/.initial=Yes,
/ConsideredUncertainAlphaTemp487/.initial=Yes,
/ConsideredUncertainAlphaTemp488/.initial=Yes,
/ConsideredUncertainAlphaTemp489/.initial=Yes,
/ConsideredUncertainAlphaTemp490/.initial=Yes,
/ConsideredUncertainAlphaTemp491/.initial=Yes,
/ConsideredUncertainAlphaTemp492/.initial=Yes,
/ConsideredUncertainAlphaTemp493/.initial=Yes,
/ConsideredUncertainAlphaTemp494/.initial=Yes,
/ConsideredUncertainAlphaTemp495/.initial=Yes,
/ConsideredUncertainAlphaTemp496/.initial=Yes,
/ConsideredUncertainAlphaTemp497/.initial=Yes,
/ConsideredUncertainAlphaTemp498/.initial=Yes,
/ConsideredUncertainAlphaTemp499/.initial=Yes,
/ConsideredUncertainAlphaTemp500/.initial=Yes,
/ConsideredUncertainAlphaTemp501/.initial=Yes,
/ConsideredUncertainAlphaTemp502/.initial=Yes,
/ConsideredUncertainAlphaTemp503/.initial=Yes,
/ConsideredUncertainAlphaTemp504/.initial=Yes,
/ConsideredUncertainAlphaTemp505/.initial=Yes,
/ConsideredUncertainAlphaTemp506/.initial=Yes,
/ConsideredUncertainAlphaTemp507/.initial=Yes,
/ConsideredUncertainAlphaTemp508/.initial=Yes,
/ConsideredUncertainAlphaTemp509/.initial=Yes,
/ConsideredUncertainAlphaTemp510/.initial=Yes,
/ConsideredUncertainAlphaTemp511/.initial=Yes,
/ConsideredUncertainAlphaTemp512/.initial=Yes,
/ConsideredUncertainAlphaTemp513/.initial=Yes,
/ConsideredUncertainAlphaTemp514/.initial=Yes,
/ConsideredUncertainAlphaTemp515/.initial=Yes,
/ConsideredUncertainAlphaTemp516/.initial=Yes,
/ConsideredUncertainAlphaTemp517/.initial=Yes,
/ConsideredUncertainAlphaTemp518/.initial=Yes,
/ConsideredUncertainAlphaTemp519/.initial=Yes,
/ConsideredUncertainAlphaTemp520/.initial=Yes,
/ConsideredUncertainAlphaTemp521/.initial=Yes,
/ConsideredUncertainAlphaTemp522/.initial=Yes,
/ConsideredUncertainAlphaTemp523/.initial=Yes,
/ConsideredUncertainAlphaTemp524/.initial=Yes,
/ConsideredUncertainAlphaTemp525/.initial=Yes,
/ConsideredUncertainAlphaTemp526/.initial=Yes,
/ConsideredUncertainAlphaTemp527/.initial=Yes,
/ConsideredUncertainAlphaTemp528/.initial=Yes,
/ConsideredUncertainAlphaTemp529/.initial=Yes,
/ConsideredUncertainAlphaTemp530/.initial=Yes,
/ConsideredUncertainAlphaTemp531/.initial=Yes,
/ConsideredUncertainAlphaTemp532/.initial=Yes,
/ConsideredUncertainAlphaTemp533/.initial=Yes,
/ConsideredUncertainAlphaTemp534/.initial=Yes,
/ConsideredUncertainAlphaTemp535/.initial=Yes,
/ConsideredUncertainAlphaTemp536/.initial=Yes,
/ConsideredUncertainAlphaTemp537/.initial=Yes,
/ConsideredUncertainAlphaTemp538/.initial=Yes,
/ConsideredUncertainAlphaTemp539/.initial=Yes,
/ConsideredUncertainAlphaTemp540/.initial=Yes,
/ConsideredUncertainAlphaTemp541/.initial=Yes,
/ConsideredUncertainAlphaTemp542/.initial=Yes,
/ConsideredUncertainAlphaTemp543/.initial=Yes,
/ConsideredUncertainAlphaTemp544/.initial=Yes,
/ConsideredUncertainAlphaTemp545/.initial=Yes,
/ConsideredUncertainAlphaTemp546/.initial=Yes,
/ConsideredUncertainAlphaTemp547/.initial=Yes,
/ConsideredUncertainAlphaTemp548/.initial=Yes,
/ConsideredUncertainAlphaTemp549/.initial=Yes,
/ConsideredUncertainAlphaTemp550/.initial=Yes,
/ConsideredUncertainAlphaTemp551/.initial=Yes,
/ConsideredUncertainAlphaTemp552/.initial=Yes,
/ConsideredUncertainAlphaTemp553/.initial=Yes,
/ConsideredUncertainAlphaTemp554/.initial=Yes,
/ConsideredUncertainAlphaTemp555/.initial=Yes,
/ConsideredUncertainAlphaTemp556/.initial=Yes,
/ConsideredUncertainAlphaTemp557/.initial=Yes,
/ConsideredUncertainAlphaTemp558/.initial=Yes,
/ConsideredUncertainAlphaTemp559/.initial=Yes,
/ConsideredUncertainAlphaTemp560/.initial=Yes,
/ConsideredUncertainAlphaTemp561/.initial=Yes,
/ConsideredUncertainAlphaTemp562/.initial=Yes,
/ConsideredUncertainAlphaTemp563/.initial=Yes,
/ConsideredUncertainAlphaTemp564/.initial=Yes,
/ConsideredUncertainAlphaTemp565/.initial=Yes,
/ConsideredUncertainAlphaTemp566/.initial=Yes,
/ConsideredUncertainAlphaTemp567/.initial=Yes,
/ConsideredUncertainAlphaTemp568/.initial=Yes,
/ConsideredUncertainAlphaTemp569/.initial=Yes,
/ConsideredUncertainAlphaTemp570/.initial=Yes,
/ConsideredUncertainAlphaTemp571/.initial=Yes,
/ConsideredUncertainAlphaTemp572/.initial=Yes,
/ConsideredUncertainAlphaTemp573/.initial=Yes,
/ConsideredUncertainAlphaTemp574/.initial=Yes,
/ConsideredUncertainAlphaTemp575/.initial=Yes,
/ConsideredUncertainAlphaTemp576/.initial=Yes,
/ConsideredUncertainAlphaTemp577/.initial=Yes,
/ConsideredUncertainAlphaTemp578/.initial=Yes,
/ConsideredUncertainAlphaTemp579/.initial=Yes,
/ConsideredUncertainAlphaTemp580/.initial=Yes,
/ConsideredUncertainAlphaTemp581/.initial=Yes,
/ConsideredUncertainAlphaTemp582/.initial=Yes,
/ConsideredUncertainAlphaTemp583/.initial=Yes,
/ConsideredUncertainAlphaTemp584/.initial=Yes,
/ConsideredUncertainAlphaTemp585/.initial=Yes,
/ConsideredUncertainAlphaTemp586/.initial=Yes,
/ConsideredUncertainAlphaTemp587/.initial=Yes,
/ConsideredUncertainAlphaTemp588/.initial=Yes,
/ConsideredUncertainAlphaTemp589/.initial=Yes,
/ConsideredUncertainAlphaTemp590/.initial=Yes,
/ConsideredUncertainAlphaTemp591/.initial=Yes,
/ConsideredUncertainAlphaTemp592/.initial=Yes,
/ConsideredUncertainAlphaTemp593/.initial=Yes,
/ConsideredUncertainAlphaTemp594/.initial=Yes,
/ConsideredUncertainAlphaTemp595/.initial=Yes,
/ConsideredUncertainAlphaTemp596/.initial=Yes,
/ConsideredUncertainAlphaTemp597/.initial=Yes,
/ConsideredUncertainAlphaTemp598/.initial=Yes,
/ConsideredUncertainAlphaTemp599/.initial=Yes,
/ConsideredUncertainAlphaTemp600/.initial=Yes,
/ConsideredUncertainAlphaTemp601/.initial=Yes,
/ConsideredUncertainAlphaTemp602/.initial=Yes,
/ConsideredUncertainAlphaTemp603/.initial=Yes,
/ConsideredUncertainAlphaTemp604/.initial=Yes,
/ConsideredUncertainAlphaTemp605/.initial=Yes,
/ConsideredUncertainAlphaTemp606/.initial=Yes,
/ConsideredUncertainAlphaTemp607/.initial=Yes,
/ConsideredUncertainAlphaTemp608/.initial=Yes,
/ConsideredUncertainAlphaTemp609/.initial=Yes,
/ConsideredUncertainAlphaTemp610/.initial=Yes,
/ConsideredUncertainAlphaTemp611/.initial=Yes,
/ConsideredUncertainAlphaTemp612/.initial=Yes,
/ConsideredUncertainAlphaTemp613/.initial=Yes,
/ConsideredUncertainAlphaTemp614/.initial=Yes,
/ConsideredUncertainAlphaTemp615/.initial=Yes,
/ConsideredUncertainAlphaTemp616/.initial=Yes,
/ConsideredUncertainAlphaTemp617/.initial=Yes,
/ConsideredUncertainAlphaTemp618/.initial=Yes,
/ConsideredUncertainAlphaTemp619/.initial=Yes,
/ConsideredUncertainAlphaTemp620/.initial=Yes,
/ConsideredUncertainAlphaTemp621/.initial=Yes,
/ConsideredUncertainAlphaTemp622/.initial=Yes,
/ConsideredUncertainAlphaTemp623/.initial=Yes,
/ConsideredUncertainAlphaTemp624/.initial=Yes,
/ConsideredUncertainAlphaTemp625/.initial=Yes,
/ConsideredUncertainAlphaTemp626/.initial=Yes,
/ConsideredUncertainAlphaTemp627/.initial=Yes,
/ConsideredUncertainAlphaTemp628/.initial=Yes,
/ConsideredUncertainAlphaTemp629/.initial=Yes,
/ConsideredUncertainAlphaTemp630/.initial=Yes,
/ConsideredUncertainAlphaTemp631/.initial=Yes,
/ConsideredUncertainAlphaTemp632/.initial=Yes,
/ConsideredUncertainAlphaTemp633/.initial=Yes,
/ConsideredUncertainAlphaTemp634/.initial=Yes,
/ConsideredUncertainAlphaTemp635/.initial=Yes,
/ConsideredUncertainAlphaTemp636/.initial=Yes,
/ConsideredUncertainAlphaTemp637/.initial=Yes,
/ConsideredUncertainAlphaTemp638/.initial=Yes,
/ConsideredUncertainAlphaTemp639/.initial=Yes,
/ConsideredUncertainAlphaTemp640/.initial=Yes,
/ConsideredUncertainAlphaTemp641/.initial=Yes,
/ConsideredUncertainAlphaTemp642/.initial=Yes,
/ConsideredUncertainAlphaTemp643/.initial=Yes,
/ConsideredUncertainAlphaTemp644/.initial=Yes,
/ConsideredUncertainAlphaTemp645/.initial=Yes,
/ConsideredUncertainAlphaTemp646/.initial=Yes,
/ConsideredUncertainAlphaTemp647/.initial=Yes,
/ConsideredUncertainAlphaTemp648/.initial=Yes,
/ConsideredUncertainAlphaTemp649/.initial=Yes,
/ConsideredUncertainAlphaTemp650/.initial=Yes,
/ConsideredUncertainAlphaTemp651/.initial=Yes,
/ConsideredUncertainAlphaTemp652/.initial=Yes,
/ConsideredUncertainAlphaTemp653/.initial=Yes,
/ConsideredUncertainAlphaTemp654/.initial=Yes,
/ConsideredUncertainAlphaTemp655/.initial=Yes,
/ConsideredUncertainAlphaTemp656/.initial=Yes,
/ConsideredUncertainAlphaTemp657/.initial=Yes,
/ConsideredUncertainAlphaTemp658/.initial=Yes,
/ConsideredUncertainAlphaTemp659/.initial=Yes,
/ConsideredUncertainAlphaTemp660/.initial=Yes,
/ConsideredUncertainAlphaTemp661/.initial=Yes,
/ConsideredUncertainAlphaTemp662/.initial=Yes,
/ConsideredUncertainAlphaTemp663/.initial=Yes,
/ConsideredUncertainAlphaTemp664/.initial=Yes,
/ConsideredUncertainAlphaTemp665/.initial=Yes,
/ConsideredUncertainAlphaTemp666/.initial=Yes,
/ConsideredUncertainAlphaTemp667/.initial=Yes,
/ConsideredUncertainAlphaTemp668/.initial=Yes,
/ConsideredUncertainAlphaTemp669/.initial=Yes,
/ConsideredUncertainAlphaTemp670/.initial=Yes,
/ConsideredUncertainAlphaTemp671/.initial=Yes,
/ConsideredUncertainAlphaTemp672/.initial=Yes,
/ConsideredUncertainAlphaTemp673/.initial=Yes,
/ConsideredUncertainAlphaTemp674/.initial=Yes,
/ConsideredUncertainAlphaTemp675/.initial=Yes,
/ConsideredUncertainAlphaTemp676/.initial=Yes,
/ConsideredUncertainAlphaTemp677/.initial=Yes,
/ConsideredUncertainAlphaTemp678/.initial=Yes,
/ConsideredUncertainAlphaTemp679/.initial=Yes,
/ConsideredUncertainAlphaTemp680/.initial=Yes,
/ConsideredUncertainAlphaTemp681/.initial=Yes,
/ConsideredUncertainAlphaTemp682/.initial=Yes,
/ConsideredUncertainAlphaTemp683/.initial=Yes,
/ConsideredUncertainAlphaTemp684/.initial=Yes,
/ConsideredUncertainAlphaTemp685/.initial=Yes,
/ConsideredUncertainAlphaTemp686/.initial=Yes,
/ConsideredUncertainAlphaTemp687/.initial=Yes,
/ConsideredUncertainAlphaTemp688/.initial=Yes,
/ConsideredUncertainAlphaTemp689/.initial=Yes,
/ConsideredUncertainAlphaTemp690/.initial=Yes,
/ConsideredUncertainAlphaTemp691/.initial=Yes,
/ConsideredUncertainAlphaTemp692/.initial=Yes,
/ConsideredUncertainAlphaTemp693/.initial=Yes,
/ConsideredUncertainAlphaTemp694/.initial=Yes,
/ConsideredUncertainAlphaTemp695/.initial=Yes,
/ConsideredUncertainAlphaTemp696/.initial=Yes,
/ConsideredUncertainAlphaTemp697/.initial=Yes,
/ConsideredUncertainAlphaTemp698/.initial=Yes,
/ConsideredUncertainAlphaTemp699/.initial=Yes,
/ConsideredUncertainAlphaTemp700/.initial=Yes,
/ConsideredUncertainAlphaTemp701/.initial=Yes,
/ConsideredUncertainAlphaTemp702/.initial=Yes,
/ConsideredUncertainAlphaTemp703/.initial=Yes,
/ConsideredUncertainAlphaTemp704/.initial=Yes,
/ConsideredUncertainAlphaTemp705/.initial=Yes,
/ConsideredUncertainAlphaTemp706/.initial=Yes,
/ConsideredUncertainAlphaTemp707/.initial=Yes,
/ConsideredUncertainAlphaTemp708/.initial=Yes,
/ConsideredUncertainAlphaTemp709/.initial=Yes,
/ConsideredUncertainAlphaTemp710/.initial=Yes,
/ConsideredUncertainAlphaTemp711/.initial=Yes,
/ConsideredUncertainAlphaTemp712/.initial=Yes,
/ConsideredUncertainAlphaTemp713/.initial=Yes,
/ConsideredUncertainAlphaTemp714/.initial=Yes,
/ConsideredUncertainAlphaTemp715/.initial=Yes,
/ConsideredUncertainAlphaTemp716/.initial=Yes,
/ConsideredUncertainAlphaTemp717/.initial=Yes,
/ConsideredUncertainAlphaTemp718/.initial=Yes,
/ConsideredUncertainAlphaTemp719/.initial=Yes,
/ConsideredUncertainAlphaTemp720/.initial=Yes,
/ConsideredUncertainAlphaTemp721/.initial=Yes,
/ConsideredUncertainAlphaTemp722/.initial=Yes,
/ConsideredUncertainAlphaTemp723/.initial=Yes,
/ConsideredUncertainAlphaTemp724/.initial=Yes,
/ConsideredUncertainAlphaTemp725/.initial=Yes,
/ConsideredUncertainAlphaTemp726/.initial=Yes,
/ConsideredUncertainAlphaTemp727/.initial=Yes,
/ConsideredUncertainAlphaTemp728/.initial=Yes,
/ConsideredUncertainAlphaTemp729/.initial=Yes,
/ConsideredUncertainAlphaTemp730/.initial=Yes,
/ConsideredUncertainAlphaTemp731/.initial=Yes,
/ConsideredUncertainAlphaTemp732/.initial=Yes,
/ConsideredUncertainAlphaTemp733/.initial=Yes,
/ConsideredUncertainAlphaTemp734/.initial=Yes,
/ConsideredUncertainAlphaTemp735/.initial=Yes,
/ConsideredUncertainAlphaTemp736/.initial=Yes,
/ConsideredUncertainAlphaTemp737/.initial=Yes,
/ConsideredUncertainAlphaTemp738/.initial=Yes,
/ConsideredUncertainAlphaTemp739/.initial=Yes,
/ConsideredUncertainAlphaTemp740/.initial=Yes,
/ConsideredUncertainAlphaTemp741/.initial=Yes,
/ConsideredUncertainAlphaTemp742/.initial=Yes,
/ConsideredUncertainAlphaTemp743/.initial=Yes,
/ConsideredUncertainAlphaTemp744/.initial=Yes,
/ConsideredUncertainAlphaTemp745/.initial=Yes,
/ConsideredUncertainAlphaTemp746/.initial=Yes,
/ConsideredUncertainAlphaTemp747/.initial=Yes,
/ConsideredUncertainAlphaTemp748/.initial=Yes,
/ConsideredUncertainAlphaTemp749/.initial=Yes,
/ConsideredUncertainAlphaTemp750/.initial=Yes,
/ConsideredUncertainAlphaTemp751/.initial=Yes,
/ConsideredUncertainAlphaTemp752/.initial=Yes,
/ConsideredUncertainAlphaTemp753/.initial=Yes,
/ConsideredUncertainAlphaTemp754/.initial=Yes,
/ConsideredUncertainAlphaTemp755/.initial=Yes,
/ConsideredUncertainAlphaTemp756/.initial=Yes,
/ConsideredUncertainAlphaTemp757/.initial=Yes,
/ConsideredUncertainAlphaTemp758/.initial=Yes,
/ConsideredUncertainAlphaTemp759/.initial=Yes,
/ConsideredUncertainAlphaTemp760/.initial=Yes,
/ConsideredUncertainAlphaTemp761/.initial=Yes,
/ConsideredUncertainAlphaTemp762/.initial=Yes,
/ConsideredUncertainAlphaTemp763/.initial=Yes,
/ConsideredUncertainAlphaTemp764/.initial=Yes,
/ConsideredUncertainAlphaTemp765/.initial=Yes,
/ConsideredUncertainAlphaTemp766/.initial=Yes,
/ConsideredUncertainAlphaTemp767/.initial=Yes,
/ConsideredUncertainAlphaTemp0InMem/.initial=Yes,
/ConsideredUncertainAlphaTemp1InMem/.initial=Yes,
/ConsideredUncertainAlphaTemp2InMem/.initial=Yes,
/ConsideredUncertainAlphaTemp3InMem/.initial=Yes,
/ConsideredUncertainAlphaTemp4InMem/.initial=Yes,
/ConsideredUncertainAlphaTemp5InMem/.initial=Yes,
/ConsideredUncertainAlphaTemp6InMem/.initial=Yes,
/ConsideredUncertainAlphaTemp7InMem/.initial=Yes,
/ConsideredUncertainAlphaTemp8InMem/.initial=Yes,
/ConsideredUncertainAlphaTemp9InMem/.initial=Yes,
/ConsideredUncertainAlphaTemp10InMem/.initial=Yes,
/ConsideredUncertainAlphaTemp11InMem/.initial=Yes,
/ConsideredUncertainAlphaTemp12InMem/.initial=Yes,
/ConsideredUncertainAlphaTemp13InMem/.initial=Yes,
/ConsideredUncertainAlphaTemp14InMem/.initial=Yes,
/ConsideredUncertainAlphaTemp15InMem/.initial=Yes,
/ConsideredUncertainAlphaTemp16InMem/.initial=Yes,
/ConsideredUncertainAlphaTemp17InMem/.initial=Yes,
/ConsideredUncertainAlphaTemp18InMem/.initial=Yes,
/ConsideredUncertainAlphaTemp19InMem/.initial=Yes,
/ConsideredUncertainAlphaTemp20InMem/.initial=Yes,
/ConsideredUncertainAlphaTemp21InMem/.initial=Yes,
/ConsideredUncertainAlphaTemp22InMem/.initial=Yes,
/ConsideredUncertainAlphaTemp23InMem/.initial=Yes,
/ConsideredUncertainAlphaTemp24InMem/.initial=Yes,
/ConsideredUncertainAlphaTemp25InMem/.initial=Yes,
/ConsideredUncertainAlphaTemp26InMem/.initial=Yes,
/ConsideredUncertainAlphaTemp27InMem/.initial=Yes,
/ConsideredUncertainAlphaTemp28InMem/.initial=Yes,
/ConsideredUncertainAlphaTemp29InMem/.initial=Yes,
/ConsideredUncertainAlphaTemp30InMem/.initial=Yes,
/ConsideredUncertainAlphaTemp31InMem/.initial=Yes,
/ConsideredUncertainAlphaTemp32InMem/.initial=Yes,
/ConsideredUncertainAlphaTemp33InMem/.initial=Yes,
/ConsideredUncertainAlphaTemp34InMem/.initial=Yes,
/ConsideredUncertainAlphaTemp35InMem/.initial=Yes,
/ConsideredUncertainAlphaTemp36InMem/.initial=Yes,
/ConsideredUncertainAlphaTemp37InMem/.initial=Yes,
/ConsideredUncertainAlphaTemp38InMem/.initial=Yes,
/ConsideredUncertainAlphaTemp39InMem/.initial=Yes,
/ConsideredUncertainAlphaTemp40InMem/.initial=Yes,
/ConsideredUncertainAlphaTemp41InMem/.initial=Yes,
/ConsideredUncertainAlphaTemp42InMem/.initial=Yes,
/ConsideredUncertainAlphaTemp43InMem/.initial=Yes,
/ConsideredUncertainAlphaTemp44InMem/.initial=Yes,
/ConsideredUncertainAlphaTemp45InMem/.initial=Yes,
/ConsideredUncertainAlphaTemp46InMem/.initial=Yes,
/ConsideredUncertainAlphaTemp47InMem/.initial=Yes,
/ConsideredUncertainAlphaTemp48InMem/.initial=Yes,
/ConsideredUncertainAlphaTemp49InMem/.initial=Yes,
/ConsideredUncertainAlphaTemp50InMem/.initial=Yes,
/ConsideredUncertainAlphaTemp51InMem/.initial=Yes,
/ConsideredUncertainAlphaTemp52InMem/.initial=Yes,
/ConsideredUncertainAlphaTemp53InMem/.initial=Yes,
/ConsideredUncertainAlphaTemp54InMem/.initial=Yes,
/ConsideredUncertainAlphaTemp55InMem/.initial=Yes,
/ConsideredUncertainAlphaTemp56InMem/.initial=Yes,
/ConsideredUncertainAlphaTemp57InMem/.initial=Yes,
/ConsideredUncertainAlphaTemp58InMem/.initial=Yes,
/ConsideredUncertainAlphaTemp59InMem/.initial=Yes,
/ConsideredUncertainAlphaTemp60InMem/.initial=Yes,
/ConsideredUncertainAlphaTemp61InMem/.initial=Yes,
/ConsideredUncertainAlphaTemp62InMem/.initial=Yes,
/ConsideredUncertainAlphaTemp63InMem/.initial=Yes,
/ConsideredUncertainAlphaTemp64InMem/.initial=Yes,
/ConsideredUncertainAlphaTemp65InMem/.initial=Yes,
/ConsideredUncertainAlphaTemp66InMem/.initial=Yes,
/ConsideredUncertainAlphaTemp67InMem/.initial=Yes,
/ConsideredUncertainAlphaTemp68InMem/.initial=Yes,
/ConsideredUncertainAlphaTemp69InMem/.initial=Yes,
/ConsideredUncertainAlphaTemp70InMem/.initial=Yes,
/ConsideredUncertainAlphaTemp71InMem/.initial=Yes,
/ConsideredUncertainAlphaTemp72InMem/.initial=Yes,
/ConsideredUncertainAlphaTemp73InMem/.initial=Yes,
/ConsideredUncertainAlphaTemp74InMem/.initial=Yes,
/ConsideredUncertainAlphaTemp75InMem/.initial=Yes,
/ConsideredUncertainAlphaTemp76InMem/.initial=Yes,
/ConsideredUncertainAlphaTemp77InMem/.initial=Yes,
/ConsideredUncertainAlphaTemp78InMem/.initial=Yes,
/ConsideredUncertainAlphaTemp79InMem/.initial=Yes,
/ConsideredUncertainAlphaTemp80InMem/.initial=Yes,
/ConsideredUncertainAlphaTemp81InMem/.initial=Yes,
/ConsideredUncertainAlphaTemp82InMem/.initial=Yes,
/ConsideredUncertainAlphaTemp83InMem/.initial=Yes,
/ConsideredUncertainAlphaTemp84InMem/.initial=Yes,
/ConsideredUncertainAlphaTemp85InMem/.initial=Yes,
/ConsideredUncertainAlphaTemp86InMem/.initial=Yes,
/ConsideredUncertainAlphaTemp87InMem/.initial=Yes,
/ConsideredUncertainAlphaTemp88InMem/.initial=Yes,
/ConsideredUncertainAlphaTemp89InMem/.initial=Yes,
/ConsideredUncertainAlphaTemp90InMem/.initial=Yes,
/ConsideredUncertainAlphaTemp91InMem/.initial=Yes,
/ConsideredUncertainAlphaTemp92InMem/.initial=Yes,
/ConsideredUncertainAlphaTemp93InMem/.initial=Yes,
/ConsideredUncertainAlphaTemp94InMem/.initial=Yes,
/ConsideredUncertainAlphaTemp95InMem/.initial=Yes,
/ConsideredUncertainAlphaTemp96InMem/.initial=Yes,
/ConsideredUncertainAlphaTemp97InMem/.initial=Yes,
/ConsideredUncertainAlphaTemp98InMem/.initial=Yes,
/ConsideredUncertainAlphaTemp99InMem/.initial=Yes,
/ConsideredUncertainAlphaTemp100InMem/.initial=Yes,
/ConsideredUncertainAlphaTemp101InMem/.initial=Yes,
/ConsideredUncertainAlphaTemp102InMem/.initial=Yes,
/ConsideredUncertainAlphaTemp103InMem/.initial=Yes,
/ConsideredUncertainAlphaTemp104InMem/.initial=Yes,
/ConsideredUncertainAlphaTemp105InMem/.initial=Yes,
/ConsideredUncertainAlphaTemp106InMem/.initial=Yes,
/ConsideredUncertainAlphaTemp107InMem/.initial=Yes,
/ConsideredUncertainAlphaTemp108InMem/.initial=Yes,
/ConsideredUncertainAlphaTemp109InMem/.initial=Yes,
/ConsideredUncertainAlphaTemp110InMem/.initial=Yes,
/ConsideredUncertainAlphaTemp111InMem/.initial=Yes,
/ConsideredUncertainAlphaTemp112InMem/.initial=Yes,
/ConsideredUncertainAlphaTemp113InMem/.initial=Yes,
/ConsideredUncertainAlphaTemp114InMem/.initial=Yes,
/ConsideredUncertainAlphaTemp115InMem/.initial=Yes,
/ConsideredUncertainAlphaTemp116InMem/.initial=Yes,
/ConsideredUncertainAlphaTemp117InMem/.initial=Yes,
/ConsideredUncertainAlphaTemp118InMem/.initial=Yes,
/ConsideredUncertainAlphaTemp119InMem/.initial=Yes,
/ConsideredUncertainAlphaTemp120InMem/.initial=Yes,
/ConsideredUncertainAlphaTemp121InMem/.initial=Yes,
/ConsideredUncertainAlphaTemp122InMem/.initial=Yes,
/ConsideredUncertainAlphaTemp123InMem/.initial=Yes,
/ConsideredUncertainAlphaTemp124InMem/.initial=Yes,
/ConsideredUncertainAlphaTemp125InMem/.initial=Yes,
/ConsideredUncertainAlphaTemp126InMem/.initial=Yes,
/ConsideredUncertainAlphaTemp127InMem/.initial=Yes,
/ConsideredUncertainAlphaTemp128InMem/.initial=Yes,
/ConsideredUncertainAlphaTemp129InMem/.initial=Yes,
/ConsideredUncertainAlphaTemp130InMem/.initial=Yes,
/ConsideredUncertainAlphaTemp131InMem/.initial=Yes,
/ConsideredUncertainAlphaTemp132InMem/.initial=Yes,
/ConsideredUncertainAlphaTemp133InMem/.initial=Yes,
/ConsideredUncertainAlphaTemp134InMem/.initial=Yes,
/ConsideredUncertainAlphaTemp135InMem/.initial=Yes,
/ConsideredUncertainAlphaTemp136InMem/.initial=Yes,
/ConsideredUncertainAlphaTemp137InMem/.initial=Yes,
/ConsideredUncertainAlphaTemp138InMem/.initial=Yes,
/ConsideredUncertainAlphaTemp139InMem/.initial=Yes,
/ConsideredUncertainAlphaTemp140InMem/.initial=Yes,
/ConsideredUncertainAlphaTemp141InMem/.initial=Yes,
/ConsideredUncertainAlphaTemp142InMem/.initial=Yes,
/ConsideredUncertainAlphaTemp143InMem/.initial=Yes,
/ConsideredUncertainAlphaTemp144InMem/.initial=Yes,
/ConsideredUncertainAlphaTemp145InMem/.initial=Yes,
/ConsideredUncertainAlphaTemp146InMem/.initial=Yes,
/ConsideredUncertainAlphaTemp147InMem/.initial=Yes,
/ConsideredUncertainAlphaTemp148InMem/.initial=Yes,
/ConsideredUncertainAlphaTemp149InMem/.initial=Yes,
/ConsideredUncertainAlphaTemp150InMem/.initial=Yes,
/ConsideredUncertainAlphaTemp151InMem/.initial=Yes,
/ConsideredUncertainAlphaTemp152InMem/.initial=Yes,
/ConsideredUncertainAlphaTemp153InMem/.initial=Yes,
/ConsideredUncertainAlphaTemp154InMem/.initial=Yes,
/ConsideredUncertainAlphaTemp155InMem/.initial=Yes,
/ConsideredUncertainAlphaTemp156InMem/.initial=Yes,
/ConsideredUncertainAlphaTemp157InMem/.initial=Yes,
/ConsideredUncertainAlphaTemp158InMem/.initial=Yes,
/ConsideredUncertainAlphaTemp159InMem/.initial=Yes,
/ConsideredUncertainAlphaTemp160InMem/.initial=Yes,
/ConsideredUncertainAlphaTemp161InMem/.initial=Yes,
/ConsideredUncertainAlphaTemp162InMem/.initial=Yes,
/ConsideredUncertainAlphaTemp163InMem/.initial=Yes,
/ConsideredUncertainAlphaTemp164InMem/.initial=Yes,
/ConsideredUncertainAlphaTemp165InMem/.initial=Yes,
/ConsideredUncertainAlphaTemp166InMem/.initial=Yes,
/ConsideredUncertainAlphaTemp167InMem/.initial=Yes,
/ConsideredUncertainAlphaTemp168InMem/.initial=Yes,
/ConsideredUncertainAlphaTemp169InMem/.initial=Yes,
/ConsideredUncertainAlphaTemp170InMem/.initial=Yes,
/ConsideredUncertainAlphaTemp171InMem/.initial=Yes,
/ConsideredUncertainAlphaTemp172InMem/.initial=Yes,
/ConsideredUncertainAlphaTemp173InMem/.initial=Yes,
/ConsideredUncertainAlphaTemp174InMem/.initial=Yes,
/ConsideredUncertainAlphaTemp175InMem/.initial=Yes,
/ConsideredUncertainAlphaTemp176InMem/.initial=Yes,
/ConsideredUncertainAlphaTemp177InMem/.initial=Yes,
/ConsideredUncertainAlphaTemp178InMem/.initial=Yes,
/ConsideredUncertainAlphaTemp179InMem/.initial=Yes,
/ConsideredUncertainAlphaTemp180InMem/.initial=Yes,
/ConsideredUncertainAlphaTemp181InMem/.initial=Yes,
/ConsideredUncertainAlphaTemp182InMem/.initial=Yes,
/ConsideredUncertainAlphaTemp183InMem/.initial=Yes,
/ConsideredUncertainAlphaTemp184InMem/.initial=Yes,
/ConsideredUncertainAlphaTemp185InMem/.initial=Yes,
/ConsideredUncertainAlphaTemp186InMem/.initial=Yes,
/ConsideredUncertainAlphaTemp187InMem/.initial=Yes,
/ConsideredUncertainAlphaTemp188InMem/.initial=Yes,
/ConsideredUncertainAlphaTemp189InMem/.initial=Yes,
/ConsideredUncertainAlphaTemp190InMem/.initial=Yes,
/ConsideredUncertainAlphaTemp191InMem/.initial=Yes,
/ConsideredUncertainAlphaTemp192InMem/.initial=Yes,
/ConsideredUncertainAlphaTemp193InMem/.initial=Yes,
/ConsideredUncertainAlphaTemp194InMem/.initial=Yes,
/ConsideredUncertainAlphaTemp195InMem/.initial=Yes,
/ConsideredUncertainAlphaTemp196InMem/.initial=Yes,
/ConsideredUncertainAlphaTemp197InMem/.initial=Yes,
/ConsideredUncertainAlphaTemp198InMem/.initial=Yes,
/ConsideredUncertainAlphaTemp199InMem/.initial=Yes,
/ConsideredUncertainAlphaTemp200InMem/.initial=Yes,
/ConsideredUncertainAlphaTemp201InMem/.initial=Yes,
/ConsideredUncertainAlphaTemp202InMem/.initial=Yes,
/ConsideredUncertainAlphaTemp203InMem/.initial=Yes,
/ConsideredUncertainAlphaTemp204InMem/.initial=Yes,
/ConsideredUncertainAlphaTemp205InMem/.initial=Yes,
/ConsideredUncertainAlphaTemp206InMem/.initial=Yes,
/ConsideredUncertainAlphaTemp207InMem/.initial=Yes,
/ConsideredUncertainAlphaTemp208InMem/.initial=Yes,
/ConsideredUncertainAlphaTemp209InMem/.initial=Yes,
/ConsideredUncertainAlphaTemp210InMem/.initial=Yes,
/ConsideredUncertainAlphaTemp211InMem/.initial=Yes,
/ConsideredUncertainAlphaTemp212InMem/.initial=Yes,
/ConsideredUncertainAlphaTemp213InMem/.initial=Yes,
/ConsideredUncertainAlphaTemp214InMem/.initial=Yes,
/ConsideredUncertainAlphaTemp215InMem/.initial=Yes,
/ConsideredUncertainAlphaTemp216InMem/.initial=Yes,
/ConsideredUncertainAlphaTemp217InMem/.initial=Yes,
/ConsideredUncertainAlphaTemp218InMem/.initial=Yes,
/ConsideredUncertainAlphaTemp219InMem/.initial=Yes,
/ConsideredUncertainAlphaTemp220InMem/.initial=Yes,
/ConsideredUncertainAlphaTemp221InMem/.initial=Yes,
/ConsideredUncertainAlphaTemp222InMem/.initial=Yes,
/ConsideredUncertainAlphaTemp223InMem/.initial=Yes,
/ConsideredUncertainAlphaTemp224InMem/.initial=Yes,
/ConsideredUncertainAlphaTemp225InMem/.initial=Yes,
/ConsideredUncertainAlphaTemp226InMem/.initial=Yes,
/ConsideredUncertainAlphaTemp227InMem/.initial=Yes,
/ConsideredUncertainAlphaTemp228InMem/.initial=Yes,
/ConsideredUncertainAlphaTemp229InMem/.initial=Yes,
/ConsideredUncertainAlphaTemp230InMem/.initial=Yes,
/ConsideredUncertainAlphaTemp231InMem/.initial=Yes,
/ConsideredUncertainAlphaTemp232InMem/.initial=Yes,
/ConsideredUncertainAlphaTemp233InMem/.initial=Yes,
/ConsideredUncertainAlphaTemp234InMem/.initial=Yes,
/ConsideredUncertainAlphaTemp235InMem/.initial=Yes,
/ConsideredUncertainAlphaTemp236InMem/.initial=Yes,
/ConsideredUncertainAlphaTemp237InMem/.initial=Yes,
/ConsideredUncertainAlphaTemp238InMem/.initial=Yes,
/ConsideredUncertainAlphaTemp239InMem/.initial=Yes,
/ConsideredUncertainAlphaTemp240InMem/.initial=Yes,
/ConsideredUncertainAlphaTemp241InMem/.initial=Yes,
/ConsideredUncertainAlphaTemp242InMem/.initial=Yes,
/ConsideredUncertainAlphaTemp243InMem/.initial=Yes,
/ConsideredUncertainAlphaTemp244InMem/.initial=Yes,
/ConsideredUncertainAlphaTemp245InMem/.initial=Yes,
/ConsideredUncertainAlphaTemp246InMem/.initial=Yes,
/ConsideredUncertainAlphaTemp247InMem/.initial=Yes,
/ConsideredUncertainAlphaTemp248InMem/.initial=Yes,
/ConsideredUncertainAlphaTemp249InMem/.initial=Yes,
/ConsideredUncertainAlphaTemp250InMem/.initial=Yes,
/ConsideredUncertainAlphaTemp251InMem/.initial=Yes,
/ConsideredUncertainAlphaTemp252InMem/.initial=Yes,
/ConsideredUncertainAlphaTemp253InMem/.initial=Yes,
/ConsideredUncertainAlphaTemp254InMem/.initial=Yes,
/ConsideredUncertainAlphaTemp255InMem/.initial=Yes,
/ConsideredUncertainAlphaTemp256InMem/.initial=Yes,
/ConsideredUncertainAlphaTemp257InMem/.initial=Yes,
/ConsideredUncertainAlphaTemp258InMem/.initial=Yes,
/ConsideredUncertainAlphaTemp259InMem/.initial=Yes,
/ConsideredUncertainAlphaTemp260InMem/.initial=Yes,
/ConsideredUncertainAlphaTemp261InMem/.initial=Yes,
/ConsideredUncertainAlphaTemp262InMem/.initial=Yes,
/ConsideredUncertainAlphaTemp263InMem/.initial=Yes,
/ConsideredUncertainAlphaTemp264InMem/.initial=Yes,
/ConsideredUncertainAlphaTemp265InMem/.initial=Yes,
/ConsideredUncertainAlphaTemp266InMem/.initial=Yes,
/ConsideredUncertainAlphaTemp267InMem/.initial=Yes,
/ConsideredUncertainAlphaTemp268InMem/.initial=Yes,
/ConsideredUncertainAlphaTemp269InMem/.initial=Yes,
/ConsideredUncertainAlphaTemp270InMem/.initial=Yes,
/ConsideredUncertainAlphaTemp271InMem/.initial=Yes,
/ConsideredUncertainAlphaTemp272InMem/.initial=Yes,
/ConsideredUncertainAlphaTemp273InMem/.initial=Yes,
/ConsideredUncertainAlphaTemp274InMem/.initial=Yes,
/ConsideredUncertainAlphaTemp275InMem/.initial=Yes,
/ConsideredUncertainAlphaTemp276InMem/.initial=Yes,
/ConsideredUncertainAlphaTemp277InMem/.initial=Yes,
/ConsideredUncertainAlphaTemp278InMem/.initial=Yes,
/ConsideredUncertainAlphaTemp279InMem/.initial=Yes,
/ConsideredUncertainAlphaTemp280InMem/.initial=Yes,
/ConsideredUncertainAlphaTemp281InMem/.initial=Yes,
/ConsideredUncertainAlphaTemp282InMem/.initial=Yes,
/ConsideredUncertainAlphaTemp283InMem/.initial=Yes,
/ConsideredUncertainAlphaTemp284InMem/.initial=Yes,
/ConsideredUncertainAlphaTemp285InMem/.initial=Yes,
/ConsideredUncertainAlphaTemp286InMem/.initial=Yes,
/ConsideredUncertainAlphaTemp287InMem/.initial=Yes,
/ConsideredUncertainAlphaTemp288InMem/.initial=Yes,
/ConsideredUncertainAlphaTemp289InMem/.initial=Yes,
/ConsideredUncertainAlphaTemp290InMem/.initial=Yes,
/ConsideredUncertainAlphaTemp291InMem/.initial=Yes,
/ConsideredUncertainAlphaTemp292InMem/.initial=Yes,
/ConsideredUncertainAlphaTemp293InMem/.initial=Yes,
/ConsideredUncertainAlphaTemp294InMem/.initial=Yes,
/ConsideredUncertainAlphaTemp295InMem/.initial=Yes,
/ConsideredUncertainAlphaTemp296InMem/.initial=Yes,
/ConsideredUncertainAlphaTemp297InMem/.initial=Yes,
/ConsideredUncertainAlphaTemp298InMem/.initial=Yes,
/ConsideredUncertainAlphaTemp299InMem/.initial=Yes,
/ConsideredUncertainAlphaTemp300InMem/.initial=Yes,
/ConsideredUncertainAlphaTemp301InMem/.initial=Yes,
/ConsideredUncertainAlphaTemp302InMem/.initial=Yes,
/ConsideredUncertainAlphaTemp303InMem/.initial=Yes,
/ConsideredUncertainAlphaTemp304InMem/.initial=Yes,
/ConsideredUncertainAlphaTemp305InMem/.initial=Yes,
/ConsideredUncertainAlphaTemp306InMem/.initial=Yes,
/ConsideredUncertainAlphaTemp307InMem/.initial=Yes,
/ConsideredUncertainAlphaTemp308InMem/.initial=Yes,
/ConsideredUncertainAlphaTemp309InMem/.initial=Yes,
/ConsideredUncertainAlphaTemp310InMem/.initial=Yes,
/ConsideredUncertainAlphaTemp311InMem/.initial=Yes,
/ConsideredUncertainAlphaTemp312InMem/.initial=Yes,
/ConsideredUncertainAlphaTemp313InMem/.initial=Yes,
/ConsideredUncertainAlphaTemp314InMem/.initial=Yes,
/ConsideredUncertainAlphaTemp315InMem/.initial=Yes,
/ConsideredUncertainAlphaTemp316InMem/.initial=Yes,
/ConsideredUncertainAlphaTemp317InMem/.initial=Yes,
/ConsideredUncertainAlphaTemp318InMem/.initial=Yes,
/ConsideredUncertainAlphaTemp319InMem/.initial=Yes,
/ConsideredUncertainAlphaTemp320InMem/.initial=Yes,
/ConsideredUncertainAlphaTemp321InMem/.initial=Yes,
/ConsideredUncertainAlphaTemp322InMem/.initial=Yes,
/ConsideredUncertainAlphaTemp323InMem/.initial=Yes,
/ConsideredUncertainAlphaTemp324InMem/.initial=Yes,
/ConsideredUncertainAlphaTemp325InMem/.initial=Yes,
/ConsideredUncertainAlphaTemp326InMem/.initial=Yes,
/ConsideredUncertainAlphaTemp327InMem/.initial=Yes,
/ConsideredUncertainAlphaTemp328InMem/.initial=Yes,
/ConsideredUncertainAlphaTemp329InMem/.initial=Yes,
/ConsideredUncertainAlphaTemp330InMem/.initial=Yes,
/ConsideredUncertainAlphaTemp331InMem/.initial=Yes,
/ConsideredUncertainAlphaTemp332InMem/.initial=Yes,
/ConsideredUncertainAlphaTemp333InMem/.initial=Yes,
/ConsideredUncertainAlphaTemp334InMem/.initial=Yes,
/ConsideredUncertainAlphaTemp335InMem/.initial=Yes,
/ConsideredUncertainAlphaTemp336InMem/.initial=Yes,
/ConsideredUncertainAlphaTemp337InMem/.initial=Yes,
/ConsideredUncertainAlphaTemp338InMem/.initial=Yes,
/ConsideredUncertainAlphaTemp339InMem/.initial=Yes,
/ConsideredUncertainAlphaTemp340InMem/.initial=Yes,
/ConsideredUncertainAlphaTemp341InMem/.initial=Yes,
/ConsideredUncertainAlphaTemp342InMem/.initial=Yes,
/ConsideredUncertainAlphaTemp343InMem/.initial=Yes,
/ConsideredUncertainAlphaTemp344InMem/.initial=Yes,
/ConsideredUncertainAlphaTemp345InMem/.initial=Yes,
/ConsideredUncertainAlphaTemp346InMem/.initial=Yes,
/ConsideredUncertainAlphaTemp347InMem/.initial=Yes,
/ConsideredUncertainAlphaTemp348InMem/.initial=Yes,
/ConsideredUncertainAlphaTemp349InMem/.initial=Yes,
/ConsideredUncertainAlphaTemp350InMem/.initial=Yes,
/ConsideredUncertainAlphaTemp351InMem/.initial=Yes,
/ConsideredUncertainAlphaTemp352InMem/.initial=Yes,
/ConsideredUncertainAlphaTemp353InMem/.initial=Yes,
/ConsideredUncertainAlphaTemp354InMem/.initial=Yes,
/ConsideredUncertainAlphaTemp355InMem/.initial=Yes,
/ConsideredUncertainAlphaTemp356InMem/.initial=Yes,
/ConsideredUncertainAlphaTemp357InMem/.initial=Yes,
/ConsideredUncertainAlphaTemp358InMem/.initial=Yes,
/ConsideredUncertainAlphaTemp359InMem/.initial=Yes,
/ConsideredUncertainAlphaTemp360InMem/.initial=Yes,
/ConsideredUncertainAlphaTemp361InMem/.initial=Yes,
/ConsideredUncertainAlphaTemp362InMem/.initial=Yes,
/ConsideredUncertainAlphaTemp363InMem/.initial=Yes,
/ConsideredUncertainAlphaTemp364InMem/.initial=Yes,
/ConsideredUncertainAlphaTemp365InMem/.initial=Yes,
/ConsideredUncertainAlphaTemp366InMem/.initial=Yes,
/ConsideredUncertainAlphaTemp367InMem/.initial=Yes,
/ConsideredUncertainAlphaTemp368InMem/.initial=Yes,
/ConsideredUncertainAlphaTemp369InMem/.initial=Yes,
/ConsideredUncertainAlphaTemp370InMem/.initial=Yes,
/ConsideredUncertainAlphaTemp371InMem/.initial=Yes,
/ConsideredUncertainAlphaTemp372InMem/.initial=Yes,
/ConsideredUncertainAlphaTemp373InMem/.initial=Yes,
/ConsideredUncertainAlphaTemp374InMem/.initial=Yes,
/ConsideredUncertainAlphaTemp375InMem/.initial=Yes,
/ConsideredUncertainAlphaTemp376InMem/.initial=Yes,
/ConsideredUncertainAlphaTemp377InMem/.initial=Yes,
/ConsideredUncertainAlphaTemp378InMem/.initial=Yes,
/ConsideredUncertainAlphaTemp379InMem/.initial=Yes,
/ConsideredUncertainAlphaTemp380InMem/.initial=Yes,
/ConsideredUncertainAlphaTemp381InMem/.initial=Yes,
/ConsideredUncertainAlphaTemp382InMem/.initial=Yes,
/ConsideredUncertainAlphaTemp383InMem/.initial=Yes,
/ConsideredUncertainAlphaTemp384InMem/.initial=Yes,
/ConsideredUncertainAlphaTemp385InMem/.initial=Yes,
/ConsideredUncertainAlphaTemp386InMem/.initial=Yes,
/ConsideredUncertainAlphaTemp387InMem/.initial=Yes,
/ConsideredUncertainAlphaTemp388InMem/.initial=Yes,
/ConsideredUncertainAlphaTemp389InMem/.initial=Yes,
/ConsideredUncertainAlphaTemp390InMem/.initial=Yes,
/ConsideredUncertainAlphaTemp391InMem/.initial=Yes,
/ConsideredUncertainAlphaTemp392InMem/.initial=Yes,
/ConsideredUncertainAlphaTemp393InMem/.initial=Yes,
/ConsideredUncertainAlphaTemp394InMem/.initial=Yes,
/ConsideredUncertainAlphaTemp395InMem/.initial=Yes,
/ConsideredUncertainAlphaTemp396InMem/.initial=Yes,
/ConsideredUncertainAlphaTemp397InMem/.initial=Yes,
/ConsideredUncertainAlphaTemp398InMem/.initial=Yes,
/ConsideredUncertainAlphaTemp399InMem/.initial=Yes,
/ConsideredUncertainAlphaTemp400InMem/.initial=Yes,
/ConsideredUncertainAlphaTemp401InMem/.initial=Yes,
/ConsideredUncertainAlphaTemp402InMem/.initial=Yes,
/ConsideredUncertainAlphaTemp403InMem/.initial=Yes,
/ConsideredUncertainAlphaTemp404InMem/.initial=Yes,
/ConsideredUncertainAlphaTemp405InMem/.initial=Yes,
/ConsideredUncertainAlphaTemp406InMem/.initial=Yes,
/ConsideredUncertainAlphaTemp407InMem/.initial=Yes,
/ConsideredUncertainAlphaTemp408InMem/.initial=Yes,
/ConsideredUncertainAlphaTemp409InMem/.initial=Yes,
/ConsideredUncertainAlphaTemp410InMem/.initial=Yes,
/ConsideredUncertainAlphaTemp411InMem/.initial=Yes,
/ConsideredUncertainAlphaTemp412InMem/.initial=Yes,
/ConsideredUncertainAlphaTemp413InMem/.initial=Yes,
/ConsideredUncertainAlphaTemp414InMem/.initial=Yes,
/ConsideredUncertainAlphaTemp415InMem/.initial=Yes,
/ConsideredUncertainAlphaTemp416InMem/.initial=Yes,
/ConsideredUncertainAlphaTemp417InMem/.initial=Yes,
/ConsideredUncertainAlphaTemp418InMem/.initial=Yes,
/ConsideredUncertainAlphaTemp419InMem/.initial=Yes,
/ConsideredUncertainAlphaTemp420InMem/.initial=Yes,
/ConsideredUncertainAlphaTemp421InMem/.initial=Yes,
/ConsideredUncertainAlphaTemp422InMem/.initial=Yes,
/ConsideredUncertainAlphaTemp423InMem/.initial=Yes,
/ConsideredUncertainAlphaTemp424InMem/.initial=Yes,
/ConsideredUncertainAlphaTemp425InMem/.initial=Yes,
/ConsideredUncertainAlphaTemp426InMem/.initial=Yes,
/ConsideredUncertainAlphaTemp427InMem/.initial=Yes,
/ConsideredUncertainAlphaTemp428InMem/.initial=Yes,
/ConsideredUncertainAlphaTemp429InMem/.initial=Yes,
/ConsideredUncertainAlphaTemp430InMem/.initial=Yes,
/ConsideredUncertainAlphaTemp431InMem/.initial=Yes,
/ConsideredUncertainAlphaTemp432InMem/.initial=Yes,
/ConsideredUncertainAlphaTemp433InMem/.initial=Yes,
/ConsideredUncertainAlphaTemp434InMem/.initial=Yes,
/ConsideredUncertainAlphaTemp435InMem/.initial=Yes,
/ConsideredUncertainAlphaTemp436InMem/.initial=Yes,
/ConsideredUncertainAlphaTemp437InMem/.initial=Yes,
/ConsideredUncertainAlphaTemp438InMem/.initial=Yes,
/ConsideredUncertainAlphaTemp439InMem/.initial=Yes,
/ConsideredUncertainAlphaTemp440InMem/.initial=Yes,
/ConsideredUncertainAlphaTemp441InMem/.initial=Yes,
/ConsideredUncertainAlphaTemp442InMem/.initial=Yes,
/ConsideredUncertainAlphaTemp443InMem/.initial=Yes,
/ConsideredUncertainAlphaTemp444InMem/.initial=Yes,
/ConsideredUncertainAlphaTemp445InMem/.initial=Yes,
/ConsideredUncertainAlphaTemp446InMem/.initial=Yes,
/ConsideredUncertainAlphaTemp447InMem/.initial=Yes,
/ConsideredUncertainAlphaTemp448InMem/.initial=Yes,
/ConsideredUncertainAlphaTemp449InMem/.initial=Yes,
/ConsideredUncertainAlphaTemp450InMem/.initial=Yes,
/ConsideredUncertainAlphaTemp451InMem/.initial=Yes,
/ConsideredUncertainAlphaTemp452InMem/.initial=Yes,
/ConsideredUncertainAlphaTemp453InMem/.initial=Yes,
/ConsideredUncertainAlphaTemp454InMem/.initial=Yes,
/ConsideredUncertainAlphaTemp455InMem/.initial=Yes,
/ConsideredUncertainAlphaTemp456InMem/.initial=Yes,
/ConsideredUncertainAlphaTemp457InMem/.initial=Yes,
/ConsideredUncertainAlphaTemp458InMem/.initial=Yes,
/ConsideredUncertainAlphaTemp459InMem/.initial=Yes,
/ConsideredUncertainAlphaTemp460InMem/.initial=Yes,
/ConsideredUncertainAlphaTemp461InMem/.initial=Yes,
/ConsideredUncertainAlphaTemp462InMem/.initial=Yes,
/ConsideredUncertainAlphaTemp463InMem/.initial=Yes,
/ConsideredUncertainAlphaTemp464InMem/.initial=Yes,
/ConsideredUncertainAlphaTemp465InMem/.initial=Yes,
/ConsideredUncertainAlphaTemp466InMem/.initial=Yes,
/ConsideredUncertainAlphaTemp467InMem/.initial=Yes,
/ConsideredUncertainAlphaTemp468InMem/.initial=Yes,
/ConsideredUncertainAlphaTemp469InMem/.initial=Yes,
/ConsideredUncertainAlphaTemp470InMem/.initial=Yes,
/ConsideredUncertainAlphaTemp471InMem/.initial=Yes,
/ConsideredUncertainAlphaTemp472InMem/.initial=Yes,
/ConsideredUncertainAlphaTemp473InMem/.initial=Yes,
/ConsideredUncertainAlphaTemp474InMem/.initial=Yes,
/ConsideredUncertainAlphaTemp475InMem/.initial=Yes,
/ConsideredUncertainAlphaTemp476InMem/.initial=Yes,
/ConsideredUncertainAlphaTemp477InMem/.initial=Yes,
/ConsideredUncertainAlphaTemp478InMem/.initial=Yes,
/ConsideredUncertainAlphaTemp479InMem/.initial=Yes,
/ConsideredUncertainAlphaTemp480InMem/.initial=Yes,
/ConsideredUncertainAlphaTemp481InMem/.initial=Yes,
/ConsideredUncertainAlphaTemp482InMem/.initial=Yes,
/ConsideredUncertainAlphaTemp483InMem/.initial=Yes,
/ConsideredUncertainAlphaTemp484InMem/.initial=Yes,
/ConsideredUncertainAlphaTemp485InMem/.initial=Yes,
/ConsideredUncertainAlphaTemp486InMem/.initial=Yes,
/ConsideredUncertainAlphaTemp487InMem/.initial=Yes,
/ConsideredUncertainAlphaTemp488InMem/.initial=Yes,
/ConsideredUncertainAlphaTemp489InMem/.initial=Yes,
/ConsideredUncertainAlphaTemp490InMem/.initial=Yes,
/ConsideredUncertainAlphaTemp491InMem/.initial=Yes,
/ConsideredUncertainAlphaTemp492InMem/.initial=Yes,
/ConsideredUncertainAlphaTemp493InMem/.initial=Yes,
/ConsideredUncertainAlphaTemp494InMem/.initial=Yes,
/ConsideredUncertainAlphaTemp495InMem/.initial=Yes,
/ConsideredUncertainAlphaTemp496InMem/.initial=Yes,
/ConsideredUncertainAlphaTemp497InMem/.initial=Yes,
/ConsideredUncertainAlphaTemp498InMem/.initial=Yes,
/ConsideredUncertainAlphaTemp499InMem/.initial=Yes,
/ConsideredUncertainAlphaTemp500InMem/.initial=Yes,
/ConsideredUncertainAlphaTemp501InMem/.initial=Yes,
/ConsideredUncertainAlphaTemp502InMem/.initial=Yes,
/ConsideredUncertainAlphaTemp503InMem/.initial=Yes,
/ConsideredUncertainAlphaTemp504InMem/.initial=Yes,
/ConsideredUncertainAlphaTemp505InMem/.initial=Yes,
/ConsideredUncertainAlphaTemp506InMem/.initial=Yes,
/ConsideredUncertainAlphaTemp507InMem/.initial=Yes,
/ConsideredUncertainAlphaTemp508InMem/.initial=Yes,
/ConsideredUncertainAlphaTemp509InMem/.initial=Yes,
/ConsideredUncertainAlphaTemp510InMem/.initial=Yes,
/ConsideredUncertainAlphaTemp511InMem/.initial=Yes,
/ConsideredUncertainAlphaTemp512InMem/.initial=Yes,
/ConsideredUncertainAlphaTemp513InMem/.initial=Yes,
/ConsideredUncertainAlphaTemp514InMem/.initial=Yes,
/ConsideredUncertainAlphaTemp515InMem/.initial=Yes,
/ConsideredUncertainAlphaTemp516InMem/.initial=Yes,
/ConsideredUncertainAlphaTemp517InMem/.initial=Yes,
/ConsideredUncertainAlphaTemp518InMem/.initial=Yes,
/ConsideredUncertainAlphaTemp519InMem/.initial=Yes,
/ConsideredUncertainAlphaTemp520InMem/.initial=Yes,
/ConsideredUncertainAlphaTemp521InMem/.initial=Yes,
/ConsideredUncertainAlphaTemp522InMem/.initial=Yes,
/ConsideredUncertainAlphaTemp523InMem/.initial=Yes,
/ConsideredUncertainAlphaTemp524InMem/.initial=Yes,
/ConsideredUncertainAlphaTemp525InMem/.initial=Yes,
/ConsideredUncertainAlphaTemp526InMem/.initial=Yes,
/ConsideredUncertainAlphaTemp527InMem/.initial=Yes,
/ConsideredUncertainAlphaTemp528InMem/.initial=Yes,
/ConsideredUncertainAlphaTemp529InMem/.initial=Yes,
/ConsideredUncertainAlphaTemp530InMem/.initial=Yes,
/ConsideredUncertainAlphaTemp531InMem/.initial=Yes,
/ConsideredUncertainAlphaTemp532InMem/.initial=Yes,
/ConsideredUncertainAlphaTemp533InMem/.initial=Yes,
/ConsideredUncertainAlphaTemp534InMem/.initial=Yes,
/ConsideredUncertainAlphaTemp535InMem/.initial=Yes,
/ConsideredUncertainAlphaTemp536InMem/.initial=Yes,
/ConsideredUncertainAlphaTemp537InMem/.initial=Yes,
/ConsideredUncertainAlphaTemp538InMem/.initial=Yes,
/ConsideredUncertainAlphaTemp539InMem/.initial=Yes,
/ConsideredUncertainAlphaTemp540InMem/.initial=Yes,
/ConsideredUncertainAlphaTemp541InMem/.initial=Yes,
/ConsideredUncertainAlphaTemp542InMem/.initial=Yes,
/ConsideredUncertainAlphaTemp543InMem/.initial=Yes,
/ConsideredUncertainAlphaTemp544InMem/.initial=Yes,
/ConsideredUncertainAlphaTemp545InMem/.initial=Yes,
/ConsideredUncertainAlphaTemp546InMem/.initial=Yes,
/ConsideredUncertainAlphaTemp547InMem/.initial=Yes,
/ConsideredUncertainAlphaTemp548InMem/.initial=Yes,
/ConsideredUncertainAlphaTemp549InMem/.initial=Yes,
/ConsideredUncertainAlphaTemp550InMem/.initial=Yes,
/ConsideredUncertainAlphaTemp551InMem/.initial=Yes,
/ConsideredUncertainAlphaTemp552InMem/.initial=Yes,
/ConsideredUncertainAlphaTemp553InMem/.initial=Yes,
/ConsideredUncertainAlphaTemp554InMem/.initial=Yes,
/ConsideredUncertainAlphaTemp555InMem/.initial=Yes,
/ConsideredUncertainAlphaTemp556InMem/.initial=Yes,
/ConsideredUncertainAlphaTemp557InMem/.initial=Yes,
/ConsideredUncertainAlphaTemp558InMem/.initial=Yes,
/ConsideredUncertainAlphaTemp559InMem/.initial=Yes,
/ConsideredUncertainAlphaTemp560InMem/.initial=Yes,
/ConsideredUncertainAlphaTemp561InMem/.initial=Yes,
/ConsideredUncertainAlphaTemp562InMem/.initial=Yes,
/ConsideredUncertainAlphaTemp563InMem/.initial=Yes,
/ConsideredUncertainAlphaTemp564InMem/.initial=Yes,
/ConsideredUncertainAlphaTemp565InMem/.initial=Yes,
/ConsideredUncertainAlphaTemp566InMem/.initial=Yes,
/ConsideredUncertainAlphaTemp567InMem/.initial=Yes,
/ConsideredUncertainAlphaTemp568InMem/.initial=Yes,
/ConsideredUncertainAlphaTemp569InMem/.initial=Yes,
/ConsideredUncertainAlphaTemp570InMem/.initial=Yes,
/ConsideredUncertainAlphaTemp571InMem/.initial=Yes,
/ConsideredUncertainAlphaTemp572InMem/.initial=Yes,
/ConsideredUncertainAlphaTemp573InMem/.initial=Yes,
/ConsideredUncertainAlphaTemp574InMem/.initial=Yes,
/ConsideredUncertainAlphaTemp575InMem/.initial=Yes,
/ConsideredUncertainAlphaTemp576InMem/.initial=Yes,
/ConsideredUncertainAlphaTemp577InMem/.initial=Yes,
/ConsideredUncertainAlphaTemp578InMem/.initial=Yes,
/ConsideredUncertainAlphaTemp579InMem/.initial=Yes,
/ConsideredUncertainAlphaTemp580InMem/.initial=Yes,
/ConsideredUncertainAlphaTemp581InMem/.initial=Yes,
/ConsideredUncertainAlphaTemp582InMem/.initial=Yes,
/ConsideredUncertainAlphaTemp583InMem/.initial=Yes,
/ConsideredUncertainAlphaTemp584InMem/.initial=Yes,
/ConsideredUncertainAlphaTemp585InMem/.initial=Yes,
/ConsideredUncertainAlphaTemp586InMem/.initial=Yes,
/ConsideredUncertainAlphaTemp587InMem/.initial=Yes,
/ConsideredUncertainAlphaTemp588InMem/.initial=Yes,
/ConsideredUncertainAlphaTemp589InMem/.initial=Yes,
/ConsideredUncertainAlphaTemp590InMem/.initial=Yes,
/ConsideredUncertainAlphaTemp591InMem/.initial=Yes,
/ConsideredUncertainAlphaTemp592InMem/.initial=Yes,
/ConsideredUncertainAlphaTemp593InMem/.initial=Yes,
/ConsideredUncertainAlphaTemp594InMem/.initial=Yes,
/ConsideredUncertainAlphaTemp595InMem/.initial=Yes,
/ConsideredUncertainAlphaTemp596InMem/.initial=Yes,
/ConsideredUncertainAlphaTemp597InMem/.initial=Yes,
/ConsideredUncertainAlphaTemp598InMem/.initial=Yes,
/ConsideredUncertainAlphaTemp599InMem/.initial=Yes,
/ConsideredUncertainAlphaTemp600InMem/.initial=Yes,
/ConsideredUncertainAlphaTemp601InMem/.initial=Yes,
/ConsideredUncertainAlphaTemp602InMem/.initial=Yes,
/ConsideredUncertainAlphaTemp603InMem/.initial=Yes,
/ConsideredUncertainAlphaTemp604InMem/.initial=Yes,
/ConsideredUncertainAlphaTemp605InMem/.initial=Yes,
/ConsideredUncertainAlphaTemp606InMem/.initial=Yes,
/ConsideredUncertainAlphaTemp607InMem/.initial=Yes,
/ConsideredUncertainAlphaTemp608InMem/.initial=Yes,
/ConsideredUncertainAlphaTemp609InMem/.initial=Yes,
/ConsideredUncertainAlphaTemp610InMem/.initial=Yes,
/ConsideredUncertainAlphaTemp611InMem/.initial=Yes,
/ConsideredUncertainAlphaTemp612InMem/.initial=Yes,
/ConsideredUncertainAlphaTemp613InMem/.initial=Yes,
/ConsideredUncertainAlphaTemp614InMem/.initial=Yes,
/ConsideredUncertainAlphaTemp615InMem/.initial=Yes,
/ConsideredUncertainAlphaTemp616InMem/.initial=Yes,
/ConsideredUncertainAlphaTemp617InMem/.initial=Yes,
/ConsideredUncertainAlphaTemp618InMem/.initial=Yes,
/ConsideredUncertainAlphaTemp619InMem/.initial=Yes,
/ConsideredUncertainAlphaTemp620InMem/.initial=Yes,
/ConsideredUncertainAlphaTemp621InMem/.initial=Yes,
/ConsideredUncertainAlphaTemp622InMem/.initial=Yes,
/ConsideredUncertainAlphaTemp623InMem/.initial=Yes,
/ConsideredUncertainAlphaTemp624InMem/.initial=Yes,
/ConsideredUncertainAlphaTemp625InMem/.initial=Yes,
/ConsideredUncertainAlphaTemp626InMem/.initial=Yes,
/ConsideredUncertainAlphaTemp627InMem/.initial=Yes,
/ConsideredUncertainAlphaTemp628InMem/.initial=Yes,
/ConsideredUncertainAlphaTemp629InMem/.initial=Yes,
/ConsideredUncertainAlphaTemp630InMem/.initial=Yes,
/ConsideredUncertainAlphaTemp631InMem/.initial=Yes,
/ConsideredUncertainAlphaTemp632InMem/.initial=Yes,
/ConsideredUncertainAlphaTemp633InMem/.initial=Yes,
/ConsideredUncertainAlphaTemp634InMem/.initial=Yes,
/ConsideredUncertainAlphaTemp635InMem/.initial=Yes,
/ConsideredUncertainAlphaTemp636InMem/.initial=Yes,
/ConsideredUncertainAlphaTemp637InMem/.initial=Yes,
/ConsideredUncertainAlphaTemp638InMem/.initial=Yes,
/ConsideredUncertainAlphaTemp639InMem/.initial=Yes,
/ConsideredUncertainAlphaTemp640InMem/.initial=Yes,
/ConsideredUncertainAlphaTemp641InMem/.initial=Yes,
/ConsideredUncertainAlphaTemp642InMem/.initial=Yes,
/ConsideredUncertainAlphaTemp643InMem/.initial=Yes,
/ConsideredUncertainAlphaTemp644InMem/.initial=Yes,
/ConsideredUncertainAlphaTemp645InMem/.initial=Yes,
/ConsideredUncertainAlphaTemp646InMem/.initial=Yes,
/ConsideredUncertainAlphaTemp647InMem/.initial=Yes,
/ConsideredUncertainAlphaTemp648InMem/.initial=Yes,
/ConsideredUncertainAlphaTemp649InMem/.initial=Yes,
/ConsideredUncertainAlphaTemp650InMem/.initial=Yes,
/ConsideredUncertainAlphaTemp651InMem/.initial=Yes,
/ConsideredUncertainAlphaTemp652InMem/.initial=Yes,
/ConsideredUncertainAlphaTemp653InMem/.initial=Yes,
/ConsideredUncertainAlphaTemp654InMem/.initial=Yes,
/ConsideredUncertainAlphaTemp655InMem/.initial=Yes,
/ConsideredUncertainAlphaTemp656InMem/.initial=Yes,
/ConsideredUncertainAlphaTemp657InMem/.initial=Yes,
/ConsideredUncertainAlphaTemp658InMem/.initial=Yes,
/ConsideredUncertainAlphaTemp659InMem/.initial=Yes,
/ConsideredUncertainAlphaTemp660InMem/.initial=Yes,
/ConsideredUncertainAlphaTemp661InMem/.initial=Yes,
/ConsideredUncertainAlphaTemp662InMem/.initial=Yes,
/ConsideredUncertainAlphaTemp663InMem/.initial=Yes,
/ConsideredUncertainAlphaTemp664InMem/.initial=Yes,
/ConsideredUncertainAlphaTemp665InMem/.initial=Yes,
/ConsideredUncertainAlphaTemp666InMem/.initial=Yes,
/ConsideredUncertainAlphaTemp667InMem/.initial=Yes,
/ConsideredUncertainAlphaTemp668InMem/.initial=Yes,
/ConsideredUncertainAlphaTemp669InMem/.initial=Yes,
/ConsideredUncertainAlphaTemp670InMem/.initial=Yes,
/ConsideredUncertainAlphaTemp671InMem/.initial=Yes,
/ConsideredUncertainAlphaTemp672InMem/.initial=Yes,
/ConsideredUncertainAlphaTemp673InMem/.initial=Yes,
/ConsideredUncertainAlphaTemp674InMem/.initial=Yes,
/ConsideredUncertainAlphaTemp675InMem/.initial=Yes,
/ConsideredUncertainAlphaTemp676InMem/.initial=Yes,
/ConsideredUncertainAlphaTemp677InMem/.initial=Yes,
/ConsideredUncertainAlphaTemp678InMem/.initial=Yes,
/ConsideredUncertainAlphaTemp679InMem/.initial=Yes,
/ConsideredUncertainAlphaTemp680InMem/.initial=Yes,
/ConsideredUncertainAlphaTemp681InMem/.initial=Yes,
/ConsideredUncertainAlphaTemp682InMem/.initial=Yes,
/ConsideredUncertainAlphaTemp683InMem/.initial=Yes,
/ConsideredUncertainAlphaTemp684InMem/.initial=Yes,
/ConsideredUncertainAlphaTemp685InMem/.initial=Yes,
/ConsideredUncertainAlphaTemp686InMem/.initial=Yes,
/ConsideredUncertainAlphaTemp687InMem/.initial=Yes,
/ConsideredUncertainAlphaTemp688InMem/.initial=Yes,
/ConsideredUncertainAlphaTemp689InMem/.initial=Yes,
/ConsideredUncertainAlphaTemp690InMem/.initial=Yes,
/ConsideredUncertainAlphaTemp691InMem/.initial=Yes,
/ConsideredUncertainAlphaTemp692InMem/.initial=Yes,
/ConsideredUncertainAlphaTemp693InMem/.initial=Yes,
/ConsideredUncertainAlphaTemp694InMem/.initial=Yes,
/ConsideredUncertainAlphaTemp695InMem/.initial=Yes,
/ConsideredUncertainAlphaTemp696InMem/.initial=Yes,
/ConsideredUncertainAlphaTemp697InMem/.initial=Yes,
/ConsideredUncertainAlphaTemp698InMem/.initial=Yes,
/ConsideredUncertainAlphaTemp699InMem/.initial=Yes,
/ConsideredUncertainAlphaTemp700InMem/.initial=Yes,
/ConsideredUncertainAlphaTemp701InMem/.initial=Yes,
/ConsideredUncertainAlphaTemp702InMem/.initial=Yes,
/ConsideredUncertainAlphaTemp703InMem/.initial=Yes,
/ConsideredUncertainAlphaTemp704InMem/.initial=Yes,
/ConsideredUncertainAlphaTemp705InMem/.initial=Yes,
/ConsideredUncertainAlphaTemp706InMem/.initial=Yes,
/ConsideredUncertainAlphaTemp707InMem/.initial=Yes,
/ConsideredUncertainAlphaTemp708InMem/.initial=Yes,
/ConsideredUncertainAlphaTemp709InMem/.initial=Yes,
/ConsideredUncertainAlphaTemp710InMem/.initial=Yes,
/ConsideredUncertainAlphaTemp711InMem/.initial=Yes,
/ConsideredUncertainAlphaTemp712InMem/.initial=Yes,
/ConsideredUncertainAlphaTemp713InMem/.initial=Yes,
/ConsideredUncertainAlphaTemp714InMem/.initial=Yes,
/ConsideredUncertainAlphaTemp715InMem/.initial=Yes,
/ConsideredUncertainAlphaTemp716InMem/.initial=Yes,
/ConsideredUncertainAlphaTemp717InMem/.initial=Yes,
/ConsideredUncertainAlphaTemp718InMem/.initial=Yes,
/ConsideredUncertainAlphaTemp719InMem/.initial=Yes,
/ConsideredUncertainAlphaTemp720InMem/.initial=Yes,
/ConsideredUncertainAlphaTemp721InMem/.initial=Yes,
/ConsideredUncertainAlphaTemp722InMem/.initial=Yes,
/ConsideredUncertainAlphaTemp723InMem/.initial=Yes,
/ConsideredUncertainAlphaTemp724InMem/.initial=Yes,
/ConsideredUncertainAlphaTemp725InMem/.initial=Yes,
/ConsideredUncertainAlphaTemp726InMem/.initial=Yes,
/ConsideredUncertainAlphaTemp727InMem/.initial=Yes,
/ConsideredUncertainAlphaTemp728InMem/.initial=Yes,
/ConsideredUncertainAlphaTemp729InMem/.initial=Yes,
/ConsideredUncertainAlphaTemp730InMem/.initial=Yes,
/ConsideredUncertainAlphaTemp731InMem/.initial=Yes,
/ConsideredUncertainAlphaTemp732InMem/.initial=Yes,
/ConsideredUncertainAlphaTemp733InMem/.initial=Yes,
/ConsideredUncertainAlphaTemp734InMem/.initial=Yes,
/ConsideredUncertainAlphaTemp735InMem/.initial=Yes,
/ConsideredUncertainAlphaTemp736InMem/.initial=Yes,
/ConsideredUncertainAlphaTemp737InMem/.initial=Yes,
/ConsideredUncertainAlphaTemp738InMem/.initial=Yes,
/ConsideredUncertainAlphaTemp739InMem/.initial=Yes,
/ConsideredUncertainAlphaTemp740InMem/.initial=Yes,
/ConsideredUncertainAlphaTemp741InMem/.initial=Yes,
/ConsideredUncertainAlphaTemp742InMem/.initial=Yes,
/ConsideredUncertainAlphaTemp743InMem/.initial=Yes,
/ConsideredUncertainAlphaTemp744InMem/.initial=Yes,
/ConsideredUncertainAlphaTemp745InMem/.initial=Yes,
/ConsideredUncertainAlphaTemp746InMem/.initial=Yes,
/ConsideredUncertainAlphaTemp747InMem/.initial=Yes,
/ConsideredUncertainAlphaTemp748InMem/.initial=Yes,
/ConsideredUncertainAlphaTemp749InMem/.initial=Yes,
/ConsideredUncertainAlphaTemp750InMem/.initial=Yes,
/ConsideredUncertainAlphaTemp751InMem/.initial=Yes,
/ConsideredUncertainAlphaTemp752InMem/.initial=Yes,
/ConsideredUncertainAlphaTemp753InMem/.initial=Yes,
/ConsideredUncertainAlphaTemp754InMem/.initial=Yes,
/ConsideredUncertainAlphaTemp755InMem/.initial=Yes,
/ConsideredUncertainAlphaTemp756InMem/.initial=Yes,
/ConsideredUncertainAlphaTemp757InMem/.initial=Yes,
/ConsideredUncertainAlphaTemp758InMem/.initial=Yes,
/ConsideredUncertainAlphaTemp759InMem/.initial=Yes,
/ConsideredUncertainAlphaTemp760InMem/.initial=Yes,
/ConsideredUncertainAlphaTemp761InMem/.initial=Yes,
/ConsideredUncertainAlphaTemp762InMem/.initial=Yes,
/ConsideredUncertainAlphaTemp763InMem/.initial=Yes,
/ConsideredUncertainAlphaTemp764InMem/.initial=Yes,
/ConsideredUncertainAlphaTemp765InMem/.initial=Yes,
/ConsideredUncertainAlphaTemp766InMem/.initial=Yes,
/ConsideredUncertainAlphaTemp767InMem/.initial=Yes,
/ConsideredUncertainOccRemScale/.initial=No,
/ConsideredUncertainOccColumnScale/.initial=No,
/ConsideredUncertainOccRowScale/.initial=No,
/ConsideredUncertainOffsetRef/.initial=Yes,
/ConsideredUncertainOccRow0/.initial=Yes,
/ConsideredUncertainOccRow1/.initial=Yes,
/ConsideredUncertainOccRow2/.initial=Yes,
/ConsideredUncertainOccRow3/.initial=Yes,
/ConsideredUncertainOccRow4/.initial=Yes,
/ConsideredUncertainOccRow5/.initial=Yes,
/ConsideredUncertainOccRow6/.initial=Yes,
/ConsideredUncertainOccRow7/.initial=Yes,
/ConsideredUncertainOccRow8/.initial=Yes,
/ConsideredUncertainOccRow9/.initial=Yes,
/ConsideredUncertainOccRow10/.initial=Yes,
/ConsideredUncertainOccRow11/.initial=Yes,
/ConsideredUncertainOccRow12/.initial=Yes,
/ConsideredUncertainOccRow13/.initial=Yes,
/ConsideredUncertainOccRow14/.initial=Yes,
/ConsideredUncertainOccRow15/.initial=Yes,
/ConsideredUncertainOccRow16/.initial=Yes,
/ConsideredUncertainOccRow17/.initial=Yes,
/ConsideredUncertainOccRow18/.initial=Yes,
/ConsideredUncertainOccRow19/.initial=Yes,
/ConsideredUncertainOccRow20/.initial=Yes,
/ConsideredUncertainOccRow21/.initial=Yes,
/ConsideredUncertainOccRow22/.initial=Yes,
/ConsideredUncertainOccRow23/.initial=Yes,
/ConsideredUncertainOccColumn0/.initial=Yes,
/ConsideredUncertainOccColumn1/.initial=Yes,
/ConsideredUncertainOccColumn2/.initial=Yes,
/ConsideredUncertainOccColumn3/.initial=Yes,
/ConsideredUncertainOccColumn4/.initial=Yes,
/ConsideredUncertainOccColumn5/.initial=Yes,
/ConsideredUncertainOccColumn6/.initial=Yes,
/ConsideredUncertainOccColumn7/.initial=Yes,
/ConsideredUncertainOccColumn8/.initial=Yes,
/ConsideredUncertainOccColumn9/.initial=Yes,
/ConsideredUncertainOccColumn10/.initial=Yes,
/ConsideredUncertainOccColumn11/.initial=Yes,
/ConsideredUncertainOccColumn12/.initial=Yes,
/ConsideredUncertainOccColumn13/.initial=Yes,
/ConsideredUncertainOccColumn14/.initial=Yes,
/ConsideredUncertainOccColumn15/.initial=Yes,
/ConsideredUncertainOccColumn16/.initial=Yes,
/ConsideredUncertainOccColumn17/.initial=Yes,
/ConsideredUncertainOccColumn18/.initial=Yes,
/ConsideredUncertainOccColumn19/.initial=Yes,
/ConsideredUncertainOccColumn20/.initial=Yes,
/ConsideredUncertainOccColumn21/.initial=Yes,
/ConsideredUncertainOccColumn22/.initial=Yes,
/ConsideredUncertainOccColumn23/.initial=Yes,
/ConsideredUncertainOccColumn24/.initial=Yes,
/ConsideredUncertainOccColumn25/.initial=Yes,
/ConsideredUncertainOccColumn26/.initial=Yes,
/ConsideredUncertainOccColumn27/.initial=Yes,
/ConsideredUncertainOccColumn28/.initial=Yes,
/ConsideredUncertainOccColumn29/.initial=Yes,
/ConsideredUncertainOccColumn30/.initial=Yes,
/ConsideredUncertainOccColumn31/.initial=Yes,
/ConsideredUncertainOffset0/.initial=Yes,
/ConsideredUncertainOffset1/.initial=Yes,
/ConsideredUncertainOffset2/.initial=Yes,
/ConsideredUncertainOffset3/.initial=Yes,
/ConsideredUncertainOffset4/.initial=Yes,
/ConsideredUncertainOffset5/.initial=Yes,
/ConsideredUncertainOffset6/.initial=Yes,
/ConsideredUncertainOffset7/.initial=Yes,
/ConsideredUncertainOffset8/.initial=Yes,
/ConsideredUncertainOffset9/.initial=Yes,
/ConsideredUncertainOffset10/.initial=Yes,
/ConsideredUncertainOffset11/.initial=Yes,
/ConsideredUncertainOffset12/.initial=Yes,
/ConsideredUncertainOffset13/.initial=Yes,
/ConsideredUncertainOffset14/.initial=Yes,
/ConsideredUncertainOffset15/.initial=Yes,
/ConsideredUncertainOffset16/.initial=Yes,
/ConsideredUncertainOffset17/.initial=Yes,
/ConsideredUncertainOffset18/.initial=Yes,
/ConsideredUncertainOffset19/.initial=Yes,
/ConsideredUncertainOffset20/.initial=Yes,
/ConsideredUncertainOffset21/.initial=Yes,
/ConsideredUncertainOffset22/.initial=Yes,
/ConsideredUncertainOffset23/.initial=Yes,
/ConsideredUncertainOffset24/.initial=Yes,
/ConsideredUncertainOffset25/.initial=Yes,
/ConsideredUncertainOffset26/.initial=Yes,
/ConsideredUncertainOffset27/.initial=Yes,
/ConsideredUncertainOffset28/.initial=Yes,
/ConsideredUncertainOffset29/.initial=Yes,
/ConsideredUncertainOffset30/.initial=Yes,
/ConsideredUncertainOffset31/.initial=Yes,
/ConsideredUncertainOffset32/.initial=Yes,
/ConsideredUncertainOffset33/.initial=Yes,
/ConsideredUncertainOffset34/.initial=Yes,
/ConsideredUncertainOffset35/.initial=Yes,
/ConsideredUncertainOffset36/.initial=Yes,
/ConsideredUncertainOffset37/.initial=Yes,
/ConsideredUncertainOffset38/.initial=Yes,
/ConsideredUncertainOffset39/.initial=Yes,
/ConsideredUncertainOffset40/.initial=Yes,
/ConsideredUncertainOffset41/.initial=Yes,
/ConsideredUncertainOffset42/.initial=Yes,
/ConsideredUncertainOffset43/.initial=Yes,
/ConsideredUncertainOffset44/.initial=Yes,
/ConsideredUncertainOffset45/.initial=Yes,
/ConsideredUncertainOffset46/.initial=Yes,
/ConsideredUncertainOffset47/.initial=Yes,
/ConsideredUncertainOffset48/.initial=Yes,
/ConsideredUncertainOffset49/.initial=Yes,
/ConsideredUncertainOffset50/.initial=Yes,
/ConsideredUncertainOffset51/.initial=Yes,
/ConsideredUncertainOffset52/.initial=Yes,
/ConsideredUncertainOffset53/.initial=Yes,
/ConsideredUncertainOffset54/.initial=Yes,
/ConsideredUncertainOffset55/.initial=Yes,
/ConsideredUncertainOffset56/.initial=Yes,
/ConsideredUncertainOffset57/.initial=Yes,
/ConsideredUncertainOffset58/.initial=Yes,
/ConsideredUncertainOffset59/.initial=Yes,
/ConsideredUncertainOffset60/.initial=Yes,
/ConsideredUncertainOffset61/.initial=Yes,
/ConsideredUncertainOffset62/.initial=Yes,
/ConsideredUncertainOffset63/.initial=Yes,
/ConsideredUncertainOffset64/.initial=Yes,
/ConsideredUncertainOffset65/.initial=Yes,
/ConsideredUncertainOffset66/.initial=Yes,
/ConsideredUncertainOffset67/.initial=Yes,
/ConsideredUncertainOffset68/.initial=Yes,
/ConsideredUncertainOffset69/.initial=Yes,
/ConsideredUncertainOffset70/.initial=Yes,
/ConsideredUncertainOffset71/.initial=Yes,
/ConsideredUncertainOffset72/.initial=Yes,
/ConsideredUncertainOffset73/.initial=Yes,
/ConsideredUncertainOffset74/.initial=Yes,
/ConsideredUncertainOffset75/.initial=Yes,
/ConsideredUncertainOffset76/.initial=Yes,
/ConsideredUncertainOffset77/.initial=Yes,
/ConsideredUncertainOffset78/.initial=Yes,
/ConsideredUncertainOffset79/.initial=Yes,
/ConsideredUncertainOffset80/.initial=Yes,
/ConsideredUncertainOffset81/.initial=Yes,
/ConsideredUncertainOffset82/.initial=Yes,
/ConsideredUncertainOffset83/.initial=Yes,
/ConsideredUncertainOffset84/.initial=Yes,
/ConsideredUncertainOffset85/.initial=Yes,
/ConsideredUncertainOffset86/.initial=Yes,
/ConsideredUncertainOffset87/.initial=Yes,
/ConsideredUncertainOffset88/.initial=Yes,
/ConsideredUncertainOffset89/.initial=Yes,
/ConsideredUncertainOffset90/.initial=Yes,
/ConsideredUncertainOffset91/.initial=Yes,
/ConsideredUncertainOffset92/.initial=Yes,
/ConsideredUncertainOffset93/.initial=Yes,
/ConsideredUncertainOffset94/.initial=Yes,
/ConsideredUncertainOffset95/.initial=Yes,
/ConsideredUncertainOffset96/.initial=Yes,
/ConsideredUncertainOffset97/.initial=Yes,
/ConsideredUncertainOffset98/.initial=Yes,
/ConsideredUncertainOffset99/.initial=Yes,
/ConsideredUncertainOffset100/.initial=Yes,
/ConsideredUncertainOffset101/.initial=Yes,
/ConsideredUncertainOffset102/.initial=Yes,
/ConsideredUncertainOffset103/.initial=Yes,
/ConsideredUncertainOffset104/.initial=Yes,
/ConsideredUncertainOffset105/.initial=Yes,
/ConsideredUncertainOffset106/.initial=Yes,
/ConsideredUncertainOffset107/.initial=Yes,
/ConsideredUncertainOffset108/.initial=Yes,
/ConsideredUncertainOffset109/.initial=Yes,
/ConsideredUncertainOffset110/.initial=Yes,
/ConsideredUncertainOffset111/.initial=Yes,
/ConsideredUncertainOffset112/.initial=Yes,
/ConsideredUncertainOffset113/.initial=Yes,
/ConsideredUncertainOffset114/.initial=Yes,
/ConsideredUncertainOffset115/.initial=Yes,
/ConsideredUncertainOffset116/.initial=Yes,
/ConsideredUncertainOffset117/.initial=Yes,
/ConsideredUncertainOffset118/.initial=Yes,
/ConsideredUncertainOffset119/.initial=Yes,
/ConsideredUncertainOffset120/.initial=Yes,
/ConsideredUncertainOffset121/.initial=Yes,
/ConsideredUncertainOffset122/.initial=Yes,
/ConsideredUncertainOffset123/.initial=Yes,
/ConsideredUncertainOffset124/.initial=Yes,
/ConsideredUncertainOffset125/.initial=Yes,
/ConsideredUncertainOffset126/.initial=Yes,
/ConsideredUncertainOffset127/.initial=Yes,
/ConsideredUncertainOffset128/.initial=Yes,
/ConsideredUncertainOffset129/.initial=Yes,
/ConsideredUncertainOffset130/.initial=Yes,
/ConsideredUncertainOffset131/.initial=Yes,
/ConsideredUncertainOffset132/.initial=Yes,
/ConsideredUncertainOffset133/.initial=Yes,
/ConsideredUncertainOffset134/.initial=Yes,
/ConsideredUncertainOffset135/.initial=Yes,
/ConsideredUncertainOffset136/.initial=Yes,
/ConsideredUncertainOffset137/.initial=Yes,
/ConsideredUncertainOffset138/.initial=Yes,
/ConsideredUncertainOffset139/.initial=Yes,
/ConsideredUncertainOffset140/.initial=Yes,
/ConsideredUncertainOffset141/.initial=Yes,
/ConsideredUncertainOffset142/.initial=Yes,
/ConsideredUncertainOffset143/.initial=Yes,
/ConsideredUncertainOffset144/.initial=Yes,
/ConsideredUncertainOffset145/.initial=Yes,
/ConsideredUncertainOffset146/.initial=Yes,
/ConsideredUncertainOffset147/.initial=Yes,
/ConsideredUncertainOffset148/.initial=Yes,
/ConsideredUncertainOffset149/.initial=Yes,
/ConsideredUncertainOffset150/.initial=Yes,
/ConsideredUncertainOffset151/.initial=Yes,
/ConsideredUncertainOffset152/.initial=Yes,
/ConsideredUncertainOffset153/.initial=Yes,
/ConsideredUncertainOffset154/.initial=Yes,
/ConsideredUncertainOffset155/.initial=Yes,
/ConsideredUncertainOffset156/.initial=Yes,
/ConsideredUncertainOffset157/.initial=Yes,
/ConsideredUncertainOffset158/.initial=Yes,
/ConsideredUncertainOffset159/.initial=Yes,
/ConsideredUncertainOffset160/.initial=Yes,
/ConsideredUncertainOffset161/.initial=Yes,
/ConsideredUncertainOffset162/.initial=Yes,
/ConsideredUncertainOffset163/.initial=Yes,
/ConsideredUncertainOffset164/.initial=Yes,
/ConsideredUncertainOffset165/.initial=Yes,
/ConsideredUncertainOffset166/.initial=Yes,
/ConsideredUncertainOffset167/.initial=Yes,
/ConsideredUncertainOffset168/.initial=Yes,
/ConsideredUncertainOffset169/.initial=Yes,
/ConsideredUncertainOffset170/.initial=Yes,
/ConsideredUncertainOffset171/.initial=Yes,
/ConsideredUncertainOffset172/.initial=Yes,
/ConsideredUncertainOffset173/.initial=Yes,
/ConsideredUncertainOffset174/.initial=Yes,
/ConsideredUncertainOffset175/.initial=Yes,
/ConsideredUncertainOffset176/.initial=Yes,
/ConsideredUncertainOffset177/.initial=Yes,
/ConsideredUncertainOffset178/.initial=Yes,
/ConsideredUncertainOffset179/.initial=Yes,
/ConsideredUncertainOffset180/.initial=Yes,
/ConsideredUncertainOffset181/.initial=Yes,
/ConsideredUncertainOffset182/.initial=Yes,
/ConsideredUncertainOffset183/.initial=Yes,
/ConsideredUncertainOffset184/.initial=Yes,
/ConsideredUncertainOffset185/.initial=Yes,
/ConsideredUncertainOffset186/.initial=Yes,
/ConsideredUncertainOffset187/.initial=Yes,
/ConsideredUncertainOffset188/.initial=Yes,
/ConsideredUncertainOffset189/.initial=Yes,
/ConsideredUncertainOffset190/.initial=Yes,
/ConsideredUncertainOffset191/.initial=Yes,
/ConsideredUncertainOffset192/.initial=Yes,
/ConsideredUncertainOffset193/.initial=Yes,
/ConsideredUncertainOffset194/.initial=Yes,
/ConsideredUncertainOffset195/.initial=Yes,
/ConsideredUncertainOffset196/.initial=Yes,
/ConsideredUncertainOffset197/.initial=Yes,
/ConsideredUncertainOffset198/.initial=Yes,
/ConsideredUncertainOffset199/.initial=Yes,
/ConsideredUncertainOffset200/.initial=Yes,
/ConsideredUncertainOffset201/.initial=Yes,
/ConsideredUncertainOffset202/.initial=Yes,
/ConsideredUncertainOffset203/.initial=Yes,
/ConsideredUncertainOffset204/.initial=Yes,
/ConsideredUncertainOffset205/.initial=Yes,
/ConsideredUncertainOffset206/.initial=Yes,
/ConsideredUncertainOffset207/.initial=Yes,
/ConsideredUncertainOffset208/.initial=Yes,
/ConsideredUncertainOffset209/.initial=Yes,
/ConsideredUncertainOffset210/.initial=Yes,
/ConsideredUncertainOffset211/.initial=Yes,
/ConsideredUncertainOffset212/.initial=Yes,
/ConsideredUncertainOffset213/.initial=Yes,
/ConsideredUncertainOffset214/.initial=Yes,
/ConsideredUncertainOffset215/.initial=Yes,
/ConsideredUncertainOffset216/.initial=Yes,
/ConsideredUncertainOffset217/.initial=Yes,
/ConsideredUncertainOffset218/.initial=Yes,
/ConsideredUncertainOffset219/.initial=Yes,
/ConsideredUncertainOffset220/.initial=Yes,
/ConsideredUncertainOffset221/.initial=Yes,
/ConsideredUncertainOffset222/.initial=Yes,
/ConsideredUncertainOffset223/.initial=Yes,
/ConsideredUncertainOffset224/.initial=Yes,
/ConsideredUncertainOffset225/.initial=Yes,
/ConsideredUncertainOffset226/.initial=Yes,
/ConsideredUncertainOffset227/.initial=Yes,
/ConsideredUncertainOffset228/.initial=Yes,
/ConsideredUncertainOffset229/.initial=Yes,
/ConsideredUncertainOffset230/.initial=Yes,
/ConsideredUncertainOffset231/.initial=Yes,
/ConsideredUncertainOffset232/.initial=Yes,
/ConsideredUncertainOffset233/.initial=Yes,
/ConsideredUncertainOffset234/.initial=Yes,
/ConsideredUncertainOffset235/.initial=Yes,
/ConsideredUncertainOffset236/.initial=Yes,
/ConsideredUncertainOffset237/.initial=Yes,
/ConsideredUncertainOffset238/.initial=Yes,
/ConsideredUncertainOffset239/.initial=Yes,
/ConsideredUncertainOffset240/.initial=Yes,
/ConsideredUncertainOffset241/.initial=Yes,
/ConsideredUncertainOffset242/.initial=Yes,
/ConsideredUncertainOffset243/.initial=Yes,
/ConsideredUncertainOffset244/.initial=Yes,
/ConsideredUncertainOffset245/.initial=Yes,
/ConsideredUncertainOffset246/.initial=Yes,
/ConsideredUncertainOffset247/.initial=Yes,
/ConsideredUncertainOffset248/.initial=Yes,
/ConsideredUncertainOffset249/.initial=Yes,
/ConsideredUncertainOffset250/.initial=Yes,
/ConsideredUncertainOffset251/.initial=Yes,
/ConsideredUncertainOffset252/.initial=Yes,
/ConsideredUncertainOffset253/.initial=Yes,
/ConsideredUncertainOffset254/.initial=Yes,
/ConsideredUncertainOffset255/.initial=Yes,
/ConsideredUncertainOffset256/.initial=Yes,
/ConsideredUncertainOffset257/.initial=Yes,
/ConsideredUncertainOffset258/.initial=Yes,
/ConsideredUncertainOffset259/.initial=Yes,
/ConsideredUncertainOffset260/.initial=Yes,
/ConsideredUncertainOffset261/.initial=Yes,
/ConsideredUncertainOffset262/.initial=Yes,
/ConsideredUncertainOffset263/.initial=Yes,
/ConsideredUncertainOffset264/.initial=Yes,
/ConsideredUncertainOffset265/.initial=Yes,
/ConsideredUncertainOffset266/.initial=Yes,
/ConsideredUncertainOffset267/.initial=Yes,
/ConsideredUncertainOffset268/.initial=Yes,
/ConsideredUncertainOffset269/.initial=Yes,
/ConsideredUncertainOffset270/.initial=Yes,
/ConsideredUncertainOffset271/.initial=Yes,
/ConsideredUncertainOffset272/.initial=Yes,
/ConsideredUncertainOffset273/.initial=Yes,
/ConsideredUncertainOffset274/.initial=Yes,
/ConsideredUncertainOffset275/.initial=Yes,
/ConsideredUncertainOffset276/.initial=Yes,
/ConsideredUncertainOffset277/.initial=Yes,
/ConsideredUncertainOffset278/.initial=Yes,
/ConsideredUncertainOffset279/.initial=Yes,
/ConsideredUncertainOffset280/.initial=Yes,
/ConsideredUncertainOffset281/.initial=Yes,
/ConsideredUncertainOffset282/.initial=Yes,
/ConsideredUncertainOffset283/.initial=Yes,
/ConsideredUncertainOffset284/.initial=Yes,
/ConsideredUncertainOffset285/.initial=Yes,
/ConsideredUncertainOffset286/.initial=Yes,
/ConsideredUncertainOffset287/.initial=Yes,
/ConsideredUncertainOffset288/.initial=Yes,
/ConsideredUncertainOffset289/.initial=Yes,
/ConsideredUncertainOffset290/.initial=Yes,
/ConsideredUncertainOffset291/.initial=Yes,
/ConsideredUncertainOffset292/.initial=Yes,
/ConsideredUncertainOffset293/.initial=Yes,
/ConsideredUncertainOffset294/.initial=Yes,
/ConsideredUncertainOffset295/.initial=Yes,
/ConsideredUncertainOffset296/.initial=Yes,
/ConsideredUncertainOffset297/.initial=Yes,
/ConsideredUncertainOffset298/.initial=Yes,
/ConsideredUncertainOffset299/.initial=Yes,
/ConsideredUncertainOffset300/.initial=Yes,
/ConsideredUncertainOffset301/.initial=Yes,
/ConsideredUncertainOffset302/.initial=Yes,
/ConsideredUncertainOffset303/.initial=Yes,
/ConsideredUncertainOffset304/.initial=Yes,
/ConsideredUncertainOffset305/.initial=Yes,
/ConsideredUncertainOffset306/.initial=Yes,
/ConsideredUncertainOffset307/.initial=Yes,
/ConsideredUncertainOffset308/.initial=Yes,
/ConsideredUncertainOffset309/.initial=Yes,
/ConsideredUncertainOffset310/.initial=Yes,
/ConsideredUncertainOffset311/.initial=Yes,
/ConsideredUncertainOffset312/.initial=Yes,
/ConsideredUncertainOffset313/.initial=Yes,
/ConsideredUncertainOffset314/.initial=Yes,
/ConsideredUncertainOffset315/.initial=Yes,
/ConsideredUncertainOffset316/.initial=Yes,
/ConsideredUncertainOffset317/.initial=Yes,
/ConsideredUncertainOffset318/.initial=Yes,
/ConsideredUncertainOffset319/.initial=Yes,
/ConsideredUncertainOffset320/.initial=Yes,
/ConsideredUncertainOffset321/.initial=Yes,
/ConsideredUncertainOffset322/.initial=Yes,
/ConsideredUncertainOffset323/.initial=Yes,
/ConsideredUncertainOffset324/.initial=Yes,
/ConsideredUncertainOffset325/.initial=Yes,
/ConsideredUncertainOffset326/.initial=Yes,
/ConsideredUncertainOffset327/.initial=Yes,
/ConsideredUncertainOffset328/.initial=Yes,
/ConsideredUncertainOffset329/.initial=Yes,
/ConsideredUncertainOffset330/.initial=Yes,
/ConsideredUncertainOffset331/.initial=Yes,
/ConsideredUncertainOffset332/.initial=Yes,
/ConsideredUncertainOffset333/.initial=Yes,
/ConsideredUncertainOffset334/.initial=Yes,
/ConsideredUncertainOffset335/.initial=Yes,
/ConsideredUncertainOffset336/.initial=Yes,
/ConsideredUncertainOffset337/.initial=Yes,
/ConsideredUncertainOffset338/.initial=Yes,
/ConsideredUncertainOffset339/.initial=Yes,
/ConsideredUncertainOffset340/.initial=Yes,
/ConsideredUncertainOffset341/.initial=Yes,
/ConsideredUncertainOffset342/.initial=Yes,
/ConsideredUncertainOffset343/.initial=Yes,
/ConsideredUncertainOffset344/.initial=Yes,
/ConsideredUncertainOffset345/.initial=Yes,
/ConsideredUncertainOffset346/.initial=Yes,
/ConsideredUncertainOffset347/.initial=Yes,
/ConsideredUncertainOffset348/.initial=Yes,
/ConsideredUncertainOffset349/.initial=Yes,
/ConsideredUncertainOffset350/.initial=Yes,
/ConsideredUncertainOffset351/.initial=Yes,
/ConsideredUncertainOffset352/.initial=Yes,
/ConsideredUncertainOffset353/.initial=Yes,
/ConsideredUncertainOffset354/.initial=Yes,
/ConsideredUncertainOffset355/.initial=Yes,
/ConsideredUncertainOffset356/.initial=Yes,
/ConsideredUncertainOffset357/.initial=Yes,
/ConsideredUncertainOffset358/.initial=Yes,
/ConsideredUncertainOffset359/.initial=Yes,
/ConsideredUncertainOffset360/.initial=Yes,
/ConsideredUncertainOffset361/.initial=Yes,
/ConsideredUncertainOffset362/.initial=Yes,
/ConsideredUncertainOffset363/.initial=Yes,
/ConsideredUncertainOffset364/.initial=Yes,
/ConsideredUncertainOffset365/.initial=Yes,
/ConsideredUncertainOffset366/.initial=Yes,
/ConsideredUncertainOffset367/.initial=Yes,
/ConsideredUncertainOffset368/.initial=Yes,
/ConsideredUncertainOffset369/.initial=Yes,
/ConsideredUncertainOffset370/.initial=Yes,
/ConsideredUncertainOffset371/.initial=Yes,
/ConsideredUncertainOffset372/.initial=Yes,
/ConsideredUncertainOffset373/.initial=Yes,
/ConsideredUncertainOffset374/.initial=Yes,
/ConsideredUncertainOffset375/.initial=Yes,
/ConsideredUncertainOffset376/.initial=Yes,
/ConsideredUncertainOffset377/.initial=Yes,
/ConsideredUncertainOffset378/.initial=Yes,
/ConsideredUncertainOffset379/.initial=Yes,
/ConsideredUncertainOffset380/.initial=Yes,
/ConsideredUncertainOffset381/.initial=Yes,
/ConsideredUncertainOffset382/.initial=Yes,
/ConsideredUncertainOffset383/.initial=Yes,
/ConsideredUncertainOffset384/.initial=Yes,
/ConsideredUncertainOffset385/.initial=Yes,
/ConsideredUncertainOffset386/.initial=Yes,
/ConsideredUncertainOffset387/.initial=Yes,
/ConsideredUncertainOffset388/.initial=Yes,
/ConsideredUncertainOffset389/.initial=Yes,
/ConsideredUncertainOffset390/.initial=Yes,
/ConsideredUncertainOffset391/.initial=Yes,
/ConsideredUncertainOffset392/.initial=Yes,
/ConsideredUncertainOffset393/.initial=Yes,
/ConsideredUncertainOffset394/.initial=Yes,
/ConsideredUncertainOffset395/.initial=Yes,
/ConsideredUncertainOffset396/.initial=Yes,
/ConsideredUncertainOffset397/.initial=Yes,
/ConsideredUncertainOffset398/.initial=Yes,
/ConsideredUncertainOffset399/.initial=Yes,
/ConsideredUncertainOffset400/.initial=Yes,
/ConsideredUncertainOffset401/.initial=Yes,
/ConsideredUncertainOffset402/.initial=Yes,
/ConsideredUncertainOffset403/.initial=Yes,
/ConsideredUncertainOffset404/.initial=Yes,
/ConsideredUncertainOffset405/.initial=Yes,
/ConsideredUncertainOffset406/.initial=Yes,
/ConsideredUncertainOffset407/.initial=Yes,
/ConsideredUncertainOffset408/.initial=Yes,
/ConsideredUncertainOffset409/.initial=Yes,
/ConsideredUncertainOffset410/.initial=Yes,
/ConsideredUncertainOffset411/.initial=Yes,
/ConsideredUncertainOffset412/.initial=Yes,
/ConsideredUncertainOffset413/.initial=Yes,
/ConsideredUncertainOffset414/.initial=Yes,
/ConsideredUncertainOffset415/.initial=Yes,
/ConsideredUncertainOffset416/.initial=Yes,
/ConsideredUncertainOffset417/.initial=Yes,
/ConsideredUncertainOffset418/.initial=Yes,
/ConsideredUncertainOffset419/.initial=Yes,
/ConsideredUncertainOffset420/.initial=Yes,
/ConsideredUncertainOffset421/.initial=Yes,
/ConsideredUncertainOffset422/.initial=Yes,
/ConsideredUncertainOffset423/.initial=Yes,
/ConsideredUncertainOffset424/.initial=Yes,
/ConsideredUncertainOffset425/.initial=Yes,
/ConsideredUncertainOffset426/.initial=Yes,
/ConsideredUncertainOffset427/.initial=Yes,
/ConsideredUncertainOffset428/.initial=Yes,
/ConsideredUncertainOffset429/.initial=Yes,
/ConsideredUncertainOffset430/.initial=Yes,
/ConsideredUncertainOffset431/.initial=Yes,
/ConsideredUncertainOffset432/.initial=Yes,
/ConsideredUncertainOffset433/.initial=Yes,
/ConsideredUncertainOffset434/.initial=Yes,
/ConsideredUncertainOffset435/.initial=Yes,
/ConsideredUncertainOffset436/.initial=Yes,
/ConsideredUncertainOffset437/.initial=Yes,
/ConsideredUncertainOffset438/.initial=Yes,
/ConsideredUncertainOffset439/.initial=Yes,
/ConsideredUncertainOffset440/.initial=Yes,
/ConsideredUncertainOffset441/.initial=Yes,
/ConsideredUncertainOffset442/.initial=Yes,
/ConsideredUncertainOffset443/.initial=Yes,
/ConsideredUncertainOffset444/.initial=Yes,
/ConsideredUncertainOffset445/.initial=Yes,
/ConsideredUncertainOffset446/.initial=Yes,
/ConsideredUncertainOffset447/.initial=Yes,
/ConsideredUncertainOffset448/.initial=Yes,
/ConsideredUncertainOffset449/.initial=Yes,
/ConsideredUncertainOffset450/.initial=Yes,
/ConsideredUncertainOffset451/.initial=Yes,
/ConsideredUncertainOffset452/.initial=Yes,
/ConsideredUncertainOffset453/.initial=Yes,
/ConsideredUncertainOffset454/.initial=Yes,
/ConsideredUncertainOffset455/.initial=Yes,
/ConsideredUncertainOffset456/.initial=Yes,
/ConsideredUncertainOffset457/.initial=Yes,
/ConsideredUncertainOffset458/.initial=Yes,
/ConsideredUncertainOffset459/.initial=Yes,
/ConsideredUncertainOffset460/.initial=Yes,
/ConsideredUncertainOffset461/.initial=Yes,
/ConsideredUncertainOffset462/.initial=Yes,
/ConsideredUncertainOffset463/.initial=Yes,
/ConsideredUncertainOffset464/.initial=Yes,
/ConsideredUncertainOffset465/.initial=Yes,
/ConsideredUncertainOffset466/.initial=Yes,
/ConsideredUncertainOffset467/.initial=Yes,
/ConsideredUncertainOffset468/.initial=Yes,
/ConsideredUncertainOffset469/.initial=Yes,
/ConsideredUncertainOffset470/.initial=Yes,
/ConsideredUncertainOffset471/.initial=Yes,
/ConsideredUncertainOffset472/.initial=Yes,
/ConsideredUncertainOffset473/.initial=Yes,
/ConsideredUncertainOffset474/.initial=Yes,
/ConsideredUncertainOffset475/.initial=Yes,
/ConsideredUncertainOffset476/.initial=Yes,
/ConsideredUncertainOffset477/.initial=Yes,
/ConsideredUncertainOffset478/.initial=Yes,
/ConsideredUncertainOffset479/.initial=Yes,
/ConsideredUncertainOffset480/.initial=Yes,
/ConsideredUncertainOffset481/.initial=Yes,
/ConsideredUncertainOffset482/.initial=Yes,
/ConsideredUncertainOffset483/.initial=Yes,
/ConsideredUncertainOffset484/.initial=Yes,
/ConsideredUncertainOffset485/.initial=Yes,
/ConsideredUncertainOffset486/.initial=Yes,
/ConsideredUncertainOffset487/.initial=Yes,
/ConsideredUncertainOffset488/.initial=Yes,
/ConsideredUncertainOffset489/.initial=Yes,
/ConsideredUncertainOffset490/.initial=Yes,
/ConsideredUncertainOffset491/.initial=Yes,
/ConsideredUncertainOffset492/.initial=Yes,
/ConsideredUncertainOffset493/.initial=Yes,
/ConsideredUncertainOffset494/.initial=Yes,
/ConsideredUncertainOffset495/.initial=Yes,
/ConsideredUncertainOffset496/.initial=Yes,
/ConsideredUncertainOffset497/.initial=Yes,
/ConsideredUncertainOffset498/.initial=Yes,
/ConsideredUncertainOffset499/.initial=Yes,
/ConsideredUncertainOffset500/.initial=Yes,
/ConsideredUncertainOffset501/.initial=Yes,
/ConsideredUncertainOffset502/.initial=Yes,
/ConsideredUncertainOffset503/.initial=Yes,
/ConsideredUncertainOffset504/.initial=Yes,
/ConsideredUncertainOffset505/.initial=Yes,
/ConsideredUncertainOffset506/.initial=Yes,
/ConsideredUncertainOffset507/.initial=Yes,
/ConsideredUncertainOffset508/.initial=Yes,
/ConsideredUncertainOffset509/.initial=Yes,
/ConsideredUncertainOffset510/.initial=Yes,
/ConsideredUncertainOffset511/.initial=Yes,
/ConsideredUncertainOffset512/.initial=Yes,
/ConsideredUncertainOffset513/.initial=Yes,
/ConsideredUncertainOffset514/.initial=Yes,
/ConsideredUncertainOffset515/.initial=Yes,
/ConsideredUncertainOffset516/.initial=Yes,
/ConsideredUncertainOffset517/.initial=Yes,
/ConsideredUncertainOffset518/.initial=Yes,
/ConsideredUncertainOffset519/.initial=Yes,
/ConsideredUncertainOffset520/.initial=Yes,
/ConsideredUncertainOffset521/.initial=Yes,
/ConsideredUncertainOffset522/.initial=Yes,
/ConsideredUncertainOffset523/.initial=Yes,
/ConsideredUncertainOffset524/.initial=Yes,
/ConsideredUncertainOffset525/.initial=Yes,
/ConsideredUncertainOffset526/.initial=Yes,
/ConsideredUncertainOffset527/.initial=Yes,
/ConsideredUncertainOffset528/.initial=Yes,
/ConsideredUncertainOffset529/.initial=Yes,
/ConsideredUncertainOffset530/.initial=Yes,
/ConsideredUncertainOffset531/.initial=Yes,
/ConsideredUncertainOffset532/.initial=Yes,
/ConsideredUncertainOffset533/.initial=Yes,
/ConsideredUncertainOffset534/.initial=Yes,
/ConsideredUncertainOffset535/.initial=Yes,
/ConsideredUncertainOffset536/.initial=Yes,
/ConsideredUncertainOffset537/.initial=Yes,
/ConsideredUncertainOffset538/.initial=Yes,
/ConsideredUncertainOffset539/.initial=Yes,
/ConsideredUncertainOffset540/.initial=Yes,
/ConsideredUncertainOffset541/.initial=Yes,
/ConsideredUncertainOffset542/.initial=Yes,
/ConsideredUncertainOffset543/.initial=Yes,
/ConsideredUncertainOffset544/.initial=Yes,
/ConsideredUncertainOffset545/.initial=Yes,
/ConsideredUncertainOffset546/.initial=Yes,
/ConsideredUncertainOffset547/.initial=Yes,
/ConsideredUncertainOffset548/.initial=Yes,
/ConsideredUncertainOffset549/.initial=Yes,
/ConsideredUncertainOffset550/.initial=Yes,
/ConsideredUncertainOffset551/.initial=Yes,
/ConsideredUncertainOffset552/.initial=Yes,
/ConsideredUncertainOffset553/.initial=Yes,
/ConsideredUncertainOffset554/.initial=Yes,
/ConsideredUncertainOffset555/.initial=Yes,
/ConsideredUncertainOffset556/.initial=Yes,
/ConsideredUncertainOffset557/.initial=Yes,
/ConsideredUncertainOffset558/.initial=Yes,
/ConsideredUncertainOffset559/.initial=Yes,
/ConsideredUncertainOffset560/.initial=Yes,
/ConsideredUncertainOffset561/.initial=Yes,
/ConsideredUncertainOffset562/.initial=Yes,
/ConsideredUncertainOffset563/.initial=Yes,
/ConsideredUncertainOffset564/.initial=Yes,
/ConsideredUncertainOffset565/.initial=Yes,
/ConsideredUncertainOffset566/.initial=Yes,
/ConsideredUncertainOffset567/.initial=Yes,
/ConsideredUncertainOffset568/.initial=Yes,
/ConsideredUncertainOffset569/.initial=Yes,
/ConsideredUncertainOffset570/.initial=Yes,
/ConsideredUncertainOffset571/.initial=Yes,
/ConsideredUncertainOffset572/.initial=Yes,
/ConsideredUncertainOffset573/.initial=Yes,
/ConsideredUncertainOffset574/.initial=Yes,
/ConsideredUncertainOffset575/.initial=Yes,
/ConsideredUncertainOffset576/.initial=Yes,
/ConsideredUncertainOffset577/.initial=Yes,
/ConsideredUncertainOffset578/.initial=Yes,
/ConsideredUncertainOffset579/.initial=Yes,
/ConsideredUncertainOffset580/.initial=Yes,
/ConsideredUncertainOffset581/.initial=Yes,
/ConsideredUncertainOffset582/.initial=Yes,
/ConsideredUncertainOffset583/.initial=Yes,
/ConsideredUncertainOffset584/.initial=Yes,
/ConsideredUncertainOffset585/.initial=Yes,
/ConsideredUncertainOffset586/.initial=Yes,
/ConsideredUncertainOffset587/.initial=Yes,
/ConsideredUncertainOffset588/.initial=Yes,
/ConsideredUncertainOffset589/.initial=Yes,
/ConsideredUncertainOffset590/.initial=Yes,
/ConsideredUncertainOffset591/.initial=Yes,
/ConsideredUncertainOffset592/.initial=Yes,
/ConsideredUncertainOffset593/.initial=Yes,
/ConsideredUncertainOffset594/.initial=Yes,
/ConsideredUncertainOffset595/.initial=Yes,
/ConsideredUncertainOffset596/.initial=Yes,
/ConsideredUncertainOffset597/.initial=Yes,
/ConsideredUncertainOffset598/.initial=Yes,
/ConsideredUncertainOffset599/.initial=Yes,
/ConsideredUncertainOffset600/.initial=Yes,
/ConsideredUncertainOffset601/.initial=Yes,
/ConsideredUncertainOffset602/.initial=Yes,
/ConsideredUncertainOffset603/.initial=Yes,
/ConsideredUncertainOffset604/.initial=Yes,
/ConsideredUncertainOffset605/.initial=Yes,
/ConsideredUncertainOffset606/.initial=Yes,
/ConsideredUncertainOffset607/.initial=Yes,
/ConsideredUncertainOffset608/.initial=Yes,
/ConsideredUncertainOffset609/.initial=Yes,
/ConsideredUncertainOffset610/.initial=Yes,
/ConsideredUncertainOffset611/.initial=Yes,
/ConsideredUncertainOffset612/.initial=Yes,
/ConsideredUncertainOffset613/.initial=Yes,
/ConsideredUncertainOffset614/.initial=Yes,
/ConsideredUncertainOffset615/.initial=Yes,
/ConsideredUncertainOffset616/.initial=Yes,
/ConsideredUncertainOffset617/.initial=Yes,
/ConsideredUncertainOffset618/.initial=Yes,
/ConsideredUncertainOffset619/.initial=Yes,
/ConsideredUncertainOffset620/.initial=Yes,
/ConsideredUncertainOffset621/.initial=Yes,
/ConsideredUncertainOffset622/.initial=Yes,
/ConsideredUncertainOffset623/.initial=Yes,
/ConsideredUncertainOffset624/.initial=Yes,
/ConsideredUncertainOffset625/.initial=Yes,
/ConsideredUncertainOffset626/.initial=Yes,
/ConsideredUncertainOffset627/.initial=Yes,
/ConsideredUncertainOffset628/.initial=Yes,
/ConsideredUncertainOffset629/.initial=Yes,
/ConsideredUncertainOffset630/.initial=Yes,
/ConsideredUncertainOffset631/.initial=Yes,
/ConsideredUncertainOffset632/.initial=Yes,
/ConsideredUncertainOffset633/.initial=Yes,
/ConsideredUncertainOffset634/.initial=Yes,
/ConsideredUncertainOffset635/.initial=Yes,
/ConsideredUncertainOffset636/.initial=Yes,
/ConsideredUncertainOffset637/.initial=Yes,
/ConsideredUncertainOffset638/.initial=Yes,
/ConsideredUncertainOffset639/.initial=Yes,
/ConsideredUncertainOffset640/.initial=Yes,
/ConsideredUncertainOffset641/.initial=Yes,
/ConsideredUncertainOffset642/.initial=Yes,
/ConsideredUncertainOffset643/.initial=Yes,
/ConsideredUncertainOffset644/.initial=Yes,
/ConsideredUncertainOffset645/.initial=Yes,
/ConsideredUncertainOffset646/.initial=Yes,
/ConsideredUncertainOffset647/.initial=Yes,
/ConsideredUncertainOffset648/.initial=Yes,
/ConsideredUncertainOffset649/.initial=Yes,
/ConsideredUncertainOffset650/.initial=Yes,
/ConsideredUncertainOffset651/.initial=Yes,
/ConsideredUncertainOffset652/.initial=Yes,
/ConsideredUncertainOffset653/.initial=Yes,
/ConsideredUncertainOffset654/.initial=Yes,
/ConsideredUncertainOffset655/.initial=Yes,
/ConsideredUncertainOffset656/.initial=Yes,
/ConsideredUncertainOffset657/.initial=Yes,
/ConsideredUncertainOffset658/.initial=Yes,
/ConsideredUncertainOffset659/.initial=Yes,
/ConsideredUncertainOffset660/.initial=Yes,
/ConsideredUncertainOffset661/.initial=Yes,
/ConsideredUncertainOffset662/.initial=Yes,
/ConsideredUncertainOffset663/.initial=Yes,
/ConsideredUncertainOffset664/.initial=Yes,
/ConsideredUncertainOffset665/.initial=Yes,
/ConsideredUncertainOffset666/.initial=Yes,
/ConsideredUncertainOffset667/.initial=Yes,
/ConsideredUncertainOffset668/.initial=Yes,
/ConsideredUncertainOffset669/.initial=Yes,
/ConsideredUncertainOffset670/.initial=Yes,
/ConsideredUncertainOffset671/.initial=Yes,
/ConsideredUncertainOffset672/.initial=Yes,
/ConsideredUncertainOffset673/.initial=Yes,
/ConsideredUncertainOffset674/.initial=Yes,
/ConsideredUncertainOffset675/.initial=Yes,
/ConsideredUncertainOffset676/.initial=Yes,
/ConsideredUncertainOffset677/.initial=Yes,
/ConsideredUncertainOffset678/.initial=Yes,
/ConsideredUncertainOffset679/.initial=Yes,
/ConsideredUncertainOffset680/.initial=Yes,
/ConsideredUncertainOffset681/.initial=Yes,
/ConsideredUncertainOffset682/.initial=Yes,
/ConsideredUncertainOffset683/.initial=Yes,
/ConsideredUncertainOffset684/.initial=Yes,
/ConsideredUncertainOffset685/.initial=Yes,
/ConsideredUncertainOffset686/.initial=Yes,
/ConsideredUncertainOffset687/.initial=Yes,
/ConsideredUncertainOffset688/.initial=Yes,
/ConsideredUncertainOffset689/.initial=Yes,
/ConsideredUncertainOffset690/.initial=Yes,
/ConsideredUncertainOffset691/.initial=Yes,
/ConsideredUncertainOffset692/.initial=Yes,
/ConsideredUncertainOffset693/.initial=Yes,
/ConsideredUncertainOffset694/.initial=Yes,
/ConsideredUncertainOffset695/.initial=Yes,
/ConsideredUncertainOffset696/.initial=Yes,
/ConsideredUncertainOffset697/.initial=Yes,
/ConsideredUncertainOffset698/.initial=Yes,
/ConsideredUncertainOffset699/.initial=Yes,
/ConsideredUncertainOffset700/.initial=Yes,
/ConsideredUncertainOffset701/.initial=Yes,
/ConsideredUncertainOffset702/.initial=Yes,
/ConsideredUncertainOffset703/.initial=Yes,
/ConsideredUncertainOffset704/.initial=Yes,
/ConsideredUncertainOffset705/.initial=Yes,
/ConsideredUncertainOffset706/.initial=Yes,
/ConsideredUncertainOffset707/.initial=Yes,
/ConsideredUncertainOffset708/.initial=Yes,
/ConsideredUncertainOffset709/.initial=Yes,
/ConsideredUncertainOffset710/.initial=Yes,
/ConsideredUncertainOffset711/.initial=Yes,
/ConsideredUncertainOffset712/.initial=Yes,
/ConsideredUncertainOffset713/.initial=Yes,
/ConsideredUncertainOffset714/.initial=Yes,
/ConsideredUncertainOffset715/.initial=Yes,
/ConsideredUncertainOffset716/.initial=Yes,
/ConsideredUncertainOffset717/.initial=Yes,
/ConsideredUncertainOffset718/.initial=Yes,
/ConsideredUncertainOffset719/.initial=Yes,
/ConsideredUncertainOffset720/.initial=Yes,
/ConsideredUncertainOffset721/.initial=Yes,
/ConsideredUncertainOffset722/.initial=Yes,
/ConsideredUncertainOffset723/.initial=Yes,
/ConsideredUncertainOffset724/.initial=Yes,
/ConsideredUncertainOffset725/.initial=Yes,
/ConsideredUncertainOffset726/.initial=Yes,
/ConsideredUncertainOffset727/.initial=Yes,
/ConsideredUncertainOffset728/.initial=Yes,
/ConsideredUncertainOffset729/.initial=Yes,
/ConsideredUncertainOffset730/.initial=Yes,
/ConsideredUncertainOffset731/.initial=Yes,
/ConsideredUncertainOffset732/.initial=Yes,
/ConsideredUncertainOffset733/.initial=Yes,
/ConsideredUncertainOffset734/.initial=Yes,
/ConsideredUncertainOffset735/.initial=Yes,
/ConsideredUncertainOffset736/.initial=Yes,
/ConsideredUncertainOffset737/.initial=Yes,
/ConsideredUncertainOffset738/.initial=Yes,
/ConsideredUncertainOffset739/.initial=Yes,
/ConsideredUncertainOffset740/.initial=Yes,
/ConsideredUncertainOffset741/.initial=Yes,
/ConsideredUncertainOffset742/.initial=Yes,
/ConsideredUncertainOffset743/.initial=Yes,
/ConsideredUncertainOffset744/.initial=Yes,
/ConsideredUncertainOffset745/.initial=Yes,
/ConsideredUncertainOffset746/.initial=Yes,
/ConsideredUncertainOffset747/.initial=Yes,
/ConsideredUncertainOffset748/.initial=Yes,
/ConsideredUncertainOffset749/.initial=Yes,
/ConsideredUncertainOffset750/.initial=Yes,
/ConsideredUncertainOffset751/.initial=Yes,
/ConsideredUncertainOffset752/.initial=Yes,
/ConsideredUncertainOffset753/.initial=Yes,
/ConsideredUncertainOffset754/.initial=Yes,
/ConsideredUncertainOffset755/.initial=Yes,
/ConsideredUncertainOffset756/.initial=Yes,
/ConsideredUncertainOffset757/.initial=Yes,
/ConsideredUncertainOffset758/.initial=Yes,
/ConsideredUncertainOffset759/.initial=Yes,
/ConsideredUncertainOffset760/.initial=Yes,
/ConsideredUncertainOffset761/.initial=Yes,
/ConsideredUncertainOffset762/.initial=Yes,
/ConsideredUncertainOffset763/.initial=Yes,
/ConsideredUncertainOffset764/.initial=Yes,
/ConsideredUncertainOffset765/.initial=Yes,
/ConsideredUncertainOffset766/.initial=Yes,
/ConsideredUncertainOffset767/.initial=Yes,
/ConsideredUncertainKtaRoCo/.initial=Yes,
/ConsideredUncertainKtaReCo/.initial=Yes,
/ConsideredUncertainKtaRoCe/.initial=Yes,
/ConsideredUncertainKtaReCe/.initial=Yes,
/ConsideredUncertainKtaScale2/.initial=No,
/ConsideredUncertainKtaTemp0/.initial=Yes,
/ConsideredUncertainKtaTemp1/.initial=Yes,
/ConsideredUncertainKtaTemp2/.initial=Yes,
/ConsideredUncertainKtaTemp3/.initial=Yes,
/ConsideredUncertainKtaTemp4/.initial=Yes,
/ConsideredUncertainKtaTemp5/.initial=Yes,
/ConsideredUncertainKtaTemp6/.initial=Yes,
/ConsideredUncertainKtaTemp7/.initial=Yes,
/ConsideredUncertainKtaTemp8/.initial=Yes,
/ConsideredUncertainKtaTemp9/.initial=Yes,
/ConsideredUncertainKtaTemp10/.initial=Yes,
/ConsideredUncertainKtaTemp11/.initial=Yes,
/ConsideredUncertainKtaTemp12/.initial=Yes,
/ConsideredUncertainKtaTemp13/.initial=Yes,
/ConsideredUncertainKtaTemp14/.initial=Yes,
/ConsideredUncertainKtaTemp15/.initial=Yes,
/ConsideredUncertainKtaTemp16/.initial=Yes,
/ConsideredUncertainKtaTemp17/.initial=Yes,
/ConsideredUncertainKtaTemp18/.initial=Yes,
/ConsideredUncertainKtaTemp19/.initial=Yes,
/ConsideredUncertainKtaTemp20/.initial=Yes,
/ConsideredUncertainKtaTemp21/.initial=Yes,
/ConsideredUncertainKtaTemp22/.initial=Yes,
/ConsideredUncertainKtaTemp23/.initial=Yes,
/ConsideredUncertainKtaTemp24/.initial=Yes,
/ConsideredUncertainKtaTemp25/.initial=Yes,
/ConsideredUncertainKtaTemp26/.initial=Yes,
/ConsideredUncertainKtaTemp27/.initial=Yes,
/ConsideredUncertainKtaTemp28/.initial=Yes,
/ConsideredUncertainKtaTemp29/.initial=Yes,
/ConsideredUncertainKtaTemp30/.initial=Yes,
/ConsideredUncertainKtaTemp31/.initial=Yes,
/ConsideredUncertainKtaTemp32/.initial=Yes,
/ConsideredUncertainKtaTemp33/.initial=Yes,
/ConsideredUncertainKtaTemp34/.initial=Yes,
/ConsideredUncertainKtaTemp35/.initial=Yes,
/ConsideredUncertainKtaTemp36/.initial=Yes,
/ConsideredUncertainKtaTemp37/.initial=Yes,
/ConsideredUncertainKtaTemp38/.initial=Yes,
/ConsideredUncertainKtaTemp39/.initial=Yes,
/ConsideredUncertainKtaTemp40/.initial=Yes,
/ConsideredUncertainKtaTemp41/.initial=Yes,
/ConsideredUncertainKtaTemp42/.initial=Yes,
/ConsideredUncertainKtaTemp43/.initial=Yes,
/ConsideredUncertainKtaTemp44/.initial=Yes,
/ConsideredUncertainKtaTemp45/.initial=Yes,
/ConsideredUncertainKtaTemp46/.initial=Yes,
/ConsideredUncertainKtaTemp47/.initial=Yes,
/ConsideredUncertainKtaTemp48/.initial=Yes,
/ConsideredUncertainKtaTemp49/.initial=Yes,
/ConsideredUncertainKtaTemp50/.initial=Yes,
/ConsideredUncertainKtaTemp51/.initial=Yes,
/ConsideredUncertainKtaTemp52/.initial=Yes,
/ConsideredUncertainKtaTemp53/.initial=Yes,
/ConsideredUncertainKtaTemp54/.initial=Yes,
/ConsideredUncertainKtaTemp55/.initial=Yes,
/ConsideredUncertainKtaTemp56/.initial=Yes,
/ConsideredUncertainKtaTemp57/.initial=Yes,
/ConsideredUncertainKtaTemp58/.initial=Yes,
/ConsideredUncertainKtaTemp59/.initial=Yes,
/ConsideredUncertainKtaTemp60/.initial=Yes,
/ConsideredUncertainKtaTemp61/.initial=Yes,
/ConsideredUncertainKtaTemp62/.initial=Yes,
/ConsideredUncertainKtaTemp63/.initial=Yes,
/ConsideredUncertainKtaTemp64/.initial=Yes,
/ConsideredUncertainKtaTemp65/.initial=Yes,
/ConsideredUncertainKtaTemp66/.initial=Yes,
/ConsideredUncertainKtaTemp67/.initial=Yes,
/ConsideredUncertainKtaTemp68/.initial=Yes,
/ConsideredUncertainKtaTemp69/.initial=Yes,
/ConsideredUncertainKtaTemp70/.initial=Yes,
/ConsideredUncertainKtaTemp71/.initial=Yes,
/ConsideredUncertainKtaTemp72/.initial=Yes,
/ConsideredUncertainKtaTemp73/.initial=Yes,
/ConsideredUncertainKtaTemp74/.initial=Yes,
/ConsideredUncertainKtaTemp75/.initial=Yes,
/ConsideredUncertainKtaTemp76/.initial=Yes,
/ConsideredUncertainKtaTemp77/.initial=Yes,
/ConsideredUncertainKtaTemp78/.initial=Yes,
/ConsideredUncertainKtaTemp79/.initial=Yes,
/ConsideredUncertainKtaTemp80/.initial=Yes,
/ConsideredUncertainKtaTemp81/.initial=Yes,
/ConsideredUncertainKtaTemp82/.initial=Yes,
/ConsideredUncertainKtaTemp83/.initial=Yes,
/ConsideredUncertainKtaTemp84/.initial=Yes,
/ConsideredUncertainKtaTemp85/.initial=Yes,
/ConsideredUncertainKtaTemp86/.initial=Yes,
/ConsideredUncertainKtaTemp87/.initial=Yes,
/ConsideredUncertainKtaTemp88/.initial=Yes,
/ConsideredUncertainKtaTemp89/.initial=Yes,
/ConsideredUncertainKtaTemp90/.initial=Yes,
/ConsideredUncertainKtaTemp91/.initial=Yes,
/ConsideredUncertainKtaTemp92/.initial=Yes,
/ConsideredUncertainKtaTemp93/.initial=Yes,
/ConsideredUncertainKtaTemp94/.initial=Yes,
/ConsideredUncertainKtaTemp95/.initial=Yes,
/ConsideredUncertainKtaTemp96/.initial=Yes,
/ConsideredUncertainKtaTemp97/.initial=Yes,
/ConsideredUncertainKtaTemp98/.initial=Yes,
/ConsideredUncertainKtaTemp99/.initial=Yes,
/ConsideredUncertainKtaTemp100/.initial=Yes,
/ConsideredUncertainKtaTemp101/.initial=Yes,
/ConsideredUncertainKtaTemp102/.initial=Yes,
/ConsideredUncertainKtaTemp103/.initial=Yes,
/ConsideredUncertainKtaTemp104/.initial=Yes,
/ConsideredUncertainKtaTemp105/.initial=Yes,
/ConsideredUncertainKtaTemp106/.initial=Yes,
/ConsideredUncertainKtaTemp107/.initial=Yes,
/ConsideredUncertainKtaTemp108/.initial=Yes,
/ConsideredUncertainKtaTemp109/.initial=Yes,
/ConsideredUncertainKtaTemp110/.initial=Yes,
/ConsideredUncertainKtaTemp111/.initial=Yes,
/ConsideredUncertainKtaTemp112/.initial=Yes,
/ConsideredUncertainKtaTemp113/.initial=Yes,
/ConsideredUncertainKtaTemp114/.initial=Yes,
/ConsideredUncertainKtaTemp115/.initial=Yes,
/ConsideredUncertainKtaTemp116/.initial=Yes,
/ConsideredUncertainKtaTemp117/.initial=Yes,
/ConsideredUncertainKtaTemp118/.initial=Yes,
/ConsideredUncertainKtaTemp119/.initial=Yes,
/ConsideredUncertainKtaTemp120/.initial=Yes,
/ConsideredUncertainKtaTemp121/.initial=Yes,
/ConsideredUncertainKtaTemp122/.initial=Yes,
/ConsideredUncertainKtaTemp123/.initial=Yes,
/ConsideredUncertainKtaTemp124/.initial=Yes,
/ConsideredUncertainKtaTemp125/.initial=Yes,
/ConsideredUncertainKtaTemp126/.initial=Yes,
/ConsideredUncertainKtaTemp127/.initial=Yes,
/ConsideredUncertainKtaTemp128/.initial=Yes,
/ConsideredUncertainKtaTemp129/.initial=Yes,
/ConsideredUncertainKtaTemp130/.initial=Yes,
/ConsideredUncertainKtaTemp131/.initial=Yes,
/ConsideredUncertainKtaTemp132/.initial=Yes,
/ConsideredUncertainKtaTemp133/.initial=Yes,
/ConsideredUncertainKtaTemp134/.initial=Yes,
/ConsideredUncertainKtaTemp135/.initial=Yes,
/ConsideredUncertainKtaTemp136/.initial=Yes,
/ConsideredUncertainKtaTemp137/.initial=Yes,
/ConsideredUncertainKtaTemp138/.initial=Yes,
/ConsideredUncertainKtaTemp139/.initial=Yes,
/ConsideredUncertainKtaTemp140/.initial=Yes,
/ConsideredUncertainKtaTemp141/.initial=Yes,
/ConsideredUncertainKtaTemp142/.initial=Yes,
/ConsideredUncertainKtaTemp143/.initial=Yes,
/ConsideredUncertainKtaTemp144/.initial=Yes,
/ConsideredUncertainKtaTemp145/.initial=Yes,
/ConsideredUncertainKtaTemp146/.initial=Yes,
/ConsideredUncertainKtaTemp147/.initial=Yes,
/ConsideredUncertainKtaTemp148/.initial=Yes,
/ConsideredUncertainKtaTemp149/.initial=Yes,
/ConsideredUncertainKtaTemp150/.initial=Yes,
/ConsideredUncertainKtaTemp151/.initial=Yes,
/ConsideredUncertainKtaTemp152/.initial=Yes,
/ConsideredUncertainKtaTemp153/.initial=Yes,
/ConsideredUncertainKtaTemp154/.initial=Yes,
/ConsideredUncertainKtaTemp155/.initial=Yes,
/ConsideredUncertainKtaTemp156/.initial=Yes,
/ConsideredUncertainKtaTemp157/.initial=Yes,
/ConsideredUncertainKtaTemp158/.initial=Yes,
/ConsideredUncertainKtaTemp159/.initial=Yes,
/ConsideredUncertainKtaTemp160/.initial=Yes,
/ConsideredUncertainKtaTemp161/.initial=Yes,
/ConsideredUncertainKtaTemp162/.initial=Yes,
/ConsideredUncertainKtaTemp163/.initial=Yes,
/ConsideredUncertainKtaTemp164/.initial=Yes,
/ConsideredUncertainKtaTemp165/.initial=Yes,
/ConsideredUncertainKtaTemp166/.initial=Yes,
/ConsideredUncertainKtaTemp167/.initial=Yes,
/ConsideredUncertainKtaTemp168/.initial=Yes,
/ConsideredUncertainKtaTemp169/.initial=Yes,
/ConsideredUncertainKtaTemp170/.initial=Yes,
/ConsideredUncertainKtaTemp171/.initial=Yes,
/ConsideredUncertainKtaTemp172/.initial=Yes,
/ConsideredUncertainKtaTemp173/.initial=Yes,
/ConsideredUncertainKtaTemp174/.initial=Yes,
/ConsideredUncertainKtaTemp175/.initial=Yes,
/ConsideredUncertainKtaTemp176/.initial=Yes,
/ConsideredUncertainKtaTemp177/.initial=Yes,
/ConsideredUncertainKtaTemp178/.initial=Yes,
/ConsideredUncertainKtaTemp179/.initial=Yes,
/ConsideredUncertainKtaTemp180/.initial=Yes,
/ConsideredUncertainKtaTemp181/.initial=Yes,
/ConsideredUncertainKtaTemp182/.initial=Yes,
/ConsideredUncertainKtaTemp183/.initial=Yes,
/ConsideredUncertainKtaTemp184/.initial=Yes,
/ConsideredUncertainKtaTemp185/.initial=Yes,
/ConsideredUncertainKtaTemp186/.initial=Yes,
/ConsideredUncertainKtaTemp187/.initial=Yes,
/ConsideredUncertainKtaTemp188/.initial=Yes,
/ConsideredUncertainKtaTemp189/.initial=Yes,
/ConsideredUncertainKtaTemp190/.initial=Yes,
/ConsideredUncertainKtaTemp191/.initial=Yes,
/ConsideredUncertainKtaTemp192/.initial=Yes,
/ConsideredUncertainKtaTemp193/.initial=Yes,
/ConsideredUncertainKtaTemp194/.initial=Yes,
/ConsideredUncertainKtaTemp195/.initial=Yes,
/ConsideredUncertainKtaTemp196/.initial=Yes,
/ConsideredUncertainKtaTemp197/.initial=Yes,
/ConsideredUncertainKtaTemp198/.initial=Yes,
/ConsideredUncertainKtaTemp199/.initial=Yes,
/ConsideredUncertainKtaTemp200/.initial=Yes,
/ConsideredUncertainKtaTemp201/.initial=Yes,
/ConsideredUncertainKtaTemp202/.initial=Yes,
/ConsideredUncertainKtaTemp203/.initial=Yes,
/ConsideredUncertainKtaTemp204/.initial=Yes,
/ConsideredUncertainKtaTemp205/.initial=Yes,
/ConsideredUncertainKtaTemp206/.initial=Yes,
/ConsideredUncertainKtaTemp207/.initial=Yes,
/ConsideredUncertainKtaTemp208/.initial=Yes,
/ConsideredUncertainKtaTemp209/.initial=Yes,
/ConsideredUncertainKtaTemp210/.initial=Yes,
/ConsideredUncertainKtaTemp211/.initial=Yes,
/ConsideredUncertainKtaTemp212/.initial=Yes,
/ConsideredUncertainKtaTemp213/.initial=Yes,
/ConsideredUncertainKtaTemp214/.initial=Yes,
/ConsideredUncertainKtaTemp215/.initial=Yes,
/ConsideredUncertainKtaTemp216/.initial=Yes,
/ConsideredUncertainKtaTemp217/.initial=Yes,
/ConsideredUncertainKtaTemp218/.initial=Yes,
/ConsideredUncertainKtaTemp219/.initial=Yes,
/ConsideredUncertainKtaTemp220/.initial=Yes,
/ConsideredUncertainKtaTemp221/.initial=Yes,
/ConsideredUncertainKtaTemp222/.initial=Yes,
/ConsideredUncertainKtaTemp223/.initial=Yes,
/ConsideredUncertainKtaTemp224/.initial=Yes,
/ConsideredUncertainKtaTemp225/.initial=Yes,
/ConsideredUncertainKtaTemp226/.initial=Yes,
/ConsideredUncertainKtaTemp227/.initial=Yes,
/ConsideredUncertainKtaTemp228/.initial=Yes,
/ConsideredUncertainKtaTemp229/.initial=Yes,
/ConsideredUncertainKtaTemp230/.initial=Yes,
/ConsideredUncertainKtaTemp231/.initial=Yes,
/ConsideredUncertainKtaTemp232/.initial=Yes,
/ConsideredUncertainKtaTemp233/.initial=Yes,
/ConsideredUncertainKtaTemp234/.initial=Yes,
/ConsideredUncertainKtaTemp235/.initial=Yes,
/ConsideredUncertainKtaTemp236/.initial=Yes,
/ConsideredUncertainKtaTemp237/.initial=Yes,
/ConsideredUncertainKtaTemp238/.initial=Yes,
/ConsideredUncertainKtaTemp239/.initial=Yes,
/ConsideredUncertainKtaTemp240/.initial=Yes,
/ConsideredUncertainKtaTemp241/.initial=Yes,
/ConsideredUncertainKtaTemp242/.initial=Yes,
/ConsideredUncertainKtaTemp243/.initial=Yes,
/ConsideredUncertainKtaTemp244/.initial=Yes,
/ConsideredUncertainKtaTemp245/.initial=Yes,
/ConsideredUncertainKtaTemp246/.initial=Yes,
/ConsideredUncertainKtaTemp247/.initial=Yes,
/ConsideredUncertainKtaTemp248/.initial=Yes,
/ConsideredUncertainKtaTemp249/.initial=Yes,
/ConsideredUncertainKtaTemp250/.initial=Yes,
/ConsideredUncertainKtaTemp251/.initial=Yes,
/ConsideredUncertainKtaTemp252/.initial=Yes,
/ConsideredUncertainKtaTemp253/.initial=Yes,
/ConsideredUncertainKtaTemp254/.initial=Yes,
/ConsideredUncertainKtaTemp255/.initial=Yes,
/ConsideredUncertainKtaTemp256/.initial=Yes,
/ConsideredUncertainKtaTemp257/.initial=Yes,
/ConsideredUncertainKtaTemp258/.initial=Yes,
/ConsideredUncertainKtaTemp259/.initial=Yes,
/ConsideredUncertainKtaTemp260/.initial=Yes,
/ConsideredUncertainKtaTemp261/.initial=Yes,
/ConsideredUncertainKtaTemp262/.initial=Yes,
/ConsideredUncertainKtaTemp263/.initial=Yes,
/ConsideredUncertainKtaTemp264/.initial=Yes,
/ConsideredUncertainKtaTemp265/.initial=Yes,
/ConsideredUncertainKtaTemp266/.initial=Yes,
/ConsideredUncertainKtaTemp267/.initial=Yes,
/ConsideredUncertainKtaTemp268/.initial=Yes,
/ConsideredUncertainKtaTemp269/.initial=Yes,
/ConsideredUncertainKtaTemp270/.initial=Yes,
/ConsideredUncertainKtaTemp271/.initial=Yes,
/ConsideredUncertainKtaTemp272/.initial=Yes,
/ConsideredUncertainKtaTemp273/.initial=Yes,
/ConsideredUncertainKtaTemp274/.initial=Yes,
/ConsideredUncertainKtaTemp275/.initial=Yes,
/ConsideredUncertainKtaTemp276/.initial=Yes,
/ConsideredUncertainKtaTemp277/.initial=Yes,
/ConsideredUncertainKtaTemp278/.initial=Yes,
/ConsideredUncertainKtaTemp279/.initial=Yes,
/ConsideredUncertainKtaTemp280/.initial=Yes,
/ConsideredUncertainKtaTemp281/.initial=Yes,
/ConsideredUncertainKtaTemp282/.initial=Yes,
/ConsideredUncertainKtaTemp283/.initial=Yes,
/ConsideredUncertainKtaTemp284/.initial=Yes,
/ConsideredUncertainKtaTemp285/.initial=Yes,
/ConsideredUncertainKtaTemp286/.initial=Yes,
/ConsideredUncertainKtaTemp287/.initial=Yes,
/ConsideredUncertainKtaTemp288/.initial=Yes,
/ConsideredUncertainKtaTemp289/.initial=Yes,
/ConsideredUncertainKtaTemp290/.initial=Yes,
/ConsideredUncertainKtaTemp291/.initial=Yes,
/ConsideredUncertainKtaTemp292/.initial=Yes,
/ConsideredUncertainKtaTemp293/.initial=Yes,
/ConsideredUncertainKtaTemp294/.initial=Yes,
/ConsideredUncertainKtaTemp295/.initial=Yes,
/ConsideredUncertainKtaTemp296/.initial=Yes,
/ConsideredUncertainKtaTemp297/.initial=Yes,
/ConsideredUncertainKtaTemp298/.initial=Yes,
/ConsideredUncertainKtaTemp299/.initial=Yes,
/ConsideredUncertainKtaTemp300/.initial=Yes,
/ConsideredUncertainKtaTemp301/.initial=Yes,
/ConsideredUncertainKtaTemp302/.initial=Yes,
/ConsideredUncertainKtaTemp303/.initial=Yes,
/ConsideredUncertainKtaTemp304/.initial=Yes,
/ConsideredUncertainKtaTemp305/.initial=Yes,
/ConsideredUncertainKtaTemp306/.initial=Yes,
/ConsideredUncertainKtaTemp307/.initial=Yes,
/ConsideredUncertainKtaTemp308/.initial=Yes,
/ConsideredUncertainKtaTemp309/.initial=Yes,
/ConsideredUncertainKtaTemp310/.initial=Yes,
/ConsideredUncertainKtaTemp311/.initial=Yes,
/ConsideredUncertainKtaTemp312/.initial=Yes,
/ConsideredUncertainKtaTemp313/.initial=Yes,
/ConsideredUncertainKtaTemp314/.initial=Yes,
/ConsideredUncertainKtaTemp315/.initial=Yes,
/ConsideredUncertainKtaTemp316/.initial=Yes,
/ConsideredUncertainKtaTemp317/.initial=Yes,
/ConsideredUncertainKtaTemp318/.initial=Yes,
/ConsideredUncertainKtaTemp319/.initial=Yes,
/ConsideredUncertainKtaTemp320/.initial=Yes,
/ConsideredUncertainKtaTemp321/.initial=Yes,
/ConsideredUncertainKtaTemp322/.initial=Yes,
/ConsideredUncertainKtaTemp323/.initial=Yes,
/ConsideredUncertainKtaTemp324/.initial=Yes,
/ConsideredUncertainKtaTemp325/.initial=Yes,
/ConsideredUncertainKtaTemp326/.initial=Yes,
/ConsideredUncertainKtaTemp327/.initial=Yes,
/ConsideredUncertainKtaTemp328/.initial=Yes,
/ConsideredUncertainKtaTemp329/.initial=Yes,
/ConsideredUncertainKtaTemp330/.initial=Yes,
/ConsideredUncertainKtaTemp331/.initial=Yes,
/ConsideredUncertainKtaTemp332/.initial=Yes,
/ConsideredUncertainKtaTemp333/.initial=Yes,
/ConsideredUncertainKtaTemp334/.initial=Yes,
/ConsideredUncertainKtaTemp335/.initial=Yes,
/ConsideredUncertainKtaTemp336/.initial=Yes,
/ConsideredUncertainKtaTemp337/.initial=Yes,
/ConsideredUncertainKtaTemp338/.initial=Yes,
/ConsideredUncertainKtaTemp339/.initial=Yes,
/ConsideredUncertainKtaTemp340/.initial=Yes,
/ConsideredUncertainKtaTemp341/.initial=Yes,
/ConsideredUncertainKtaTemp342/.initial=Yes,
/ConsideredUncertainKtaTemp343/.initial=Yes,
/ConsideredUncertainKtaTemp344/.initial=Yes,
/ConsideredUncertainKtaTemp345/.initial=Yes,
/ConsideredUncertainKtaTemp346/.initial=Yes,
/ConsideredUncertainKtaTemp347/.initial=Yes,
/ConsideredUncertainKtaTemp348/.initial=Yes,
/ConsideredUncertainKtaTemp349/.initial=Yes,
/ConsideredUncertainKtaTemp350/.initial=Yes,
/ConsideredUncertainKtaTemp351/.initial=Yes,
/ConsideredUncertainKtaTemp352/.initial=Yes,
/ConsideredUncertainKtaTemp353/.initial=Yes,
/ConsideredUncertainKtaTemp354/.initial=Yes,
/ConsideredUncertainKtaTemp355/.initial=Yes,
/ConsideredUncertainKtaTemp356/.initial=Yes,
/ConsideredUncertainKtaTemp357/.initial=Yes,
/ConsideredUncertainKtaTemp358/.initial=Yes,
/ConsideredUncertainKtaTemp359/.initial=Yes,
/ConsideredUncertainKtaTemp360/.initial=Yes,
/ConsideredUncertainKtaTemp361/.initial=Yes,
/ConsideredUncertainKtaTemp362/.initial=Yes,
/ConsideredUncertainKtaTemp363/.initial=Yes,
/ConsideredUncertainKtaTemp364/.initial=Yes,
/ConsideredUncertainKtaTemp365/.initial=Yes,
/ConsideredUncertainKtaTemp366/.initial=Yes,
/ConsideredUncertainKtaTemp367/.initial=Yes,
/ConsideredUncertainKtaTemp368/.initial=Yes,
/ConsideredUncertainKtaTemp369/.initial=Yes,
/ConsideredUncertainKtaTemp370/.initial=Yes,
/ConsideredUncertainKtaTemp371/.initial=Yes,
/ConsideredUncertainKtaTemp372/.initial=Yes,
/ConsideredUncertainKtaTemp373/.initial=Yes,
/ConsideredUncertainKtaTemp374/.initial=Yes,
/ConsideredUncertainKtaTemp375/.initial=Yes,
/ConsideredUncertainKtaTemp376/.initial=Yes,
/ConsideredUncertainKtaTemp377/.initial=Yes,
/ConsideredUncertainKtaTemp378/.initial=Yes,
/ConsideredUncertainKtaTemp379/.initial=Yes,
/ConsideredUncertainKtaTemp380/.initial=Yes,
/ConsideredUncertainKtaTemp381/.initial=Yes,
/ConsideredUncertainKtaTemp382/.initial=Yes,
/ConsideredUncertainKtaTemp383/.initial=Yes,
/ConsideredUncertainKtaTemp384/.initial=Yes,
/ConsideredUncertainKtaTemp385/.initial=Yes,
/ConsideredUncertainKtaTemp386/.initial=Yes,
/ConsideredUncertainKtaTemp387/.initial=Yes,
/ConsideredUncertainKtaTemp388/.initial=Yes,
/ConsideredUncertainKtaTemp389/.initial=Yes,
/ConsideredUncertainKtaTemp390/.initial=Yes,
/ConsideredUncertainKtaTemp391/.initial=Yes,
/ConsideredUncertainKtaTemp392/.initial=Yes,
/ConsideredUncertainKtaTemp393/.initial=Yes,
/ConsideredUncertainKtaTemp394/.initial=Yes,
/ConsideredUncertainKtaTemp395/.initial=Yes,
/ConsideredUncertainKtaTemp396/.initial=Yes,
/ConsideredUncertainKtaTemp397/.initial=Yes,
/ConsideredUncertainKtaTemp398/.initial=Yes,
/ConsideredUncertainKtaTemp399/.initial=Yes,
/ConsideredUncertainKtaTemp400/.initial=Yes,
/ConsideredUncertainKtaTemp401/.initial=Yes,
/ConsideredUncertainKtaTemp402/.initial=Yes,
/ConsideredUncertainKtaTemp403/.initial=Yes,
/ConsideredUncertainKtaTemp404/.initial=Yes,
/ConsideredUncertainKtaTemp405/.initial=Yes,
/ConsideredUncertainKtaTemp406/.initial=Yes,
/ConsideredUncertainKtaTemp407/.initial=Yes,
/ConsideredUncertainKtaTemp408/.initial=Yes,
/ConsideredUncertainKtaTemp409/.initial=Yes,
/ConsideredUncertainKtaTemp410/.initial=Yes,
/ConsideredUncertainKtaTemp411/.initial=Yes,
/ConsideredUncertainKtaTemp412/.initial=Yes,
/ConsideredUncertainKtaTemp413/.initial=Yes,
/ConsideredUncertainKtaTemp414/.initial=Yes,
/ConsideredUncertainKtaTemp415/.initial=Yes,
/ConsideredUncertainKtaTemp416/.initial=Yes,
/ConsideredUncertainKtaTemp417/.initial=Yes,
/ConsideredUncertainKtaTemp418/.initial=Yes,
/ConsideredUncertainKtaTemp419/.initial=Yes,
/ConsideredUncertainKtaTemp420/.initial=Yes,
/ConsideredUncertainKtaTemp421/.initial=Yes,
/ConsideredUncertainKtaTemp422/.initial=Yes,
/ConsideredUncertainKtaTemp423/.initial=Yes,
/ConsideredUncertainKtaTemp424/.initial=Yes,
/ConsideredUncertainKtaTemp425/.initial=Yes,
/ConsideredUncertainKtaTemp426/.initial=Yes,
/ConsideredUncertainKtaTemp427/.initial=Yes,
/ConsideredUncertainKtaTemp428/.initial=Yes,
/ConsideredUncertainKtaTemp429/.initial=Yes,
/ConsideredUncertainKtaTemp430/.initial=Yes,
/ConsideredUncertainKtaTemp431/.initial=Yes,
/ConsideredUncertainKtaTemp432/.initial=Yes,
/ConsideredUncertainKtaTemp433/.initial=Yes,
/ConsideredUncertainKtaTemp434/.initial=Yes,
/ConsideredUncertainKtaTemp435/.initial=Yes,
/ConsideredUncertainKtaTemp436/.initial=Yes,
/ConsideredUncertainKtaTemp437/.initial=Yes,
/ConsideredUncertainKtaTemp438/.initial=Yes,
/ConsideredUncertainKtaTemp439/.initial=Yes,
/ConsideredUncertainKtaTemp440/.initial=Yes,
/ConsideredUncertainKtaTemp441/.initial=Yes,
/ConsideredUncertainKtaTemp442/.initial=Yes,
/ConsideredUncertainKtaTemp443/.initial=Yes,
/ConsideredUncertainKtaTemp444/.initial=Yes,
/ConsideredUncertainKtaTemp445/.initial=Yes,
/ConsideredUncertainKtaTemp446/.initial=Yes,
/ConsideredUncertainKtaTemp447/.initial=Yes,
/ConsideredUncertainKtaTemp448/.initial=Yes,
/ConsideredUncertainKtaTemp449/.initial=Yes,
/ConsideredUncertainKtaTemp450/.initial=Yes,
/ConsideredUncertainKtaTemp451/.initial=Yes,
/ConsideredUncertainKtaTemp452/.initial=Yes,
/ConsideredUncertainKtaTemp453/.initial=Yes,
/ConsideredUncertainKtaTemp454/.initial=Yes,
/ConsideredUncertainKtaTemp455/.initial=Yes,
/ConsideredUncertainKtaTemp456/.initial=Yes,
/ConsideredUncertainKtaTemp457/.initial=Yes,
/ConsideredUncertainKtaTemp458/.initial=Yes,
/ConsideredUncertainKtaTemp459/.initial=Yes,
/ConsideredUncertainKtaTemp460/.initial=Yes,
/ConsideredUncertainKtaTemp461/.initial=Yes,
/ConsideredUncertainKtaTemp462/.initial=Yes,
/ConsideredUncertainKtaTemp463/.initial=Yes,
/ConsideredUncertainKtaTemp464/.initial=Yes,
/ConsideredUncertainKtaTemp465/.initial=Yes,
/ConsideredUncertainKtaTemp466/.initial=Yes,
/ConsideredUncertainKtaTemp467/.initial=Yes,
/ConsideredUncertainKtaTemp468/.initial=Yes,
/ConsideredUncertainKtaTemp469/.initial=Yes,
/ConsideredUncertainKtaTemp470/.initial=Yes,
/ConsideredUncertainKtaTemp471/.initial=Yes,
/ConsideredUncertainKtaTemp472/.initial=Yes,
/ConsideredUncertainKtaTemp473/.initial=Yes,
/ConsideredUncertainKtaTemp474/.initial=Yes,
/ConsideredUncertainKtaTemp475/.initial=Yes,
/ConsideredUncertainKtaTemp476/.initial=Yes,
/ConsideredUncertainKtaTemp477/.initial=Yes,
/ConsideredUncertainKtaTemp478/.initial=Yes,
/ConsideredUncertainKtaTemp479/.initial=Yes,
/ConsideredUncertainKtaTemp480/.initial=Yes,
/ConsideredUncertainKtaTemp481/.initial=Yes,
/ConsideredUncertainKtaTemp482/.initial=Yes,
/ConsideredUncertainKtaTemp483/.initial=Yes,
/ConsideredUncertainKtaTemp484/.initial=Yes,
/ConsideredUncertainKtaTemp485/.initial=Yes,
/ConsideredUncertainKtaTemp486/.initial=Yes,
/ConsideredUncertainKtaTemp487/.initial=Yes,
/ConsideredUncertainKtaTemp488/.initial=Yes,
/ConsideredUncertainKtaTemp489/.initial=Yes,
/ConsideredUncertainKtaTemp490/.initial=Yes,
/ConsideredUncertainKtaTemp491/.initial=Yes,
/ConsideredUncertainKtaTemp492/.initial=Yes,
/ConsideredUncertainKtaTemp493/.initial=Yes,
/ConsideredUncertainKtaTemp494/.initial=Yes,
/ConsideredUncertainKtaTemp495/.initial=Yes,
/ConsideredUncertainKtaTemp496/.initial=Yes,
/ConsideredUncertainKtaTemp497/.initial=Yes,
/ConsideredUncertainKtaTemp498/.initial=Yes,
/ConsideredUncertainKtaTemp499/.initial=Yes,
/ConsideredUncertainKtaTemp500/.initial=Yes,
/ConsideredUncertainKtaTemp501/.initial=Yes,
/ConsideredUncertainKtaTemp502/.initial=Yes,
/ConsideredUncertainKtaTemp503/.initial=Yes,
/ConsideredUncertainKtaTemp504/.initial=Yes,
/ConsideredUncertainKtaTemp505/.initial=Yes,
/ConsideredUncertainKtaTemp506/.initial=Yes,
/ConsideredUncertainKtaTemp507/.initial=Yes,
/ConsideredUncertainKtaTemp508/.initial=Yes,
/ConsideredUncertainKtaTemp509/.initial=Yes,
/ConsideredUncertainKtaTemp510/.initial=Yes,
/ConsideredUncertainKtaTemp511/.initial=Yes,
/ConsideredUncertainKtaTemp512/.initial=Yes,
/ConsideredUncertainKtaTemp513/.initial=Yes,
/ConsideredUncertainKtaTemp514/.initial=Yes,
/ConsideredUncertainKtaTemp515/.initial=Yes,
/ConsideredUncertainKtaTemp516/.initial=Yes,
/ConsideredUncertainKtaTemp517/.initial=Yes,
/ConsideredUncertainKtaTemp518/.initial=Yes,
/ConsideredUncertainKtaTemp519/.initial=Yes,
/ConsideredUncertainKtaTemp520/.initial=Yes,
/ConsideredUncertainKtaTemp521/.initial=Yes,
/ConsideredUncertainKtaTemp522/.initial=Yes,
/ConsideredUncertainKtaTemp523/.initial=Yes,
/ConsideredUncertainKtaTemp524/.initial=Yes,
/ConsideredUncertainKtaTemp525/.initial=Yes,
/ConsideredUncertainKtaTemp526/.initial=Yes,
/ConsideredUncertainKtaTemp527/.initial=Yes,
/ConsideredUncertainKtaTemp528/.initial=Yes,
/ConsideredUncertainKtaTemp529/.initial=Yes,
/ConsideredUncertainKtaTemp530/.initial=Yes,
/ConsideredUncertainKtaTemp531/.initial=Yes,
/ConsideredUncertainKtaTemp532/.initial=Yes,
/ConsideredUncertainKtaTemp533/.initial=Yes,
/ConsideredUncertainKtaTemp534/.initial=Yes,
/ConsideredUncertainKtaTemp535/.initial=Yes,
/ConsideredUncertainKtaTemp536/.initial=Yes,
/ConsideredUncertainKtaTemp537/.initial=Yes,
/ConsideredUncertainKtaTemp538/.initial=Yes,
/ConsideredUncertainKtaTemp539/.initial=Yes,
/ConsideredUncertainKtaTemp540/.initial=Yes,
/ConsideredUncertainKtaTemp541/.initial=Yes,
/ConsideredUncertainKtaTemp542/.initial=Yes,
/ConsideredUncertainKtaTemp543/.initial=Yes,
/ConsideredUncertainKtaTemp544/.initial=Yes,
/ConsideredUncertainKtaTemp545/.initial=Yes,
/ConsideredUncertainKtaTemp546/.initial=Yes,
/ConsideredUncertainKtaTemp547/.initial=Yes,
/ConsideredUncertainKtaTemp548/.initial=Yes,
/ConsideredUncertainKtaTemp549/.initial=Yes,
/ConsideredUncertainKtaTemp550/.initial=Yes,
/ConsideredUncertainKtaTemp551/.initial=Yes,
/ConsideredUncertainKtaTemp552/.initial=Yes,
/ConsideredUncertainKtaTemp553/.initial=Yes,
/ConsideredUncertainKtaTemp554/.initial=Yes,
/ConsideredUncertainKtaTemp555/.initial=Yes,
/ConsideredUncertainKtaTemp556/.initial=Yes,
/ConsideredUncertainKtaTemp557/.initial=Yes,
/ConsideredUncertainKtaTemp558/.initial=Yes,
/ConsideredUncertainKtaTemp559/.initial=Yes,
/ConsideredUncertainKtaTemp560/.initial=Yes,
/ConsideredUncertainKtaTemp561/.initial=Yes,
/ConsideredUncertainKtaTemp562/.initial=Yes,
/ConsideredUncertainKtaTemp563/.initial=Yes,
/ConsideredUncertainKtaTemp564/.initial=Yes,
/ConsideredUncertainKtaTemp565/.initial=Yes,
/ConsideredUncertainKtaTemp566/.initial=Yes,
/ConsideredUncertainKtaTemp567/.initial=Yes,
/ConsideredUncertainKtaTemp568/.initial=Yes,
/ConsideredUncertainKtaTemp569/.initial=Yes,
/ConsideredUncertainKtaTemp570/.initial=Yes,
/ConsideredUncertainKtaTemp571/.initial=Yes,
/ConsideredUncertainKtaTemp572/.initial=Yes,
/ConsideredUncertainKtaTemp573/.initial=Yes,
/ConsideredUncertainKtaTemp574/.initial=Yes,
/ConsideredUncertainKtaTemp575/.initial=Yes,
/ConsideredUncertainKtaTemp576/.initial=Yes,
/ConsideredUncertainKtaTemp577/.initial=Yes,
/ConsideredUncertainKtaTemp578/.initial=Yes,
/ConsideredUncertainKtaTemp579/.initial=Yes,
/ConsideredUncertainKtaTemp580/.initial=Yes,
/ConsideredUncertainKtaTemp581/.initial=Yes,
/ConsideredUncertainKtaTemp582/.initial=Yes,
/ConsideredUncertainKtaTemp583/.initial=Yes,
/ConsideredUncertainKtaTemp584/.initial=Yes,
/ConsideredUncertainKtaTemp585/.initial=Yes,
/ConsideredUncertainKtaTemp586/.initial=Yes,
/ConsideredUncertainKtaTemp587/.initial=Yes,
/ConsideredUncertainKtaTemp588/.initial=Yes,
/ConsideredUncertainKtaTemp589/.initial=Yes,
/ConsideredUncertainKtaTemp590/.initial=Yes,
/ConsideredUncertainKtaTemp591/.initial=Yes,
/ConsideredUncertainKtaTemp592/.initial=Yes,
/ConsideredUncertainKtaTemp593/.initial=Yes,
/ConsideredUncertainKtaTemp594/.initial=Yes,
/ConsideredUncertainKtaTemp595/.initial=Yes,
/ConsideredUncertainKtaTemp596/.initial=Yes,
/ConsideredUncertainKtaTemp597/.initial=Yes,
/ConsideredUncertainKtaTemp598/.initial=Yes,
/ConsideredUncertainKtaTemp599/.initial=Yes,
/ConsideredUncertainKtaTemp600/.initial=Yes,
/ConsideredUncertainKtaTemp601/.initial=Yes,
/ConsideredUncertainKtaTemp602/.initial=Yes,
/ConsideredUncertainKtaTemp603/.initial=Yes,
/ConsideredUncertainKtaTemp604/.initial=Yes,
/ConsideredUncertainKtaTemp605/.initial=Yes,
/ConsideredUncertainKtaTemp606/.initial=Yes,
/ConsideredUncertainKtaTemp607/.initial=Yes,
/ConsideredUncertainKtaTemp608/.initial=Yes,
/ConsideredUncertainKtaTemp609/.initial=Yes,
/ConsideredUncertainKtaTemp610/.initial=Yes,
/ConsideredUncertainKtaTemp611/.initial=Yes,
/ConsideredUncertainKtaTemp612/.initial=Yes,
/ConsideredUncertainKtaTemp613/.initial=Yes,
/ConsideredUncertainKtaTemp614/.initial=Yes,
/ConsideredUncertainKtaTemp615/.initial=Yes,
/ConsideredUncertainKtaTemp616/.initial=Yes,
/ConsideredUncertainKtaTemp617/.initial=Yes,
/ConsideredUncertainKtaTemp618/.initial=Yes,
/ConsideredUncertainKtaTemp619/.initial=Yes,
/ConsideredUncertainKtaTemp620/.initial=Yes,
/ConsideredUncertainKtaTemp621/.initial=Yes,
/ConsideredUncertainKtaTemp622/.initial=Yes,
/ConsideredUncertainKtaTemp623/.initial=Yes,
/ConsideredUncertainKtaTemp624/.initial=Yes,
/ConsideredUncertainKtaTemp625/.initial=Yes,
/ConsideredUncertainKtaTemp626/.initial=Yes,
/ConsideredUncertainKtaTemp627/.initial=Yes,
/ConsideredUncertainKtaTemp628/.initial=Yes,
/ConsideredUncertainKtaTemp629/.initial=Yes,
/ConsideredUncertainKtaTemp630/.initial=Yes,
/ConsideredUncertainKtaTemp631/.initial=Yes,
/ConsideredUncertainKtaTemp632/.initial=Yes,
/ConsideredUncertainKtaTemp633/.initial=Yes,
/ConsideredUncertainKtaTemp634/.initial=Yes,
/ConsideredUncertainKtaTemp635/.initial=Yes,
/ConsideredUncertainKtaTemp636/.initial=Yes,
/ConsideredUncertainKtaTemp637/.initial=Yes,
/ConsideredUncertainKtaTemp638/.initial=Yes,
/ConsideredUncertainKtaTemp639/.initial=Yes,
/ConsideredUncertainKtaTemp640/.initial=Yes,
/ConsideredUncertainKtaTemp641/.initial=Yes,
/ConsideredUncertainKtaTemp642/.initial=Yes,
/ConsideredUncertainKtaTemp643/.initial=Yes,
/ConsideredUncertainKtaTemp644/.initial=Yes,
/ConsideredUncertainKtaTemp645/.initial=Yes,
/ConsideredUncertainKtaTemp646/.initial=Yes,
/ConsideredUncertainKtaTemp647/.initial=Yes,
/ConsideredUncertainKtaTemp648/.initial=Yes,
/ConsideredUncertainKtaTemp649/.initial=Yes,
/ConsideredUncertainKtaTemp650/.initial=Yes,
/ConsideredUncertainKtaTemp651/.initial=Yes,
/ConsideredUncertainKtaTemp652/.initial=Yes,
/ConsideredUncertainKtaTemp653/.initial=Yes,
/ConsideredUncertainKtaTemp654/.initial=Yes,
/ConsideredUncertainKtaTemp655/.initial=Yes,
/ConsideredUncertainKtaTemp656/.initial=Yes,
/ConsideredUncertainKtaTemp657/.initial=Yes,
/ConsideredUncertainKtaTemp658/.initial=Yes,
/ConsideredUncertainKtaTemp659/.initial=Yes,
/ConsideredUncertainKtaTemp660/.initial=Yes,
/ConsideredUncertainKtaTemp661/.initial=Yes,
/ConsideredUncertainKtaTemp662/.initial=Yes,
/ConsideredUncertainKtaTemp663/.initial=Yes,
/ConsideredUncertainKtaTemp664/.initial=Yes,
/ConsideredUncertainKtaTemp665/.initial=Yes,
/ConsideredUncertainKtaTemp666/.initial=Yes,
/ConsideredUncertainKtaTemp667/.initial=Yes,
/ConsideredUncertainKtaTemp668/.initial=Yes,
/ConsideredUncertainKtaTemp669/.initial=Yes,
/ConsideredUncertainKtaTemp670/.initial=Yes,
/ConsideredUncertainKtaTemp671/.initial=Yes,
/ConsideredUncertainKtaTemp672/.initial=Yes,
/ConsideredUncertainKtaTemp673/.initial=Yes,
/ConsideredUncertainKtaTemp674/.initial=Yes,
/ConsideredUncertainKtaTemp675/.initial=Yes,
/ConsideredUncertainKtaTemp676/.initial=Yes,
/ConsideredUncertainKtaTemp677/.initial=Yes,
/ConsideredUncertainKtaTemp678/.initial=Yes,
/ConsideredUncertainKtaTemp679/.initial=Yes,
/ConsideredUncertainKtaTemp680/.initial=Yes,
/ConsideredUncertainKtaTemp681/.initial=Yes,
/ConsideredUncertainKtaTemp682/.initial=Yes,
/ConsideredUncertainKtaTemp683/.initial=Yes,
/ConsideredUncertainKtaTemp684/.initial=Yes,
/ConsideredUncertainKtaTemp685/.initial=Yes,
/ConsideredUncertainKtaTemp686/.initial=Yes,
/ConsideredUncertainKtaTemp687/.initial=Yes,
/ConsideredUncertainKtaTemp688/.initial=Yes,
/ConsideredUncertainKtaTemp689/.initial=Yes,
/ConsideredUncertainKtaTemp690/.initial=Yes,
/ConsideredUncertainKtaTemp691/.initial=Yes,
/ConsideredUncertainKtaTemp692/.initial=Yes,
/ConsideredUncertainKtaTemp693/.initial=Yes,
/ConsideredUncertainKtaTemp694/.initial=Yes,
/ConsideredUncertainKtaTemp695/.initial=Yes,
/ConsideredUncertainKtaTemp696/.initial=Yes,
/ConsideredUncertainKtaTemp697/.initial=Yes,
/ConsideredUncertainKtaTemp698/.initial=Yes,
/ConsideredUncertainKtaTemp699/.initial=Yes,
/ConsideredUncertainKtaTemp700/.initial=Yes,
/ConsideredUncertainKtaTemp701/.initial=Yes,
/ConsideredUncertainKtaTemp702/.initial=Yes,
/ConsideredUncertainKtaTemp703/.initial=Yes,
/ConsideredUncertainKtaTemp704/.initial=Yes,
/ConsideredUncertainKtaTemp705/.initial=Yes,
/ConsideredUncertainKtaTemp706/.initial=Yes,
/ConsideredUncertainKtaTemp707/.initial=Yes,
/ConsideredUncertainKtaTemp708/.initial=Yes,
/ConsideredUncertainKtaTemp709/.initial=Yes,
/ConsideredUncertainKtaTemp710/.initial=Yes,
/ConsideredUncertainKtaTemp711/.initial=Yes,
/ConsideredUncertainKtaTemp712/.initial=Yes,
/ConsideredUncertainKtaTemp713/.initial=Yes,
/ConsideredUncertainKtaTemp714/.initial=Yes,
/ConsideredUncertainKtaTemp715/.initial=Yes,
/ConsideredUncertainKtaTemp716/.initial=Yes,
/ConsideredUncertainKtaTemp717/.initial=Yes,
/ConsideredUncertainKtaTemp718/.initial=Yes,
/ConsideredUncertainKtaTemp719/.initial=Yes,
/ConsideredUncertainKtaTemp720/.initial=Yes,
/ConsideredUncertainKtaTemp721/.initial=Yes,
/ConsideredUncertainKtaTemp722/.initial=Yes,
/ConsideredUncertainKtaTemp723/.initial=Yes,
/ConsideredUncertainKtaTemp724/.initial=Yes,
/ConsideredUncertainKtaTemp725/.initial=Yes,
/ConsideredUncertainKtaTemp726/.initial=Yes,
/ConsideredUncertainKtaTemp727/.initial=Yes,
/ConsideredUncertainKtaTemp728/.initial=Yes,
/ConsideredUncertainKtaTemp729/.initial=Yes,
/ConsideredUncertainKtaTemp730/.initial=Yes,
/ConsideredUncertainKtaTemp731/.initial=Yes,
/ConsideredUncertainKtaTemp732/.initial=Yes,
/ConsideredUncertainKtaTemp733/.initial=Yes,
/ConsideredUncertainKtaTemp734/.initial=Yes,
/ConsideredUncertainKtaTemp735/.initial=Yes,
/ConsideredUncertainKtaTemp736/.initial=Yes,
/ConsideredUncertainKtaTemp737/.initial=Yes,
/ConsideredUncertainKtaTemp738/.initial=Yes,
/ConsideredUncertainKtaTemp739/.initial=Yes,
/ConsideredUncertainKtaTemp740/.initial=Yes,
/ConsideredUncertainKtaTemp741/.initial=Yes,
/ConsideredUncertainKtaTemp742/.initial=Yes,
/ConsideredUncertainKtaTemp743/.initial=Yes,
/ConsideredUncertainKtaTemp744/.initial=Yes,
/ConsideredUncertainKtaTemp745/.initial=Yes,
/ConsideredUncertainKtaTemp746/.initial=Yes,
/ConsideredUncertainKtaTemp747/.initial=Yes,
/ConsideredUncertainKtaTemp748/.initial=Yes,
/ConsideredUncertainKtaTemp749/.initial=Yes,
/ConsideredUncertainKtaTemp750/.initial=Yes,
/ConsideredUncertainKtaTemp751/.initial=Yes,
/ConsideredUncertainKtaTemp752/.initial=Yes,
/ConsideredUncertainKtaTemp753/.initial=Yes,
/ConsideredUncertainKtaTemp754/.initial=Yes,
/ConsideredUncertainKtaTemp755/.initial=Yes,
/ConsideredUncertainKtaTemp756/.initial=Yes,
/ConsideredUncertainKtaTemp757/.initial=Yes,
/ConsideredUncertainKtaTemp758/.initial=Yes,
/ConsideredUncertainKtaTemp759/.initial=Yes,
/ConsideredUncertainKtaTemp760/.initial=Yes,
/ConsideredUncertainKtaTemp761/.initial=Yes,
/ConsideredUncertainKtaTemp762/.initial=Yes,
/ConsideredUncertainKtaTemp763/.initial=Yes,
/ConsideredUncertainKtaTemp764/.initial=Yes,
/ConsideredUncertainKtaTemp765/.initial=Yes,
/ConsideredUncertainKtaTemp766/.initial=Yes,
/ConsideredUncertainKtaTemp767/.initial=Yes,
/ConsideredUncertainKtaTemp0InMem/.initial=Yes,
/ConsideredUncertainKtaTemp1InMem/.initial=Yes,
/ConsideredUncertainKtaTemp2InMem/.initial=Yes,
/ConsideredUncertainKtaTemp3InMem/.initial=Yes,
/ConsideredUncertainKtaTemp4InMem/.initial=Yes,
/ConsideredUncertainKtaTemp5InMem/.initial=Yes,
/ConsideredUncertainKtaTemp6InMem/.initial=Yes,
/ConsideredUncertainKtaTemp7InMem/.initial=Yes,
/ConsideredUncertainKtaTemp8InMem/.initial=Yes,
/ConsideredUncertainKtaTemp9InMem/.initial=Yes,
/ConsideredUncertainKtaTemp10InMem/.initial=Yes,
/ConsideredUncertainKtaTemp11InMem/.initial=Yes,
/ConsideredUncertainKtaTemp12InMem/.initial=Yes,
/ConsideredUncertainKtaTemp13InMem/.initial=Yes,
/ConsideredUncertainKtaTemp14InMem/.initial=Yes,
/ConsideredUncertainKtaTemp15InMem/.initial=Yes,
/ConsideredUncertainKtaTemp16InMem/.initial=Yes,
/ConsideredUncertainKtaTemp17InMem/.initial=Yes,
/ConsideredUncertainKtaTemp18InMem/.initial=Yes,
/ConsideredUncertainKtaTemp19InMem/.initial=Yes,
/ConsideredUncertainKtaTemp20InMem/.initial=Yes,
/ConsideredUncertainKtaTemp21InMem/.initial=Yes,
/ConsideredUncertainKtaTemp22InMem/.initial=Yes,
/ConsideredUncertainKtaTemp23InMem/.initial=Yes,
/ConsideredUncertainKtaTemp24InMem/.initial=Yes,
/ConsideredUncertainKtaTemp25InMem/.initial=Yes,
/ConsideredUncertainKtaTemp26InMem/.initial=Yes,
/ConsideredUncertainKtaTemp27InMem/.initial=Yes,
/ConsideredUncertainKtaTemp28InMem/.initial=Yes,
/ConsideredUncertainKtaTemp29InMem/.initial=Yes,
/ConsideredUncertainKtaTemp30InMem/.initial=Yes,
/ConsideredUncertainKtaTemp31InMem/.initial=Yes,
/ConsideredUncertainKtaTemp32InMem/.initial=Yes,
/ConsideredUncertainKtaTemp33InMem/.initial=Yes,
/ConsideredUncertainKtaTemp34InMem/.initial=Yes,
/ConsideredUncertainKtaTemp35InMem/.initial=Yes,
/ConsideredUncertainKtaTemp36InMem/.initial=Yes,
/ConsideredUncertainKtaTemp37InMem/.initial=Yes,
/ConsideredUncertainKtaTemp38InMem/.initial=Yes,
/ConsideredUncertainKtaTemp39InMem/.initial=Yes,
/ConsideredUncertainKtaTemp40InMem/.initial=Yes,
/ConsideredUncertainKtaTemp41InMem/.initial=Yes,
/ConsideredUncertainKtaTemp42InMem/.initial=Yes,
/ConsideredUncertainKtaTemp43InMem/.initial=Yes,
/ConsideredUncertainKtaTemp44InMem/.initial=Yes,
/ConsideredUncertainKtaTemp45InMem/.initial=Yes,
/ConsideredUncertainKtaTemp46InMem/.initial=Yes,
/ConsideredUncertainKtaTemp47InMem/.initial=Yes,
/ConsideredUncertainKtaTemp48InMem/.initial=Yes,
/ConsideredUncertainKtaTemp49InMem/.initial=Yes,
/ConsideredUncertainKtaTemp50InMem/.initial=Yes,
/ConsideredUncertainKtaTemp51InMem/.initial=Yes,
/ConsideredUncertainKtaTemp52InMem/.initial=Yes,
/ConsideredUncertainKtaTemp53InMem/.initial=Yes,
/ConsideredUncertainKtaTemp54InMem/.initial=Yes,
/ConsideredUncertainKtaTemp55InMem/.initial=Yes,
/ConsideredUncertainKtaTemp56InMem/.initial=Yes,
/ConsideredUncertainKtaTemp57InMem/.initial=Yes,
/ConsideredUncertainKtaTemp58InMem/.initial=Yes,
/ConsideredUncertainKtaTemp59InMem/.initial=Yes,
/ConsideredUncertainKtaTemp60InMem/.initial=Yes,
/ConsideredUncertainKtaTemp61InMem/.initial=Yes,
/ConsideredUncertainKtaTemp62InMem/.initial=Yes,
/ConsideredUncertainKtaTemp63InMem/.initial=Yes,
/ConsideredUncertainKtaTemp64InMem/.initial=Yes,
/ConsideredUncertainKtaTemp65InMem/.initial=Yes,
/ConsideredUncertainKtaTemp66InMem/.initial=Yes,
/ConsideredUncertainKtaTemp67InMem/.initial=Yes,
/ConsideredUncertainKtaTemp68InMem/.initial=Yes,
/ConsideredUncertainKtaTemp69InMem/.initial=Yes,
/ConsideredUncertainKtaTemp70InMem/.initial=Yes,
/ConsideredUncertainKtaTemp71InMem/.initial=Yes,
/ConsideredUncertainKtaTemp72InMem/.initial=Yes,
/ConsideredUncertainKtaTemp73InMem/.initial=Yes,
/ConsideredUncertainKtaTemp74InMem/.initial=Yes,
/ConsideredUncertainKtaTemp75InMem/.initial=Yes,
/ConsideredUncertainKtaTemp76InMem/.initial=Yes,
/ConsideredUncertainKtaTemp77InMem/.initial=Yes,
/ConsideredUncertainKtaTemp78InMem/.initial=Yes,
/ConsideredUncertainKtaTemp79InMem/.initial=Yes,
/ConsideredUncertainKtaTemp80InMem/.initial=Yes,
/ConsideredUncertainKtaTemp81InMem/.initial=Yes,
/ConsideredUncertainKtaTemp82InMem/.initial=Yes,
/ConsideredUncertainKtaTemp83InMem/.initial=Yes,
/ConsideredUncertainKtaTemp84InMem/.initial=Yes,
/ConsideredUncertainKtaTemp85InMem/.initial=Yes,
/ConsideredUncertainKtaTemp86InMem/.initial=Yes,
/ConsideredUncertainKtaTemp87InMem/.initial=Yes,
/ConsideredUncertainKtaTemp88InMem/.initial=Yes,
/ConsideredUncertainKtaTemp89InMem/.initial=Yes,
/ConsideredUncertainKtaTemp90InMem/.initial=Yes,
/ConsideredUncertainKtaTemp91InMem/.initial=Yes,
/ConsideredUncertainKtaTemp92InMem/.initial=Yes,
/ConsideredUncertainKtaTemp93InMem/.initial=Yes,
/ConsideredUncertainKtaTemp94InMem/.initial=Yes,
/ConsideredUncertainKtaTemp95InMem/.initial=Yes,
/ConsideredUncertainKtaTemp96InMem/.initial=Yes,
/ConsideredUncertainKtaTemp97InMem/.initial=Yes,
/ConsideredUncertainKtaTemp98InMem/.initial=Yes,
/ConsideredUncertainKtaTemp99InMem/.initial=Yes,
/ConsideredUncertainKtaTemp100InMem/.initial=Yes,
/ConsideredUncertainKtaTemp101InMem/.initial=Yes,
/ConsideredUncertainKtaTemp102InMem/.initial=Yes,
/ConsideredUncertainKtaTemp103InMem/.initial=Yes,
/ConsideredUncertainKtaTemp104InMem/.initial=Yes,
/ConsideredUncertainKtaTemp105InMem/.initial=Yes,
/ConsideredUncertainKtaTemp106InMem/.initial=Yes,
/ConsideredUncertainKtaTemp107InMem/.initial=Yes,
/ConsideredUncertainKtaTemp108InMem/.initial=Yes,
/ConsideredUncertainKtaTemp109InMem/.initial=Yes,
/ConsideredUncertainKtaTemp110InMem/.initial=Yes,
/ConsideredUncertainKtaTemp111InMem/.initial=Yes,
/ConsideredUncertainKtaTemp112InMem/.initial=Yes,
/ConsideredUncertainKtaTemp113InMem/.initial=Yes,
/ConsideredUncertainKtaTemp114InMem/.initial=Yes,
/ConsideredUncertainKtaTemp115InMem/.initial=Yes,
/ConsideredUncertainKtaTemp116InMem/.initial=Yes,
/ConsideredUncertainKtaTemp117InMem/.initial=Yes,
/ConsideredUncertainKtaTemp118InMem/.initial=Yes,
/ConsideredUncertainKtaTemp119InMem/.initial=Yes,
/ConsideredUncertainKtaTemp120InMem/.initial=Yes,
/ConsideredUncertainKtaTemp121InMem/.initial=Yes,
/ConsideredUncertainKtaTemp122InMem/.initial=Yes,
/ConsideredUncertainKtaTemp123InMem/.initial=Yes,
/ConsideredUncertainKtaTemp124InMem/.initial=Yes,
/ConsideredUncertainKtaTemp125InMem/.initial=Yes,
/ConsideredUncertainKtaTemp126InMem/.initial=Yes,
/ConsideredUncertainKtaTemp127InMem/.initial=Yes,
/ConsideredUncertainKtaTemp128InMem/.initial=Yes,
/ConsideredUncertainKtaTemp129InMem/.initial=Yes,
/ConsideredUncertainKtaTemp130InMem/.initial=Yes,
/ConsideredUncertainKtaTemp131InMem/.initial=Yes,
/ConsideredUncertainKtaTemp132InMem/.initial=Yes,
/ConsideredUncertainKtaTemp133InMem/.initial=Yes,
/ConsideredUncertainKtaTemp134InMem/.initial=Yes,
/ConsideredUncertainKtaTemp135InMem/.initial=Yes,
/ConsideredUncertainKtaTemp136InMem/.initial=Yes,
/ConsideredUncertainKtaTemp137InMem/.initial=Yes,
/ConsideredUncertainKtaTemp138InMem/.initial=Yes,
/ConsideredUncertainKtaTemp139InMem/.initial=Yes,
/ConsideredUncertainKtaTemp140InMem/.initial=Yes,
/ConsideredUncertainKtaTemp141InMem/.initial=Yes,
/ConsideredUncertainKtaTemp142InMem/.initial=Yes,
/ConsideredUncertainKtaTemp143InMem/.initial=Yes,
/ConsideredUncertainKtaTemp144InMem/.initial=Yes,
/ConsideredUncertainKtaTemp145InMem/.initial=Yes,
/ConsideredUncertainKtaTemp146InMem/.initial=Yes,
/ConsideredUncertainKtaTemp147InMem/.initial=Yes,
/ConsideredUncertainKtaTemp148InMem/.initial=Yes,
/ConsideredUncertainKtaTemp149InMem/.initial=Yes,
/ConsideredUncertainKtaTemp150InMem/.initial=Yes,
/ConsideredUncertainKtaTemp151InMem/.initial=Yes,
/ConsideredUncertainKtaTemp152InMem/.initial=Yes,
/ConsideredUncertainKtaTemp153InMem/.initial=Yes,
/ConsideredUncertainKtaTemp154InMem/.initial=Yes,
/ConsideredUncertainKtaTemp155InMem/.initial=Yes,
/ConsideredUncertainKtaTemp156InMem/.initial=Yes,
/ConsideredUncertainKtaTemp157InMem/.initial=Yes,
/ConsideredUncertainKtaTemp158InMem/.initial=Yes,
/ConsideredUncertainKtaTemp159InMem/.initial=Yes,
/ConsideredUncertainKtaTemp160InMem/.initial=Yes,
/ConsideredUncertainKtaTemp161InMem/.initial=Yes,
/ConsideredUncertainKtaTemp162InMem/.initial=Yes,
/ConsideredUncertainKtaTemp163InMem/.initial=Yes,
/ConsideredUncertainKtaTemp164InMem/.initial=Yes,
/ConsideredUncertainKtaTemp165InMem/.initial=Yes,
/ConsideredUncertainKtaTemp166InMem/.initial=Yes,
/ConsideredUncertainKtaTemp167InMem/.initial=Yes,
/ConsideredUncertainKtaTemp168InMem/.initial=Yes,
/ConsideredUncertainKtaTemp169InMem/.initial=Yes,
/ConsideredUncertainKtaTemp170InMem/.initial=Yes,
/ConsideredUncertainKtaTemp171InMem/.initial=Yes,
/ConsideredUncertainKtaTemp172InMem/.initial=Yes,
/ConsideredUncertainKtaTemp173InMem/.initial=Yes,
/ConsideredUncertainKtaTemp174InMem/.initial=Yes,
/ConsideredUncertainKtaTemp175InMem/.initial=Yes,
/ConsideredUncertainKtaTemp176InMem/.initial=Yes,
/ConsideredUncertainKtaTemp177InMem/.initial=Yes,
/ConsideredUncertainKtaTemp178InMem/.initial=Yes,
/ConsideredUncertainKtaTemp179InMem/.initial=Yes,
/ConsideredUncertainKtaTemp180InMem/.initial=Yes,
/ConsideredUncertainKtaTemp181InMem/.initial=Yes,
/ConsideredUncertainKtaTemp182InMem/.initial=Yes,
/ConsideredUncertainKtaTemp183InMem/.initial=Yes,
/ConsideredUncertainKtaTemp184InMem/.initial=Yes,
/ConsideredUncertainKtaTemp185InMem/.initial=Yes,
/ConsideredUncertainKtaTemp186InMem/.initial=Yes,
/ConsideredUncertainKtaTemp187InMem/.initial=Yes,
/ConsideredUncertainKtaTemp188InMem/.initial=Yes,
/ConsideredUncertainKtaTemp189InMem/.initial=Yes,
/ConsideredUncertainKtaTemp190InMem/.initial=Yes,
/ConsideredUncertainKtaTemp191InMem/.initial=Yes,
/ConsideredUncertainKtaTemp192InMem/.initial=Yes,
/ConsideredUncertainKtaTemp193InMem/.initial=Yes,
/ConsideredUncertainKtaTemp194InMem/.initial=Yes,
/ConsideredUncertainKtaTemp195InMem/.initial=Yes,
/ConsideredUncertainKtaTemp196InMem/.initial=Yes,
/ConsideredUncertainKtaTemp197InMem/.initial=Yes,
/ConsideredUncertainKtaTemp198InMem/.initial=Yes,
/ConsideredUncertainKtaTemp199InMem/.initial=Yes,
/ConsideredUncertainKtaTemp200InMem/.initial=Yes,
/ConsideredUncertainKtaTemp201InMem/.initial=Yes,
/ConsideredUncertainKtaTemp202InMem/.initial=Yes,
/ConsideredUncertainKtaTemp203InMem/.initial=Yes,
/ConsideredUncertainKtaTemp204InMem/.initial=Yes,
/ConsideredUncertainKtaTemp205InMem/.initial=Yes,
/ConsideredUncertainKtaTemp206InMem/.initial=Yes,
/ConsideredUncertainKtaTemp207InMem/.initial=Yes,
/ConsideredUncertainKtaTemp208InMem/.initial=Yes,
/ConsideredUncertainKtaTemp209InMem/.initial=Yes,
/ConsideredUncertainKtaTemp210InMem/.initial=Yes,
/ConsideredUncertainKtaTemp211InMem/.initial=Yes,
/ConsideredUncertainKtaTemp212InMem/.initial=Yes,
/ConsideredUncertainKtaTemp213InMem/.initial=Yes,
/ConsideredUncertainKtaTemp214InMem/.initial=Yes,
/ConsideredUncertainKtaTemp215InMem/.initial=Yes,
/ConsideredUncertainKtaTemp216InMem/.initial=Yes,
/ConsideredUncertainKtaTemp217InMem/.initial=Yes,
/ConsideredUncertainKtaTemp218InMem/.initial=Yes,
/ConsideredUncertainKtaTemp219InMem/.initial=Yes,
/ConsideredUncertainKtaTemp220InMem/.initial=Yes,
/ConsideredUncertainKtaTemp221InMem/.initial=Yes,
/ConsideredUncertainKtaTemp222InMem/.initial=Yes,
/ConsideredUncertainKtaTemp223InMem/.initial=Yes,
/ConsideredUncertainKtaTemp224InMem/.initial=Yes,
/ConsideredUncertainKtaTemp225InMem/.initial=Yes,
/ConsideredUncertainKtaTemp226InMem/.initial=Yes,
/ConsideredUncertainKtaTemp227InMem/.initial=Yes,
/ConsideredUncertainKtaTemp228InMem/.initial=Yes,
/ConsideredUncertainKtaTemp229InMem/.initial=Yes,
/ConsideredUncertainKtaTemp230InMem/.initial=Yes,
/ConsideredUncertainKtaTemp231InMem/.initial=Yes,
/ConsideredUncertainKtaTemp232InMem/.initial=Yes,
/ConsideredUncertainKtaTemp233InMem/.initial=Yes,
/ConsideredUncertainKtaTemp234InMem/.initial=Yes,
/ConsideredUncertainKtaTemp235InMem/.initial=Yes,
/ConsideredUncertainKtaTemp236InMem/.initial=Yes,
/ConsideredUncertainKtaTemp237InMem/.initial=Yes,
/ConsideredUncertainKtaTemp238InMem/.initial=Yes,
/ConsideredUncertainKtaTemp239InMem/.initial=Yes,
/ConsideredUncertainKtaTemp240InMem/.initial=Yes,
/ConsideredUncertainKtaTemp241InMem/.initial=Yes,
/ConsideredUncertainKtaTemp242InMem/.initial=Yes,
/ConsideredUncertainKtaTemp243InMem/.initial=Yes,
/ConsideredUncertainKtaTemp244InMem/.initial=Yes,
/ConsideredUncertainKtaTemp245InMem/.initial=Yes,
/ConsideredUncertainKtaTemp246InMem/.initial=Yes,
/ConsideredUncertainKtaTemp247InMem/.initial=Yes,
/ConsideredUncertainKtaTemp248InMem/.initial=Yes,
/ConsideredUncertainKtaTemp249InMem/.initial=Yes,
/ConsideredUncertainKtaTemp250InMem/.initial=Yes,
/ConsideredUncertainKtaTemp251InMem/.initial=Yes,
/ConsideredUncertainKtaTemp252InMem/.initial=Yes,
/ConsideredUncertainKtaTemp253InMem/.initial=Yes,
/ConsideredUncertainKtaTemp254InMem/.initial=Yes,
/ConsideredUncertainKtaTemp255InMem/.initial=Yes,
/ConsideredUncertainKtaTemp256InMem/.initial=Yes,
/ConsideredUncertainKtaTemp257InMem/.initial=Yes,
/ConsideredUncertainKtaTemp258InMem/.initial=Yes,
/ConsideredUncertainKtaTemp259InMem/.initial=Yes,
/ConsideredUncertainKtaTemp260InMem/.initial=Yes,
/ConsideredUncertainKtaTemp261InMem/.initial=Yes,
/ConsideredUncertainKtaTemp262InMem/.initial=Yes,
/ConsideredUncertainKtaTemp263InMem/.initial=Yes,
/ConsideredUncertainKtaTemp264InMem/.initial=Yes,
/ConsideredUncertainKtaTemp265InMem/.initial=Yes,
/ConsideredUncertainKtaTemp266InMem/.initial=Yes,
/ConsideredUncertainKtaTemp267InMem/.initial=Yes,
/ConsideredUncertainKtaTemp268InMem/.initial=Yes,
/ConsideredUncertainKtaTemp269InMem/.initial=Yes,
/ConsideredUncertainKtaTemp270InMem/.initial=Yes,
/ConsideredUncertainKtaTemp271InMem/.initial=Yes,
/ConsideredUncertainKtaTemp272InMem/.initial=Yes,
/ConsideredUncertainKtaTemp273InMem/.initial=Yes,
/ConsideredUncertainKtaTemp274InMem/.initial=Yes,
/ConsideredUncertainKtaTemp275InMem/.initial=Yes,
/ConsideredUncertainKtaTemp276InMem/.initial=Yes,
/ConsideredUncertainKtaTemp277InMem/.initial=Yes,
/ConsideredUncertainKtaTemp278InMem/.initial=Yes,
/ConsideredUncertainKtaTemp279InMem/.initial=Yes,
/ConsideredUncertainKtaTemp280InMem/.initial=Yes,
/ConsideredUncertainKtaTemp281InMem/.initial=Yes,
/ConsideredUncertainKtaTemp282InMem/.initial=Yes,
/ConsideredUncertainKtaTemp283InMem/.initial=Yes,
/ConsideredUncertainKtaTemp284InMem/.initial=Yes,
/ConsideredUncertainKtaTemp285InMem/.initial=Yes,
/ConsideredUncertainKtaTemp286InMem/.initial=Yes,
/ConsideredUncertainKtaTemp287InMem/.initial=Yes,
/ConsideredUncertainKtaTemp288InMem/.initial=Yes,
/ConsideredUncertainKtaTemp289InMem/.initial=Yes,
/ConsideredUncertainKtaTemp290InMem/.initial=Yes,
/ConsideredUncertainKtaTemp291InMem/.initial=Yes,
/ConsideredUncertainKtaTemp292InMem/.initial=Yes,
/ConsideredUncertainKtaTemp293InMem/.initial=Yes,
/ConsideredUncertainKtaTemp294InMem/.initial=Yes,
/ConsideredUncertainKtaTemp295InMem/.initial=Yes,
/ConsideredUncertainKtaTemp296InMem/.initial=Yes,
/ConsideredUncertainKtaTemp297InMem/.initial=Yes,
/ConsideredUncertainKtaTemp298InMem/.initial=Yes,
/ConsideredUncertainKtaTemp299InMem/.initial=Yes,
/ConsideredUncertainKtaTemp300InMem/.initial=Yes,
/ConsideredUncertainKtaTemp301InMem/.initial=Yes,
/ConsideredUncertainKtaTemp302InMem/.initial=Yes,
/ConsideredUncertainKtaTemp303InMem/.initial=Yes,
/ConsideredUncertainKtaTemp304InMem/.initial=Yes,
/ConsideredUncertainKtaTemp305InMem/.initial=Yes,
/ConsideredUncertainKtaTemp306InMem/.initial=Yes,
/ConsideredUncertainKtaTemp307InMem/.initial=Yes,
/ConsideredUncertainKtaTemp308InMem/.initial=Yes,
/ConsideredUncertainKtaTemp309InMem/.initial=Yes,
/ConsideredUncertainKtaTemp310InMem/.initial=Yes,
/ConsideredUncertainKtaTemp311InMem/.initial=Yes,
/ConsideredUncertainKtaTemp312InMem/.initial=Yes,
/ConsideredUncertainKtaTemp313InMem/.initial=Yes,
/ConsideredUncertainKtaTemp314InMem/.initial=Yes,
/ConsideredUncertainKtaTemp315InMem/.initial=Yes,
/ConsideredUncertainKtaTemp316InMem/.initial=Yes,
/ConsideredUncertainKtaTemp317InMem/.initial=Yes,
/ConsideredUncertainKtaTemp318InMem/.initial=Yes,
/ConsideredUncertainKtaTemp319InMem/.initial=Yes,
/ConsideredUncertainKtaTemp320InMem/.initial=Yes,
/ConsideredUncertainKtaTemp321InMem/.initial=Yes,
/ConsideredUncertainKtaTemp322InMem/.initial=Yes,
/ConsideredUncertainKtaTemp323InMem/.initial=Yes,
/ConsideredUncertainKtaTemp324InMem/.initial=Yes,
/ConsideredUncertainKtaTemp325InMem/.initial=Yes,
/ConsideredUncertainKtaTemp326InMem/.initial=Yes,
/ConsideredUncertainKtaTemp327InMem/.initial=Yes,
/ConsideredUncertainKtaTemp328InMem/.initial=Yes,
/ConsideredUncertainKtaTemp329InMem/.initial=Yes,
/ConsideredUncertainKtaTemp330InMem/.initial=Yes,
/ConsideredUncertainKtaTemp331InMem/.initial=Yes,
/ConsideredUncertainKtaTemp332InMem/.initial=Yes,
/ConsideredUncertainKtaTemp333InMem/.initial=Yes,
/ConsideredUncertainKtaTemp334InMem/.initial=Yes,
/ConsideredUncertainKtaTemp335InMem/.initial=Yes,
/ConsideredUncertainKtaTemp336InMem/.initial=Yes,
/ConsideredUncertainKtaTemp337InMem/.initial=Yes,
/ConsideredUncertainKtaTemp338InMem/.initial=Yes,
/ConsideredUncertainKtaTemp339InMem/.initial=Yes,
/ConsideredUncertainKtaTemp340InMem/.initial=Yes,
/ConsideredUncertainKtaTemp341InMem/.initial=Yes,
/ConsideredUncertainKtaTemp342InMem/.initial=Yes,
/ConsideredUncertainKtaTemp343InMem/.initial=Yes,
/ConsideredUncertainKtaTemp344InMem/.initial=Yes,
/ConsideredUncertainKtaTemp345InMem/.initial=Yes,
/ConsideredUncertainKtaTemp346InMem/.initial=Yes,
/ConsideredUncertainKtaTemp347InMem/.initial=Yes,
/ConsideredUncertainKtaTemp348InMem/.initial=Yes,
/ConsideredUncertainKtaTemp349InMem/.initial=Yes,
/ConsideredUncertainKtaTemp350InMem/.initial=Yes,
/ConsideredUncertainKtaTemp351InMem/.initial=Yes,
/ConsideredUncertainKtaTemp352InMem/.initial=Yes,
/ConsideredUncertainKtaTemp353InMem/.initial=Yes,
/ConsideredUncertainKtaTemp354InMem/.initial=Yes,
/ConsideredUncertainKtaTemp355InMem/.initial=Yes,
/ConsideredUncertainKtaTemp356InMem/.initial=Yes,
/ConsideredUncertainKtaTemp357InMem/.initial=Yes,
/ConsideredUncertainKtaTemp358InMem/.initial=Yes,
/ConsideredUncertainKtaTemp359InMem/.initial=Yes,
/ConsideredUncertainKtaTemp360InMem/.initial=Yes,
/ConsideredUncertainKtaTemp361InMem/.initial=Yes,
/ConsideredUncertainKtaTemp362InMem/.initial=Yes,
/ConsideredUncertainKtaTemp363InMem/.initial=Yes,
/ConsideredUncertainKtaTemp364InMem/.initial=Yes,
/ConsideredUncertainKtaTemp365InMem/.initial=Yes,
/ConsideredUncertainKtaTemp366InMem/.initial=Yes,
/ConsideredUncertainKtaTemp367InMem/.initial=Yes,
/ConsideredUncertainKtaTemp368InMem/.initial=Yes,
/ConsideredUncertainKtaTemp369InMem/.initial=Yes,
/ConsideredUncertainKtaTemp370InMem/.initial=Yes,
/ConsideredUncertainKtaTemp371InMem/.initial=Yes,
/ConsideredUncertainKtaTemp372InMem/.initial=Yes,
/ConsideredUncertainKtaTemp373InMem/.initial=Yes,
/ConsideredUncertainKtaTemp374InMem/.initial=Yes,
/ConsideredUncertainKtaTemp375InMem/.initial=Yes,
/ConsideredUncertainKtaTemp376InMem/.initial=Yes,
/ConsideredUncertainKtaTemp377InMem/.initial=Yes,
/ConsideredUncertainKtaTemp378InMem/.initial=Yes,
/ConsideredUncertainKtaTemp379InMem/.initial=Yes,
/ConsideredUncertainKtaTemp380InMem/.initial=Yes,
/ConsideredUncertainKtaTemp381InMem/.initial=Yes,
/ConsideredUncertainKtaTemp382InMem/.initial=Yes,
/ConsideredUncertainKtaTemp383InMem/.initial=Yes,
/ConsideredUncertainKtaTemp384InMem/.initial=Yes,
/ConsideredUncertainKtaTemp385InMem/.initial=Yes,
/ConsideredUncertainKtaTemp386InMem/.initial=Yes,
/ConsideredUncertainKtaTemp387InMem/.initial=Yes,
/ConsideredUncertainKtaTemp388InMem/.initial=Yes,
/ConsideredUncertainKtaTemp389InMem/.initial=Yes,
/ConsideredUncertainKtaTemp390InMem/.initial=Yes,
/ConsideredUncertainKtaTemp391InMem/.initial=Yes,
/ConsideredUncertainKtaTemp392InMem/.initial=Yes,
/ConsideredUncertainKtaTemp393InMem/.initial=Yes,
/ConsideredUncertainKtaTemp394InMem/.initial=Yes,
/ConsideredUncertainKtaTemp395InMem/.initial=Yes,
/ConsideredUncertainKtaTemp396InMem/.initial=Yes,
/ConsideredUncertainKtaTemp397InMem/.initial=Yes,
/ConsideredUncertainKtaTemp398InMem/.initial=Yes,
/ConsideredUncertainKtaTemp399InMem/.initial=Yes,
/ConsideredUncertainKtaTemp400InMem/.initial=Yes,
/ConsideredUncertainKtaTemp401InMem/.initial=Yes,
/ConsideredUncertainKtaTemp402InMem/.initial=Yes,
/ConsideredUncertainKtaTemp403InMem/.initial=Yes,
/ConsideredUncertainKtaTemp404InMem/.initial=Yes,
/ConsideredUncertainKtaTemp405InMem/.initial=Yes,
/ConsideredUncertainKtaTemp406InMem/.initial=Yes,
/ConsideredUncertainKtaTemp407InMem/.initial=Yes,
/ConsideredUncertainKtaTemp408InMem/.initial=Yes,
/ConsideredUncertainKtaTemp409InMem/.initial=Yes,
/ConsideredUncertainKtaTemp410InMem/.initial=Yes,
/ConsideredUncertainKtaTemp411InMem/.initial=Yes,
/ConsideredUncertainKtaTemp412InMem/.initial=Yes,
/ConsideredUncertainKtaTemp413InMem/.initial=Yes,
/ConsideredUncertainKtaTemp414InMem/.initial=Yes,
/ConsideredUncertainKtaTemp415InMem/.initial=Yes,
/ConsideredUncertainKtaTemp416InMem/.initial=Yes,
/ConsideredUncertainKtaTemp417InMem/.initial=Yes,
/ConsideredUncertainKtaTemp418InMem/.initial=Yes,
/ConsideredUncertainKtaTemp419InMem/.initial=Yes,
/ConsideredUncertainKtaTemp420InMem/.initial=Yes,
/ConsideredUncertainKtaTemp421InMem/.initial=Yes,
/ConsideredUncertainKtaTemp422InMem/.initial=Yes,
/ConsideredUncertainKtaTemp423InMem/.initial=Yes,
/ConsideredUncertainKtaTemp424InMem/.initial=Yes,
/ConsideredUncertainKtaTemp425InMem/.initial=Yes,
/ConsideredUncertainKtaTemp426InMem/.initial=Yes,
/ConsideredUncertainKtaTemp427InMem/.initial=Yes,
/ConsideredUncertainKtaTemp428InMem/.initial=Yes,
/ConsideredUncertainKtaTemp429InMem/.initial=Yes,
/ConsideredUncertainKtaTemp430InMem/.initial=Yes,
/ConsideredUncertainKtaTemp431InMem/.initial=Yes,
/ConsideredUncertainKtaTemp432InMem/.initial=Yes,
/ConsideredUncertainKtaTemp433InMem/.initial=Yes,
/ConsideredUncertainKtaTemp434InMem/.initial=Yes,
/ConsideredUncertainKtaTemp435InMem/.initial=Yes,
/ConsideredUncertainKtaTemp436InMem/.initial=Yes,
/ConsideredUncertainKtaTemp437InMem/.initial=Yes,
/ConsideredUncertainKtaTemp438InMem/.initial=Yes,
/ConsideredUncertainKtaTemp439InMem/.initial=Yes,
/ConsideredUncertainKtaTemp440InMem/.initial=Yes,
/ConsideredUncertainKtaTemp441InMem/.initial=Yes,
/ConsideredUncertainKtaTemp442InMem/.initial=Yes,
/ConsideredUncertainKtaTemp443InMem/.initial=Yes,
/ConsideredUncertainKtaTemp444InMem/.initial=Yes,
/ConsideredUncertainKtaTemp445InMem/.initial=Yes,
/ConsideredUncertainKtaTemp446InMem/.initial=Yes,
/ConsideredUncertainKtaTemp447InMem/.initial=Yes,
/ConsideredUncertainKtaTemp448InMem/.initial=Yes,
/ConsideredUncertainKtaTemp449InMem/.initial=Yes,
/ConsideredUncertainKtaTemp450InMem/.initial=Yes,
/ConsideredUncertainKtaTemp451InMem/.initial=Yes,
/ConsideredUncertainKtaTemp452InMem/.initial=Yes,
/ConsideredUncertainKtaTemp453InMem/.initial=Yes,
/ConsideredUncertainKtaTemp454InMem/.initial=Yes,
/ConsideredUncertainKtaTemp455InMem/.initial=Yes,
/ConsideredUncertainKtaTemp456InMem/.initial=Yes,
/ConsideredUncertainKtaTemp457InMem/.initial=Yes,
/ConsideredUncertainKtaTemp458InMem/.initial=Yes,
/ConsideredUncertainKtaTemp459InMem/.initial=Yes,
/ConsideredUncertainKtaTemp460InMem/.initial=Yes,
/ConsideredUncertainKtaTemp461InMem/.initial=Yes,
/ConsideredUncertainKtaTemp462InMem/.initial=Yes,
/ConsideredUncertainKtaTemp463InMem/.initial=Yes,
/ConsideredUncertainKtaTemp464InMem/.initial=Yes,
/ConsideredUncertainKtaTemp465InMem/.initial=Yes,
/ConsideredUncertainKtaTemp466InMem/.initial=Yes,
/ConsideredUncertainKtaTemp467InMem/.initial=Yes,
/ConsideredUncertainKtaTemp468InMem/.initial=Yes,
/ConsideredUncertainKtaTemp469InMem/.initial=Yes,
/ConsideredUncertainKtaTemp470InMem/.initial=Yes,
/ConsideredUncertainKtaTemp471InMem/.initial=Yes,
/ConsideredUncertainKtaTemp472InMem/.initial=Yes,
/ConsideredUncertainKtaTemp473InMem/.initial=Yes,
/ConsideredUncertainKtaTemp474InMem/.initial=Yes,
/ConsideredUncertainKtaTemp475InMem/.initial=Yes,
/ConsideredUncertainKtaTemp476InMem/.initial=Yes,
/ConsideredUncertainKtaTemp477InMem/.initial=Yes,
/ConsideredUncertainKtaTemp478InMem/.initial=Yes,
/ConsideredUncertainKtaTemp479InMem/.initial=Yes,
/ConsideredUncertainKtaTemp480InMem/.initial=Yes,
/ConsideredUncertainKtaTemp481InMem/.initial=Yes,
/ConsideredUncertainKtaTemp482InMem/.initial=Yes,
/ConsideredUncertainKtaTemp483InMem/.initial=Yes,
/ConsideredUncertainKtaTemp484InMem/.initial=Yes,
/ConsideredUncertainKtaTemp485InMem/.initial=Yes,
/ConsideredUncertainKtaTemp486InMem/.initial=Yes,
/ConsideredUncertainKtaTemp487InMem/.initial=Yes,
/ConsideredUncertainKtaTemp488InMem/.initial=Yes,
/ConsideredUncertainKtaTemp489InMem/.initial=Yes,
/ConsideredUncertainKtaTemp490InMem/.initial=Yes,
/ConsideredUncertainKtaTemp491InMem/.initial=Yes,
/ConsideredUncertainKtaTemp492InMem/.initial=Yes,
/ConsideredUncertainKtaTemp493InMem/.initial=Yes,
/ConsideredUncertainKtaTemp494InMem/.initial=Yes,
/ConsideredUncertainKtaTemp495InMem/.initial=Yes,
/ConsideredUncertainKtaTemp496InMem/.initial=Yes,
/ConsideredUncertainKtaTemp497InMem/.initial=Yes,
/ConsideredUncertainKtaTemp498InMem/.initial=Yes,
/ConsideredUncertainKtaTemp499InMem/.initial=Yes,
/ConsideredUncertainKtaTemp500InMem/.initial=Yes,
/ConsideredUncertainKtaTemp501InMem/.initial=Yes,
/ConsideredUncertainKtaTemp502InMem/.initial=Yes,
/ConsideredUncertainKtaTemp503InMem/.initial=Yes,
/ConsideredUncertainKtaTemp504InMem/.initial=Yes,
/ConsideredUncertainKtaTemp505InMem/.initial=Yes,
/ConsideredUncertainKtaTemp506InMem/.initial=Yes,
/ConsideredUncertainKtaTemp507InMem/.initial=Yes,
/ConsideredUncertainKtaTemp508InMem/.initial=Yes,
/ConsideredUncertainKtaTemp509InMem/.initial=Yes,
/ConsideredUncertainKtaTemp510InMem/.initial=Yes,
/ConsideredUncertainKtaTemp511InMem/.initial=Yes,
/ConsideredUncertainKtaTemp512InMem/.initial=Yes,
/ConsideredUncertainKtaTemp513InMem/.initial=Yes,
/ConsideredUncertainKtaTemp514InMem/.initial=Yes,
/ConsideredUncertainKtaTemp515InMem/.initial=Yes,
/ConsideredUncertainKtaTemp516InMem/.initial=Yes,
/ConsideredUncertainKtaTemp517InMem/.initial=Yes,
/ConsideredUncertainKtaTemp518InMem/.initial=Yes,
/ConsideredUncertainKtaTemp519InMem/.initial=Yes,
/ConsideredUncertainKtaTemp520InMem/.initial=Yes,
/ConsideredUncertainKtaTemp521InMem/.initial=Yes,
/ConsideredUncertainKtaTemp522InMem/.initial=Yes,
/ConsideredUncertainKtaTemp523InMem/.initial=Yes,
/ConsideredUncertainKtaTemp524InMem/.initial=Yes,
/ConsideredUncertainKtaTemp525InMem/.initial=Yes,
/ConsideredUncertainKtaTemp526InMem/.initial=Yes,
/ConsideredUncertainKtaTemp527InMem/.initial=Yes,
/ConsideredUncertainKtaTemp528InMem/.initial=Yes,
/ConsideredUncertainKtaTemp529InMem/.initial=Yes,
/ConsideredUncertainKtaTemp530InMem/.initial=Yes,
/ConsideredUncertainKtaTemp531InMem/.initial=Yes,
/ConsideredUncertainKtaTemp532InMem/.initial=Yes,
/ConsideredUncertainKtaTemp533InMem/.initial=Yes,
/ConsideredUncertainKtaTemp534InMem/.initial=Yes,
/ConsideredUncertainKtaTemp535InMem/.initial=Yes,
/ConsideredUncertainKtaTemp536InMem/.initial=Yes,
/ConsideredUncertainKtaTemp537InMem/.initial=Yes,
/ConsideredUncertainKtaTemp538InMem/.initial=Yes,
/ConsideredUncertainKtaTemp539InMem/.initial=Yes,
/ConsideredUncertainKtaTemp540InMem/.initial=Yes,
/ConsideredUncertainKtaTemp541InMem/.initial=Yes,
/ConsideredUncertainKtaTemp542InMem/.initial=Yes,
/ConsideredUncertainKtaTemp543InMem/.initial=Yes,
/ConsideredUncertainKtaTemp544InMem/.initial=Yes,
/ConsideredUncertainKtaTemp545InMem/.initial=Yes,
/ConsideredUncertainKtaTemp546InMem/.initial=Yes,
/ConsideredUncertainKtaTemp547InMem/.initial=Yes,
/ConsideredUncertainKtaTemp548InMem/.initial=Yes,
/ConsideredUncertainKtaTemp549InMem/.initial=Yes,
/ConsideredUncertainKtaTemp550InMem/.initial=Yes,
/ConsideredUncertainKtaTemp551InMem/.initial=Yes,
/ConsideredUncertainKtaTemp552InMem/.initial=Yes,
/ConsideredUncertainKtaTemp553InMem/.initial=Yes,
/ConsideredUncertainKtaTemp554InMem/.initial=Yes,
/ConsideredUncertainKtaTemp555InMem/.initial=Yes,
/ConsideredUncertainKtaTemp556InMem/.initial=Yes,
/ConsideredUncertainKtaTemp557InMem/.initial=Yes,
/ConsideredUncertainKtaTemp558InMem/.initial=Yes,
/ConsideredUncertainKtaTemp559InMem/.initial=Yes,
/ConsideredUncertainKtaTemp560InMem/.initial=Yes,
/ConsideredUncertainKtaTemp561InMem/.initial=Yes,
/ConsideredUncertainKtaTemp562InMem/.initial=Yes,
/ConsideredUncertainKtaTemp563InMem/.initial=Yes,
/ConsideredUncertainKtaTemp564InMem/.initial=Yes,
/ConsideredUncertainKtaTemp565InMem/.initial=Yes,
/ConsideredUncertainKtaTemp566InMem/.initial=Yes,
/ConsideredUncertainKtaTemp567InMem/.initial=Yes,
/ConsideredUncertainKtaTemp568InMem/.initial=Yes,
/ConsideredUncertainKtaTemp569InMem/.initial=Yes,
/ConsideredUncertainKtaTemp570InMem/.initial=Yes,
/ConsideredUncertainKtaTemp571InMem/.initial=Yes,
/ConsideredUncertainKtaTemp572InMem/.initial=Yes,
/ConsideredUncertainKtaTemp573InMem/.initial=Yes,
/ConsideredUncertainKtaTemp574InMem/.initial=Yes,
/ConsideredUncertainKtaTemp575InMem/.initial=Yes,
/ConsideredUncertainKtaTemp576InMem/.initial=Yes,
/ConsideredUncertainKtaTemp577InMem/.initial=Yes,
/ConsideredUncertainKtaTemp578InMem/.initial=Yes,
/ConsideredUncertainKtaTemp579InMem/.initial=Yes,
/ConsideredUncertainKtaTemp580InMem/.initial=Yes,
/ConsideredUncertainKtaTemp581InMem/.initial=Yes,
/ConsideredUncertainKtaTemp582InMem/.initial=Yes,
/ConsideredUncertainKtaTemp583InMem/.initial=Yes,
/ConsideredUncertainKtaTemp584InMem/.initial=Yes,
/ConsideredUncertainKtaTemp585InMem/.initial=Yes,
/ConsideredUncertainKtaTemp586InMem/.initial=Yes,
/ConsideredUncertainKtaTemp587InMem/.initial=Yes,
/ConsideredUncertainKtaTemp588InMem/.initial=Yes,
/ConsideredUncertainKtaTemp589InMem/.initial=Yes,
/ConsideredUncertainKtaTemp590InMem/.initial=Yes,
/ConsideredUncertainKtaTemp591InMem/.initial=Yes,
/ConsideredUncertainKtaTemp592InMem/.initial=Yes,
/ConsideredUncertainKtaTemp593InMem/.initial=Yes,
/ConsideredUncertainKtaTemp594InMem/.initial=Yes,
/ConsideredUncertainKtaTemp595InMem/.initial=Yes,
/ConsideredUncertainKtaTemp596InMem/.initial=Yes,
/ConsideredUncertainKtaTemp597InMem/.initial=Yes,
/ConsideredUncertainKtaTemp598InMem/.initial=Yes,
/ConsideredUncertainKtaTemp599InMem/.initial=Yes,
/ConsideredUncertainKtaTemp600InMem/.initial=Yes,
/ConsideredUncertainKtaTemp601InMem/.initial=Yes,
/ConsideredUncertainKtaTemp602InMem/.initial=Yes,
/ConsideredUncertainKtaTemp603InMem/.initial=Yes,
/ConsideredUncertainKtaTemp604InMem/.initial=Yes,
/ConsideredUncertainKtaTemp605InMem/.initial=Yes,
/ConsideredUncertainKtaTemp606InMem/.initial=Yes,
/ConsideredUncertainKtaTemp607InMem/.initial=Yes,
/ConsideredUncertainKtaTemp608InMem/.initial=Yes,
/ConsideredUncertainKtaTemp609InMem/.initial=Yes,
/ConsideredUncertainKtaTemp610InMem/.initial=Yes,
/ConsideredUncertainKtaTemp611InMem/.initial=Yes,
/ConsideredUncertainKtaTemp612InMem/.initial=Yes,
/ConsideredUncertainKtaTemp613InMem/.initial=Yes,
/ConsideredUncertainKtaTemp614InMem/.initial=Yes,
/ConsideredUncertainKtaTemp615InMem/.initial=Yes,
/ConsideredUncertainKtaTemp616InMem/.initial=Yes,
/ConsideredUncertainKtaTemp617InMem/.initial=Yes,
/ConsideredUncertainKtaTemp618InMem/.initial=Yes,
/ConsideredUncertainKtaTemp619InMem/.initial=Yes,
/ConsideredUncertainKtaTemp620InMem/.initial=Yes,
/ConsideredUncertainKtaTemp621InMem/.initial=Yes,
/ConsideredUncertainKtaTemp622InMem/.initial=Yes,
/ConsideredUncertainKtaTemp623InMem/.initial=Yes,
/ConsideredUncertainKtaTemp624InMem/.initial=Yes,
/ConsideredUncertainKtaTemp625InMem/.initial=Yes,
/ConsideredUncertainKtaTemp626InMem/.initial=Yes,
/ConsideredUncertainKtaTemp627InMem/.initial=Yes,
/ConsideredUncertainKtaTemp628InMem/.initial=Yes,
/ConsideredUncertainKtaTemp629InMem/.initial=Yes,
/ConsideredUncertainKtaTemp630InMem/.initial=Yes,
/ConsideredUncertainKtaTemp631InMem/.initial=Yes,
/ConsideredUncertainKtaTemp632InMem/.initial=Yes,
/ConsideredUncertainKtaTemp633InMem/.initial=Yes,
/ConsideredUncertainKtaTemp634InMem/.initial=Yes,
/ConsideredUncertainKtaTemp635InMem/.initial=Yes,
/ConsideredUncertainKtaTemp636InMem/.initial=Yes,
/ConsideredUncertainKtaTemp637InMem/.initial=Yes,
/ConsideredUncertainKtaTemp638InMem/.initial=Yes,
/ConsideredUncertainKtaTemp639InMem/.initial=Yes,
/ConsideredUncertainKtaTemp640InMem/.initial=Yes,
/ConsideredUncertainKtaTemp641InMem/.initial=Yes,
/ConsideredUncertainKtaTemp642InMem/.initial=Yes,
/ConsideredUncertainKtaTemp643InMem/.initial=Yes,
/ConsideredUncertainKtaTemp644InMem/.initial=Yes,
/ConsideredUncertainKtaTemp645InMem/.initial=Yes,
/ConsideredUncertainKtaTemp646InMem/.initial=Yes,
/ConsideredUncertainKtaTemp647InMem/.initial=Yes,
/ConsideredUncertainKtaTemp648InMem/.initial=Yes,
/ConsideredUncertainKtaTemp649InMem/.initial=Yes,
/ConsideredUncertainKtaTemp650InMem/.initial=Yes,
/ConsideredUncertainKtaTemp651InMem/.initial=Yes,
/ConsideredUncertainKtaTemp652InMem/.initial=Yes,
/ConsideredUncertainKtaTemp653InMem/.initial=Yes,
/ConsideredUncertainKtaTemp654InMem/.initial=Yes,
/ConsideredUncertainKtaTemp655InMem/.initial=Yes,
/ConsideredUncertainKtaTemp656InMem/.initial=Yes,
/ConsideredUncertainKtaTemp657InMem/.initial=Yes,
/ConsideredUncertainKtaTemp658InMem/.initial=Yes,
/ConsideredUncertainKtaTemp659InMem/.initial=Yes,
/ConsideredUncertainKtaTemp660InMem/.initial=Yes,
/ConsideredUncertainKtaTemp661InMem/.initial=Yes,
/ConsideredUncertainKtaTemp662InMem/.initial=Yes,
/ConsideredUncertainKtaTemp663InMem/.initial=Yes,
/ConsideredUncertainKtaTemp664InMem/.initial=Yes,
/ConsideredUncertainKtaTemp665InMem/.initial=Yes,
/ConsideredUncertainKtaTemp666InMem/.initial=Yes,
/ConsideredUncertainKtaTemp667InMem/.initial=Yes,
/ConsideredUncertainKtaTemp668InMem/.initial=Yes,
/ConsideredUncertainKtaTemp669InMem/.initial=Yes,
/ConsideredUncertainKtaTemp670InMem/.initial=Yes,
/ConsideredUncertainKtaTemp671InMem/.initial=Yes,
/ConsideredUncertainKtaTemp672InMem/.initial=Yes,
/ConsideredUncertainKtaTemp673InMem/.initial=Yes,
/ConsideredUncertainKtaTemp674InMem/.initial=Yes,
/ConsideredUncertainKtaTemp675InMem/.initial=Yes,
/ConsideredUncertainKtaTemp676InMem/.initial=Yes,
/ConsideredUncertainKtaTemp677InMem/.initial=Yes,
/ConsideredUncertainKtaTemp678InMem/.initial=Yes,
/ConsideredUncertainKtaTemp679InMem/.initial=Yes,
/ConsideredUncertainKtaTemp680InMem/.initial=Yes,
/ConsideredUncertainKtaTemp681InMem/.initial=Yes,
/ConsideredUncertainKtaTemp682InMem/.initial=Yes,
/ConsideredUncertainKtaTemp683InMem/.initial=Yes,
/ConsideredUncertainKtaTemp684InMem/.initial=Yes,
/ConsideredUncertainKtaTemp685InMem/.initial=Yes,
/ConsideredUncertainKtaTemp686InMem/.initial=Yes,
/ConsideredUncertainKtaTemp687InMem/.initial=Yes,
/ConsideredUncertainKtaTemp688InMem/.initial=Yes,
/ConsideredUncertainKtaTemp689InMem/.initial=Yes,
/ConsideredUncertainKtaTemp690InMem/.initial=Yes,
/ConsideredUncertainKtaTemp691InMem/.initial=Yes,
/ConsideredUncertainKtaTemp692InMem/.initial=Yes,
/ConsideredUncertainKtaTemp693InMem/.initial=Yes,
/ConsideredUncertainKtaTemp694InMem/.initial=Yes,
/ConsideredUncertainKtaTemp695InMem/.initial=Yes,
/ConsideredUncertainKtaTemp696InMem/.initial=Yes,
/ConsideredUncertainKtaTemp697InMem/.initial=Yes,
/ConsideredUncertainKtaTemp698InMem/.initial=Yes,
/ConsideredUncertainKtaTemp699InMem/.initial=Yes,
/ConsideredUncertainKtaTemp700InMem/.initial=Yes,
/ConsideredUncertainKtaTemp701InMem/.initial=Yes,
/ConsideredUncertainKtaTemp702InMem/.initial=Yes,
/ConsideredUncertainKtaTemp703InMem/.initial=Yes,
/ConsideredUncertainKtaTemp704InMem/.initial=Yes,
/ConsideredUncertainKtaTemp705InMem/.initial=Yes,
/ConsideredUncertainKtaTemp706InMem/.initial=Yes,
/ConsideredUncertainKtaTemp707InMem/.initial=Yes,
/ConsideredUncertainKtaTemp708InMem/.initial=Yes,
/ConsideredUncertainKtaTemp709InMem/.initial=Yes,
/ConsideredUncertainKtaTemp710InMem/.initial=Yes,
/ConsideredUncertainKtaTemp711InMem/.initial=Yes,
/ConsideredUncertainKtaTemp712InMem/.initial=Yes,
/ConsideredUncertainKtaTemp713InMem/.initial=Yes,
/ConsideredUncertainKtaTemp714InMem/.initial=Yes,
/ConsideredUncertainKtaTemp715InMem/.initial=Yes,
/ConsideredUncertainKtaTemp716InMem/.initial=Yes,
/ConsideredUncertainKtaTemp717InMem/.initial=Yes,
/ConsideredUncertainKtaTemp718InMem/.initial=Yes,
/ConsideredUncertainKtaTemp719InMem/.initial=Yes,
/ConsideredUncertainKtaTemp720InMem/.initial=Yes,
/ConsideredUncertainKtaTemp721InMem/.initial=Yes,
/ConsideredUncertainKtaTemp722InMem/.initial=Yes,
/ConsideredUncertainKtaTemp723InMem/.initial=Yes,
/ConsideredUncertainKtaTemp724InMem/.initial=Yes,
/ConsideredUncertainKtaTemp725InMem/.initial=Yes,
/ConsideredUncertainKtaTemp726InMem/.initial=Yes,
/ConsideredUncertainKtaTemp727InMem/.initial=Yes,
/ConsideredUncertainKtaTemp728InMem/.initial=Yes,
/ConsideredUncertainKtaTemp729InMem/.initial=Yes,
/ConsideredUncertainKtaTemp730InMem/.initial=Yes,
/ConsideredUncertainKtaTemp731InMem/.initial=Yes,
/ConsideredUncertainKtaTemp732InMem/.initial=Yes,
/ConsideredUncertainKtaTemp733InMem/.initial=Yes,
/ConsideredUncertainKtaTemp734InMem/.initial=Yes,
/ConsideredUncertainKtaTemp735InMem/.initial=Yes,
/ConsideredUncertainKtaTemp736InMem/.initial=Yes,
/ConsideredUncertainKtaTemp737InMem/.initial=Yes,
/ConsideredUncertainKtaTemp738InMem/.initial=Yes,
/ConsideredUncertainKtaTemp739InMem/.initial=Yes,
/ConsideredUncertainKtaTemp740InMem/.initial=Yes,
/ConsideredUncertainKtaTemp741InMem/.initial=Yes,
/ConsideredUncertainKtaTemp742InMem/.initial=Yes,
/ConsideredUncertainKtaTemp743InMem/.initial=Yes,
/ConsideredUncertainKtaTemp744InMem/.initial=Yes,
/ConsideredUncertainKtaTemp745InMem/.initial=Yes,
/ConsideredUncertainKtaTemp746InMem/.initial=Yes,
/ConsideredUncertainKtaTemp747InMem/.initial=Yes,
/ConsideredUncertainKtaTemp748InMem/.initial=Yes,
/ConsideredUncertainKtaTemp749InMem/.initial=Yes,
/ConsideredUncertainKtaTemp750InMem/.initial=Yes,
/ConsideredUncertainKtaTemp751InMem/.initial=Yes,
/ConsideredUncertainKtaTemp752InMem/.initial=Yes,
/ConsideredUncertainKtaTemp753InMem/.initial=Yes,
/ConsideredUncertainKtaTemp754InMem/.initial=Yes,
/ConsideredUncertainKtaTemp755InMem/.initial=Yes,
/ConsideredUncertainKtaTemp756InMem/.initial=Yes,
/ConsideredUncertainKtaTemp757InMem/.initial=Yes,
/ConsideredUncertainKtaTemp758InMem/.initial=Yes,
/ConsideredUncertainKtaTemp759InMem/.initial=Yes,
/ConsideredUncertainKtaTemp760InMem/.initial=Yes,
/ConsideredUncertainKtaTemp761InMem/.initial=Yes,
/ConsideredUncertainKtaTemp762InMem/.initial=Yes,
/ConsideredUncertainKtaTemp763InMem/.initial=Yes,
/ConsideredUncertainKtaTemp764InMem/.initial=Yes,
/ConsideredUncertainKtaTemp765InMem/.initial=Yes,
/ConsideredUncertainKtaTemp766InMem/.initial=Yes,
/ConsideredUncertainKtaTemp767InMem/.initial=Yes,
/ConsideredUncertainKvRoCo/.initial=Yes,
/ConsideredUncertainKvReCo/.initial=Yes,
/ConsideredUncertainKvRoCe/.initial=Yes,
/ConsideredUncertainKvReCe/.initial=Yes,
/ConsideredUncertainKvTemp0InMem/.initial=Yes,
/ConsideredUncertainKvTemp1InMem/.initial=Yes,
/ConsideredUncertainKvTemp2InMem/.initial=Yes,
/ConsideredUncertainKvTemp3InMem/.initial=Yes,
/ConsideredUncertainKvTemp4InMem/.initial=Yes,
/ConsideredUncertainKvTemp5InMem/.initial=Yes,
/ConsideredUncertainKvTemp6InMem/.initial=Yes,
/ConsideredUncertainKvTemp7InMem/.initial=Yes,
/ConsideredUncertainKvTemp8InMem/.initial=Yes,
/ConsideredUncertainKvTemp9InMem/.initial=Yes,
/ConsideredUncertainKvTemp10InMem/.initial=Yes,
/ConsideredUncertainKvTemp11InMem/.initial=Yes,
/ConsideredUncertainKvTemp12InMem/.initial=Yes,
/ConsideredUncertainKvTemp13InMem/.initial=Yes,
/ConsideredUncertainKvTemp14InMem/.initial=Yes,
/ConsideredUncertainKvTemp15InMem/.initial=Yes,
/ConsideredUncertainKvTemp16InMem/.initial=Yes,
/ConsideredUncertainKvTemp17InMem/.initial=Yes,
/ConsideredUncertainKvTemp18InMem/.initial=Yes,
/ConsideredUncertainKvTemp19InMem/.initial=Yes,
/ConsideredUncertainKvTemp20InMem/.initial=Yes,
/ConsideredUncertainKvTemp21InMem/.initial=Yes,
/ConsideredUncertainKvTemp22InMem/.initial=Yes,
/ConsideredUncertainKvTemp23InMem/.initial=Yes,
/ConsideredUncertainKvTemp24InMem/.initial=Yes,
/ConsideredUncertainKvTemp25InMem/.initial=Yes,
/ConsideredUncertainKvTemp26InMem/.initial=Yes,
/ConsideredUncertainKvTemp27InMem/.initial=Yes,
/ConsideredUncertainKvTemp28InMem/.initial=Yes,
/ConsideredUncertainKvTemp29InMem/.initial=Yes,
/ConsideredUncertainKvTemp30InMem/.initial=Yes,
/ConsideredUncertainKvTemp31InMem/.initial=Yes,
/ConsideredUncertainKvTemp32InMem/.initial=Yes,
/ConsideredUncertainKvTemp33InMem/.initial=Yes,
/ConsideredUncertainKvTemp34InMem/.initial=Yes,
/ConsideredUncertainKvTemp35InMem/.initial=Yes,
/ConsideredUncertainKvTemp36InMem/.initial=Yes,
/ConsideredUncertainKvTemp37InMem/.initial=Yes,
/ConsideredUncertainKvTemp38InMem/.initial=Yes,
/ConsideredUncertainKvTemp39InMem/.initial=Yes,
/ConsideredUncertainKvTemp40InMem/.initial=Yes,
/ConsideredUncertainKvTemp41InMem/.initial=Yes,
/ConsideredUncertainKvTemp42InMem/.initial=Yes,
/ConsideredUncertainKvTemp43InMem/.initial=Yes,
/ConsideredUncertainKvTemp44InMem/.initial=Yes,
/ConsideredUncertainKvTemp45InMem/.initial=Yes,
/ConsideredUncertainKvTemp46InMem/.initial=Yes,
/ConsideredUncertainKvTemp47InMem/.initial=Yes,
/ConsideredUncertainKvTemp48InMem/.initial=Yes,
/ConsideredUncertainKvTemp49InMem/.initial=Yes,
/ConsideredUncertainKvTemp50InMem/.initial=Yes,
/ConsideredUncertainKvTemp51InMem/.initial=Yes,
/ConsideredUncertainKvTemp52InMem/.initial=Yes,
/ConsideredUncertainKvTemp53InMem/.initial=Yes,
/ConsideredUncertainKvTemp54InMem/.initial=Yes,
/ConsideredUncertainKvTemp55InMem/.initial=Yes,
/ConsideredUncertainKvTemp56InMem/.initial=Yes,
/ConsideredUncertainKvTemp57InMem/.initial=Yes,
/ConsideredUncertainKvTemp58InMem/.initial=Yes,
/ConsideredUncertainKvTemp59InMem/.initial=Yes,
/ConsideredUncertainKvTemp60InMem/.initial=Yes,
/ConsideredUncertainKvTemp61InMem/.initial=Yes,
/ConsideredUncertainKvTemp62InMem/.initial=Yes,
/ConsideredUncertainKvTemp63InMem/.initial=Yes,
/ConsideredUncertainKvTemp64InMem/.initial=Yes,
/ConsideredUncertainKvTemp65InMem/.initial=Yes,
/ConsideredUncertainKvTemp66InMem/.initial=Yes,
/ConsideredUncertainKvTemp67InMem/.initial=Yes,
/ConsideredUncertainKvTemp68InMem/.initial=Yes,
/ConsideredUncertainKvTemp69InMem/.initial=Yes,
/ConsideredUncertainKvTemp70InMem/.initial=Yes,
/ConsideredUncertainKvTemp71InMem/.initial=Yes,
/ConsideredUncertainKvTemp72InMem/.initial=Yes,
/ConsideredUncertainKvTemp73InMem/.initial=Yes,
/ConsideredUncertainKvTemp74InMem/.initial=Yes,
/ConsideredUncertainKvTemp75InMem/.initial=Yes,
/ConsideredUncertainKvTemp76InMem/.initial=Yes,
/ConsideredUncertainKvTemp77InMem/.initial=Yes,
/ConsideredUncertainKvTemp78InMem/.initial=Yes,
/ConsideredUncertainKvTemp79InMem/.initial=Yes,
/ConsideredUncertainKvTemp80InMem/.initial=Yes,
/ConsideredUncertainKvTemp81InMem/.initial=Yes,
/ConsideredUncertainKvTemp82InMem/.initial=Yes,
/ConsideredUncertainKvTemp83InMem/.initial=Yes,
/ConsideredUncertainKvTemp84InMem/.initial=Yes,
/ConsideredUncertainKvTemp85InMem/.initial=Yes,
/ConsideredUncertainKvTemp86InMem/.initial=Yes,
/ConsideredUncertainKvTemp87InMem/.initial=Yes,
/ConsideredUncertainKvTemp88InMem/.initial=Yes,
/ConsideredUncertainKvTemp89InMem/.initial=Yes,
/ConsideredUncertainKvTemp90InMem/.initial=Yes,
/ConsideredUncertainKvTemp91InMem/.initial=Yes,
/ConsideredUncertainKvTemp92InMem/.initial=Yes,
/ConsideredUncertainKvTemp93InMem/.initial=Yes,
/ConsideredUncertainKvTemp94InMem/.initial=Yes,
/ConsideredUncertainKvTemp95InMem/.initial=Yes,
/ConsideredUncertainKvTemp96InMem/.initial=Yes,
/ConsideredUncertainKvTemp97InMem/.initial=Yes,
/ConsideredUncertainKvTemp98InMem/.initial=Yes,
/ConsideredUncertainKvTemp99InMem/.initial=Yes,
/ConsideredUncertainKvTemp100InMem/.initial=Yes,
/ConsideredUncertainKvTemp101InMem/.initial=Yes,
/ConsideredUncertainKvTemp102InMem/.initial=Yes,
/ConsideredUncertainKvTemp103InMem/.initial=Yes,
/ConsideredUncertainKvTemp104InMem/.initial=Yes,
/ConsideredUncertainKvTemp105InMem/.initial=Yes,
/ConsideredUncertainKvTemp106InMem/.initial=Yes,
/ConsideredUncertainKvTemp107InMem/.initial=Yes,
/ConsideredUncertainKvTemp108InMem/.initial=Yes,
/ConsideredUncertainKvTemp109InMem/.initial=Yes,
/ConsideredUncertainKvTemp110InMem/.initial=Yes,
/ConsideredUncertainKvTemp111InMem/.initial=Yes,
/ConsideredUncertainKvTemp112InMem/.initial=Yes,
/ConsideredUncertainKvTemp113InMem/.initial=Yes,
/ConsideredUncertainKvTemp114InMem/.initial=Yes,
/ConsideredUncertainKvTemp115InMem/.initial=Yes,
/ConsideredUncertainKvTemp116InMem/.initial=Yes,
/ConsideredUncertainKvTemp117InMem/.initial=Yes,
/ConsideredUncertainKvTemp118InMem/.initial=Yes,
/ConsideredUncertainKvTemp119InMem/.initial=Yes,
/ConsideredUncertainKvTemp120InMem/.initial=Yes,
/ConsideredUncertainKvTemp121InMem/.initial=Yes,
/ConsideredUncertainKvTemp122InMem/.initial=Yes,
/ConsideredUncertainKvTemp123InMem/.initial=Yes,
/ConsideredUncertainKvTemp124InMem/.initial=Yes,
/ConsideredUncertainKvTemp125InMem/.initial=Yes,
/ConsideredUncertainKvTemp126InMem/.initial=Yes,
/ConsideredUncertainKvTemp127InMem/.initial=Yes,
/ConsideredUncertainKvTemp128InMem/.initial=Yes,
/ConsideredUncertainKvTemp129InMem/.initial=Yes,
/ConsideredUncertainKvTemp130InMem/.initial=Yes,
/ConsideredUncertainKvTemp131InMem/.initial=Yes,
/ConsideredUncertainKvTemp132InMem/.initial=Yes,
/ConsideredUncertainKvTemp133InMem/.initial=Yes,
/ConsideredUncertainKvTemp134InMem/.initial=Yes,
/ConsideredUncertainKvTemp135InMem/.initial=Yes,
/ConsideredUncertainKvTemp136InMem/.initial=Yes,
/ConsideredUncertainKvTemp137InMem/.initial=Yes,
/ConsideredUncertainKvTemp138InMem/.initial=Yes,
/ConsideredUncertainKvTemp139InMem/.initial=Yes,
/ConsideredUncertainKvTemp140InMem/.initial=Yes,
/ConsideredUncertainKvTemp141InMem/.initial=Yes,
/ConsideredUncertainKvTemp142InMem/.initial=Yes,
/ConsideredUncertainKvTemp143InMem/.initial=Yes,
/ConsideredUncertainKvTemp144InMem/.initial=Yes,
/ConsideredUncertainKvTemp145InMem/.initial=Yes,
/ConsideredUncertainKvTemp146InMem/.initial=Yes,
/ConsideredUncertainKvTemp147InMem/.initial=Yes,
/ConsideredUncertainKvTemp148InMem/.initial=Yes,
/ConsideredUncertainKvTemp149InMem/.initial=Yes,
/ConsideredUncertainKvTemp150InMem/.initial=Yes,
/ConsideredUncertainKvTemp151InMem/.initial=Yes,
/ConsideredUncertainKvTemp152InMem/.initial=Yes,
/ConsideredUncertainKvTemp153InMem/.initial=Yes,
/ConsideredUncertainKvTemp154InMem/.initial=Yes,
/ConsideredUncertainKvTemp155InMem/.initial=Yes,
/ConsideredUncertainKvTemp156InMem/.initial=Yes,
/ConsideredUncertainKvTemp157InMem/.initial=Yes,
/ConsideredUncertainKvTemp158InMem/.initial=Yes,
/ConsideredUncertainKvTemp159InMem/.initial=Yes,
/ConsideredUncertainKvTemp160InMem/.initial=Yes,
/ConsideredUncertainKvTemp161InMem/.initial=Yes,
/ConsideredUncertainKvTemp162InMem/.initial=Yes,
/ConsideredUncertainKvTemp163InMem/.initial=Yes,
/ConsideredUncertainKvTemp164InMem/.initial=Yes,
/ConsideredUncertainKvTemp165InMem/.initial=Yes,
/ConsideredUncertainKvTemp166InMem/.initial=Yes,
/ConsideredUncertainKvTemp167InMem/.initial=Yes,
/ConsideredUncertainKvTemp168InMem/.initial=Yes,
/ConsideredUncertainKvTemp169InMem/.initial=Yes,
/ConsideredUncertainKvTemp170InMem/.initial=Yes,
/ConsideredUncertainKvTemp171InMem/.initial=Yes,
/ConsideredUncertainKvTemp172InMem/.initial=Yes,
/ConsideredUncertainKvTemp173InMem/.initial=Yes,
/ConsideredUncertainKvTemp174InMem/.initial=Yes,
/ConsideredUncertainKvTemp175InMem/.initial=Yes,
/ConsideredUncertainKvTemp176InMem/.initial=Yes,
/ConsideredUncertainKvTemp177InMem/.initial=Yes,
/ConsideredUncertainKvTemp178InMem/.initial=Yes,
/ConsideredUncertainKvTemp179InMem/.initial=Yes,
/ConsideredUncertainKvTemp180InMem/.initial=Yes,
/ConsideredUncertainKvTemp181InMem/.initial=Yes,
/ConsideredUncertainKvTemp182InMem/.initial=Yes,
/ConsideredUncertainKvTemp183InMem/.initial=Yes,
/ConsideredUncertainKvTemp184InMem/.initial=Yes,
/ConsideredUncertainKvTemp185InMem/.initial=Yes,
/ConsideredUncertainKvTemp186InMem/.initial=Yes,
/ConsideredUncertainKvTemp187InMem/.initial=Yes,
/ConsideredUncertainKvTemp188InMem/.initial=Yes,
/ConsideredUncertainKvTemp189InMem/.initial=Yes,
/ConsideredUncertainKvTemp190InMem/.initial=Yes,
/ConsideredUncertainKvTemp191InMem/.initial=Yes,
/ConsideredUncertainKvTemp192InMem/.initial=Yes,
/ConsideredUncertainKvTemp193InMem/.initial=Yes,
/ConsideredUncertainKvTemp194InMem/.initial=Yes,
/ConsideredUncertainKvTemp195InMem/.initial=Yes,
/ConsideredUncertainKvTemp196InMem/.initial=Yes,
/ConsideredUncertainKvTemp197InMem/.initial=Yes,
/ConsideredUncertainKvTemp198InMem/.initial=Yes,
/ConsideredUncertainKvTemp199InMem/.initial=Yes,
/ConsideredUncertainKvTemp200InMem/.initial=Yes,
/ConsideredUncertainKvTemp201InMem/.initial=Yes,
/ConsideredUncertainKvTemp202InMem/.initial=Yes,
/ConsideredUncertainKvTemp203InMem/.initial=Yes,
/ConsideredUncertainKvTemp204InMem/.initial=Yes,
/ConsideredUncertainKvTemp205InMem/.initial=Yes,
/ConsideredUncertainKvTemp206InMem/.initial=Yes,
/ConsideredUncertainKvTemp207InMem/.initial=Yes,
/ConsideredUncertainKvTemp208InMem/.initial=Yes,
/ConsideredUncertainKvTemp209InMem/.initial=Yes,
/ConsideredUncertainKvTemp210InMem/.initial=Yes,
/ConsideredUncertainKvTemp211InMem/.initial=Yes,
/ConsideredUncertainKvTemp212InMem/.initial=Yes,
/ConsideredUncertainKvTemp213InMem/.initial=Yes,
/ConsideredUncertainKvTemp214InMem/.initial=Yes,
/ConsideredUncertainKvTemp215InMem/.initial=Yes,
/ConsideredUncertainKvTemp216InMem/.initial=Yes,
/ConsideredUncertainKvTemp217InMem/.initial=Yes,
/ConsideredUncertainKvTemp218InMem/.initial=Yes,
/ConsideredUncertainKvTemp219InMem/.initial=Yes,
/ConsideredUncertainKvTemp220InMem/.initial=Yes,
/ConsideredUncertainKvTemp221InMem/.initial=Yes,
/ConsideredUncertainKvTemp222InMem/.initial=Yes,
/ConsideredUncertainKvTemp223InMem/.initial=Yes,
/ConsideredUncertainKvTemp224InMem/.initial=Yes,
/ConsideredUncertainKvTemp225InMem/.initial=Yes,
/ConsideredUncertainKvTemp226InMem/.initial=Yes,
/ConsideredUncertainKvTemp227InMem/.initial=Yes,
/ConsideredUncertainKvTemp228InMem/.initial=Yes,
/ConsideredUncertainKvTemp229InMem/.initial=Yes,
/ConsideredUncertainKvTemp230InMem/.initial=Yes,
/ConsideredUncertainKvTemp231InMem/.initial=Yes,
/ConsideredUncertainKvTemp232InMem/.initial=Yes,
/ConsideredUncertainKvTemp233InMem/.initial=Yes,
/ConsideredUncertainKvTemp234InMem/.initial=Yes,
/ConsideredUncertainKvTemp235InMem/.initial=Yes,
/ConsideredUncertainKvTemp236InMem/.initial=Yes,
/ConsideredUncertainKvTemp237InMem/.initial=Yes,
/ConsideredUncertainKvTemp238InMem/.initial=Yes,
/ConsideredUncertainKvTemp239InMem/.initial=Yes,
/ConsideredUncertainKvTemp240InMem/.initial=Yes,
/ConsideredUncertainKvTemp241InMem/.initial=Yes,
/ConsideredUncertainKvTemp242InMem/.initial=Yes,
/ConsideredUncertainKvTemp243InMem/.initial=Yes,
/ConsideredUncertainKvTemp244InMem/.initial=Yes,
/ConsideredUncertainKvTemp245InMem/.initial=Yes,
/ConsideredUncertainKvTemp246InMem/.initial=Yes,
/ConsideredUncertainKvTemp247InMem/.initial=Yes,
/ConsideredUncertainKvTemp248InMem/.initial=Yes,
/ConsideredUncertainKvTemp249InMem/.initial=Yes,
/ConsideredUncertainKvTemp250InMem/.initial=Yes,
/ConsideredUncertainKvTemp251InMem/.initial=Yes,
/ConsideredUncertainKvTemp252InMem/.initial=Yes,
/ConsideredUncertainKvTemp253InMem/.initial=Yes,
/ConsideredUncertainKvTemp254InMem/.initial=Yes,
/ConsideredUncertainKvTemp255InMem/.initial=Yes,
/ConsideredUncertainKvTemp256InMem/.initial=Yes,
/ConsideredUncertainKvTemp257InMem/.initial=Yes,
/ConsideredUncertainKvTemp258InMem/.initial=Yes,
/ConsideredUncertainKvTemp259InMem/.initial=Yes,
/ConsideredUncertainKvTemp260InMem/.initial=Yes,
/ConsideredUncertainKvTemp261InMem/.initial=Yes,
/ConsideredUncertainKvTemp262InMem/.initial=Yes,
/ConsideredUncertainKvTemp263InMem/.initial=Yes,
/ConsideredUncertainKvTemp264InMem/.initial=Yes,
/ConsideredUncertainKvTemp265InMem/.initial=Yes,
/ConsideredUncertainKvTemp266InMem/.initial=Yes,
/ConsideredUncertainKvTemp267InMem/.initial=Yes,
/ConsideredUncertainKvTemp268InMem/.initial=Yes,
/ConsideredUncertainKvTemp269InMem/.initial=Yes,
/ConsideredUncertainKvTemp270InMem/.initial=Yes,
/ConsideredUncertainKvTemp271InMem/.initial=Yes,
/ConsideredUncertainKvTemp272InMem/.initial=Yes,
/ConsideredUncertainKvTemp273InMem/.initial=Yes,
/ConsideredUncertainKvTemp274InMem/.initial=Yes,
/ConsideredUncertainKvTemp275InMem/.initial=Yes,
/ConsideredUncertainKvTemp276InMem/.initial=Yes,
/ConsideredUncertainKvTemp277InMem/.initial=Yes,
/ConsideredUncertainKvTemp278InMem/.initial=Yes,
/ConsideredUncertainKvTemp279InMem/.initial=Yes,
/ConsideredUncertainKvTemp280InMem/.initial=Yes,
/ConsideredUncertainKvTemp281InMem/.initial=Yes,
/ConsideredUncertainKvTemp282InMem/.initial=Yes,
/ConsideredUncertainKvTemp283InMem/.initial=Yes,
/ConsideredUncertainKvTemp284InMem/.initial=Yes,
/ConsideredUncertainKvTemp285InMem/.initial=Yes,
/ConsideredUncertainKvTemp286InMem/.initial=Yes,
/ConsideredUncertainKvTemp287InMem/.initial=Yes,
/ConsideredUncertainKvTemp288InMem/.initial=Yes,
/ConsideredUncertainKvTemp289InMem/.initial=Yes,
/ConsideredUncertainKvTemp290InMem/.initial=Yes,
/ConsideredUncertainKvTemp291InMem/.initial=Yes,
/ConsideredUncertainKvTemp292InMem/.initial=Yes,
/ConsideredUncertainKvTemp293InMem/.initial=Yes,
/ConsideredUncertainKvTemp294InMem/.initial=Yes,
/ConsideredUncertainKvTemp295InMem/.initial=Yes,
/ConsideredUncertainKvTemp296InMem/.initial=Yes,
/ConsideredUncertainKvTemp297InMem/.initial=Yes,
/ConsideredUncertainKvTemp298InMem/.initial=Yes,
/ConsideredUncertainKvTemp299InMem/.initial=Yes,
/ConsideredUncertainKvTemp300InMem/.initial=Yes,
/ConsideredUncertainKvTemp301InMem/.initial=Yes,
/ConsideredUncertainKvTemp302InMem/.initial=Yes,
/ConsideredUncertainKvTemp303InMem/.initial=Yes,
/ConsideredUncertainKvTemp304InMem/.initial=Yes,
/ConsideredUncertainKvTemp305InMem/.initial=Yes,
/ConsideredUncertainKvTemp306InMem/.initial=Yes,
/ConsideredUncertainKvTemp307InMem/.initial=Yes,
/ConsideredUncertainKvTemp308InMem/.initial=Yes,
/ConsideredUncertainKvTemp309InMem/.initial=Yes,
/ConsideredUncertainKvTemp310InMem/.initial=Yes,
/ConsideredUncertainKvTemp311InMem/.initial=Yes,
/ConsideredUncertainKvTemp312InMem/.initial=Yes,
/ConsideredUncertainKvTemp313InMem/.initial=Yes,
/ConsideredUncertainKvTemp314InMem/.initial=Yes,
/ConsideredUncertainKvTemp315InMem/.initial=Yes,
/ConsideredUncertainKvTemp316InMem/.initial=Yes,
/ConsideredUncertainKvTemp317InMem/.initial=Yes,
/ConsideredUncertainKvTemp318InMem/.initial=Yes,
/ConsideredUncertainKvTemp319InMem/.initial=Yes,
/ConsideredUncertainKvTemp320InMem/.initial=Yes,
/ConsideredUncertainKvTemp321InMem/.initial=Yes,
/ConsideredUncertainKvTemp322InMem/.initial=Yes,
/ConsideredUncertainKvTemp323InMem/.initial=Yes,
/ConsideredUncertainKvTemp324InMem/.initial=Yes,
/ConsideredUncertainKvTemp325InMem/.initial=Yes,
/ConsideredUncertainKvTemp326InMem/.initial=Yes,
/ConsideredUncertainKvTemp327InMem/.initial=Yes,
/ConsideredUncertainKvTemp328InMem/.initial=Yes,
/ConsideredUncertainKvTemp329InMem/.initial=Yes,
/ConsideredUncertainKvTemp330InMem/.initial=Yes,
/ConsideredUncertainKvTemp331InMem/.initial=Yes,
/ConsideredUncertainKvTemp332InMem/.initial=Yes,
/ConsideredUncertainKvTemp333InMem/.initial=Yes,
/ConsideredUncertainKvTemp334InMem/.initial=Yes,
/ConsideredUncertainKvTemp335InMem/.initial=Yes,
/ConsideredUncertainKvTemp336InMem/.initial=Yes,
/ConsideredUncertainKvTemp337InMem/.initial=Yes,
/ConsideredUncertainKvTemp338InMem/.initial=Yes,
/ConsideredUncertainKvTemp339InMem/.initial=Yes,
/ConsideredUncertainKvTemp340InMem/.initial=Yes,
/ConsideredUncertainKvTemp341InMem/.initial=Yes,
/ConsideredUncertainKvTemp342InMem/.initial=Yes,
/ConsideredUncertainKvTemp343InMem/.initial=Yes,
/ConsideredUncertainKvTemp344InMem/.initial=Yes,
/ConsideredUncertainKvTemp345InMem/.initial=Yes,
/ConsideredUncertainKvTemp346InMem/.initial=Yes,
/ConsideredUncertainKvTemp347InMem/.initial=Yes,
/ConsideredUncertainKvTemp348InMem/.initial=Yes,
/ConsideredUncertainKvTemp349InMem/.initial=Yes,
/ConsideredUncertainKvTemp350InMem/.initial=Yes,
/ConsideredUncertainKvTemp351InMem/.initial=Yes,
/ConsideredUncertainKvTemp352InMem/.initial=Yes,
/ConsideredUncertainKvTemp353InMem/.initial=Yes,
/ConsideredUncertainKvTemp354InMem/.initial=Yes,
/ConsideredUncertainKvTemp355InMem/.initial=Yes,
/ConsideredUncertainKvTemp356InMem/.initial=Yes,
/ConsideredUncertainKvTemp357InMem/.initial=Yes,
/ConsideredUncertainKvTemp358InMem/.initial=Yes,
/ConsideredUncertainKvTemp359InMem/.initial=Yes,
/ConsideredUncertainKvTemp360InMem/.initial=Yes,
/ConsideredUncertainKvTemp361InMem/.initial=Yes,
/ConsideredUncertainKvTemp362InMem/.initial=Yes,
/ConsideredUncertainKvTemp363InMem/.initial=Yes,
/ConsideredUncertainKvTemp364InMem/.initial=Yes,
/ConsideredUncertainKvTemp365InMem/.initial=Yes,
/ConsideredUncertainKvTemp366InMem/.initial=Yes,
/ConsideredUncertainKvTemp367InMem/.initial=Yes,
/ConsideredUncertainKvTemp368InMem/.initial=Yes,
/ConsideredUncertainKvTemp369InMem/.initial=Yes,
/ConsideredUncertainKvTemp370InMem/.initial=Yes,
/ConsideredUncertainKvTemp371InMem/.initial=Yes,
/ConsideredUncertainKvTemp372InMem/.initial=Yes,
/ConsideredUncertainKvTemp373InMem/.initial=Yes,
/ConsideredUncertainKvTemp374InMem/.initial=Yes,
/ConsideredUncertainKvTemp375InMem/.initial=Yes,
/ConsideredUncertainKvTemp376InMem/.initial=Yes,
/ConsideredUncertainKvTemp377InMem/.initial=Yes,
/ConsideredUncertainKvTemp378InMem/.initial=Yes,
/ConsideredUncertainKvTemp379InMem/.initial=Yes,
/ConsideredUncertainKvTemp380InMem/.initial=Yes,
/ConsideredUncertainKvTemp381InMem/.initial=Yes,
/ConsideredUncertainKvTemp382InMem/.initial=Yes,
/ConsideredUncertainKvTemp383InMem/.initial=Yes,
/ConsideredUncertainKvTemp384InMem/.initial=Yes,
/ConsideredUncertainKvTemp385InMem/.initial=Yes,
/ConsideredUncertainKvTemp386InMem/.initial=Yes,
/ConsideredUncertainKvTemp387InMem/.initial=Yes,
/ConsideredUncertainKvTemp388InMem/.initial=Yes,
/ConsideredUncertainKvTemp389InMem/.initial=Yes,
/ConsideredUncertainKvTemp390InMem/.initial=Yes,
/ConsideredUncertainKvTemp391InMem/.initial=Yes,
/ConsideredUncertainKvTemp392InMem/.initial=Yes,
/ConsideredUncertainKvTemp393InMem/.initial=Yes,
/ConsideredUncertainKvTemp394InMem/.initial=Yes,
/ConsideredUncertainKvTemp395InMem/.initial=Yes,
/ConsideredUncertainKvTemp396InMem/.initial=Yes,
/ConsideredUncertainKvTemp397InMem/.initial=Yes,
/ConsideredUncertainKvTemp398InMem/.initial=Yes,
/ConsideredUncertainKvTemp399InMem/.initial=Yes,
/ConsideredUncertainKvTemp400InMem/.initial=Yes,
/ConsideredUncertainKvTemp401InMem/.initial=Yes,
/ConsideredUncertainKvTemp402InMem/.initial=Yes,
/ConsideredUncertainKvTemp403InMem/.initial=Yes,
/ConsideredUncertainKvTemp404InMem/.initial=Yes,
/ConsideredUncertainKvTemp405InMem/.initial=Yes,
/ConsideredUncertainKvTemp406InMem/.initial=Yes,
/ConsideredUncertainKvTemp407InMem/.initial=Yes,
/ConsideredUncertainKvTemp408InMem/.initial=Yes,
/ConsideredUncertainKvTemp409InMem/.initial=Yes,
/ConsideredUncertainKvTemp410InMem/.initial=Yes,
/ConsideredUncertainKvTemp411InMem/.initial=Yes,
/ConsideredUncertainKvTemp412InMem/.initial=Yes,
/ConsideredUncertainKvTemp413InMem/.initial=Yes,
/ConsideredUncertainKvTemp414InMem/.initial=Yes,
/ConsideredUncertainKvTemp415InMem/.initial=Yes,
/ConsideredUncertainKvTemp416InMem/.initial=Yes,
/ConsideredUncertainKvTemp417InMem/.initial=Yes,
/ConsideredUncertainKvTemp418InMem/.initial=Yes,
/ConsideredUncertainKvTemp419InMem/.initial=Yes,
/ConsideredUncertainKvTemp420InMem/.initial=Yes,
/ConsideredUncertainKvTemp421InMem/.initial=Yes,
/ConsideredUncertainKvTemp422InMem/.initial=Yes,
/ConsideredUncertainKvTemp423InMem/.initial=Yes,
/ConsideredUncertainKvTemp424InMem/.initial=Yes,
/ConsideredUncertainKvTemp425InMem/.initial=Yes,
/ConsideredUncertainKvTemp426InMem/.initial=Yes,
/ConsideredUncertainKvTemp427InMem/.initial=Yes,
/ConsideredUncertainKvTemp428InMem/.initial=Yes,
/ConsideredUncertainKvTemp429InMem/.initial=Yes,
/ConsideredUncertainKvTemp430InMem/.initial=Yes,
/ConsideredUncertainKvTemp431InMem/.initial=Yes,
/ConsideredUncertainKvTemp432InMem/.initial=Yes,
/ConsideredUncertainKvTemp433InMem/.initial=Yes,
/ConsideredUncertainKvTemp434InMem/.initial=Yes,
/ConsideredUncertainKvTemp435InMem/.initial=Yes,
/ConsideredUncertainKvTemp436InMem/.initial=Yes,
/ConsideredUncertainKvTemp437InMem/.initial=Yes,
/ConsideredUncertainKvTemp438InMem/.initial=Yes,
/ConsideredUncertainKvTemp439InMem/.initial=Yes,
/ConsideredUncertainKvTemp440InMem/.initial=Yes,
/ConsideredUncertainKvTemp441InMem/.initial=Yes,
/ConsideredUncertainKvTemp442InMem/.initial=Yes,
/ConsideredUncertainKvTemp443InMem/.initial=Yes,
/ConsideredUncertainKvTemp444InMem/.initial=Yes,
/ConsideredUncertainKvTemp445InMem/.initial=Yes,
/ConsideredUncertainKvTemp446InMem/.initial=Yes,
/ConsideredUncertainKvTemp447InMem/.initial=Yes,
/ConsideredUncertainKvTemp448InMem/.initial=Yes,
/ConsideredUncertainKvTemp449InMem/.initial=Yes,
/ConsideredUncertainKvTemp450InMem/.initial=Yes,
/ConsideredUncertainKvTemp451InMem/.initial=Yes,
/ConsideredUncertainKvTemp452InMem/.initial=Yes,
/ConsideredUncertainKvTemp453InMem/.initial=Yes,
/ConsideredUncertainKvTemp454InMem/.initial=Yes,
/ConsideredUncertainKvTemp455InMem/.initial=Yes,
/ConsideredUncertainKvTemp456InMem/.initial=Yes,
/ConsideredUncertainKvTemp457InMem/.initial=Yes,
/ConsideredUncertainKvTemp458InMem/.initial=Yes,
/ConsideredUncertainKvTemp459InMem/.initial=Yes,
/ConsideredUncertainKvTemp460InMem/.initial=Yes,
/ConsideredUncertainKvTemp461InMem/.initial=Yes,
/ConsideredUncertainKvTemp462InMem/.initial=Yes,
/ConsideredUncertainKvTemp463InMem/.initial=Yes,
/ConsideredUncertainKvTemp464InMem/.initial=Yes,
/ConsideredUncertainKvTemp465InMem/.initial=Yes,
/ConsideredUncertainKvTemp466InMem/.initial=Yes,
/ConsideredUncertainKvTemp467InMem/.initial=Yes,
/ConsideredUncertainKvTemp468InMem/.initial=Yes,
/ConsideredUncertainKvTemp469InMem/.initial=Yes,
/ConsideredUncertainKvTemp470InMem/.initial=Yes,
/ConsideredUncertainKvTemp471InMem/.initial=Yes,
/ConsideredUncertainKvTemp472InMem/.initial=Yes,
/ConsideredUncertainKvTemp473InMem/.initial=Yes,
/ConsideredUncertainKvTemp474InMem/.initial=Yes,
/ConsideredUncertainKvTemp475InMem/.initial=Yes,
/ConsideredUncertainKvTemp476InMem/.initial=Yes,
/ConsideredUncertainKvTemp477InMem/.initial=Yes,
/ConsideredUncertainKvTemp478InMem/.initial=Yes,
/ConsideredUncertainKvTemp479InMem/.initial=Yes,
/ConsideredUncertainKvTemp480InMem/.initial=Yes,
/ConsideredUncertainKvTemp481InMem/.initial=Yes,
/ConsideredUncertainKvTemp482InMem/.initial=Yes,
/ConsideredUncertainKvTemp483InMem/.initial=Yes,
/ConsideredUncertainKvTemp484InMem/.initial=Yes,
/ConsideredUncertainKvTemp485InMem/.initial=Yes,
/ConsideredUncertainKvTemp486InMem/.initial=Yes,
/ConsideredUncertainKvTemp487InMem/.initial=Yes,
/ConsideredUncertainKvTemp488InMem/.initial=Yes,
/ConsideredUncertainKvTemp489InMem/.initial=Yes,
/ConsideredUncertainKvTemp490InMem/.initial=Yes,
/ConsideredUncertainKvTemp491InMem/.initial=Yes,
/ConsideredUncertainKvTemp492InMem/.initial=Yes,
/ConsideredUncertainKvTemp493InMem/.initial=Yes,
/ConsideredUncertainKvTemp494InMem/.initial=Yes,
/ConsideredUncertainKvTemp495InMem/.initial=Yes,
/ConsideredUncertainKvTemp496InMem/.initial=Yes,
/ConsideredUncertainKvTemp497InMem/.initial=Yes,
/ConsideredUncertainKvTemp498InMem/.initial=Yes,
/ConsideredUncertainKvTemp499InMem/.initial=Yes,
/ConsideredUncertainKvTemp500InMem/.initial=Yes,
/ConsideredUncertainKvTemp501InMem/.initial=Yes,
/ConsideredUncertainKvTemp502InMem/.initial=Yes,
/ConsideredUncertainKvTemp503InMem/.initial=Yes,
/ConsideredUncertainKvTemp504InMem/.initial=Yes,
/ConsideredUncertainKvTemp505InMem/.initial=Yes,
/ConsideredUncertainKvTemp506InMem/.initial=Yes,
/ConsideredUncertainKvTemp507InMem/.initial=Yes,
/ConsideredUncertainKvTemp508InMem/.initial=Yes,
/ConsideredUncertainKvTemp509InMem/.initial=Yes,
/ConsideredUncertainKvTemp510InMem/.initial=Yes,
/ConsideredUncertainKvTemp511InMem/.initial=Yes,
/ConsideredUncertainKvTemp512InMem/.initial=Yes,
/ConsideredUncertainKvTemp513InMem/.initial=Yes,
/ConsideredUncertainKvTemp514InMem/.initial=Yes,
/ConsideredUncertainKvTemp515InMem/.initial=Yes,
/ConsideredUncertainKvTemp516InMem/.initial=Yes,
/ConsideredUncertainKvTemp517InMem/.initial=Yes,
/ConsideredUncertainKvTemp518InMem/.initial=Yes,
/ConsideredUncertainKvTemp519InMem/.initial=Yes,
/ConsideredUncertainKvTemp520InMem/.initial=Yes,
/ConsideredUncertainKvTemp521InMem/.initial=Yes,
/ConsideredUncertainKvTemp522InMem/.initial=Yes,
/ConsideredUncertainKvTemp523InMem/.initial=Yes,
/ConsideredUncertainKvTemp524InMem/.initial=Yes,
/ConsideredUncertainKvTemp525InMem/.initial=Yes,
/ConsideredUncertainKvTemp526InMem/.initial=Yes,
/ConsideredUncertainKvTemp527InMem/.initial=Yes,
/ConsideredUncertainKvTemp528InMem/.initial=Yes,
/ConsideredUncertainKvTemp529InMem/.initial=Yes,
/ConsideredUncertainKvTemp530InMem/.initial=Yes,
/ConsideredUncertainKvTemp531InMem/.initial=Yes,
/ConsideredUncertainKvTemp532InMem/.initial=Yes,
/ConsideredUncertainKvTemp533InMem/.initial=Yes,
/ConsideredUncertainKvTemp534InMem/.initial=Yes,
/ConsideredUncertainKvTemp535InMem/.initial=Yes,
/ConsideredUncertainKvTemp536InMem/.initial=Yes,
/ConsideredUncertainKvTemp537InMem/.initial=Yes,
/ConsideredUncertainKvTemp538InMem/.initial=Yes,
/ConsideredUncertainKvTemp539InMem/.initial=Yes,
/ConsideredUncertainKvTemp540InMem/.initial=Yes,
/ConsideredUncertainKvTemp541InMem/.initial=Yes,
/ConsideredUncertainKvTemp542InMem/.initial=Yes,
/ConsideredUncertainKvTemp543InMem/.initial=Yes,
/ConsideredUncertainKvTemp544InMem/.initial=Yes,
/ConsideredUncertainKvTemp545InMem/.initial=Yes,
/ConsideredUncertainKvTemp546InMem/.initial=Yes,
/ConsideredUncertainKvTemp547InMem/.initial=Yes,
/ConsideredUncertainKvTemp548InMem/.initial=Yes,
/ConsideredUncertainKvTemp549InMem/.initial=Yes,
/ConsideredUncertainKvTemp550InMem/.initial=Yes,
/ConsideredUncertainKvTemp551InMem/.initial=Yes,
/ConsideredUncertainKvTemp552InMem/.initial=Yes,
/ConsideredUncertainKvTemp553InMem/.initial=Yes,
/ConsideredUncertainKvTemp554InMem/.initial=Yes,
/ConsideredUncertainKvTemp555InMem/.initial=Yes,
/ConsideredUncertainKvTemp556InMem/.initial=Yes,
/ConsideredUncertainKvTemp557InMem/.initial=Yes,
/ConsideredUncertainKvTemp558InMem/.initial=Yes,
/ConsideredUncertainKvTemp559InMem/.initial=Yes,
/ConsideredUncertainKvTemp560InMem/.initial=Yes,
/ConsideredUncertainKvTemp561InMem/.initial=Yes,
/ConsideredUncertainKvTemp562InMem/.initial=Yes,
/ConsideredUncertainKvTemp563InMem/.initial=Yes,
/ConsideredUncertainKvTemp564InMem/.initial=Yes,
/ConsideredUncertainKvTemp565InMem/.initial=Yes,
/ConsideredUncertainKvTemp566InMem/.initial=Yes,
/ConsideredUncertainKvTemp567InMem/.initial=Yes,
/ConsideredUncertainKvTemp568InMem/.initial=Yes,
/ConsideredUncertainKvTemp569InMem/.initial=Yes,
/ConsideredUncertainKvTemp570InMem/.initial=Yes,
/ConsideredUncertainKvTemp571InMem/.initial=Yes,
/ConsideredUncertainKvTemp572InMem/.initial=Yes,
/ConsideredUncertainKvTemp573InMem/.initial=Yes,
/ConsideredUncertainKvTemp574InMem/.initial=Yes,
/ConsideredUncertainKvTemp575InMem/.initial=Yes,
/ConsideredUncertainKvTemp576InMem/.initial=Yes,
/ConsideredUncertainKvTemp577InMem/.initial=Yes,
/ConsideredUncertainKvTemp578InMem/.initial=Yes,
/ConsideredUncertainKvTemp579InMem/.initial=Yes,
/ConsideredUncertainKvTemp580InMem/.initial=Yes,
/ConsideredUncertainKvTemp581InMem/.initial=Yes,
/ConsideredUncertainKvTemp582InMem/.initial=Yes,
/ConsideredUncertainKvTemp583InMem/.initial=Yes,
/ConsideredUncertainKvTemp584InMem/.initial=Yes,
/ConsideredUncertainKvTemp585InMem/.initial=Yes,
/ConsideredUncertainKvTemp586InMem/.initial=Yes,
/ConsideredUncertainKvTemp587InMem/.initial=Yes,
/ConsideredUncertainKvTemp588InMem/.initial=Yes,
/ConsideredUncertainKvTemp589InMem/.initial=Yes,
/ConsideredUncertainKvTemp590InMem/.initial=Yes,
/ConsideredUncertainKvTemp591InMem/.initial=Yes,
/ConsideredUncertainKvTemp592InMem/.initial=Yes,
/ConsideredUncertainKvTemp593InMem/.initial=Yes,
/ConsideredUncertainKvTemp594InMem/.initial=Yes,
/ConsideredUncertainKvTemp595InMem/.initial=Yes,
/ConsideredUncertainKvTemp596InMem/.initial=Yes,
/ConsideredUncertainKvTemp597InMem/.initial=Yes,
/ConsideredUncertainKvTemp598InMem/.initial=Yes,
/ConsideredUncertainKvTemp599InMem/.initial=Yes,
/ConsideredUncertainKvTemp600InMem/.initial=Yes,
/ConsideredUncertainKvTemp601InMem/.initial=Yes,
/ConsideredUncertainKvTemp602InMem/.initial=Yes,
/ConsideredUncertainKvTemp603InMem/.initial=Yes,
/ConsideredUncertainKvTemp604InMem/.initial=Yes,
/ConsideredUncertainKvTemp605InMem/.initial=Yes,
/ConsideredUncertainKvTemp606InMem/.initial=Yes,
/ConsideredUncertainKvTemp607InMem/.initial=Yes,
/ConsideredUncertainKvTemp608InMem/.initial=Yes,
/ConsideredUncertainKvTemp609InMem/.initial=Yes,
/ConsideredUncertainKvTemp610InMem/.initial=Yes,
/ConsideredUncertainKvTemp611InMem/.initial=Yes,
/ConsideredUncertainKvTemp612InMem/.initial=Yes,
/ConsideredUncertainKvTemp613InMem/.initial=Yes,
/ConsideredUncertainKvTemp614InMem/.initial=Yes,
/ConsideredUncertainKvTemp615InMem/.initial=Yes,
/ConsideredUncertainKvTemp616InMem/.initial=Yes,
/ConsideredUncertainKvTemp617InMem/.initial=Yes,
/ConsideredUncertainKvTemp618InMem/.initial=Yes,
/ConsideredUncertainKvTemp619InMem/.initial=Yes,
/ConsideredUncertainKvTemp620InMem/.initial=Yes,
/ConsideredUncertainKvTemp621InMem/.initial=Yes,
/ConsideredUncertainKvTemp622InMem/.initial=Yes,
/ConsideredUncertainKvTemp623InMem/.initial=Yes,
/ConsideredUncertainKvTemp624InMem/.initial=Yes,
/ConsideredUncertainKvTemp625InMem/.initial=Yes,
/ConsideredUncertainKvTemp626InMem/.initial=Yes,
/ConsideredUncertainKvTemp627InMem/.initial=Yes,
/ConsideredUncertainKvTemp628InMem/.initial=Yes,
/ConsideredUncertainKvTemp629InMem/.initial=Yes,
/ConsideredUncertainKvTemp630InMem/.initial=Yes,
/ConsideredUncertainKvTemp631InMem/.initial=Yes,
/ConsideredUncertainKvTemp632InMem/.initial=Yes,
/ConsideredUncertainKvTemp633InMem/.initial=Yes,
/ConsideredUncertainKvTemp634InMem/.initial=Yes,
/ConsideredUncertainKvTemp635InMem/.initial=Yes,
/ConsideredUncertainKvTemp636InMem/.initial=Yes,
/ConsideredUncertainKvTemp637InMem/.initial=Yes,
/ConsideredUncertainKvTemp638InMem/.initial=Yes,
/ConsideredUncertainKvTemp639InMem/.initial=Yes,
/ConsideredUncertainKvTemp640InMem/.initial=Yes,
/ConsideredUncertainKvTemp641InMem/.initial=Yes,
/ConsideredUncertainKvTemp642InMem/.initial=Yes,
/ConsideredUncertainKvTemp643InMem/.initial=Yes,
/ConsideredUncertainKvTemp644InMem/.initial=Yes,
/ConsideredUncertainKvTemp645InMem/.initial=Yes,
/ConsideredUncertainKvTemp646InMem/.initial=Yes,
/ConsideredUncertainKvTemp647InMem/.initial=Yes,
/ConsideredUncertainKvTemp648InMem/.initial=Yes,
/ConsideredUncertainKvTemp649InMem/.initial=Yes,
/ConsideredUncertainKvTemp650InMem/.initial=Yes,
/ConsideredUncertainKvTemp651InMem/.initial=Yes,
/ConsideredUncertainKvTemp652InMem/.initial=Yes,
/ConsideredUncertainKvTemp653InMem/.initial=Yes,
/ConsideredUncertainKvTemp654InMem/.initial=Yes,
/ConsideredUncertainKvTemp655InMem/.initial=Yes,
/ConsideredUncertainKvTemp656InMem/.initial=Yes,
/ConsideredUncertainKvTemp657InMem/.initial=Yes,
/ConsideredUncertainKvTemp658InMem/.initial=Yes,
/ConsideredUncertainKvTemp659InMem/.initial=Yes,
/ConsideredUncertainKvTemp660InMem/.initial=Yes,
/ConsideredUncertainKvTemp661InMem/.initial=Yes,
/ConsideredUncertainKvTemp662InMem/.initial=Yes,
/ConsideredUncertainKvTemp663InMem/.initial=Yes,
/ConsideredUncertainKvTemp664InMem/.initial=Yes,
/ConsideredUncertainKvTemp665InMem/.initial=Yes,
/ConsideredUncertainKvTemp666InMem/.initial=Yes,
/ConsideredUncertainKvTemp667InMem/.initial=Yes,
/ConsideredUncertainKvTemp668InMem/.initial=Yes,
/ConsideredUncertainKvTemp669InMem/.initial=Yes,
/ConsideredUncertainKvTemp670InMem/.initial=Yes,
/ConsideredUncertainKvTemp671InMem/.initial=Yes,
/ConsideredUncertainKvTemp672InMem/.initial=Yes,
/ConsideredUncertainKvTemp673InMem/.initial=Yes,
/ConsideredUncertainKvTemp674InMem/.initial=Yes,
/ConsideredUncertainKvTemp675InMem/.initial=Yes,
/ConsideredUncertainKvTemp676InMem/.initial=Yes,
/ConsideredUncertainKvTemp677InMem/.initial=Yes,
/ConsideredUncertainKvTemp678InMem/.initial=Yes,
/ConsideredUncertainKvTemp679InMem/.initial=Yes,
/ConsideredUncertainKvTemp680InMem/.initial=Yes,
/ConsideredUncertainKvTemp681InMem/.initial=Yes,
/ConsideredUncertainKvTemp682InMem/.initial=Yes,
/ConsideredUncertainKvTemp683InMem/.initial=Yes,
/ConsideredUncertainKvTemp684InMem/.initial=Yes,
/ConsideredUncertainKvTemp685InMem/.initial=Yes,
/ConsideredUncertainKvTemp686InMem/.initial=Yes,
/ConsideredUncertainKvTemp687InMem/.initial=Yes,
/ConsideredUncertainKvTemp688InMem/.initial=Yes,
/ConsideredUncertainKvTemp689InMem/.initial=Yes,
/ConsideredUncertainKvTemp690InMem/.initial=Yes,
/ConsideredUncertainKvTemp691InMem/.initial=Yes,
/ConsideredUncertainKvTemp692InMem/.initial=Yes,
/ConsideredUncertainKvTemp693InMem/.initial=Yes,
/ConsideredUncertainKvTemp694InMem/.initial=Yes,
/ConsideredUncertainKvTemp695InMem/.initial=Yes,
/ConsideredUncertainKvTemp696InMem/.initial=Yes,
/ConsideredUncertainKvTemp697InMem/.initial=Yes,
/ConsideredUncertainKvTemp698InMem/.initial=Yes,
/ConsideredUncertainKvTemp699InMem/.initial=Yes,
/ConsideredUncertainKvTemp700InMem/.initial=Yes,
/ConsideredUncertainKvTemp701InMem/.initial=Yes,
/ConsideredUncertainKvTemp702InMem/.initial=Yes,
/ConsideredUncertainKvTemp703InMem/.initial=Yes,
/ConsideredUncertainKvTemp704InMem/.initial=Yes,
/ConsideredUncertainKvTemp705InMem/.initial=Yes,
/ConsideredUncertainKvTemp706InMem/.initial=Yes,
/ConsideredUncertainKvTemp707InMem/.initial=Yes,
/ConsideredUncertainKvTemp708InMem/.initial=Yes,
/ConsideredUncertainKvTemp709InMem/.initial=Yes,
/ConsideredUncertainKvTemp710InMem/.initial=Yes,
/ConsideredUncertainKvTemp711InMem/.initial=Yes,
/ConsideredUncertainKvTemp712InMem/.initial=Yes,
/ConsideredUncertainKvTemp713InMem/.initial=Yes,
/ConsideredUncertainKvTemp714InMem/.initial=Yes,
/ConsideredUncertainKvTemp715InMem/.initial=Yes,
/ConsideredUncertainKvTemp716InMem/.initial=Yes,
/ConsideredUncertainKvTemp717InMem/.initial=Yes,
/ConsideredUncertainKvTemp718InMem/.initial=Yes,
/ConsideredUncertainKvTemp719InMem/.initial=Yes,
/ConsideredUncertainKvTemp720InMem/.initial=Yes,
/ConsideredUncertainKvTemp721InMem/.initial=Yes,
/ConsideredUncertainKvTemp722InMem/.initial=Yes,
/ConsideredUncertainKvTemp723InMem/.initial=Yes,
/ConsideredUncertainKvTemp724InMem/.initial=Yes,
/ConsideredUncertainKvTemp725InMem/.initial=Yes,
/ConsideredUncertainKvTemp726InMem/.initial=Yes,
/ConsideredUncertainKvTemp727InMem/.initial=Yes,
/ConsideredUncertainKvTemp728InMem/.initial=Yes,
/ConsideredUncertainKvTemp729InMem/.initial=Yes,
/ConsideredUncertainKvTemp730InMem/.initial=Yes,
/ConsideredUncertainKvTemp731InMem/.initial=Yes,
/ConsideredUncertainKvTemp732InMem/.initial=Yes,
/ConsideredUncertainKvTemp733InMem/.initial=Yes,
/ConsideredUncertainKvTemp734InMem/.initial=Yes,
/ConsideredUncertainKvTemp735InMem/.initial=Yes,
/ConsideredUncertainKvTemp736InMem/.initial=Yes,
/ConsideredUncertainKvTemp737InMem/.initial=Yes,
/ConsideredUncertainKvTemp738InMem/.initial=Yes,
/ConsideredUncertainKvTemp739InMem/.initial=Yes,
/ConsideredUncertainKvTemp740InMem/.initial=Yes,
/ConsideredUncertainKvTemp741InMem/.initial=Yes,
/ConsideredUncertainKvTemp742InMem/.initial=Yes,
/ConsideredUncertainKvTemp743InMem/.initial=Yes,
/ConsideredUncertainKvTemp744InMem/.initial=Yes,
/ConsideredUncertainKvTemp745InMem/.initial=Yes,
/ConsideredUncertainKvTemp746InMem/.initial=Yes,
/ConsideredUncertainKvTemp747InMem/.initial=Yes,
/ConsideredUncertainKvTemp748InMem/.initial=Yes,
/ConsideredUncertainKvTemp749InMem/.initial=Yes,
/ConsideredUncertainKvTemp750InMem/.initial=Yes,
/ConsideredUncertainKvTemp751InMem/.initial=Yes,
/ConsideredUncertainKvTemp752InMem/.initial=Yes,
/ConsideredUncertainKvTemp753InMem/.initial=Yes,
/ConsideredUncertainKvTemp754InMem/.initial=Yes,
/ConsideredUncertainKvTemp755InMem/.initial=Yes,
/ConsideredUncertainKvTemp756InMem/.initial=Yes,
/ConsideredUncertainKvTemp757InMem/.initial=Yes,
/ConsideredUncertainKvTemp758InMem/.initial=Yes,
/ConsideredUncertainKvTemp759InMem/.initial=Yes,
/ConsideredUncertainKvTemp760InMem/.initial=Yes,
/ConsideredUncertainKvTemp761InMem/.initial=Yes,
/ConsideredUncertainKvTemp762InMem/.initial=Yes,
/ConsideredUncertainKvTemp763InMem/.initial=Yes,
/ConsideredUncertainKvTemp764InMem/.initial=Yes,
/ConsideredUncertainKvTemp765InMem/.initial=Yes,
/ConsideredUncertainKvTemp766InMem/.initial=Yes,
/ConsideredUncertainKvTemp767InMem/.initial=Yes,
/ConsideredUncertainIlChessC0/.initial=Yes,
/ConsideredUncertainIlChessC1/.initial=Yes,
/ConsideredUncertainIlChessC2/.initial=Yes,
/ConsideredUncertainVdd/.initial=Yes,
/ConsideredUncertainPtat/.initial=Yes,
/ConsideredUncertainPtatArt/.initial=Yes,
}

    \newcommand{\flashvalue}[1]{\getvariable{FlashValue#1}}
    \pgfkeys{
    /FlashValueAlphaPtat/.initial=4,
/FlashValueOccRowScale/.initial=2,
/FlashValueOccColScale/.initial=2,
/FlashValueOccRemScale/.initial=0,
/FlashValuePixOsAverage/.initial=65466,
/FlashValueAlphaScale/.initial=8,
/FlashValueAccRowScale/.initial=9,
/FlashValueAccColScale/.initial=10,
/FlashValueAccRemScale/.initial=5,
/FlashValuePixSensitivityAverage/.initial=16152,
/FlashValueGainEe/.initial=6276,
/FlashValueVPtat25/.initial=12194,
/FlashValueKvPtat/.initial=8,
/FlashValueKtPtat/.initial=342,
/FlashValueKVdd/.initial=161,
/FlashValueVdd25/.initial=110,
/FlashValueKvRoCo/.initial=4,
/FlashValueKvReCo/.initial=3,
/FlashValueKvRoCe/.initial=3,
/FlashValueKvReCe/.initial=3,
/FlashValueIlChessC2/.initial=0,
/FlashValueIlChessC1/.initial=8,
/FlashValueIlChessC0/.initial=15,
/FlashValueKtRoCo/.initial=68,
/FlashValueKtReCo/.initial=64,
/FlashValueKtRoCe/.initial=60,
/FlashValueKtReCe/.initial=58,
/FlashValueResolutionEe/.initial=2,
/FlashValueKvScale/.initial=3,
/FlashValueKtaScale1/.initial=5,
/FlashValueKtaScale2/.initial=2,
/FlashValueAlphaSp1/.initial=59,
/FlashValueAlphaSp0/.initial=126,
/FlashValueOffsetSp1/.initial=5,
/FlashValueOffsetSp0/.initial=944,
/FlashValueCpKv/.initial=3,
/FlashValueCpKta/.initial=36,
/FlashValueKsTa/.initial=246,
/FlashValueTgc/.initial=0,
/FlashValueKsTo1/.initial=197,
/FlashValueKsTo0/.initial=204,
/FlashValueKsTo3/.initial=151,
/FlashValueKsTo2/.initial=177,
/FlashValueStep/.initial=2,
/FlashValueCt3/.initial=6,
/FlashValueCt2/.initial=6,
/FlashValueKsToScale/.initial=9,
}

\usepackage{cuted}
\usepackage{capt-of}

\begin{document}

\verticaladjustment{-5pt}

\maketitle

\begin{strip}
\centering
\captionsetup{type=figure}
\includegraphics[trim={1cm 0cm 1cm 0cm},width=0.95\textwidth]{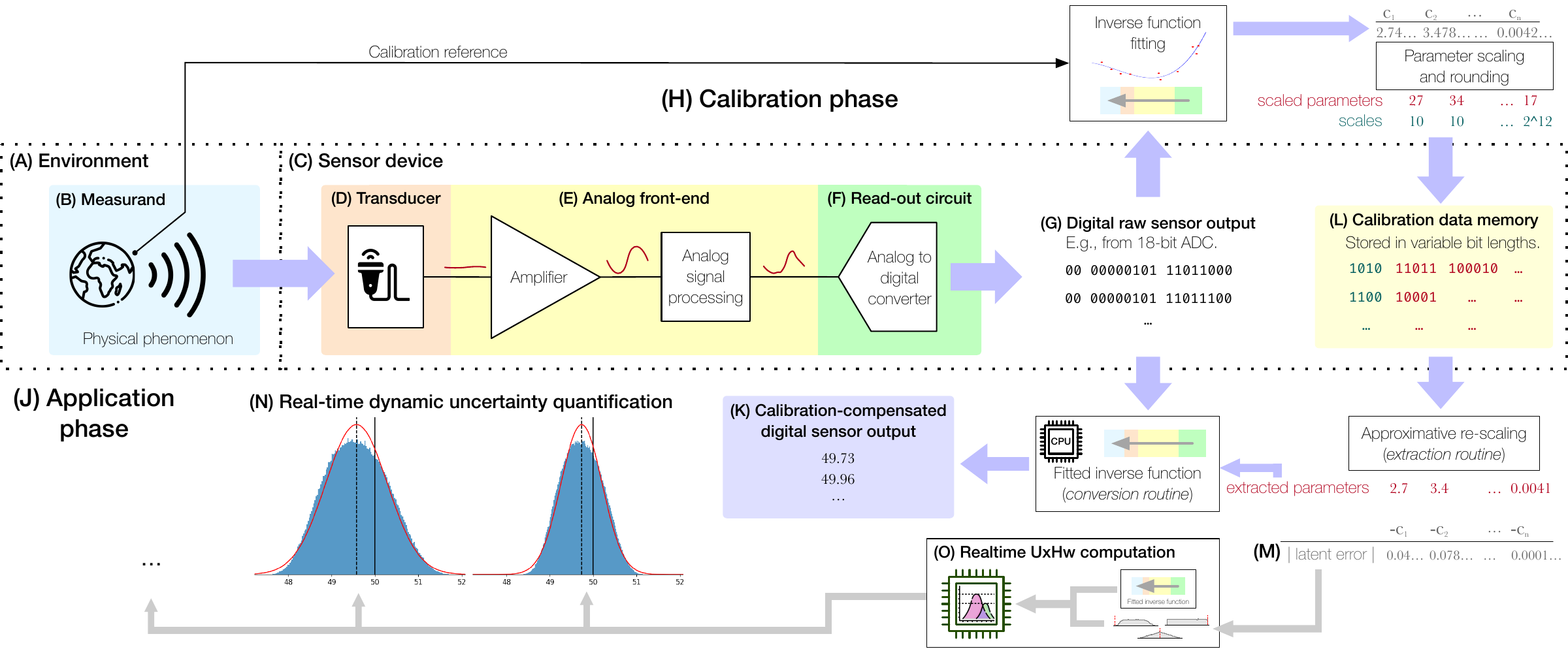}
\captionof{figure}{Simplified illustrative diagram for a calibration-compensated
sensor. (A) The environment produces a physical phenomenon as an analog,
potentially non-electrical, signal related to a quantity (B) we want to measure.
(C) On the sensor device, a component picks up the signal and transduces it to
an electrical signal (D). (E) An amplifier amplifies the electrical signal and
analog circuits process it further. (F) An analog-to-digital converter turns
measurements into digital values (G). These values do not have an obvious
correspondence to the measurand. (H) The calibration phase uses reference values
to fit a function (\emph{conversion routine}) which applications (J) will use to
compute measurand values (K) from the raw readings (G). The conversion routine
uses parameters from the calibration phase (J) that the manufacturer pre-stored
in the sensor device memory (L). Because these parameters are constrained in
precision by finite digital memories, quantization errors (M) in the extracted
parameters cause the sensor output of the conversion routines to contain
significant amounts of epistemic uncertainty (N). (O) Conventional approaches
can quantify this uncertainty in a static off-device offline approach: this
work presents a method for dynamic on-device real-time uncertainty
quantification for sensor outputs which depend on pre-stored calibration data.}
\label{fig:ThermopileInfraredSensorDiagram}
\end{strip}

\begin{abstract}
Modern data-driven applications that make real-time decisions increasingly
depend on advanced sensors which use pre-stored calibration data. In such
applications, accurate characterization of sensor output uncertainty is
important for reliable data interpretation. Here, we present a method for
real-time on-device dynamic uncertainty quantification for sensor outputs which
depend on pre-stored calibration data. We show how sensor calibration
compensation equations (essential in advanced sensing systems) propagate
uncertainties resulting from the quantization of calibration parameters to the
sensor output. We use a low-cost thermal sensor as a motivating example and
show these ideas are practical and possible on actual embedded sensor systems by
prototyping them on two commercially-available uncertainty tracking hardware
platforms. One has average power dissipation 16.7\;mW and achieves 42.9$\times$
speedup compared to the equal-accuracy Monte Carlo computation (the status quo),
and the other 147.15\;mW and achieves 94.4$\times$ speedup. We present a
proof-of-usefulness application using the quantified uncertainty in edge
detection over ten test scenes where we show accuracy and precision average
improvement by 4.97 and 40.25 percentage points, respectively, trading off
sensitivity. Another application example examines uncertainty quantification for
four different calibration-data storage scenarios and compute that a 48\%
increase in memory yields 75\% smaller uncertainty metrics over the baseline.

\end{abstract}

\thispagestyle{firststyle}
\ifthenelse{\boolean{shortarticle}}{\ifthenelse{\boolean{singlecolumn}}{\abscontentformatted}{\abscontent}}{}

\vspace{-0.1in}

Cutting-edge realtime cyber-physical systems are increasingly data-driven and
rely on ever more complex sensors \cite{chun2021artificial, yang2024sensor,
lenk2023neuromorphic, shi2022miniature} whose functionality depends on pre-stored factory
calibration data~\cite{melexis2019mlx,bosch2024bme680, bosch2024bme280,
te2017ms5611, nxp2024mpl115a2, analog2019adis16470}. For such systems, the
efficient and effective real-time operation hinges on the sensor output accuracy
and characterized uncertainty. Yet, the conventional approach to
sensing--control integration quantifies uncertainty per subsystem in an offline
manner as opposed to a holistic and real-time approach. This work demonstrates
an approach that enables real-time on-device dynamic quantification of the
uncertainty which arises due to pre-stored calibration data.

Uncertainty about the true value of a quantity is either \emph{aleatoric} or
\emph{epistemic}. Epistemic uncertainty is due to incomplete information. For
example, a 20-digit estimate of $\pi$ is epistemically uncertain beyond the 20th
digit. Aleatoric uncertainty is due to inherent randomness and no amount of
extra information reduces it---for example, the outcome of a fair coin toss.
This work shows how epistemic uncertainty in the calibration data of sensors
impacts the sensor output uncertainty and offers a practical solution compatible
with resource-constrained sensing systems.

Figure \ref{fig:ThermopileInfraredSensorDiagram} shows a simplified overview of
a calibration-compensated sensor and a simplified diagram of its calibration and
field application. In Figure~\ref{fig:ThermopileInfraredSensorDiagram}.A, a
physical phenomenon emits a signal which travels to the sensor, in
Figure~\ref{fig:ThermopileInfraredSensorDiagram}.C, where it creates a
measurable electric potential difference (voltage) across a transducer
(Figure~\ref{fig:ThermopileInfraredSensorDiagram}.D). The sensor circuits
amplify and filter this voltage signal, and then feed it to an analog-to-digital
converter (\adc) which digitizes it. This digital form is the raw output of the
sensor and not the desired physical quantity (measurand): it needs an additional
calibration-compensating mathematical transformation.

The sensor's digital output and the measurand value follow a
correspondence that depends on material- and manufacturing-related quantities.
Estimating this correspondence is an important part of sensor
calibration~\cite{fraden2016handbook}. The outputs of the calibration process
enable the construction of the inverse correspondence as a function for
converting the raw digital sensor output to an estimate of the measurand. Such functions are
commonly called \emph{conversion routines} or \emph{calibration-compensation routines}, and the
calibration outputs are called \emph{calibration parameters}.

Because of material and manufacturing variations, the exact calibration
parameters between different instances of the same sensor are typically
different. For this reason, sensor devices usually include a digital
non-volatile memory that holds information about the calibration parameters---we
refer to that information as \emph{calibration data}. Sensor driver functions
that recover calibration parameters from calibration data are called
\emph{extraction routines}.

Low-cost manufacturing and miniaturization impose strict size and power
constraints on sensors. To conserve space and energy, the calibration data is a
representation of the calibration parameters that uses less information. This
work shows how this treatment of the calibration data affects the sensor output
uncertainty in a significant measurable way.

Due to fundamental reasons of how calibration-compensated sensors work,
there is unavoidable uncertainty about the sensor outputs, even if we assume a
\emph{perfect measurement} that has \emph{zero noise}.  This uncertainty is
not due to \adc{} quantization
(Figure~\ref{fig:ThermopileInfraredSensorDiagram}.F) but because of deliberate
data loss when storing coefficients in limited memory
(Figure~\ref{fig:ThermopileInfraredSensorDiagram}.H). Because of the dynamic
nature of this uncertainty, sensor systems require a dynamic realtime approach
to accurately quantify it.

\newcommand\figlabel[1]{%
  \phantomsubcaption\label{#1}%
}

This article provides a practical solution for on-device realtime quantification
of dynamic uncertainty arising from pre-stored calibration data by making the
following contributions to the state of the art:

\begin{enumerate}
    \item[\ding{192}] Demonstration of hardware and algorithmic techniques
    for the quantification of dynamic uncertainty in sensor outputs, with
    practical demonstration on two low-power FPGA-based hardware implementations
    of processor-native uncertainty tracking~\cite{tsoutsouras2021laplace}.
    These present a way to improve the status quo, enabling more informed
    decision-making in critical applications where the reliability of sensor
    data is paramount. (Sections~\ref{section:hardwareUncertaintyTracking}
    and~\ref{SectionEvaluation}.)
    \item[\ding{193}] An analysis of epistemic uncertainty in the measurements
    of calibration-compensated sensors using a state-of-the-art infrared
    sensor (MLX90640) as a driving example. This epistemic uncertainty
    originates from the representation uncertainty of calibration numbers stored
    in-sensor and serves as the ground truth for the on-device approach.
    (Section~\ref{section:conversionCalibrationRoutines}.)

    \item[\ding{194}] Two experimental proof-of-usefulness examples: an
    edge detection application where making use of the uncertainty information
    leads to improved accuracy and precision
    (Section~\ref{section:edgeDetectionApplication}), and an application in
    sensor design for faster design feedback cycles, which examines the output
    uncertainty for four different calibration data storage scenarios 
    (Section~\ref{SectionApplicationDesignSpaceExploration}).
    \item[\ding{195}] An extensive dataset of raw infrared measurements from
    four MLX90640 instances amounting to a total of 322\,560 temperature
    readings of a high-accuracy, high-precision infrared calibration source,
    which we will make public.
\end{enumerate}

Our method and insights are applicable to sensors and devices that use
conversion routines with calibration data, or other quantized pre-stored
information. These insights are important for modern sensors, which use
increasingly complex processes for measuring phenomena and involve more
nonlinear elements~\cite{lenk2023neuromorphic, potyrailo2020extraordinary,
nakata2018wearable} while also pushing the boundaries of power consumption and
portability~\cite{maag2018survey}. Related work on sensor uncertainty
characterization often lumps representation uncertainty together with other
errors in a ``systematic errors'' category despite representation uncertainty
being straightforward and worthwhile to quantify
explicitly~\cite{fraden2016handbook, ru2022mems}, as this work shows.

\subsection*{Representation uncertainty}

Finite-precision representations map ranges of values in a dense domain to
a single stored value, which discards information and introduces
\emph{representation error} (e.g., rounding or truncating a floating-point value
to an integer). Because the mapping is many-to-one, the pre-conversion value
becomes unknowable from the stored value alone. This work models the plausible
pre-conversion values as a uniform distribution over the quantization bin, and
calls that \emph{representation uncertainty}. This uncertainty matters when
devices quantize calibration coefficients for non-volatile storage and later
reconstruct them at run time, since signal-path quantization-mitigation methods
do not add information to stored coefficients, and reuse of uncertain values in
later expressions can introduce correlation that complicates closed-form
propagation. Supplementary Section~\ref{section:discretizationErasesInformation}
further discusses representation uncertainty in calibration-compensated sensor
systems.

\section{Real-time on-device sensor uncertainty computation}
\label{section:hardwareUncertaintyTracking}

\begin{figure*}[t]
    \centering
    \begin{subfigure}[t]{0.31\linewidth}
        \centering
        \includegraphics[clip,trim={0.1cm 0.3cm 0.1cm 0.9cm},width=\linewidth]{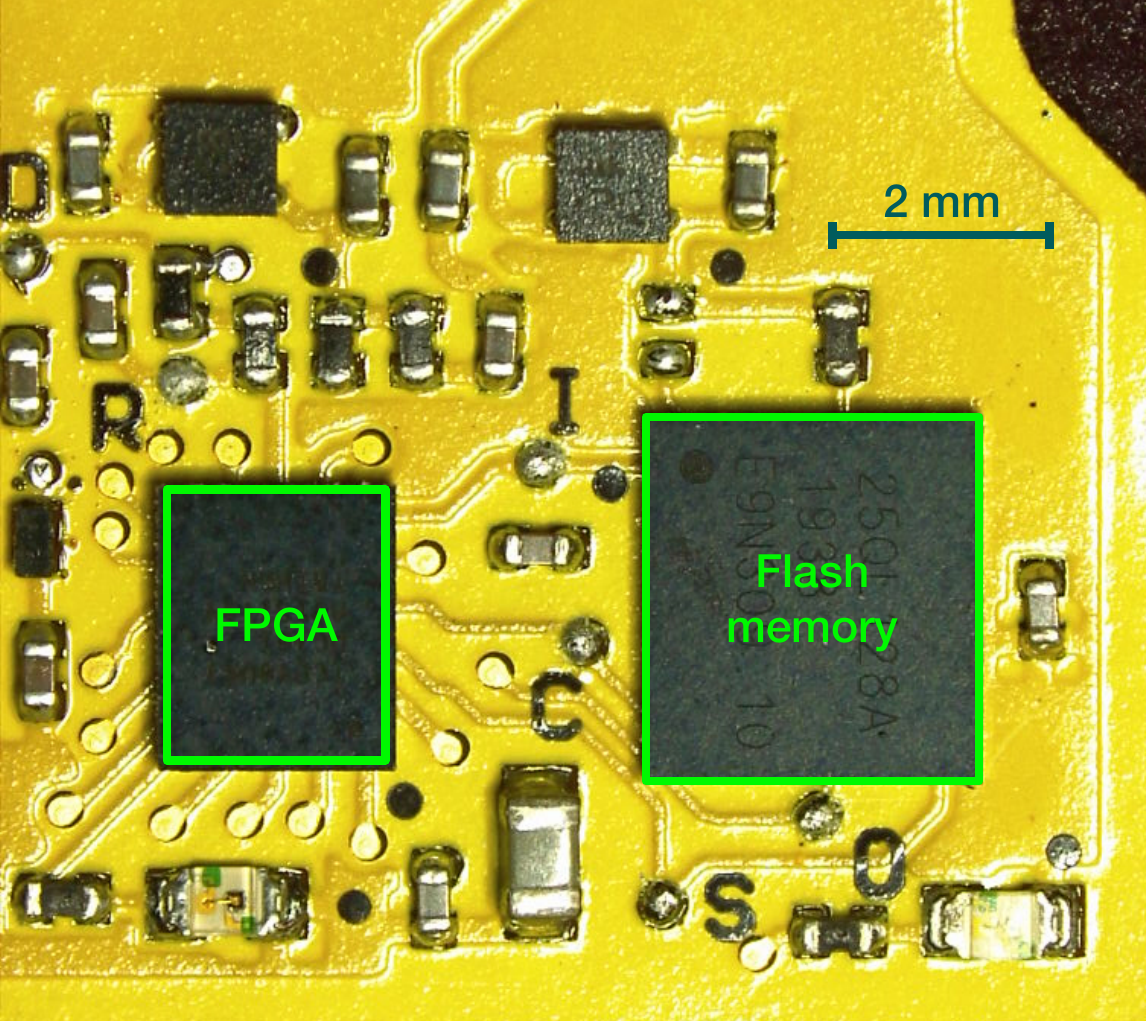}
        \caption{UxHw-FPGA-5k.}
        \label{fig:C0microSDHardware}
    \end{subfigure}
    \begin{subfigure}[t]{0.31\linewidth}
        \centering
        \includegraphics[clip,trim={0.1cm 0.2cm 0.1cm 0.9cm},width=\linewidth]{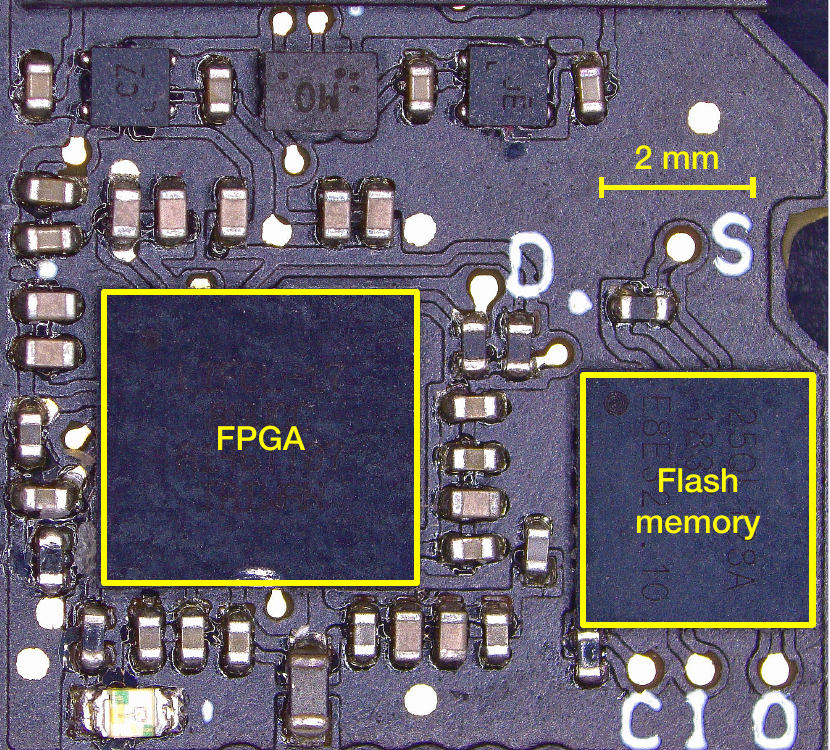}
        \caption{UxHw-FPGA-17k.}
        \label{fig:C0microSDPlusHardware}
    \end{subfigure}
    \begin{subfigure}[t]{0.31\linewidth}
        \centering
        \includegraphics[clip,trim={0.1cm 0.2cm 0.1cm 0cm},width=\linewidth]{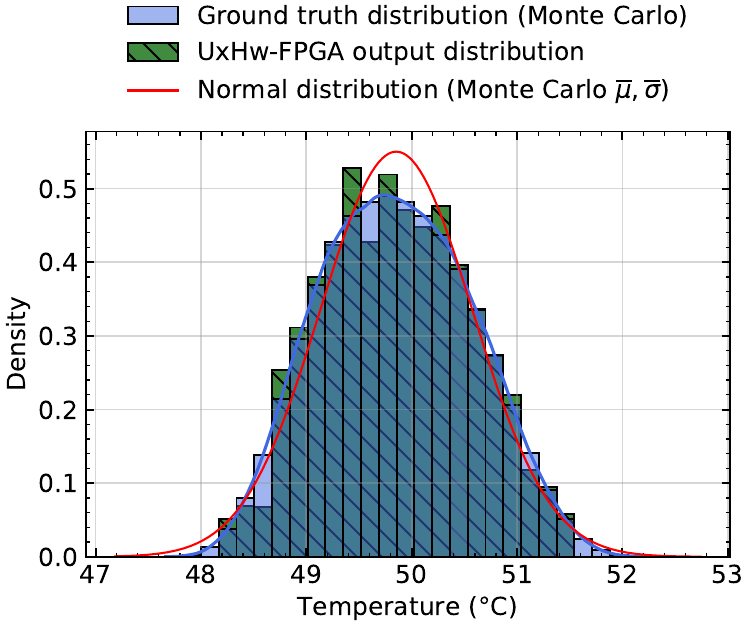}
        \caption{}
        \label{fig:NativeUncertaintyTrackingResult}
    \end{subfigure}
    \caption{\textbf{a,} Commercially-available system-on-module UxHw-FPGA-5k
    for native uncertainty tracking with \qty{12}{\mega\hertz} clock speed,
    \qty{128}{\kibi\byte} RAM, \qty{12}{\milli\watt} base power,
    \qty{10}{\milli\meter}~$\times$~\qty{15}{\milli\meter} size. \textbf{b,}
    Commercially-available system-on-module UxHw-FPGA-17k for native uncertainty
    tracking with \qty{45}{\mega\hertz} clock speed, \qty{320}{\kibi\byte}
    RAM, \qty{99}{\milli\watt} base power,
    \qty{10}{\milli\meter}~$\times$~\qty{15}{\milli\meter} size. \textbf{c,}
    Conversion routine output distribution when running on Ux-FPGA-5k and
    Ux-FPGA-17k with native uncertainty tracking. This result approximates the
    corresponding output distribution from the Monte Carlo execution with a
    Wasserstein distance of \qty{0.0189}{\degreeCelsius}, in under
    \qty{5}{\second} per pixel and with average power \qty{16.7}{\milli\watt} on
    UxHw-FPGA-5k. By contrast, to ensure same level of accuracy with 90\%
    confidence using Monte Carlo execution, the embedded system must re-execute
    the conversion routines at least 3000 times, which takes
    \qty{203}{\second}.}
\end{figure*}

The conventional approach for dealing with representation uncertainty
involves the theoretical off-line study of the subject system to determine the
uncertainty characteristics as constants which usually represent the worst case
scenario across the operating domain (or a small amount of domain regions).
Determined engineers apply compute-intensive Monte Carlo methods for
fine-grained insights about the uncertainty characteristics of the
system~\cite{zhou2025real,zhao2023uncertainty}. The Monte Carlo approach while
effective, is too compute-intensive and too slow to deploy on actual real-time
devices. 

This article presents realistic on-device real-time uncertainty
quantification in sensing systems which builds on recent advances in
deterministic arithmetic with probability distributions
(UxHw)~\cite{bilgin2025quantization}. We use two different hardware
system-on-module implementations of an uncertainty-extended processor
microarchitecture~\cite{tsoutsouras2021laplace} to benchmark the proposed UxHw
approach.
Instead of compute-intensive Monte Carlo simulations, the UxHw system uses
deterministic arithmetic on fixed-size uncertainty representations to propagate
uncertainty through computer operations, yielding deterministic runtimes and
memory usage.

In contrast with bootstrap-based approaches which apply non-parametric
bootstrap resampling to per-input ensembles of outputs produced by a
conventional estimator to compute uncertainty statistics~\cite{zhou2025real},
our proposed approach applies the UxHw system to propagate the parametric
representation uncertainty of pre-stored calibration parameters through the
calibration-extraction and calibration-compensation code, yielding an
approximate uncertainty distribution for each calibration-compensated sensor
output.

Figures~\ref{fig:C0microSDHardware} and~\ref{fig:C0microSDPlusHardware} show the
two evaluation platforms, which this article refers to as UxHw-FPGA-5k and
UxHw-FPGA-17k, respectively. UxHw-FPGA-5k builds on RISC-V RV32I and is
optimized for low power but is also slower. UxHw-FPGA-17k, which builds on
RISC-V RV32IM, is optimized for performance and is about ten times faster, but
also consumes more power. On these two hardware platforms, we perform real-time
dynamic uncertainty quantification for the calibration extraction and
calibration-compensation code. 

Section~\ref{section:conversionCalibrationRoutines} takes a closer look at the
arithmetic involved in calibration-compensation of raw sensor outputs for a
low-power infrared sensor which we use as a motivating application for our
approach. For a target object at \qty{50}{\degreeCelsius},
Figure~\ref{fig:NativeUncertaintyTrackingResult} presents the resulting output
uncertainty of a single pixel, quantified by the Monte Carlo approach (blue
histogram) and by our proposed UxHw approach (green histogram). The red line
represents a normal distribution of the same mean and variance as the Monte
Carlo data and highlights how this uncertainty follows a non-normal
distribution.

\section{Conversion or calibration routines}
\label{section:sensorComponents}
\label{section:conversionCalibrationRoutines}

Calibration-compensation routines are often provided by the sensor manufacturer
in the form of computer code. This article uses the Melexis MLX90640
thermal sensor~\cite{melexis2019mlx} as a driving example to demonstrate
real-time on-device uncertainty quantification of epistemic uncertainty which
arises in calibration-compensated sensors because of the representation
uncertainty in the calibration data.

As part of calibration compensation, the driver code must convert a raw 18-bit
\textsc{\Large adc} reading of, e.g., \codelight{00\;00000101\;11011000} (1496)
to an interpretable and useful quantity (in this case:
\qty{159.35}{\degreeCelsius}). To be able to convert the raw transducer reading
to a temperature estimate for each pixel, the code first reads pre-stored
calibration data from the non-volatile memory on the sensor device and applies
the extraction routines to get the calibration parameters. For this conversion,
the code uses 37 calibration parameters, four of which have different values for
each of the 768 pixels of the sensor.

\subsection{Motivating example using infrared sensor}

The MLX90640 has a thermal sensor array which senses radiation in the
far-infrared range (FIR; \qty{15}{\micro\meter} to \qty{1}{\milli\meter}
wavelength) with pixel resolution 32\texttimes24~\cite{melexis2019mlx}. The
pixel-sensors use thermopiles which transduce temperature differences to voltage
via the Seebeck effect~\cite{van1986thermal}.

\begin{figure*}[t]
    \centering
    \includegraphics[trim={0cm 0cm 0cm 0cm},clip,width=0.99\linewidth]{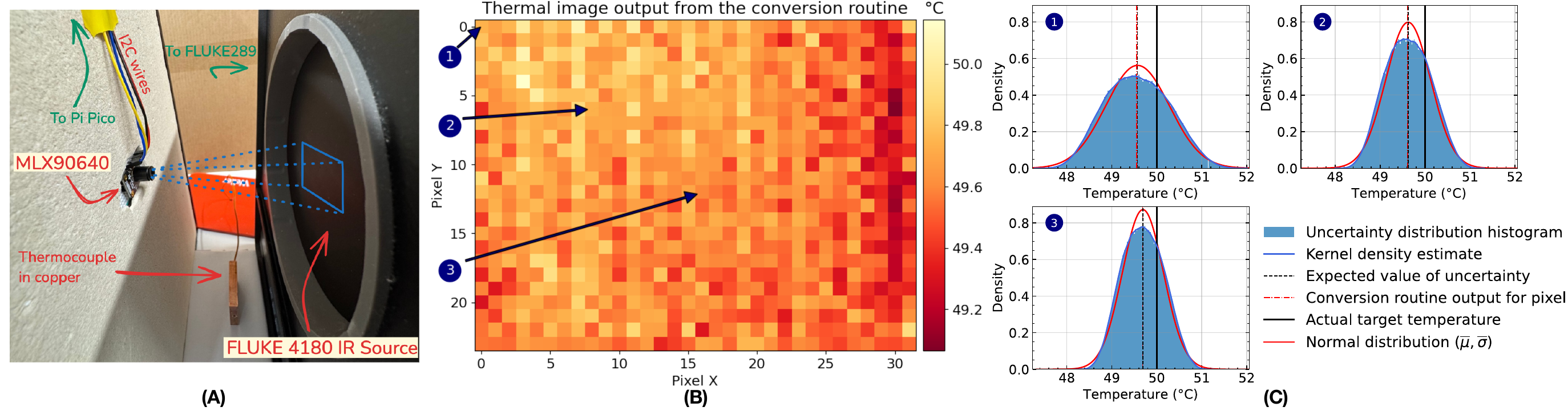}
    \label{fig:MlxSensorExperimentMegafigure}
    \caption{\textbf{A,} Experimental measurement setup: an
    MLX90640~\cite{melexis2019mlx} thermal sensor, controlled by a
    microcontroller, measures infrared radiation emanating from a FLUKE 4180
    infrared calibration source. A FLUKE 289 thermocouple monitors ambient
    temperature. Red: key items. Green: supporting items, out of frame. Blue:
    sensor field-of-view projection. \textbf{B,} The thermal image sensor outut
    of the setup in \ref{fig:MlxSensorExperimentalSetupTopView}.A, after
    conversion. \textbf{C,} Distribution of due-to-limited-precision uncertainty
    for three selected pixels of
    Figure~\ref{fig:MlxSensorExperimentalSetupScene}. Our method uncovers the
    dynamic uncertainty distribution of estimated temperature for every pixel of
    the image.
    \label{fig:MlxSensorExperimentalSetup}
    \label{fig:MlxSensorExperimentalSetupTopView}
    \label{fig:MlxSensorExperimentalSetupScene}
    \label{fig:MlxScenePixelTemperatureDistributions}}
\end{figure*}

Let $\alpha$ be the pixel sensitivity, $S$ be the thermoelectric (Seebeck)
effect coefficient slope~\cite{van1986thermal}, and let $\epsilon$ be the target
material emissivity. Let $c_0 = \qty{273.15}{\kelvin}$
(\qty{0}{\degreeCelsius}). Let $V_\mathrm{out}$ represent the analog output
voltage of a thermopile-based infrared sensor. Equation~\ref{eq:mlx90640Tout},
adapted from the sensor datasheet~\cite{melexis2019mlx}, gives the target object
temperature ($T_o$; in the Kelvin scale) for one pixel of the frame of the
sensor.
\begin{gather}
    T_o = \sqrt[4]{\frac{V_\mathrm{out}}{\alpha S (\sqrt[4]{\frac{V_\mathrm{out}}{\alpha} + T_{a-r}} - c_0)} + T_{a-r}}, \label{eq:mlx90640Tout}\\
    \mathrm{where}~T_{a-r} = - \frac{(1 - \epsilon) T_r^4}{\epsilon} - \frac{T_a^4}{\epsilon}. \nonumber
\end{gather}
Supplementary Material Section~\ref{sec:SupplementaryInfrared} provides further
context about the relevant physics as well as raw
and extracted values and associated uncertainty for all calibration parameters.

\subsubsection*{Calibration data extraction}  Each MLX90640 sensor has its own
calibration data which the manufacturer stored on the sensor. Because the
extraction routines have non-linear terms, uncertainty quantification becomes
more difficult. As an illustrative example of a parameter extraction, let
$\alpha_i$ be the calibration parameter related to the sensor sensitivity for
pixel $i$. Let $A_{\mathrm{ref}}$, $R$, $C$, and $D$ refer to calibration data
which have representation
uncertainty because they were quantized and stored in the device
memory. Let $s_R$,
$s_C$, and $s_D$ refer to respective scales (non-uncertain calibration data).
Let $\mathrm{TGC}$ represent the thermal gradient coefficient and
$\alpha_{\mathrm{CP}}$ compensation pixel values. $\mathrm{TGC}$ and
$\alpha_{\mathrm{CP}}$ are other calibration parameters with their own
extraction routines and the extraction of $\alpha$ depends on their extracted
values. Equation~\ref{eq:alphaExtractionEquation} is the mathematical equation
for the extraction routine for $\alpha_i$, where $i$ is the pixel index.
\begin{equation}
    \alpha_i = \frac{A_\mathrm{ref} + 2^{s_R} R_{\mathrm{row}(i)} + 2^{s_C} C_{\mathrm{col}(i)} + 2^{s_D} D_i}{2^{s_\alpha}} - \mathrm{TGC}\cdot\frac{\alpha_{\mathrm{CP0}} + \alpha_{\mathrm{CP1}}}{2}. \label{eq:alphaExtractionEquation}
\end{equation}

\begin{figure}[t]
    \centering

    \begin{subfigure}[t]{0.48\columnwidth}
        \centering
        \includegraphics[trim={0.1cm -0.5cm 0cm 0.1cm},clip,width=\linewidth]{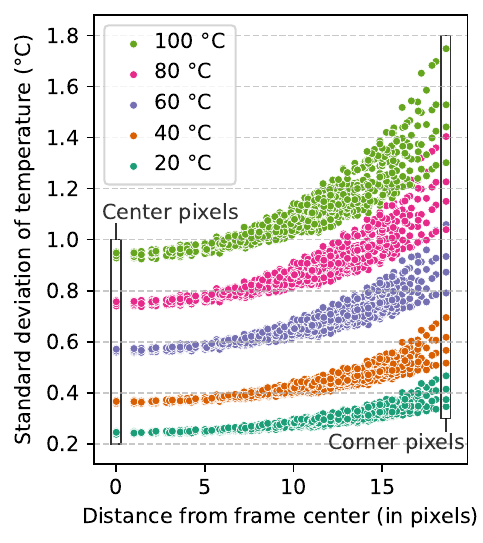}
        \caption{}
        \label{fig:MlxSceneStdTemperatureVsDistanceFromCenter}
    \end{subfigure}
    \begin{subfigure}[t]{0.4\columnwidth}
        \centering
        \includegraphics[clip,trim={0.1cm 0.3cm 0.1cm 0.1cm},width=\linewidth]{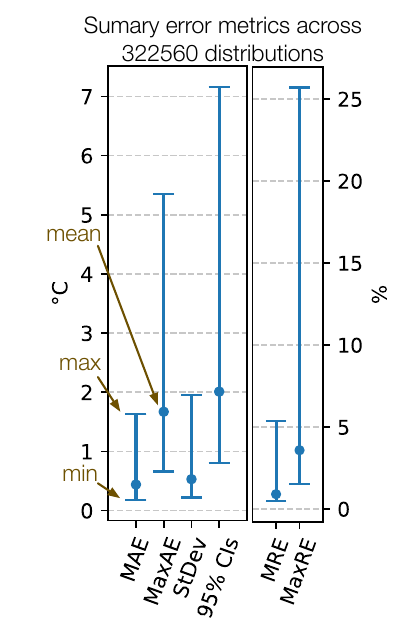}
        \caption{}
        \label{fig:MlxSensorOutputsErrorsCombined}
    \end{subfigure}
    
    \caption{\textbf{a,} The standard deviation of conversion routine output
    temperature for all 768 pixels of a single measurement of a single MLX90640
    sensor instance. For a target temperature of \qty{100}{\degreeCelsius}, the
    output uncertainty has average standard deviation \qty{0.94}{\degreeCelsius}
    for the center pixels, while for the corner pixels std average
    \qty{1.5}{\degreeCelsius} and max \qty{1.75}{\degreeCelsius}. Overall, the
    standard deviation has an increasing trend as pixels approach the frame edge
    of the sensor output and as the temperature increases. \textbf{b,} Summary
    for the probable absolute and relative errors because of representation
    uncertainty of the MLX90640 calibration data (conventional sensor output vs distribution), from all 768 pixels of four
    tested sensor instances, each tested for 21 target temperatures
    (\qty{322560} sensor output distributions). MAE: Mean Absolute Error, MaxAE:
    Max Absolute Error, StDev: Standard Deviation, 95\% CIs: Size of 95\%
    Confidence Intervals, MRE: Mean Relative Error, MaxRE: Max Relative Error.
    \label{fig:MlxSensorOutputsErrors}}
\end{figure}

\subsection{Quantifying representation uncertainty in MLX90640}
\label{sec:MLX90640Analysis}

The calibration data fundamentally represent real-valued quantities that the
calibration process converted to integers to fit them in the sensor memory.
Because of this, the calibration data are uncertain to the degree that their
integer representation cannot fully represent the before-conversion numbers.
Using floating-point representations for calibration data would partly
mitigate but not eliminate the issue. 
We quantify this representation uncertainty as a uniform distribution with unit
support length, centered on the raw calibration data.

Figure~\ref{fig:MlxSensorExperimentalSetupTopView}.A presents a physical
measurement setup analogous to the diagram in
Figure~\ref{fig:ThermopileInfraredSensorDiagram}. For one measurement using the
setup, Figure~\ref{fig:MlxSensorExperimentalSetupScene}.B shows the thermal
image output of the manufacturer's conversion routines. For three pixels along
the main diagonal of the thermal image (top-left, halfway to the center, and
center), Figure~\ref{fig:MlxScenePixelTemperatureDistributions}.C shows the
probability distribution of the pixel temperature. This distribution is due to
epistemic uncertainty that is a direct effect of the reduced information of the
calibration data and we computed it by off-line off-device Monte Carlo
simulation. Figure~\ref{fig:MlxScenePixelTemperatureDistributions}.C visually
shows how the distributions are non-uniform and different from Gaussians of the
sample mean and variance.

Figure~\ref{fig:MlxSceneStdTemperatureVsDistanceFromCenter} shows how this
uncertainty is different for each pixel and temperature and
Figure~\ref{fig:MlxSensorOutputsErrorsCombined} summarizes probable absolute and
relative error metrics due to the sensor output uncertainty.

Because the extracted calibration parameters are functions of the raw
calibration data, the parameters are also uncertain, and their uncertainty is a
function of the raw calibration data uncertainty. To quantify this effect, we
ran a 500K-iteration Monte Carlo execution of the conversion routines for each
measurement and each pixel.
Figure~\ref{fig:MlxSceneStdTemperatureVsDistanceFromCenter} shows how the
standard deviation of uncertainty of the MLX90640 conversion routine output
increases with the pixel-distance from the image center from
\qty{0.24}{\degreeCelsius} for center pixels at target temperature of
\qty{20}{\degreeCelsius}, to as high as \qty{1.75}{\degreeCelsius} for corner
pixels at target temperature of \qty{100}{\degreeCelsius}.
Figure~\ref{fig:MlxSensorOutputsErrorsCombined} presents aggregate metrics for
all \qty{322560} the MLX90640 output pixel distributions. The probable maximum
absolute error can be as high as \qty{5.35}{\degreeCelsius} (or more), and as
high as 25.7\% max relative error (or more). The standard deviation can be as
high as \qty{1.9}{\degreeCelsius} and the 95\% confidence interval can be as
high as \qty{7.16}{\degreeCelsius}. The representation uncertainty of the
calibration parameters causes sensor output uncertainty to be greater than the
nominal sensor accuracy of \textpm\qty{3}{\degreeCelsius}~\cite{melexis2019mlx}.

Supplementary Material Section~\ref{SectionSupplementaryDataAndFiguresMLX}
further discusses representation uncertainty in the calibration data of the
MLX90640.

\subsubsection*{Applicability to other sensors and systems} 

This work examines the particular thermal imaging sensor as a driving
example to show the effect of epistemic uncertainty in the sensor outputs and to
present the hardware-based framework for its on-device dynamic computation. The
presented framework extends to every sensor and system which uses calibration
data stored on the device. Calibration data which is originally real numbers and
is ultimately stored in a numerical domain of finite precision is unavoidably
subject to representation uncertainty (recall
Section~\ref{section:discretizationErasesInformation}). This even applies to
cases where the calibration-compensation is part of digital in-sensor processing
before and not in the host microcontroller. In the context of the UxHw approach
aiding device design-space exploration,
Section~\ref{SectionApplicationDesignSpaceExploration} examines how different
storage strategies affect this uncertainty which is however always there.

Modern sensors are increasingly more complex and include digital logic for
calibration componsation which depends on the sensor device storing data from
factory calibration. This same pattern shows up in environmental
sensors~\cite{bosch2024bme680,bosch2024bme280,te2017ms5611,nxp2024mpl115a2,te2017ms5837},
inertial
sensors~\cite{kasei2023ak8963,analog2023adis16550,analog2019adis16470,texas2021tmcs1123},
and others~\cite{flusso2021fls110, ams2016as7263, sensirion2011sfm3100}.

\section{Evaluation}
\label{SectionEvaluation}

We quantify the accuracy of the approximate uncertainty quantification of the
uncertainty-tracking hardware as the Wasserstein
distance~\cite{ramdas2017wasserstein} of the output distributions from the
hardware to the Monte Carlo execution output. Intuitively, the Wasserstein
distance measures the cost of transforming one distribution into another. Let
two one-dimensional empirical distributions $P$ and $Q$ and let $F_P$ and $F_Q$
be the respective cumulative density functions.
Equation~\ref{eq:WassersteinDistanceDefinition} defines the first Wasserstein
distance \( W_1(P, Q) \) with the difference of $F_P$ and $F_Q$.
\begin{equation}
    W_1(P, Q) = \int |F_P(x) - F_Q(x)| dx. \label{eq:WassersteinDistanceDefinition}
\end{equation}

The uncertainty information of both UxHw-FPGA-5k and UxHw-FPGA-17k achieves
Wasserstein distance \qty{0.0189}{\degreeCelsius} to the output distribution
from the ground-truth 500K-iteration Monte Carlo execution. We compare the
speed of the hardware uncertainty tracking of UxHw-FPGA-5k and UxHw-FPGA-17k
against the speed of a brute-force Monte Carlo execution of equal accuracy on
the same hardware but without uncertainty-tracking. Since an actual deployment
cannot check the Wasserstein distance against some large ground truth dataset at
runtime, it must make a Monte Carlo iteration count cutoff based on offline
analysis of empirical data: we report the time it takes to run the Monte Carlo
execution with iteration count cutoff which 90\% of the time yields Wasserstein
distance \qty{0.0189}{\degreeCelsius} (i.e., worse distance 10\% of the time).
We refer to these as EqMCp90-FPGA-5k and EqMCp90-FPGA-17k. A higher percentile
confidence for convergence requires even more compute time in the EqMCp90-FPGA
Monte Carlo executions.

Variant UxHw-FPGA-5k with average power \qty{16.7}{\milli\watt} runs in under
\qty{4740}{\milli\second} per pixel while the equal-accuracy brute-force
EqMCp90-FPGA-5k needs at least \qty{203509}{\milli\second} (42.9$\times$
speedup of UxHw over corresponding EqMCp90). Variant UxHw-FPGA-17k runs in
\qty{350}{\milli\second} per pixel with average power under
\qty{147.15}{\milli\watt}, while the equal-accuracy brute-force computation
needs at least \qty{33042}{\milli\second} (94.4$\times$ speedup of UxHw
over corresponding EqMCp90). Table~\ref{tab:C0MicroSDLatencyMetrics} lists workload latency
for the two hardware variants and their corresponding equal-accuracy Monte Carlo
execution, and the speedup the UxHw-FPGA variants achieve over the corresponding
EqMCp90-FPGA variant. Table~\ref{tab:C0MicroSDPowerDissipationMetrics} lists the
power dissipation of extracting the calibration parameters and their
uncertainty, and computing the pixel output temperature and its uncertainty
(excluding I2C sensor communication).

\begin{table}[t]
    \centering
    \caption{Power dissipation measurements for FPGA-5k and FPGA-17k, the base
    platforms for UxHw-FPGA-5k and UxHw-FPGA-17k and the corresponding
    EqMCp90-FPGA-5k and EqMCp90-FPGA-17k executions. The low power dissipations
    verify our method as a valid approach for uncertainty quantification for
    sensor outputs on low-power edge devices.}
    \begin{tabular}{l r r r}
        \toprule
            & \textbf{Idle power} & \textbf{Average power} & \textbf{Max power}        \\
            \midrule
            \rowcolor{a} \textbf{FPGA-5k}   & \qty{12}{\milli\watt}   & \qty{17}{\milli\watt}   & \qty{35}{\milli\watt}  \\
            \rowcolor{b} \textbf{FPGA-17k}  & \qty{99}{\milli\watt}   & \qty{147}{\milli\watt}  & \qty{208}{\milli\watt} \\
            \bottomrule
    \end{tabular}
    \label{tab:C0MicroSDPowerDissipationMetrics}
\end{table}

\begin{table}[t]
    \centering
    \caption{Iso-quality computation latency measurements for UxHw-FPGA-5k and
    UxHw-FPGA-17k and the corresponding EqMCp90-FPGA-5k and EqMCp90-FPGA-17k
    executions, for the same Wasserstein distance 0.0189\,\textdegree{C} to the
    ground truth Monte Carlo execution output. The uncertainty-tracking hardware
    variants achieve 42.9$\times$ and 94.4$\times$ speedup over the
    corresponding equal-accuracy brute-force Monte Carlo execution (on the same
    processor).}
    \begin{tabular}{l r c}
        \toprule
            & \textbf{Latency (ms)}  & \textbf{Speedup}\\
            \midrule
            \rowcolor{b} \textbf{EqMCp90-FPGA-5k}      & 203\,509 &  - \\
            \rowcolor{a} \textbf{UxHw-FPGA-5k}         & 4\,740 & 42.9$\times$\\
            \hline
            \rowcolor{b} \textbf{EqMCp90-FPGA-17k}     & 33\,042 & - \\
            \rowcolor{a} \textbf{UxHw-FPGA-17k}        & 350 & 94.4$\times$ \\
            \bottomrule
    \end{tabular}
    \label{tab:C0MicroSDLatencyMetrics}
\end{table}

\begin{figure*}[t]
    \centering
    \includegraphics[clip,width=0.95\textwidth]{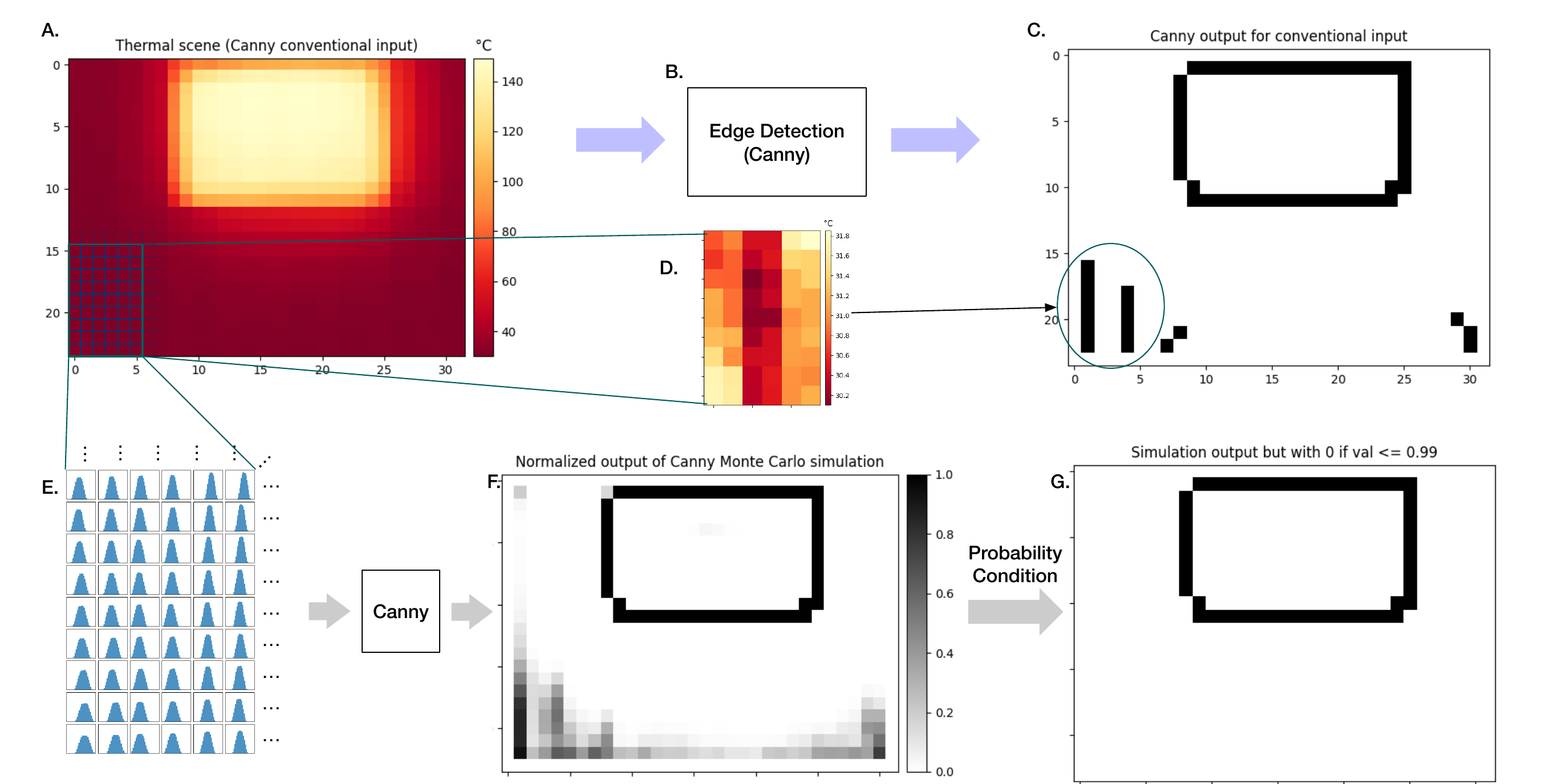}

    \caption{\textbf{The uncertainty information helps against false positives in edge detection.}
    \textbf{A,} The conventional sensor output.
    \textbf{B,} Canny edge detector.
    \textbf{C,} The result of the Canny edge detector for the conventional sensor output yields false positive edges 
    close to the bottom corners of the frame.
    \textbf{D,} Small horizontal fluctuations leads to false positive edges.
    \textbf{E,} The distribution of the temperature frame samples from Section~\ref{section:conversionCalibrationRoutines}.
    \textbf{F,} The result of the Canny execution using the temperature frame distribution yields probabilistic edges, 
    with the highest probabilities for the true positive edges. False-positive edges in every sample frame: 26.8\textpm8.
    \textbf{G,} The true edges have edge probability equal to one. We filter the data with probability lower than one and 
    the true positive edges remain.
    \label{fig:EdgeDetectionMegaFigure}}

\end{figure*}

\subsection{Application: Edge detection on uncertain data}
\label{section:edgeDetectionApplication}

Detecting object edges in an image is a common process in imaging systems and is
one of the steps necessary for object detection and
identification~\cite{shin2001comparison, bansal20212d}. As one
demonstrative application of the impact of epistemic uncertainty in
calibration-compensated sensors and the potential usefulness of tracking it in
real time, we analyze how the repersentation uncertainty in the thermal data of
the MLX90640 (recall Section~\ref{section:discretizationErasesInformation})
propagates through a common edge detection implementation: the Canny
algorithm~\cite{canny1986computational}. These effects of uncertainty all result
from the manner of storage of the calibration data during the calibration
process (recall Section~\ref{section:conversionCalibrationRoutines}).

The top part of Figure~\ref{fig:EdgeDetectionMegaFigure} shows how the
application of the Canny detector on the sensor output for an edge detection
scene (Figure~\ref{fig:EdgeDetectionMegaFigure}.A) leads to false-positive edges
in the output (Figure~\ref{fig:EdgeDetectionMegaFigure}.C). The Canny algorithm
uses a gradient-based operator and then decides which pixels are edges. The
false-positive edges occur because of small gradients in the otherwise quiet
part of the image (Figure~\ref{fig:EdgeDetectionMegaFigure}.D). 

\DeclareSIUnit{\pp}{\textup{percentage points}}

We ran a 500K-iteration Monte Carlo execution of the Canny algorithm where each
input sample is one sample-frame of 768 pixel temperatures of
Section~\ref{sec:MLX90640Analysis}---Figure~\ref{fig:EdgeDetectionMegaFigure}.E
shows this input to the Canny algorithm.
Figure~\ref{fig:EdgeDetectionMegaFigure}.F shows the normalized aggregate result
that represents the probability of Canny identifying each pixel as an edge,
across the probable input temperature frames. Applying the conventional
algorithm on different probable temperature frames leads to artifactual
false-positive edges every time. The mean count of false-positives per sample
frame is 26.8 pixels (i.e., 3.5\% of the picture is falsely characterized as
edges), with standard deviation 8.0, and max count 57 (i.e., 7.4\% of the
picture). Figure~\ref{fig:EdgeDetectionMegaFigure}.G shows how we can now filter
out the these false-positives by using a high-probability constraint, e.g.,
requiring more than 0.99 probability of edge to accept it.

We repeat this process for a total of ten scenes of varying complexity,
emissivity, and background uniformity, and we use three empirical probability
thresholds to quantify changes in the performance of the edge classification.
The results show that the utilization of the epistemic uncertainty information
as part of the Canny algorithm lead to improvements in accuracy and precision at
the expense of sensitivity. For the 80\%-probability threshold the approach
yields on average +\qty{6}{\pp} in accuracy, +\qty{46.3}{\pp} in precision, and
-\qty{16.6}{\pp} in sensitivity. Supplementary Material
Section~\ref{SectionSupplementaryDataAndFiguresMLX} lists further details about
the scenes and detailed performance metrics per scene. Supplementary
Figure~\ref{FigureEdgeDetectionSelectedScenes} contains visuals for three of the
ten scenes.

\subsection{Application: calibration memory size scenario exploration}
\label{SectionApplicationDesignSpaceExploration}

The epistemic uncertainties of
Sections~\ref{section:conversionCalibrationRoutines} and~\ref{SectionEvaluation}
arise from engineering choices made in the sensor design and calibration phase.
In particular, the representation uncertainty of the calibration data is a
direct effect of the on-device flash memory size and the rounding algorithm.
This section overviews how the UxHw approach applies to sensor design with the
potential of faster design feedback cycles by enabling faster choice--effect
approximation for design choices such as the on-board memory and the calibration
process.

The calibration data that lead to the sensor output uncertainties of
Figure~\ref{fig:MlxSensorOutputsErrorsCombined} take up \qty{1.5}{\kibi\byte} of
the \qty{1.625}{\kibi\byte} EEPROM. This section examines the effect on this
representation uncertainty when increasing the memory limit and when using
different data formats to store calibration data. We study the effect of the
four following hypothetical scenarios of using more memory for the MLX90640
calibration data. \textbf{TwoMoreBits:} Use two more bits for each piece
of calibration data: this increases the necessary memory by about 48\% to
\qty{2.225}{\kibi\byte}. \textbf{TwiceBits:} Use twice the amount of bits for
each piece of calibration data: this increases the memory by about 108\% to
\qty{3.125}{\kibi\byte}. \textbf{FP16:} Use the 16-bit IEEE-754 floating-point
data format to store each calibration data: this increases the memory by about
228\% to\qty{4.925}{\kibi\byte}. \textbf{FP32:} Use the 32-bit IEEE-754
floating-point data format (single precision floating point) for each piece of
calibration data: this increases the necessary on-device memory by about 548\%
to \qty{9.725}{\kibi\byte}.

In scenarios TwoMoreBits and TwiceBits the data format stays the same but
the quantization regions are smaller and thus the representation uncertainty is
also narrower. Scenarios FP16 and FP32 use the a uniform width across all pieces
of calibration data while the original format assigns non-uniform fidelities to
different categories of calibration data. We compare these scenarios agains the
NearestInteger baseline scenario of
Section~\ref{section:conversionCalibrationRoutines}.

\begin{figure}[t]
    \centering
    \includegraphics[width=0.9\columnwidth]{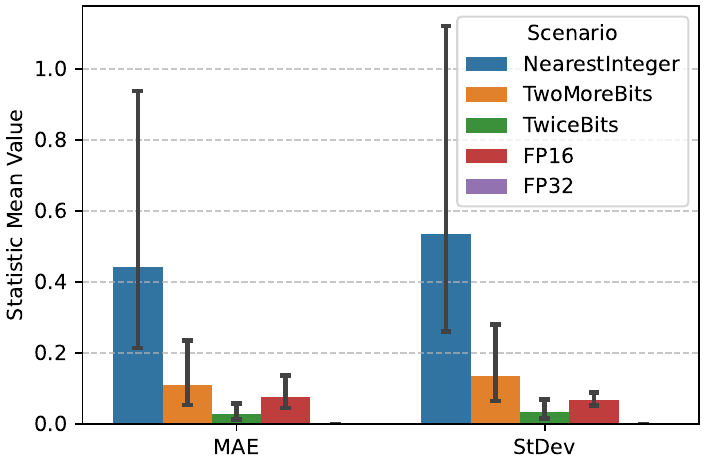}
    \caption{Mean absolute error and standard deviation statistic of the
    sensor output epistemic uncertainty distribution under the four examined
    calibration data storage scenarios. Error bars show the empirical 95\%
    percentile interval.}
    \label{FigureCalibrationDataScenarioComparison}
\end{figure}

Figure~\ref{FigureCalibrationDataScenarioComparison} shows the mean
absolute error and standard deviation statistic of the sensor output epistemic
uncertainty distribution under the four examined calibration data storage
scenarios. Even using two more bits at the cost of 48\% more memory requirements
yields smaller possible due-to-quantization errors and narrower sensor output
uncertainty distributions. Supplementary
Table~\ref{TableUncertaintyStatisticsByScenarioPctChange} presents finer-grained
percent-change data for the summary statistics of the sensor output uncertainty.
As expected, scenarios with more information storage capacity result to narrower
epistemic uncertainty in the sensor outputs. The IEEE-754 32-bit scenario yields
99.99\% smaller error statistics across all datasets. While the IEEE-754 16-bit
requires more memory than the TwiceBits scenario, it yields less stable benefit
compared to the nearest-integer rounding baseline.

These scenario simulations take hours to complete on a state-of-the-art
machine. More sophisticated approaches would check for hybrid scenarios where
calibration data is stored with non-uniform widths, as happens with the real
calibration data on the MLX90640, and this means even more resource-intensive
and time-consuming simulations. Engineers and designers can apply the UxHw
method of this article for faster approximation of the effect of
design choices on the output uncertainty and benefit from faster simulations and
less time lost on design feedback cycles.

\section{Conclusions}
\label{section:conclusions}

In calibration-compensated sensor systems, real-valued calibration-related
quantities undergo rounding as part of the calibration process that stores them
as data in the sensor memory for later extraction. This process introduces
representation uncertainty over the range of real numbers that rounding
collapses to the same digital representation. This representation uncertainty
leads to epistemic uncertainty in the sensor outputs. This uncertainty is an
inherent fact of the physics and computation relevant to the sensor and does not
come from noise in the measurement---it would be present even assuming perfect
noiseless measurement. This epistemic uncertainty of the sensor output is not of
a constant size or distribution shape. We cannot store it as an offset
distribution and re-apply it for new measurements: For each pixel and raw
temperature reading, the distribution is different. This uncertainty is also
different based on the conversion technique of the calibration process.

The correct quantification of such uncertainty in conventional systems
involves offline off-device analysis, often with compute-intensive Monte Carlo
methods. This work employs processor-native uncertainty-tracking using UxHw to
present a framework for on-device realtime dynamic uncertainty quantification.
We show practical use of the framework with two commercial uncertainty-extended
RISC-V processors and achieve uncertain quantification accuracy within
\qty{0.0189}{\degreeCelsius} Wasserstein distance to the Monte Carlo execution
output, while also achieving 42.9$\times$ and 94.4$\times$ speedup compared to
the equal-accuracy brute-force computation of using Monte Carlo execution.
The hardware platforms achieve these results with as little as
\qty{35}{\milli\watt} and \qty{208}{\milli\watt}, respectively, demonstrating
the feasibility of the approach on small scale sensing systems.

We show two proof-of-concept usecases of the UxHw method for real-life
applications to demonstrate how the uncertainty quantification framework of
this work is practical and applicable for low-power embedded systems. The
tracking of epistemic uncertainty in sensor outputs through downstream
application computations enables better automated decision making: In the edge
detection application the uncertainty information helps greatly reduce
false-positive edges (Section~\ref{section:edgeDetectionApplication}). The
dynamic quantification of this uncertainty can speed up design cycles and
efficient design-space exploration by aiding trade-off decisions of embedded
designs such as memory space
(Section~\ref{SectionApplicationDesignSpaceExploration}).

\subsubsection*{Future directions} Embedded systems with calibration-compensated
sensors can rely on field calibration with higher-precision calibration data
retention, when the sensing system can support it. Recent techniques from
quantization-aware neural networks could help producing better-quality quantized
fits directly~\cite{jacob2018quantization}, while also using block
floating-point or bfloat16 instead of unsigned integers~\cite{tye2023materials}.
Bayesian computing techniques may hold the key in uncertainty-aware sensor
calibration~\cite{berger2022bayesian}. Since sensor devices already bear unique
identifiers, manufacturers could provide digital distribution channels for
higher-precision calibration parameters.

\section{Methods}
\label{section:methods}

\subsection{Experimental measurement setup}
Four MLX90640 (Adafruit breakboard) instances take five infrared measurements of
a FLUKE 4180 high-precision calibration source from a distance of
\qty{10}{\centi\meter} at 21 temperature configurations. The calibration source
emissivity is 0.95. We start at source temperature \qty{10}{\degreeCelsius} and
measure up to \qty{110}{\degreeCelsius} with step \qty{5}{\degreeCelsius}, with
stabilization wait time \qty{10}{\minute}. After stabilization, we place each
sensor opposite across the source on an affixed vertical breadboard and take
five infrared measurements. The breadboard ensures each sensor is in the same
spot at the time of measurement. The field of view of the sensor is fully within
the \qty{15}{\centi\meter} diameter of the infrared source disk. Materials
around the sensor position are shielded from infrared radiation using Superwool
HT. A FLUKE 289 multimeter monitors the ambient temperature between the sensor
and the source using a copper-stabilized thermocouple, outside the source's
field-of-view. (The MLX90640 also reports its own ambient temperature reading.)

A Raspberry Pi Pico WH 2022 with a RP2040 microcontroller~\cite{rpi2025rp2040}
running custom MicroPython\footnote{The interpreter firmware is the official
release \code{RPI\_PICO\_W-20240602-v1.23.0.uf2}.} code drives each MLX90640
sensor over I\textsuperscript{2}C, first reading the EEPROM data and making five
raw data measurements with sampling frequency \qty{0.5}{\hertz} (highest
accuracy for MLX90640). The Pico sends the data to a MacBook Pro M1 over
Bluetooth LE. A Keithley 2281 precision DC supply powers the Pico at
\qty{5}{\volt}, which powers the MLX90640 from the \codelight{3V3} pin.

\subsection{Monte Carlo executions}
Using the uniform uncertainty centered around each calibration data model, we do
a Monte Carlo execution with \qty{500000} re-executions of the conversion
routines for all 768 pixels of four distinct MLX90640 sensor instances, and
across five different actual measurements from each, each against 21 target
temperatures settings of the infrared source. The Monte Carlo execution result
data amount to \qty{322560} empirical distributions total, each made up of
\qty{500000} samples---a total data size of 645.12 GB. We provide the input
measurements as an open dataset and also open-source code that generates the
result data in a way easily-configurable to simulate a subset of the results.
Histogram plots use the Doane binning algorithm from
NumPy~\cite{harris2020array} unless specified otherwise.

\subsubsection*{Benchmarking equipment}
The Monte Carlo executions ran on a Dell Precision 5820 workstation with Ubuntu
24.04 LTS, Linux 6.11, Python 3.12, 16-core Intel i7-7820X 4.5 GHz, and 32 GiB
RAM, as well as on a MacBook Pro with Apple M1 Pro, 32 GiB RAM, Sequoia 15,
Apple clang 16.0. The executions for
Section~\ref{section:hardwareUncertaintyTracking} ran on commercially available
FPGA implementations of the Laplace
microarchitecture~\cite{tsoutsouras2021laplace} (see
Section~\ref{sec:methodsRealTimeHardware}).

\subsubsection*{Metrics}
Let $s$ be an index over the sensor instances, $t$ over the tested temperatures,
$m$ over the per-instance per-temperature measurement repetitions, and $i$ over
the sensor pixels. We compute the aggregate metrics between the conventional
outputs of the sensors, $x_{s,t,m,i}$, and the corresponding sensor output
distribution, $Z_{s,t,m,i} \sim \{z_{s,t,m,i,j}\}_{j=1}^{500000}$, which is the
result of the Monte Carlo simulation of the conversion routines with
representation uncertainty in the calibration data. Equations~\ref{EquationMAE}
through~\ref{EquationCIs} give the metric formulas:
\begin{align}
    \text{MAE} (s,t,m,i)&= \mathbb{E}\langle |x_{s,t,m,i}-Z_{s,t,m,i}| \rangle,\label{EquationMAE}\\
    \text{MRE} (s,t,m,i)&= \mathbb{E}\langle |\frac{Z_{s,t,m,i}-x_{s,t,m,i}}{Z_{s,t,m,i}}| \rangle,\\
    \text{MaxAE} (s,t,m,i)&= \max\left(|x_{s,t,m,i}-Z_{s,t,m,i}|\right),\\
    \text{MaxRE} (s,t,m,i)&= \max\left( |\frac{Z_{s,t,m,i}-x_{s,t,m,i}}{Z_{s,t,m,i}}| \right),\\
    \text{StDev} (s,t,m,i)&= \sqrt{\frac{1}{N-1}\sum_{j=1}^{N=500000}\left(z_{s,t,m,i,j} - \mathbb{E}\langle Z_{s,t,m,i} \rangle\right)^2},\\
    \text{SizeCI}_{95\%} (s,t,m,i)&= Q_{0.975}(Z_{s,t,m,i}) - Q_{0.025}(Z_{s,t,m,i}).\label{EquationCIs}
\end{align}

\subsection{Code changes to MLX90640 driver for simulation}

We assume that the original calibration process performed by the manufacturer,
to generate the manufacturer-supplied calibration data, rounds the calibration
data using the C \codelight{round()} function which rounds half-way cases away
from zero. To compute the uncertainty of the calibration uncertainty we model
the representation uncertainty of the calibration data and then
forward-propagate it through the extraction routines. To achieve that, we
simulate uniform additive noise in the range [-0.5, 0.5), and center it around
the integer value of each calibration data out of the sensor memory. The
introduction of this additive noise must happen: \ding{202}~After \emph{logical}
bit shifting operations\footnote{Because the data type is unsigned
(\codelight{uint16\_t}), right shift operations in the C conversion routines are
always logical shifts~\cite[Sec. 6.5.7]{iso1999c99}.} that aim to place the data
in the bits of correct significance; \ding{203}~after arithmetic operations
aiming to restore the negative values domain; \ding{204}~before any operations
including variables with values dependent on other calibration parameters;
\ding{205} before other arithmetic operations of multiplication or division on
the data. (In theory, additions and subtractions with constant scalars do not
affect the representation uncertainty.)

While computing different calibration parameters, the parameter extraction code
reads some calibration data more than once. To achieve correct modelling of
variable dependencies in the simulation, our simulation code makes sure to sample
the representation uncertainty of these calibration data once and reuse the
value during the same execution.

The official C driver re-discretizes some extracted calibration parameters to
conserve dynamic memory at the time of operation. Because we are studying the
effect of representation uncertainty to the output temperature image, we change
the simulation code to treat that data as floating-point numbers and sidestep
that particular code. See supplementary material for further discussion on this.

\subsection{Canny}
The edge detection code uses the Canny operator implementation from Python
package scikit-image v0.24~\cite{padregosa2011scikit}. We configure for unit
standard deviation Gaussian filter, and \codelight{nearest} mode for filtering
the array borders. Higher standard deviation for the Gaussian filter leads to
fewer false-positive edges but also reduces the accuracy of true-positive edges.

\subsection{Real-time uncertainty quantification hardware}
\label{sec:methodsRealTimeHardware}
We use two commercial implementations of processor-native uncertainty-tracking
microarchitecture~\cite{tsoutsouras2021laplace} as systems-on-module on FPGA:
the UxHw-FPGA-5k and the UxHw-FPGA-17k.
The devices track uncertainty in computation for floating-point variables and
are based on RISC-V RV32I and RV32IM. We compile C code for the
uncertainty-tracking hardware using the publicly-provided online access to the
toolchain. We configure the uncertainty-tracking hardware for
uncertainty-tracking computation with the smallest-available uncertainty
representation (C0-microSD-N).

\subsubsection*{Accuracy-based speedup benchmarking}

To quantify the speedup of the UxHw approach we quantify how much faster it
estimates the sensor output distribution. To do this, we first quantify the
accuracy of the UxHw output in terms of the Wasserstein distance to the
\qty{500000}{} Monte Carlo execution result. Then, we perform batches of
independent Monte Carlo simulations where each stops when it achieves the
Wasserstein distance. The result is the empirical distribution of the Monte
Carlo iteration counts necessary to beat UxHw. From this distribution we pick
the 90-th percentile that we call \emph{EqMCp90} to time against UxHw on the
FPGA-5k and 17k.

\acknow{Orestis Kaparounakis was supported by an EPSRC DTP Studentship.\\
Figure~\ref{fig:ThermopileInfraredSensorDiagram} icon attributions
(flaticon.com): the earth, analog signal, transducer, and distributions icons
are by Freepik.}\showacknow

\vspace{0.25in}

\bibliography{references}

\clearpage
\pagebreak
\appendix
\renewcommand\thesection{SM}
\section{Supplementary material}
\addcontentsline{toc}{section}{Supplementary material}

Because contemporary digital computers are voltage-based, analog environmental
information such as temperature or sound entering the computer system invariably
pass through an electrical transducer to convert the original analog phenomenon
to a voltage. And because the only way conventional digital computers convert
voltage levels to digital values is to use an analog-to-digital converter
(\textsc{\Large adc}), sensors in general transduce some physical phenomenon
ultimately into a voltage.

The following sections provide further discussion and context around the core
idea of representation uncertainty as well as supplementary material for the
experimental validation using the Melexis MLX90640 far-infrared sensor array.

\subsection{Infrared radiation and sensing} \label{sec:SupplementaryInfrared} Infrared, or thermal, radiation is
electromagnetic radiation in the infrared spectrum, i.e., wavelengths from
\qty{750}{\nano\meter} to \qty{1}{\milli\meter}. It emanates from the surface of
every material and has power proportional to the degree of thermal motion of the
material's particles, colloquially known as the material's temperature.

The thermal emission power also depends on other material properties, e.g., its
chemical composition~\cite{neuer1998spectral}. Ideal materials that emit the
maximum theoretical radiation are called \emph{black bodies}. Actual materials
emit radiation at a fraction of the black-body radiation---that fraction is
called \emph{emissivity}. I.e., emissivity is a dimensionless quantity ranging
from 0 to 1 and it quantifies a material's efficiency in emitting thermal
radiation relative to an ideal black body.

Let $T$ be the temperature of an object whose material has emissivity
$\epsilon$. Let $P$ represent the total infrared radiation emission power per
unit surface area, and let $\sigma$ be the Stefan-Boltzmann constant.
Equation~\ref{eq:stefanBoltzmannLaw} states the Stefan-Boltzmann law which
describes the relationship between the temperature of the object and the
emission of infrared radiation.

\begin{equation}
    P = \epsilon \sigma T^4. \label{eq:stefanBoltzmannLaw}
\end{equation}

Equation~\ref{eq:stefanBoltzmannLaw} assumes constant emissivity across
all wavelengths and temperatures. Some materials with low emissivity, such as
commercial aluminum ($\epsilon \approx 0.04-0.09$), deviate from this
assumption~\cite{melexis2019mlx90616}. These materials predominantly reflect
incident radiation rather than emit it, leading to underestimation of their true
temperature when using infrared sensing techniques~\cite{wen2004emissivity}. In
such cases, the detected radiation may primarily originate from the surrounding
environment rather than the object itself, rendering the simple Stefan-Boltzmann
model inadequate. Accurate temperature measurement of low-emissivity materials
necessitates more sophisticated models that account for factors such as
environmental radiation reflection, wavelength-dependent emissivity, and surface
conditions.

\subsubsection*{Thermopile-based infrared radiation sensors}

Thermopile sensors capture electromagnetic energy in the infrared spectrum and
convert it to electrical energy in the form of a measurable voltage via the
thermoelectric (Seebeck) effect.

A thermopile consists of two or more thermocouples. A thermocouple is a
temperature measurement device made of two dissimilar metals joined at two
junctions. One junction receives the incoming radiation (hot junction) and the
other is at a reference temperature (cold junction). This creates a temperature
difference between the two junctions which results in a voltage potential due to
the Seebeck effect.

Let $S_{XY}$ be the Seebeck coefficient between metals $X$ and $Y$. The
temperature at the hot junction is $T_H$ while at the cold junction is $T_C$.
Equation~\ref{eq:thermocoupleVoltage} shows the voltage output  $V$ of the
thermocouple~\cite{van1986thermal}.

\begin{equation}
    V = S_{XY}(T_H -T_C).  \label{eq:thermocoupleVoltage}
\end{equation}

Individual thermocouple voltages are typically small, e.g.,
\qty{58.7}{\micro\volt\per\kelvin} for an ANSI Type E chromel-constantan
thermocouple~\cite{ansi2024thermocouples}, one of the highest coefficients in
the common junction types. Thermopiles use two or more thermocouple pairs placed
in series, and these voltages add up. To increase sensitivity and utility,
sensor package designs often include an amplification step before
analog-to-digital conversion~\cite{winsen2022mrt313, marek2021thermopile,
hamamatsu2024thermopile, analog2016ndir}.

Let $\alpha$ be the pixel sensitivity and $S$ be the Seebeck coefficient slope.
Let $c_0 = \qty{273.15}{\kelvin}$ (\qty{0}{\degreeCelsius}). Let
$V_\mathrm{out}$ represent the analog output voltage of a thermopile-based
infrared sensor. Equation~\ref{eq:mlx90640ToutSupplementary}, adapted from the
sensor datasheet~\cite{melexis2019mlx}, gives the target object temperature
($T_o$; in the Kelvin scale) for one pixel of the frame of the sensor.
\begin{equation}
    T_o = \sqrt[4]{\frac{V_\mathrm{out}}{\alpha S (\sqrt[4]{\frac{V_\mathrm{out}}{\alpha} + T_{a-r}} - c_0)} + T_{a-r}}. \label{eq:mlx90640ToutSupplementary}
\end{equation}
Where the term $T_{a-r}$ encodes the effect of ambient and reflected radiation
on the sensor (\autoref{eq:mlx90640TarSupplementary}).
\begin{equation}
    T_{a-r} = - \frac{(1 - \epsilon) T_r^4}{\epsilon} - \frac{T_a^4}{\epsilon}. \label{eq:mlx90640TarSupplementary}
\end{equation}

The manufacturer provides conversion routines to convert raw sensor readings
(Figure~\ref{fig:ThermopileInfraredSensorDiagram}.G) to per-pixel temperature
values. These routines are functions of the 768-pixel raw sensor data, 37
calibration parameters (four of which are 768-element vectors), and two dynamic
sensor-provided parameters (device temperature and voltage). Supplementary
Material Table~\ref{tab:Mlx90640CalibrationParametersExtracted} lists the
extracted value for all calibration parameters, and Supplementary Material
Table~\ref{tab:Mlx90640CalibrationMemoryData} lists the raw calibration data.

\subsection{Discretization introduces uncertainty}
\label{section:discretizationErasesInformation}

Real numbers are infinite but digital storage and memory are finite.
Conversion of a number from a dense or continuous domain to a sparser domain
maps regions of the originating domain, each to a single value in the target
domain---the target domain has lower \emph{precision}.

\subsubsection*{Representation error} Because of information loss, if after
conversion we project a lower-precision quantity back to its original domain,
the projection and the original quantity will almost certainly differ. Depending
on the scientific field and the specific numerical domains involved, the
difference between the original quantity and the projection is called
quantization error, discretization error, or rounding error. This work refers to
this error which arises due to conversion to a lower-precision representation as
\emph{representation error}. For example, in an application that requires
large amounts of parameters (e.g., a machine learning algorithm), engineers may
change the parameter representation size to reduce memory requirements, e.g.,
from \qty{64}{\bit} to \qty{16}{\bit}. This reduces the available information of
the parameters and introduces representation error. This error is not because of
noise in the original parameter values but rather an inherent principle of
quantization.

\subsubsection*{Representation uncertainty}  When converting quantities to a
lower-precision domain, different numbers in the original domain end up as the
same number in the target domain. Because conversion is a many-to-one function,
we cannot know what the original number was before conversion. We therefore
represent the probable original values as a uniform probability distribution
around the converted value.
This work refers to this probability distribution
as \emph{representation uncertainty}. For example, representation uncertainty
arises in computer code when converting a floating-point value to an integer
value via truncation or rounding.

\subsubsection*{Relation to quantization-mitigation methods}

Engineers reduce quantization error in the \emph{signal path}, e.g., at an
ADC output or in repeated fixed-point arithmetic, by increasing bit depth,
adding dither, oversampling, or using error feedback. Dithering, oversampling,
and error-feedback corrections target quantization of time-varying quantities:
they reduce deterministic artifacts and convert rounding effects into error
terms that filtering suppresses. This article studies a different mechanism also
present in sensors whereby devices quantize calibration data once to fit into
limited non-volatile memory and then reuse those fixed bytes to reconstruct
calibration parameters at run time. Applying these techniques does not increase
the information content of stored coefficients. Repeated measurements do not
reduce the resulting error unless the system performs online identification or
stores side-information. Our analysis treats each stored datum as an unknown
pre-quantization value within its quantization bin and dynamically propagates
this representation uncertainty through the calibration-extraction and
calibration-compensation code.

Section~\ref{SectionApplicationDesignSpaceExploration} examines how different
bit-depth choices for calibration data storage affect this uncertainty.

\subsubsection*{Representation uncertainty and correlation} Two or more
uncertain quantities may share a relation---this means they share \emph{mutual
information} and are \emph{correlated}.
When inputs are correlated, uncertainty
quantification must sample from the joint probability distribution of the
quantities to account for the correlations~\cite{papadopoulos2001uncertainty}.

Representation errors stemming from the discretization of different quantities
are \emph{mutually independent}~\cite{drosg2009dealing}. However, correlations
commonly arise during computation, even for independent inputs. This happens
because an uncertain quantity may take part more than once in calculating the
result---and at every operation it takes part in after the first, the operands
will share mutual information. These runtime correlations make it hard to derive
a closed-form solution for the uncertainty of computation, and analyses today
often have to rely on computationally-expensive Monte Carlo methods.

\subsection{Quantization and representation uncertainty examples}

For a simple analog-to-digital conversion example, consider an 10-bit
\textsc{\Large adc}, with 1024 levels of quantization. We use it to quantize a
voltage input that can range from \qty{0}{\volt} to \qty{1}{\volt}.
Equation~\ref{eq:quantizationMidTread} is the input-output relationship of
typical mid-tread quantization.
\begin{equation}
    Q_{\mathrm{MT}}(x) = \Delta \cdot \left\lfloor \frac{x}{\Delta} + \frac{1}{2} \right\rfloor.  \label{eq:quantizationMidTread}
\end{equation}
We adjust the \adc{} for $\Delta = 0.0009765625$, dividing the voltage range in
1024 regions. For $x = \qty{0}{\volt}$, the \adc{} gives output 0, and for $x =
\qty{0.5}{\volt}$, 512. All voltages in the same region of size $\Delta$ result
to the same integer. The digital system that receives the \adc{} output of 512
should heed that the \adc{} input could be any voltage in the region
$[\qty{0.49951171875}{\volt}, \qty{0.50048828125}{\volt})$---an epistemic
uncertainty. If a function describes the relation of a physical quantity to the
input voltage, the inverse function describes the propagation of the uncertainty
about the input voltage to the uncertainty of the physical quantity. Because
sensors transduce physical quantities to voltages, and so need a way to relate
the raw output back to the physical quantity, this property is significant in to
them.

Representation uncertainty as an effect of quantization via a type conversion
arises in computer code that stores fundamental real-values quantities,
eventually as integers in a storage device. For example, this happens during
sensor calibration processes that store calibration parameters on-device as
integers, as is the case with the MLX90640 sensor~\cite{melexis2019mlx}. Because
the difference between sequential integers is one ($\Delta=1$ in
Equation~\ref{eq:quantizationMidTread}),
Equation~\ref{eq:quantizationNearestInteger} is a representation of
nearest-integer rounding that rounds halfway cases away from zero, as does the C
function \codelight{round()}.
\begin{equation}
    Q_{\mathrm{NI}}(x) = \left\lfloor x + \frac{1}{2} \right\rfloor.  \label{eq:quantizationNearestInteger}
\end{equation}

\begin{figure}[tbh]
    \centering

    \includegraphics[trim={1.2cm 0cm 1.6cm 1cm},clip,width=0.48\textwidth]{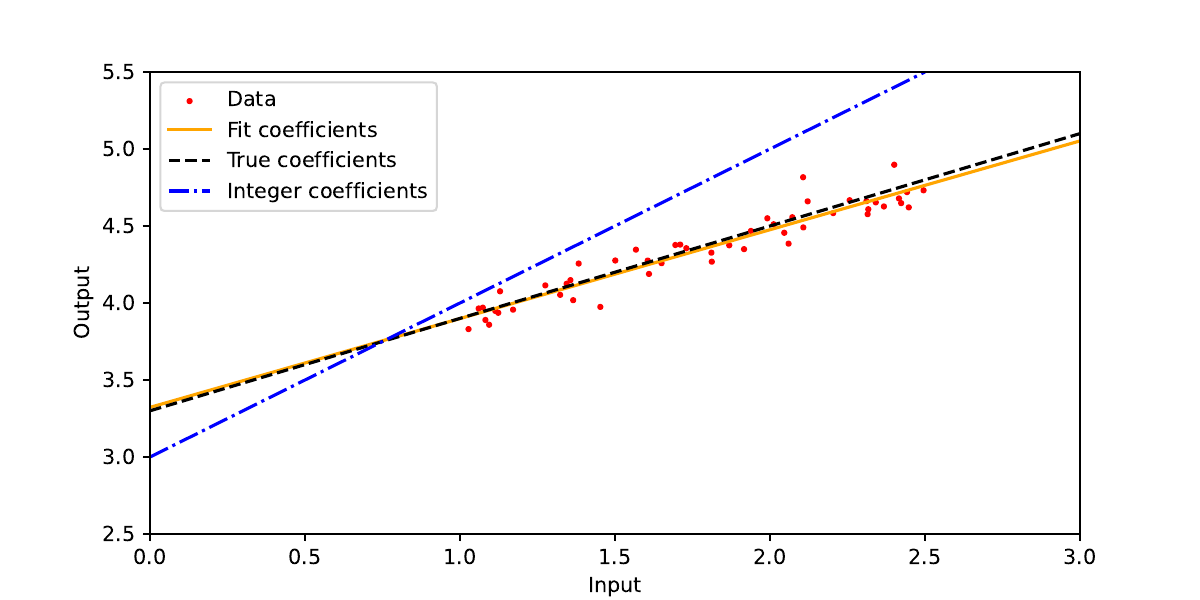}

    \caption{Simplified example of sensor calibration using coefficients of a
    linear fit. While linear regression (yellow) may be sufficiently close to
    the true model (black), using the coefficients after rounding is an
    inaccurate model (blue). Sensors uses coefficient-based calibration need to
    store device-specific numbers in the constrained bit lengths of non-volatile
    memory.}
    \label{fig:LinearFitIntegerExample}

\end{figure}

One common method of sensor calibration is fitting an expected function type to
the raw sensor data and ground truth data from a controlled calibration
environment~\cite{tatsch2023open, fraden2016handbook}. As a simple example,
consider a linear fit with two coefficients, $a x + b$. Let $a^\star$, $b^\star$
be the ideal coefficients with minimum error between ground truth and
prediction. Let $a_{\mathrm{fit}}$, $b_{\mathrm{fit}}$ be the coefficient
results of the fitting algorithm. Let $a_{\mathrm{int}}$, $b_{\mathrm{int}}$ be
the integer coefficients stored on the sensor device. For a physical model with
$a^\star = 0.6$ and $b^\star = 3.4$, a fit\footnote{Implementation with
\codelight{numpy.polyfit} of Python 3.12.8, NumPy 2.1.1.} on 100 data points
with zero-mean Gaussian noise (with standard deviation 0.1) finds
$a_{\mathrm{fit}} = 0.55782446$, $b_{\mathrm{fit}}=3.3512181$, and with
nearest-integer rounding $a_{\mathrm{int}} = 1$, $b_{\mathrm{int}} = 3$.
Figure~\ref{fig:LinearFitIntegerExample} visualizes the example and the
inaccuracy of the integer coefficients. In a more realistic scenario,
calibration processes scale and offset the data to reduce the memory footprint
of the calibration parameter and also achieve a more precise integer fit for
two-parameter linear-fit scenario. However, calibration processes perform this
space optimization for more than two parameters and for non-linear
transformations, and inevitably cannot fully avoid this error.

Sensors that leverage coefficient-based fitting as part of the calibration
process need to store the calibration data in a non-volatile memory of the
sensor package because these data are different for each instance of the sensor.
The storage encoding must be able to sufficiently represent the effective
coefficient space for this particular sensor type and sensor manufacturers may
provide coefficient extraction procedures to recover the coefficients (these
procedures need to be the exact same for all instances of the sensor).
Non-linear models with increased numbers of coefficients exacerbate this issue
as they are inherently more sensitive to smaller changes in the coefficient
values.

\subsubsection*{Rounding and truncation in C} The calibration process itself is
computer code, often a program written in the C programming language. Before
storing the calibration data in the memory of each device, the calibration
program performs arithmetic operations and ultimately, for quantities that do
not correspond to ordinal values, either casts the value to an integer, often as
an implicit conversion (implicit cast), or explicitly rounds the value to the
nearest integer. In the C language, when casting a floating-point number to an
integer the implementation discards fractional part (truncation towards zero;
Equation~\ref{eq:quantizationTruncationTowardsZero})~\cite{iso1999c99}\footnote[2]{Holds
for all standard revisions C89, C99, C11, C17, and C23 (which also includes the
decimal type).}.
\begin{equation}
    Q_{\mathrm{TTZ}}(x) = \mathrm{sgn(x)}  \left\lfloor \lvert x \rvert \right\rfloor.  \label{eq:quantizationTruncationTowardsZero}
\end{equation}

\smallskip
Sensor calibration processes and software typically explicitly defines the
rounding approach for the calibration parameters, which most often is
nearest-integer rounding (Equation~\ref{eq:quantizationNearestInteger}). For
some sensors, application engineers are able to perform their own calibration
and store application-specific calibration data back to the sensor memory. This
is the case for Bosch BNO055 where the official sensor driver includes functions
for passing calibration data back to the sensor package as integers of type
\codelight{signed\;short}. Those calibration data come from calibration
parameters which are in the real domain. So, there is representation uncertainty
when storing them as \codelight{signed\;short} (typically 16 bits in low-power
systems). Also, the manufacturer does not clarify a rounding scheme for passing
these parameters, and the different ranges of representation uncertainty
associated with different rounding schemes leads to different sensor output
uncertainty.

\subsubsection*{Tracking uncertainty across re-discretizations}

The official C driver re-discretizes some extracted calibration parameters to
conserve runtime memory usage. For example, it stores the extracted per-pixel
$\alpha$, $\mathrm{K_{Ta}}$, and $\mathrm{K_{V}}$ coefficients as integer
numbers, after explicitly rounding them from floating point to the nearest
integers. At the time of temperature conversion, the driver code performs a
second set of ad hoc extractions for these parameters and then uses them in
floating-point form. Because we are studying the effect of representation
uncertainty to the output temperature image, we change the execution code to
treat this data as floating-point numbers throughout execution, and sidestep the
re-discretization.

This memory-conservation technique introduces a second level of representation
uncertainty affecting the output of the sensor that warrants its own separate
investigation. Probationary Monte Carlo executions for quantifying the effect of
the memory-conservation technique showed that the change was much smaller
compared to the base representation uncertainty of the calibration data. For the
sake of brevity and clarity, and without real loss of generality, we bypass the
specific part of the driver and do not quantify that effect in this work.

\subsection{Supplementary data and figures for Melexis MLX90640}

\label{SectionSupplementaryDataAndFiguresMLX}

In case of sensors where use of calibration data is necessary, sensor
manufacturers provide conversion routines to convert raw sensor readings to
meaningful sensor outputs. For the MLX90640, these routines convert raw data to
per-pixel temperature values, as exemplified by
Figures~\ref{fig:ThermopileInfraredSensorDiagram}(G)
and~\ref{fig:ThermopileInfraredSensorDiagram}(L), respectively. These conversion
routines are functions of the 768-pixel raw sensor data, 37 calibration
parameters (four of which are 768-element arrays), and two dynamic
sensor-provided parameters (device temperature and voltage).
Table~\ref{tab:MLX90640OutputUncertaintyMetrics} lists aggregate metrics
for all \qty{322560} the MLX90640 output pixel distributions (recall
Figure~\ref{fig:MlxSensorOutputsErrorsCombined}).

Table~\ref{tab:Mlx90640CalibrationParametersExtracted} lists the extracted value
for each calibration parameter. The official MLX90640
driver\footnote{https://github.com/melexis/mlx90640-library} extracts the raw
calibration data from the sensor by reading the EEPROM memory. The driver firsta
reads the data from the memory as unsigned 16-bit integer numbers
(\verb|uint16_t|). Then, it performs logical and arithmetic operations on the
values read, such as bit shifts and floating-point scaling, to compute the
calibration parameters---in some sense, the calibration parameters have their
own set of conversion routines.

Table~\ref{tab:Mlx90640CalibrationMemoryData} lists the raw scalar calibration
data that is available in the sensor memory. Applications that need to extract
the calibration parameters from this data are subject to the uncertainty of the
quantized representation of the data. This is an effect of the calibration
process that converts analog quantities or digital floating-point values
ultimately to an integer to store in the sensor device memory. Using
floating-point representations for calibration data would partly mitigate but
not eliminate the issue. 

The Uncertainty column of Supplementary Material
Table~\ref{tab:Mlx90640CalibrationParametersExtractedShort} shows the rough
outline of the representation uncertainty for five scalar calibration parameters
relevant to the sensitivity calibration of the sensor. Because these
uncertainties show up through operations that sometimes reference the value of
other calibration parameters more than once, the uncertainty of calibration
parameters can share mutual information. This happens even though these
uncertainties originate from uncorrelated representation uncertainty of
calibration data due to integer conversion~\cite{drosg2009dealing}. For example,
in Equation~\ref{eq:alphaExtractionEquation} the uncertainties of $\alpha_3$ and
$\alpha_4$ share mutual information in $A_{\mathrm{ref}}$, $R_0$,
$\mathrm{TGC}$, and $\alpha_{\mathrm{CP}}$.
Figure~\ref{fig:CalibrationExtraction} further illustrates the extraction
of the calibration parameter tensor $\alpha$.

\begin{figure*}[tp]
    \centering

    \begin{subfigure}[t]{0.45\textwidth}
        \centering
        \includegraphics[clip,width=\linewidth]{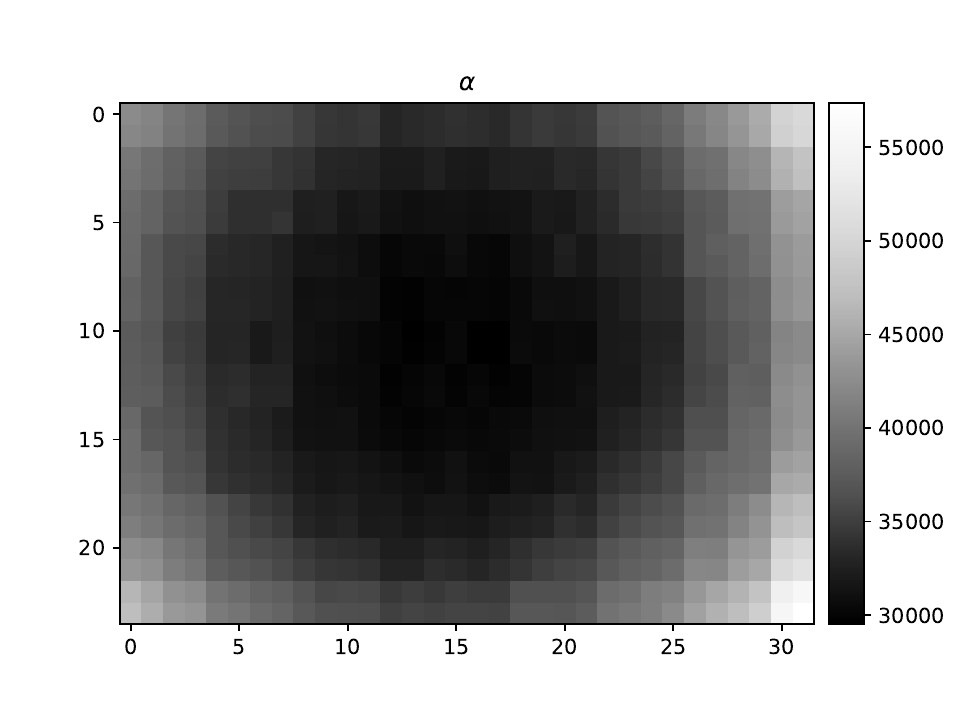}
        \caption{The sensitivity calibration parameter $\alpha$ has different
        value for each pixel. Because of the representation uncertainty in the
        calibration data each $\alpha$-pixel has a different probability
        distribution.}
        \label{fig:MlxCalibrationParameterParticleAlpha}
    \end{subfigure}
    \hspace{0.5em}
    \begin{subfigure}[t]{0.45\textwidth}
        \centering
        \includegraphics[clip,width=\linewidth]{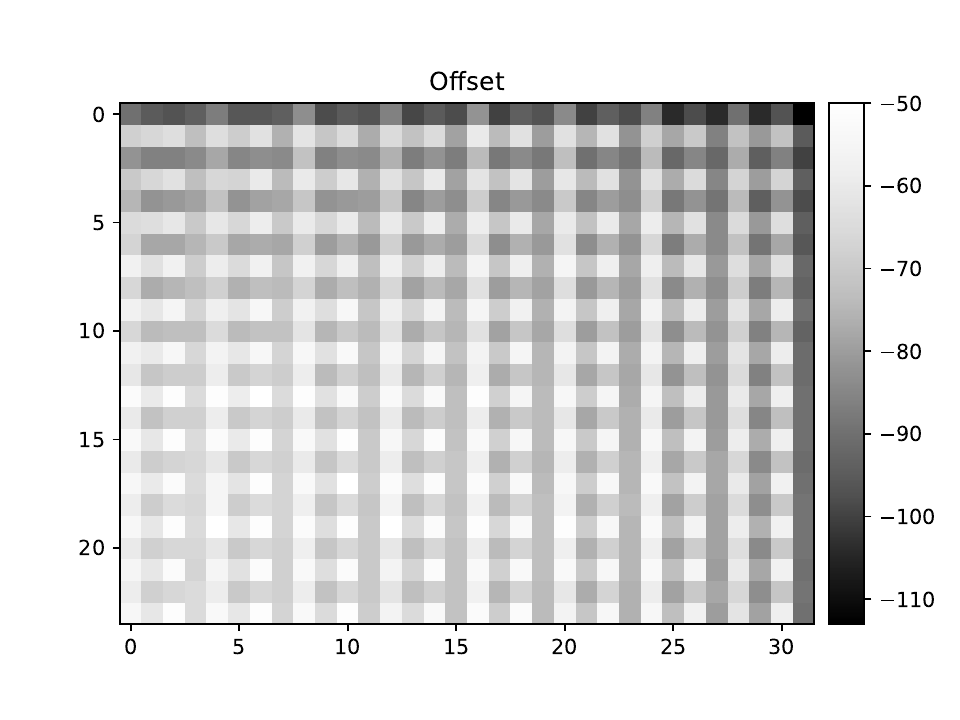}
        \caption{The offset calibration parameter has different value for each
        pixel. Because of the representation uncertainty in the calibration data
        each pixel offset has a different probability
        distribution.}
        \label{fig:MlxCalibrationParameterParticleOffset}
    \end{subfigure}
    \\[-1ex]
    \begin{subfigure}[t]{0.45\textwidth}
        \centering
        \includegraphics[clip,width=\linewidth]{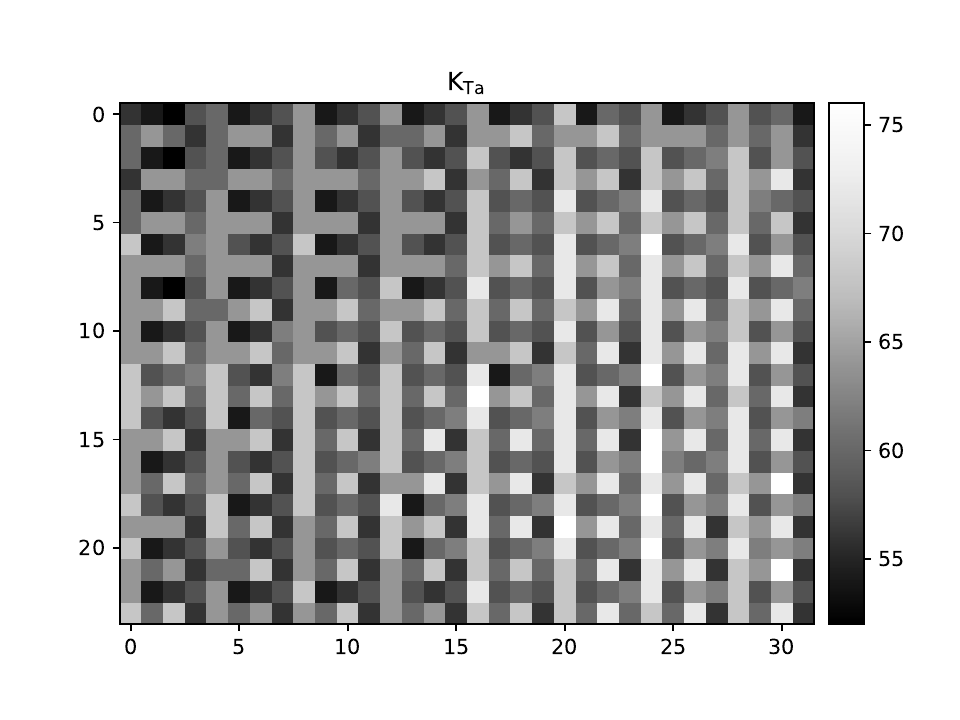}
        \caption{Per-pixel calibration parameter values for parameter
        $\mathrm{K_{Ta}}$ repeat based on a pattern. Because of the representation
        uncertainty in the calibration data, each pixel has an associated
        probability
        distribution.}
        \label{fig:MlxCalibrationParameterParticleKta}
    \end{subfigure}
    \hspace{0.5em}
    \begin{subfigure}[t]{0.45\textwidth}
        \centering
        \includegraphics[clip,width=\linewidth]{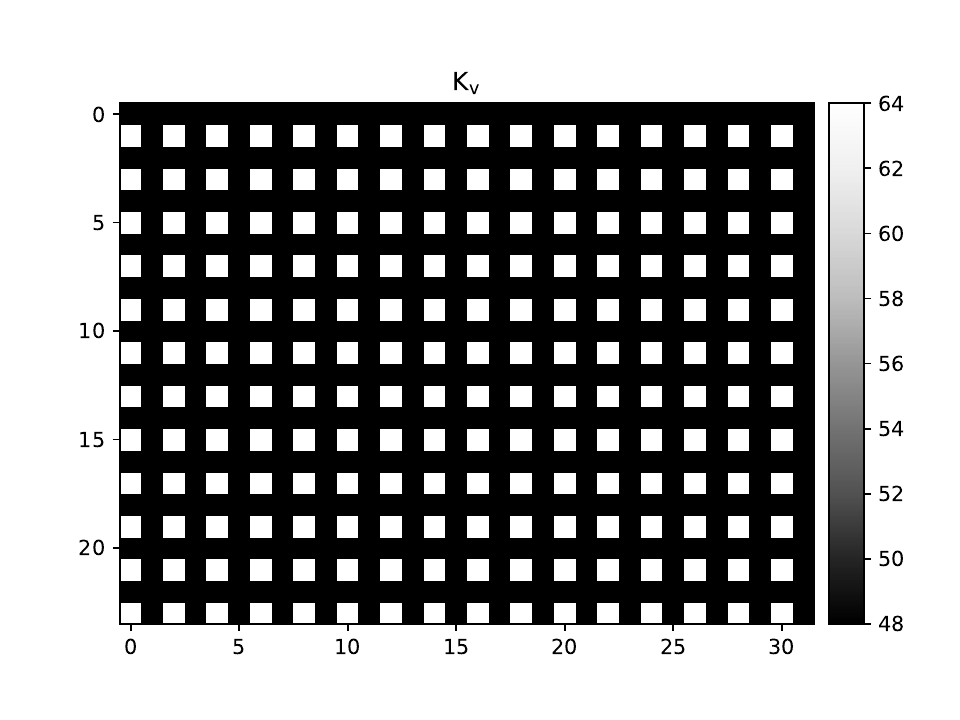}
        \caption{Per-pixel calibration parameter values for parameter
        $\mathrm{K_{V}}$ take one of two values depending on the pixel location.
        Because of the representation uncertainty in the calibration data, each of
        these has an associated probability
        distribution.}
        \label{fig:MlxCalibrationParameterParticleKv}
    \end{subfigure}

    \caption{Particle values of the per-pixel calibration parameters. Each pixel
    value has an associated probability distribution that describes the probable
    values this parameter can take due to representation uncertainty in the
    calibration data.\label{fig:SupplPerPixelCalibrationParameters}}

\end{figure*}

\begin{figure*}[tp]
    \centering

    \begin{subfigure}[t]{0.3\textwidth}
        \centering
        \includegraphics[clip,width=\linewidth]{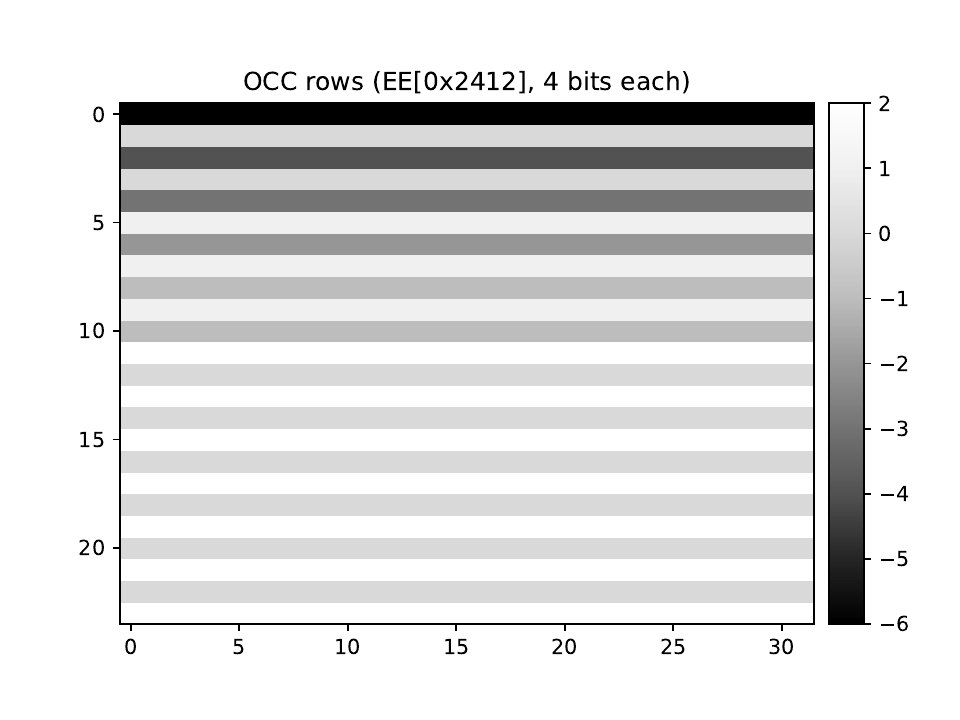}
        \caption{Per-row calibration data for offset calibration parameters.}
        \label{fig:MlxFlashDataOccRows}
    \end{subfigure}
    \hspace{0.5em}
    \begin{subfigure}[t]{0.3\textwidth}
        \centering
        \includegraphics[clip,width=\linewidth]{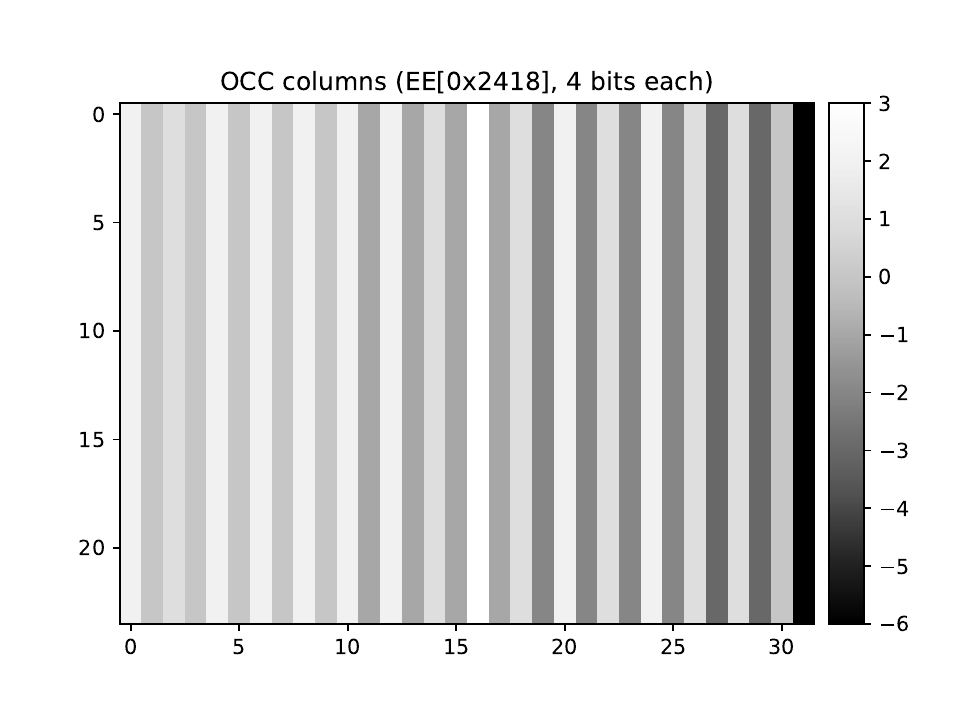}
        \caption{Per-column calibration data for offset calibration parameters.}
        \label{fig:MlxFlashDataOccCols}
    \end{subfigure}
    \hspace{0.5em}
    \begin{subfigure}[t]{0.3\textwidth}
        \centering
        \includegraphics[clip,width=\linewidth]{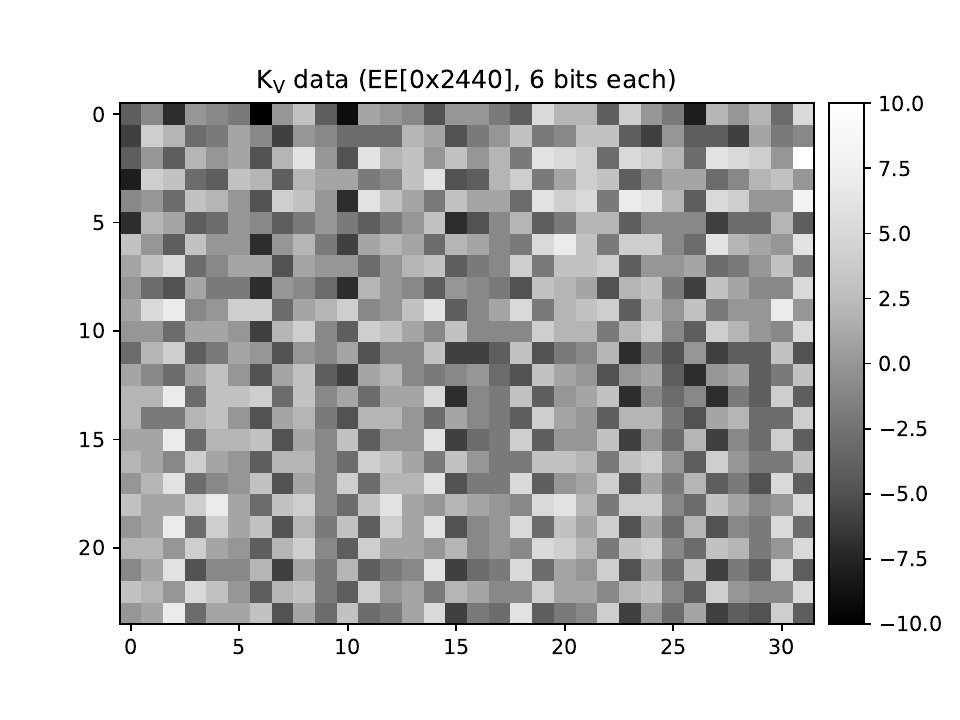}
        \caption{Per-pixel calibration data remainder for offset calibration parameters.}
        \label{fig:MlxFlashDataOffsetPixels}
    \end{subfigure}
    \\[-1ex]
    \begin{subfigure}[t]{0.3\textwidth}
        \centering
        \includegraphics[clip,width=\linewidth]{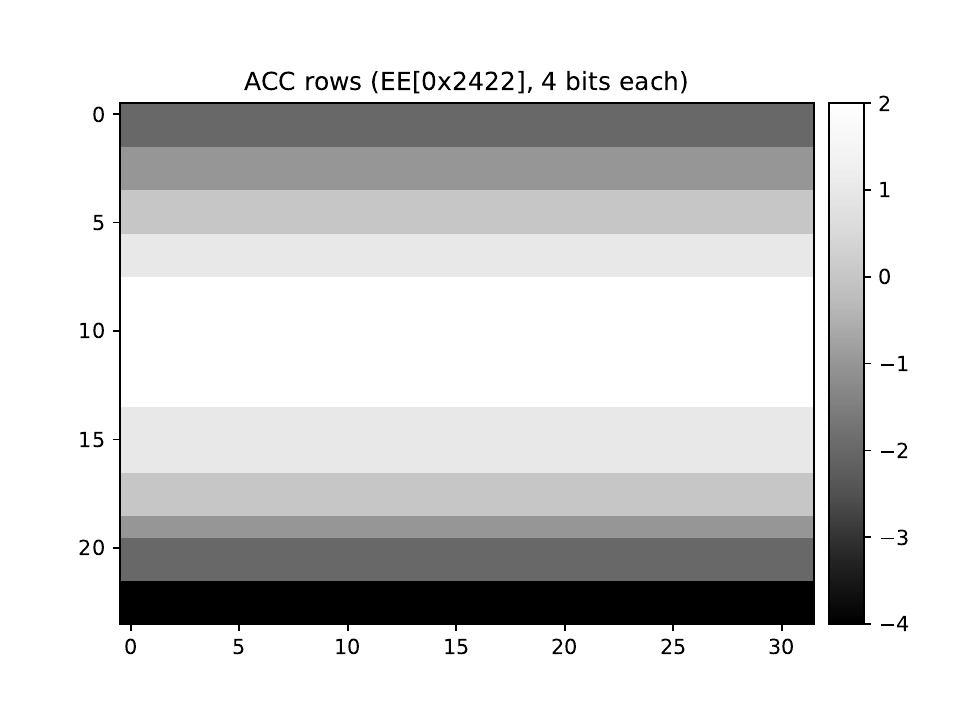}
        \caption{Per-row calibration data for $\alpha$ calibration parameters.}
        \label{fig:MlxFlashDataAccRows}
    \end{subfigure}
    \hspace{0.5em}
    \begin{subfigure}[t]{0.3\textwidth}
        \centering
        \includegraphics[clip,width=\linewidth]{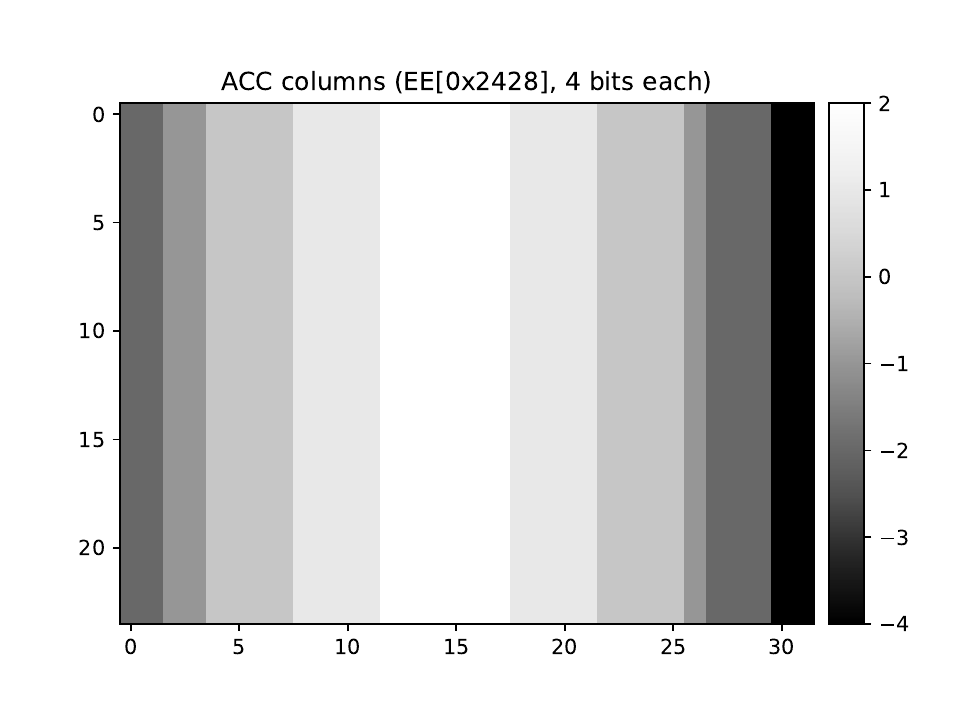}
        \caption{Per-column calibration data for $\alpha$ calibration parameters.}
        \label{fig:MlxFlashDataAccCols}
    \end{subfigure}
    \hspace{0.5em}
    \begin{subfigure}[t]{0.3\textwidth}
        \centering
        \includegraphics[clip,width=\linewidth]{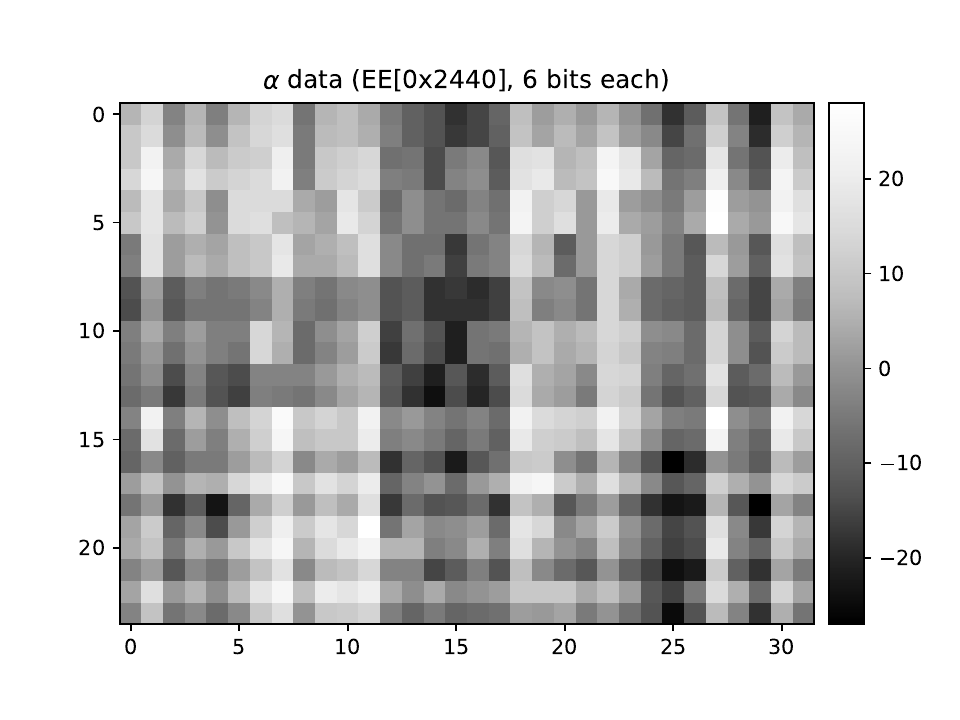}
        \caption{Per-pixel calibration data remainder for $\alpha$ calibration parameters.}
        \label{fig:MlxFlashDataAlphaPixels}
    \end{subfigure}
    \\[-1ex]
    \begin{subfigure}[t]{0.3\textwidth}
        \centering
        \includegraphics[clip,width=\linewidth]{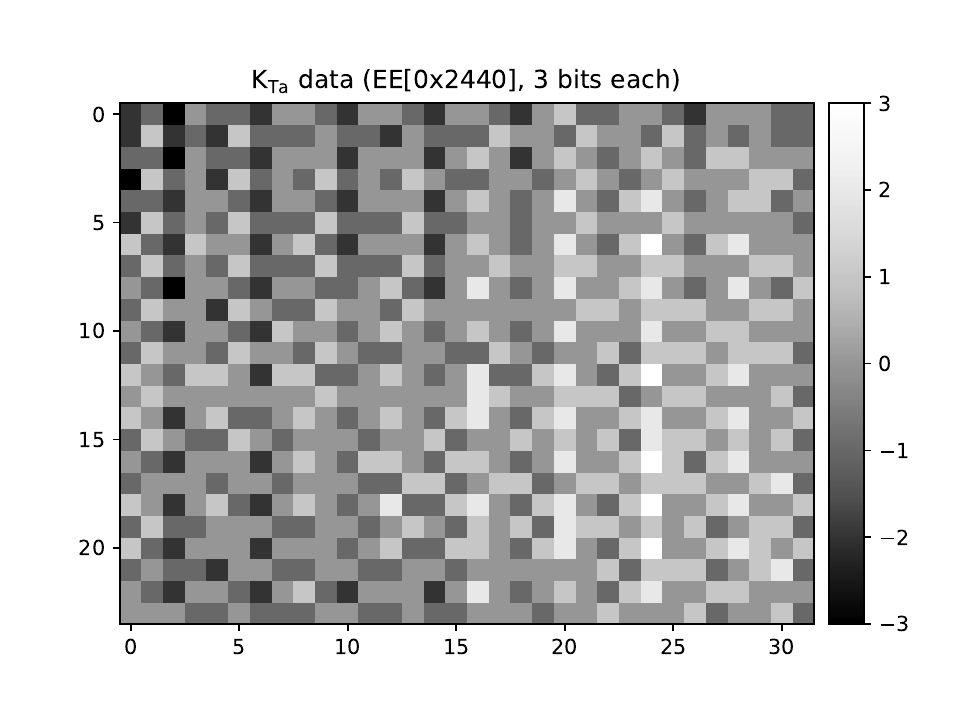}
        \caption{Per-pixel calibration data for $\mathrm{K_{Ta}}$ calibration parameters.}
        \label{fig:MlxFlashDataKtaPixels}
    \end{subfigure}

    \caption{Pixel-location-dependent calibration data for the MLX90640
    measurement conversion routines. Every calibration data integer value $X$
    from the sensor EEPROM represents a fundamentally real-valued quantity that
    could have been anywhere in the range $(X-0.5, X+0.5)$ before the
    calibration process rounded it to the integer $X$. Because the sensor driver
    constructs the $K_V$ calibration parameter by selecting between four scalar
    calibration data values based on the pixel location and scaling the
    selection, we do not show per-pixel calibration data for $K_V$
    here.\label{fig:SupplPerPixelCalibrationRawData}}

\end{figure*}

Figure~\ref{fig:SupplPerPixelCalibrationParameters} shows the conventional
values of the four extracted per-pixel calibration parameters. Each of these is
a 768-element vector that for better intuition we plot as an image of the same
dimensions as the sensor output.
Figure~\ref{fig:SupplPerPixelCalibrationRawData} presents the raw calibration
data (from the sensor memory) which is the input for the extraction routines
that compute the calibration parameters of
Figure~\ref{fig:SupplPerPixelCalibrationParameters}.

\begin{table}[tbh]
    \centering
    \caption{Metrics summary for the output uncertainty of the MLX90640 from all
    768 pixels of four tested sensor instances, each tested for 21 target
    temperatures (322\,560 sensor output distributions). This is the same data as
    Figure~\ref{fig:MlxSensorOutputsErrorsCombined}, in literal numerical
    format.}
    \begin{tabular}{lSSS}
        \toprule
        \textbf{Metric aggregates} & \textbf{Min} & \textbf{Mean} & \textbf{Max}\\
         \midrule
            \rowcolor{a} \textbf{Mean Absolute Errors (\textdegree{C})}     & 0.18    & 0.44      & 1.63      \\
            \rowcolor{b} \textbf{Mean Relative Errors}                      & 0.46\%  & 0.90\%    & 5.38\%    \\
            \rowcolor{a} \textbf{Max Absolute Errors (\textdegree{C})}      & 0.66    & 1.67      & 5.35      \\
            \rowcolor{b} \textbf{Max Relative Errors}                       & 1.50\%  & 3.58\%    & 25.70\%   \\
            \rowcolor{a} \textbf{Standard deviations (\textdegree{C})}      & 0.22    & 0.53      & 1.95      \\
            \rowcolor{b} \textbf{Size of 95\% CIs (\textdegree{C})}         & 0.81    & 2.01      & 7.16      \\
            \bottomrule\\[-6ex]
            \multicolumn{4}{p{4cm}}{\scriptsize CIs=Confidence Intervals}
    \end{tabular}
    \label{tab:MLX90640OutputUncertaintyMetrics}
\end{table}

\newcommand{\mlxinstanceid}{\codelight{0x1f15cb2d0189}}

\setcounter{rownumber}{0}
\renewcommand{\therownumber}{R\arabic{rownumber}}
\renewcommand{\rowcountstep}{\refstepcounter{rownumber}\therownumber}

\subsubsection*{Uncertainty information and edge detection}

As one demonstrative application of the impact of epistemic uncertainty in
calibration-compensated sensing, and the potential usefulness of tracking it in
real time, we examine the propagation of the representation uncertainty from the
MLX90640 conversion routines through the Canny edge detection algorithm. We
study ten different scenes of varying complexity, emissivity, and background.
For each scene, we run a Monte Carlo simulations with 500K re-executions of the
Canny algorithm. For each re-execution, the input sample is one probable
sample-frame of 768 pixel temperatures from the MLX90640 sensor (recall
Section~\ref{sec:MLX90640Analysis}). The output of each simulation is a
distribution of edge-annotated frames. 

\begin{figure*}[tbh]
    \centering
    \includegraphics[clip,width=0.95\textwidth]{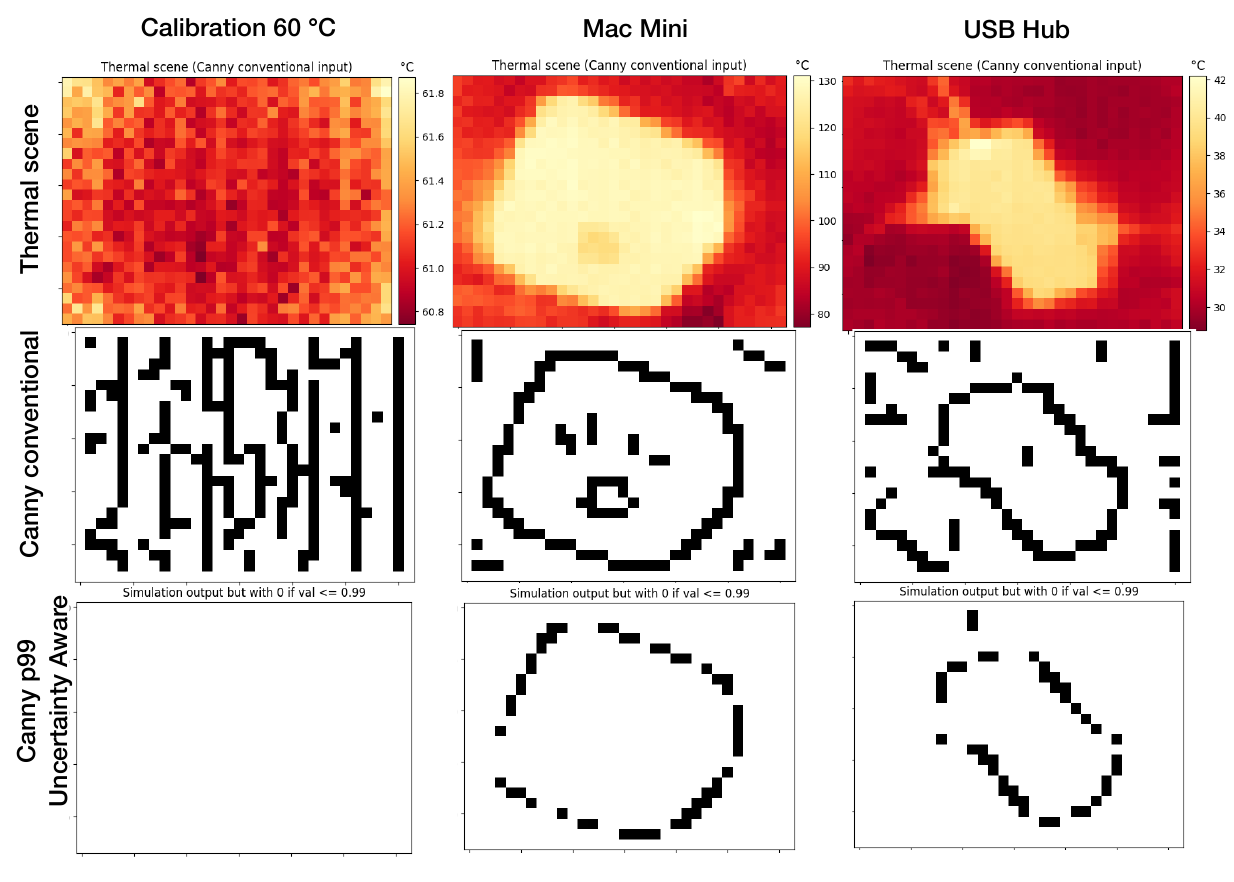}
    \caption{Edge detection thermal image output, conventional edge detection,
    result, and uncertainty-aware edge detection p99 result for the scenes
    Calibration Source \qty{60}{\degreeCelsius}, Mac Mini M4, and USB Hub. For
    all tested scenes the conventional results show artifact that relate to the
    representation uncertainty of the sensor calibration data. The
    uncertainty-aware version of the algorithm can better marginalize these
    artifacts to a cleaner image using a tunable probability threshold. For the
    p99 threshold, the uncertainty-aware approach achieves 100\%, 96\%, and
    96.8\% accuracy, as much as 5.3\% higher than the conventional output. For
    the cases of the Mac Mini M4 and the USB Hub this accuracy increase comes
    with a drop to sensitivity. Applications can tune the probability threshold to
    their accuracy, precision, and sensitivity requirements.}
    \label{FigureEdgeDetectionSelectedScenes}
\end{figure*}

The rows of Figure~\ref{FigureEdgeDetectionSelectedScenes} show the thermal
image output, the edge detection output using the conventional thermal image,
and the edge detection output using the uncertainty-aware framework of
Section~\ref{section:edgeDetectionApplication}.
Figure~\ref{FigureEdgeDetectionSelectedScenes} shows these for three testcase
scenes (columns): Calibration Source \qty{60}{\degreeCelsius} which corresponds
to an image of the infrared calibration source where all pixels are nominally
the same temperature, a Mac Mini M4 that has heated up under load, and a
USB hub which has heated up under heavy-traffic.

Table~\ref{TableEdgeDetectionMetricsFull} lists the performance metrics of the
conventional and of the uncertainty-aware approaches

\subsubsection*{Uncertainty information and design-space exploration}

Supplementary Table~\ref{TableUncertaintyStatisticsByScenarioPctChange}
presents detailed data for the percent-change differences of the sensor output
uncertainty for each of the four alternative calibration-data storage
scenarios.

\begin{table*}[tbh]
    \centering
    \caption{The calibration parameters for the Melexis MLX90640 infrared sensor
    conversion routines, after extraction from the device memory and applying
    the extraction routines. Column \emph{Row} provides a reference numbering.
    Column \emph{Calibration parameter} gives the name of each parameter. The
    conventional extracted value appears in the third column, as well as the
    rough uncertainty plots of column \emph{Uncertainty} as a vertical red
    dashed line. Column \emph{Uncertainty} documents which parameters we
    consider for epistemic uncertainty due to representation uncertainty in this
    work and its rough shape. Parameters without uncertainty are either
    categorical numbers stored without loss of precision or values the
    manufacturer has directly hardcoded in the sensor driver source code (i.e.,
    they do not actually come from the sensor device memory). The ``hardcoded''
    values have uncertainty that we are unable to confidently quantify with our
    method because the necessary information is not available. The last column
    gives the uncertainty type of the calibration parameter, where applicable.}
    \begin{tabular}{l c S c c}
        \toprule

        \textbf{Row} & \textbf{Calibration parameter} & \textbf{Extracted value} & \textbf{Uncertainty} & \textbf{Uncertainty type}\\  \hline
        \rowcolor{b} \rowcountstep & $K_{V_{\mathrm{DD}}}$ & -3040 & \includegraphics[width=1.25cm]{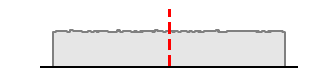} & Epistemic \\
        \rowcolor{a} \rowcountstep & $V_\mathrm{DD25}$ & -12864 & \includegraphics[width=1.25cm]{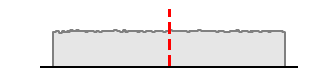} & Epistemic \\
        \rowcolor{b} \rowcountstep & $\mathrm{K}_{V_\mathrm{PTAT}}$ & 0.001953125 & \includegraphics[width=1.25cm]{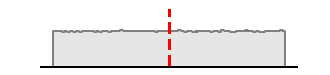} & Epistemic \\
        \rowcolor{a} \rowcountstep & $\mathrm{K}_{T_\mathrm{PTAT}}$ & 42.75 & \includegraphics[width=1.25cm]{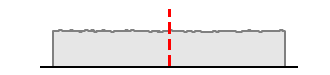} & Epistemic \\
        \rowcolor{b} \rowcountstep & $V_\mathrm{{PTAT25}}$ & 12194 & \includegraphics[width=1.25cm]{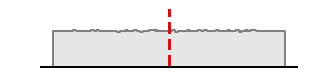} & Epistemic \\
        \rowcolor{a} \rowcountstep & $\alpha_\mathrm{{PTAT}}$ & 9 & \includegraphics[width=1.25cm]{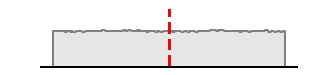} & Epistemic \\
        \rowcolor{b} \rowcountstep & $\mathrm{GAIN}$ & 6276 & \includegraphics[width=1.25cm]{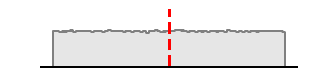} & Epistemic \\
        \rowcolor{a} \rowcountstep & $\mathrm{TGC}$ & 0 & \includegraphics[width=1.25cm]{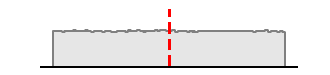} & Epistemic \\
        \rowcolor{b} \rowcountstep & $K_{V_\mathrm{CP}}$ & 0.375 & \includegraphics[width=1.25cm]{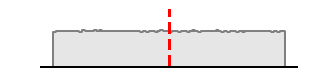} & Epistemic \\
        \rowcolor{a} \rowcountstep & $K_\mathrm{{Ta_{CP}}}$ & 0.00439453125 & \includegraphics[width=1.25cm]{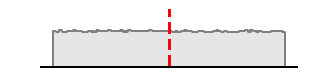} & Epistemic \\
        \rowcolor{b} \rowcountstep & $\mathrm{Resolution_{EE}}$ & 2 & No (nominal) & N/A \\
        \rowcolor{a} \rowcountstep & $\mathrm{CalibrationMode_{EE}}$ & 128 & No (nominal)& N/A \\
        \rowcolor{b} \rowcountstep & $\mathrm{Ks_{Ta}}$ & -0.001220703125 & \includegraphics[width=1.25cm]{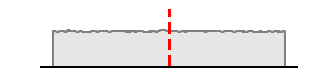} & Epistemic \\
        \rowcolor{a} \rowcountstep & $\mathrm{Ks_{To}}[0]$ & -0.000396728515625 & \includegraphics[width=1.25cm]{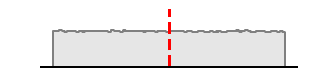} & Epistemic \\
        \rowcolor{b} \rowcountstep & $\mathrm{Ks_{To}}[1]$ & -0.00045013427734375 & \includegraphics[width=1.25cm]{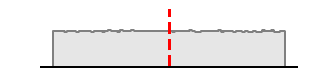} & Epistemic \\
        \rowcolor{a} \rowcountstep & $\mathrm{Ks_{To}}[2]$ & -0.00060272216796875 & \includegraphics[width=1.25cm]{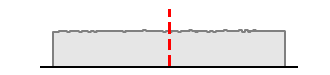} & Epistemic \\
        \rowcolor{b} \rowcountstep & $\mathrm{Ks_{To}}[3]$ & -0.00080108642578125 & \includegraphics[width=1.25cm]{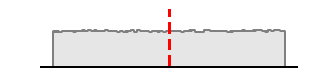} & Epistemic \\
        \rowcolor{a} \rowcountstep & $\mathrm{Ks_{To}}[4]$ & -0.0 & Not quantified (hardcoded) & Epistemic \\
        \rowcolor{b} \rowcountstep & $\mathrm{CT}[0]$ & -40 & Not quantified (hardcoded) & Epistemic \\
        \rowcolor{a} \rowcountstep & $\mathrm{CT}[1]$ & 0 & Not quantified (hardcoded) & Epistemic \\
        \rowcolor{b} \rowcountstep & $\mathrm{CT}[2]$ & 120 & \includegraphics[width=1.25cm]{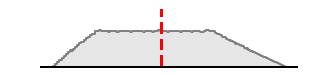} & Epistemic \\
        \rowcolor{a} \rowcountstep & $\mathrm{CT}[3]$ & 240 & \includegraphics[width=1.25cm]{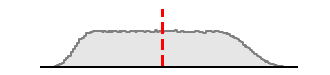} & Epistemic \\
        \rowcolor{b} \rowcountstep & $s_\alpha$ & 11 & No (ordinal) & N/A \\
        \rowcolor{a} \rowcountstep & $s_\mathrm{K_{Ta}}$ & 13 & No (ordinal) & N/A \\
        \rowcolor{b} \rowcountstep & $s_\mathrm{K_V}$ & 7 & No (ordinal) & N/A \\
        \rowcolor{a} \rowcountstep & $\alpha_{\mathrm{CP0}}$ & 3.6670826375484467e-09 & \includegraphics[width=1.25cm]{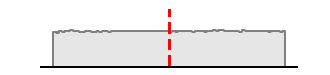} & Epistemic \\
        \rowcolor{b} \rowcountstep & $\alpha_{\mathrm{CP1}}$ & 3.5238372220192105e-09 & \includegraphics[width=1.25cm]{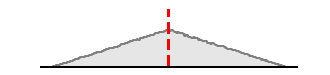} & Epistemic \\
        \rowcolor{a} \rowcountstep & $\mathrm{CP_{Offset}}[0]$ & -80 & \includegraphics[width=1.25cm]{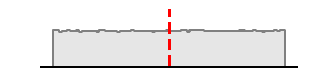} & Epistemic \\
        \rowcolor{b} \rowcountstep & $\mathrm{CP_{Offset}}[1]$ & -75 & \includegraphics[width=1.25cm]{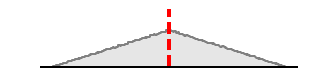} & Epistemic \\
        \rowcolor{a} \rowcountstep & $\mathrm{IL_{Chess_C}}[0]$ & 0.9375 & \includegraphics[width=1.25cm]{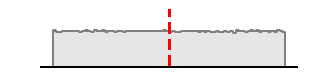} & Epistemic \\
        \rowcolor{b} \rowcountstep & $\mathrm{IL_{Chess_C}}[1]$ & 4.0 & \includegraphics[width=1.25cm]{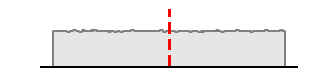} & Epistemic \\
        \rowcolor{a} \rowcountstep & $\mathrm{IL_{Chess_C}}[2]$ & 0.0 & \includegraphics[width=1.25cm]{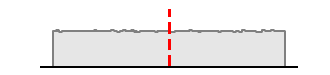} & Epistemic \\
    \end{tabular}
    
    \begin{tabular}{l p{3.2cm} p{5.2cm} p{4cm} c}
        \rowcolor{b} \rowcountstep & \centering $\mathrm{BrokenPixels[]}$  & \centering Empty  & No (ordinal) & N/A \\
        \rowcolor{a} \rowcountstep & \centering $\mathrm{OutlierPixels[]}$ &  \centering Empty & No (ordinal) & N/A \\
        \rowcolor{b} \rowcountstep & \centering $\alpha$ &  \centering Suppl. Fig.~\ref{fig:CalibrationExtraction}.A & Pixel (12,16): \includegraphics[width=1.25cm]{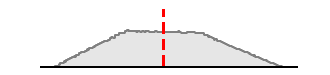}  & Epistemic \\ 
        \rowcolor{a} \rowcountstep & \centering Offset & \centering  Suppl. Fig.~\ref{fig:MlxCalibrationParameterParticleOffset} & Pixel (12,16): \includegraphics[width=1.25cm]{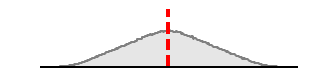} & Epistemic \\
        \rowcolor{b} \rowcountstep & \centering $K_\mathrm{Ta}$ &  \centering Suppl. Fig.~\ref{fig:MlxCalibrationParameterParticleKta} & Pixel (12,16): \includegraphics[width=1.25cm]{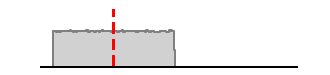} & Epistemic \\
        \rowcolor{a} \rowcountstep & \centering $K_V$ & \centering  Suppl. Fig.~\ref{fig:MlxCalibrationParameterParticleKv} & Pixel (12,16): \includegraphics[width=1.25cm]{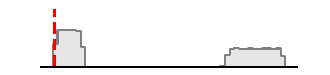} & Epistemic \\
        \bottomrule\\[-6ex]
        \multicolumn{5}{p{10cm}}{\scriptsize The values in the table above are from sensor instance with identifier \mlxinstanceid.}
    \end{tabular}

    \label{tab:Mlx90640CalibrationParametersExtracted}
\end{table*}

\begin{table*}[tbh]
    \centering
    \caption{Calibration data from the flash memory of an Melexis MLX90640
    infrared sensor device. Column \emph{Row} provides a reference numbering.
    Column \emph{Calibration data name} gives the name of the calibration data
    parameter. The same information appears in the MLX90640 datasheet as the
    ``calibration parameters memory'' table~\cite{melexis2019mlx}. The symbol
    ``\textpm'' means the number is in two's complement format. Column
    \emph{Flash value} gives the calibration data value read from the sensor
    memory and column \emph{Bit count} gives the data size in that memory.
    Column \emph{Source register} documents the origin address of the data. The
    last column documents which calibration data we consider for representation
    uncertainty: we do not consider ordinal values (scaling factors), nominal
    values (interleaving pattern mode), or hardcoded values as having
    representation uncertainty.}
    \setcounter{rownumber}{0}
    \begin{tabular}{l c S c c c}
        \toprule
        \textbf{Row} & \textbf{Calibration data name}   & \textbf{Flash value}  &  \textbf{Bit count} & \textbf{Source register} & \textbf{Considered uncertain} \\  \hline
        \rowcolor{b} \rowcountstep & (Alpha PTAT - 8)*4 & \flashvalue{AlphaPtat}         & 4     & EE[\codelight{0x2410}] & \considereduncertain{AlphaPtat} \\
        \rowcolor{a} \rowcountstep & Scale OCC Row      & \flashvalue{OccRowScale}         & 4     & EE[\codelight{0x2410}] & \considereduncertain{OccRowScale} \\
        \rowcolor{b} \rowcountstep & Scale OCC Col      & \flashvalue{OccColScale}         & 4     & EE[\codelight{0x2410}] & \considereduncertain{OccColumnScale} \\
        \rowcolor{a} \rowcountstep & Scale OCC Rem      & \flashvalue{OccRemScale}         & 4     & EE[\codelight{0x2410}] & \considereduncertain{OccRemScale} \\
        \rowcolor{b} \rowcountstep & \textpm\, Pix Os Average  & \flashvalue{PixOsAverage}  & 16  & EE[\codelight{0x2411}] & \considereduncertain{OffsetRef} \\
        \rowcolor{a} \rowcountstep & Alpha Scale - 30   & \flashvalue{AlphaScale}         & 4     & EE[\codelight{0x2420}] & \considereduncertain{AlphaScale} \\
        \rowcolor{a} \rowcountstep & Scale ACC Row      & \flashvalue{AccRowScale}         & 4     & EE[\codelight{0x2420}] & \considereduncertain{AccRowScale} \\
        \rowcolor{b} \rowcountstep & Scale ACC Col      & \flashvalue{AccColScale}         & 4     & EE[\codelight{0x2420}] & \considereduncertain{AccColumnScale} \\
        \rowcolor{a} \rowcountstep & Scale ACC Rem      & \flashvalue{AccRemScale}         & 4     & EE[\codelight{0x2420}] & \considereduncertain{AccRemScale} \\
        \rowcolor{a} \rowcountstep & Pix Sensitivity Average & 14403    & 16  & EE[\codelight{0x2421}] & Yes \\
        \rowcolor{b} \rowcountstep & \textpm\, Gain       & \flashvalue{GainEe}    & 16    & EE[\codelight{0x2430}] & \considereduncertain{GainEe} \\
        \rowcolor{a} \rowcountstep & \textpm\, PTAT 25    & \flashvalue{VPtat25}   & 16    & EE[\codelight{0x2431}] & \considereduncertain{VPtat25} \\
        \rowcolor{a} \rowcountstep & \textpm\, Kv PTAT    & \flashvalue{KvPtat}       & 6     & EE[\codelight{0x2432}] & \considereduncertain{KvPtat} \\
        \rowcolor{b} \rowcountstep & \textpm\, Kt PTAT    & \flashvalue{KtPtat}     & 10    & EE[\codelight{0x2432}] & \considereduncertain{KtPtat} \\
        \rowcolor{a} \rowcountstep & \textpm\, Kv Vdd     & \flashvalue{KVdd}     & 8     & EE[\codelight{0x2433}] & \considereduncertain{KVdd} \\
        \rowcolor{b} \rowcountstep & \textpm\, Vdd 25     & \flashvalue{Vdd25}     & 8     & EE[\codelight{0x2433}] & \considereduncertain{Vdd25} \\

        \rowcolor{a} \rowcountstep & \textpm\, Kv Avg Row-Odd Column-Odd    & \flashvalue{KvRoCo} & 4     & EE[\codelight{0x2434}] & \considereduncertain{KvRoCo} \\
        \rowcolor{b} \rowcountstep & \textpm\, Kv Avg Row-Even Column-Odd   & \flashvalue{KvReCo} & 4     & EE[\codelight{0x2434}] & \considereduncertain{KvReCo} \\
        \rowcolor{a} \rowcountstep & \textpm\, Kv Avg Row-Odd Column-Even   & \flashvalue{KvRoCe} & 4     & EE[\codelight{0x2434}] & \considereduncertain{KvRoCe} \\
        \rowcolor{b} \rowcountstep & \textpm\, Kv Avg Row-Even Column-Even  & \flashvalue{KvReCe} & 4     & EE[\codelight{0x2434}] & \considereduncertain{KvReCe} \\
        
        \rowcolor{a} \rowcountstep & \textpm\, IL Chess C3   & \flashvalue{IlChessC2}  & 5   & EE[\codelight{0x2435}] & \considereduncertain{IlChessC2} \\
        \rowcolor{b} \rowcountstep & \textpm\, IL Chess C2   & \flashvalue{IlChessC1}    & 5     & EE[\codelight{0x2435}] & \considereduncertain{IlChessC1} \\
        \rowcolor{a} \rowcountstep & \textpm\, IL Chess C1   & \flashvalue{IlChessC0}   & 6    & EE[\codelight{0x2435}] & \considereduncertain{IlChessC0} \\

        \rowcolor{b} \rowcountstep & \textpm\, Kt Avg Row-Odd Column-Odd        & \flashvalue{KtRoCo}    & 8     & EE[\codelight{0x2436}] & \considereduncertain{KtaRoCo} \\
        \rowcolor{a} \rowcountstep & \textpm\, Kt Avg Row-Odd Column-Even       & \flashvalue{KtRoCe}    & 8     & EE[\codelight{0x2436}] & \considereduncertain{KtaRoCe} \\
        \rowcolor{b} \rowcountstep & \textpm\, Kt Avg Row-Even Column-Odd       & \flashvalue{KtReCo}    & 8     & EE[\codelight{0x2437}] & \considereduncertain{KtaReCo} \\
        \rowcolor{a} \rowcountstep & \textpm\, Kt Avg Row-Even Column-Even      & \flashvalue{KtReCe}    & 8     & EE[\codelight{0x2437}] & \considereduncertain{KtaReCe} \\
        
        \rowcolor{b} \rowcountstep & Res Control Calib  & \flashvalue{ResolutionEe}    & 2      & EE[\codelight{0x2438}] & No \\    
        \rowcolor{a} \rowcountstep & Kv Scale           & \flashvalue{KvScale}          & 4      & EE[\codelight{0x2438}] & \considereduncertain{KvScale} \\
        \rowcolor{b} \rowcountstep & Kta Scale 1        & \flashvalue{KtaScale1}        & 4      & EE[\codelight{0x2438}] & \considereduncertain{KtaScale1} \\
        \rowcolor{a} \rowcountstep & Kta Scale 2        & \flashvalue{KtaScale2}        & 4      & EE[\codelight{0x2438}] & \considereduncertain{KtaScale2} \\
        
        \rowcolor{b} \rowcountstep & \textpm\, Alpha ((CP Subpage 1) / (CP Subpage 0) - 1)*2\textsuperscript{7}   & \flashvalue{AlphaSp1} & 6     & EE[\codelight{0x2439}] & \considereduncertain{AlphaSp1} \\
        \rowcolor{a} \rowcountstep & Alpha CP Subpage 0 & \flashvalue{AlphaSp0}   & 10  & EE[\codelight{0x2439}] & \considereduncertain{AlphaSp0} \\
        
        \rowcolor{b} \rowcountstep & \textpm\, Offset (CP Subpage 1 -  CP Subpage 0)  & \flashvalue{OffsetSp1}  & 4 & EE[\codelight{0x243A}] & \considereduncertain{OffsetSp1} \\
        \rowcolor{a} \rowcountstep & \textpm\, Offset CP Subpage 0                    & \flashvalue{OffsetSp0}  & 4 & EE[\codelight{0x243A}] & \considereduncertain{OffsetSp0} \\
        
        \rowcolor{b} \rowcountstep & \textpm\, Kv CP    & \flashvalue{CpKv}    & 8  & EE[\codelight{0x243B}] & \considereduncertain{CpKv} \\
        \rowcolor{a} \rowcountstep & \textpm\, Kta CP   & \flashvalue{CpKta}   & 8  & EE[\codelight{0x243B}] & \considereduncertain{CpKta} \\

        \rowcolor{b} \rowcountstep & \textpm\, KsTa*2\textsuperscript{13}       & \flashvalue{KsTa}  & 8 & EE[\codelight{0x243C}] & \considereduncertain{KsTa} \\
        \rowcolor{a} \rowcountstep & TGC (\textpm 4)*2\textsuperscript{7}       & \flashvalue{Tgc}   & 8 & EE[\codelight{0x243C}] & \considereduncertain{Tgc} \\
        
        \rowcolor{b} \rowcountstep & \textpm\, KsTo Range 2 (0\textdegree{C}\dots CT1\textdegree{C})    & \flashvalue{KsTo1}  & 8 & EE[\codelight{0x243D}] & \considereduncertain{KsTo1} \\
        \rowcolor{a} \rowcountstep & \textpm\, KsTo Range 1 (<0\textdegree{C})              & \flashvalue{KsTo0}  & 8 & EE[\codelight{0x243D}] & \considereduncertain{KsTo0} \\
        
        \rowcolor{b} \rowcountstep & \textpm\, KsTo Range 4 (CT2\textdegree{C}\dots)        & \flashvalue{KsTo3}  & 8 & EE[\codelight{0x243E}] & \considereduncertain{KsTo3} \\
        \rowcolor{a} \rowcountstep & \textpm\, KsTo Range 3 (CT1\textdegree{C}\dots CT2\textdegree{C})  & \flashvalue{KsTo2}  & 8 & EE[\codelight{0x243E}] & \considereduncertain{KsTo2} \\
        
        \rowcolor{b} \rowcountstep & Temp Step x10              & \flashvalue{Step}         & 2   & EE[\codelight{0x243F}] & \considereduncertain{Step} \\
        \rowcolor{a} \rowcountstep & CT4                        & \flashvalue{Ct3}          & 4   & EE[\codelight{0x243F}] & \considereduncertain{Ct3} \\
        \rowcolor{b} \rowcountstep & CT3                        & \flashvalue{Ct2}          & 4   & EE[\codelight{0x243F}] & \considereduncertain{Ct2} \\
        \rowcolor{a} \rowcountstep & KsTo Scale Offset - 8      & \flashvalue{KsToScale}    & 4   & EE[\codelight{0x243F}] & No \\    

        \bottomrule\\[-6ex]
        \multicolumn{5}{p{10cm}}{\scriptsize The values in the table above are from sensor instance with identifier \mlxinstanceid.}
    \end{tabular}

    \label{tab:Mlx90640CalibrationMemoryData}
\end{table*}

\begin{figure*}[tbh]
\centering
\begin{minipage}{\textwidth}
    \centering
    \captionof{table}{Representative subset of extracted calibration parameters
    for the Melexis MLX90640 infrared sensor conversion routines. The
    Uncertainty column documents which parameters we consider for epistemic
    uncertainty due to representation uncertainty in this work and its rough
    shape. The conventionally-extracted value appears in the uncertainty plots
    as a vertical red dashed line. Scale-related parameters such as $s_\alpha$
    are essentially ordinal numbers stored without loss of precision.
    Supplementary Material
    Table~\ref{tab:Mlx90640CalibrationParametersExtracted} is the full table.
    \label{tab:Mlx90640CalibrationParametersExtractedShort}}
    \begin{tabular}{l c S c c}
        \toprule

        \textbf{Row} & \textbf{Calibration parameter} & \textbf{Extracted value} & \textbf{Uncertainty} & \textbf{Uncertainty type}\\  \hline
        \rowcolor{b} R8 & $\mathrm{TGC}$ & 0 & \includegraphics[width=1.25cm]{Illustrations/MlxSensorCalibrationFlashValues/MlxCalibrationParameterTGC.pdf} & Epistemic \\
        \rowcolor{a} R23 & $s_\alpha$ & 11 & No (ordinal) & N/A \\
        \rowcolor{b} R26 & $\alpha_{\mathrm{CP0}}$ & 3.346940502524376e-09 & \includegraphics[width=1.25cm]{Illustrations/MlxSensorCalibrationFlashValues/MlxCalibrationParameterCpAlpha0.pdf} & Epistemic \\
        \rowcolor{a} R27 & $\alpha_{\mathrm{CP1}}$ & 3.425384420552291e-09 & \includegraphics[width=1.25cm]{Illustrations/MlxSensorCalibrationFlashValues/MlxCalibrationParameterCpAlpha1.pdf} & Epistemic \\
    \end{tabular}\\[-0.25ex]
    \begin{tabular}{l p{3.1cm} p{4.5cm} p{2.7cm} c}
        \rowcolor{b} R35 & \centering $\alpha$ &  \centering Suppl. Fig.~\ref{fig:CalibrationExtraction}.A & Pixel (12,16): \includegraphics[width=1.25cm]{Illustrations/MlxSensorCalibrationFlashValues/MlxCalibrationParameterAlpha524.pdf}  & Epistemic\hspace{1.8em} \\ 
        \bottomrule
    \end{tabular}
\end{minipage}

\begin{minipage}{\textwidth}
    \centering
    \includegraphics[width=0.95\textwidth]{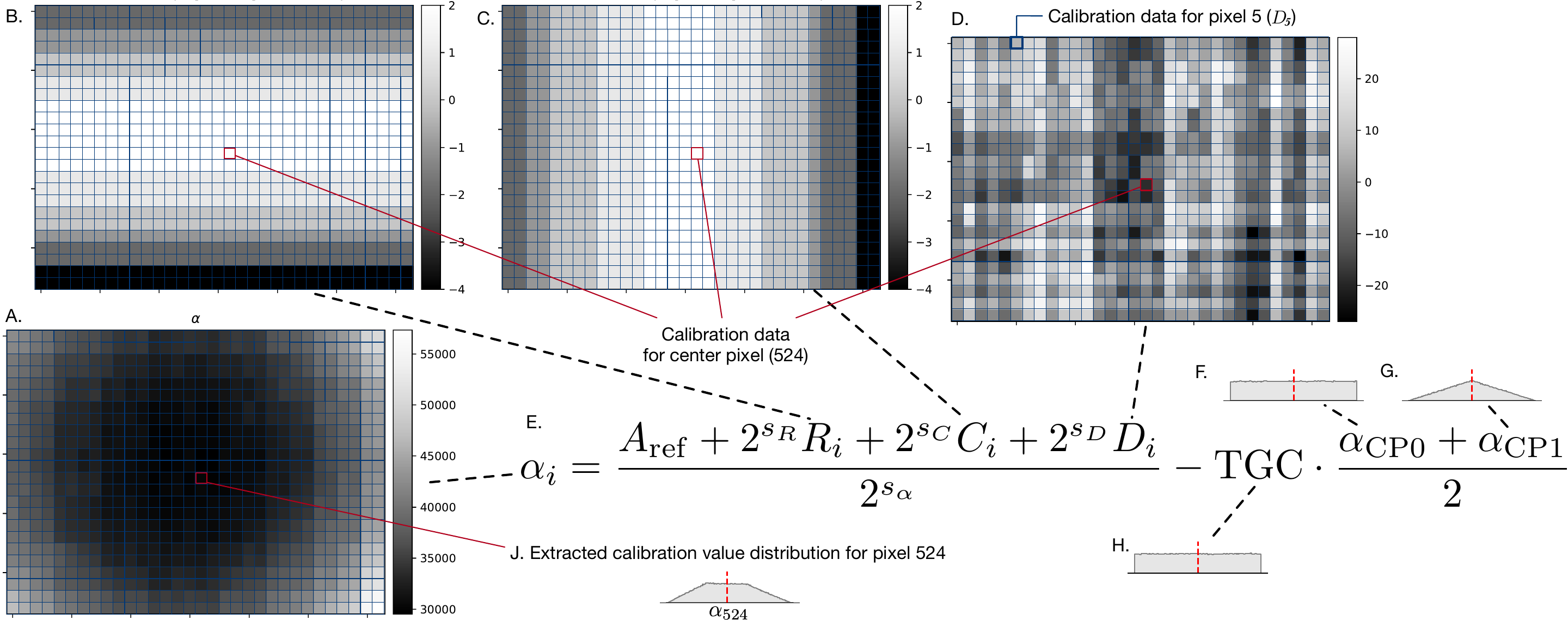}
    \caption{\textbf{Extraction of the sensitivity calibration parameter
    $\alpha$ (Table~\ref{tab:Mlx90640CalibrationParametersExtractedShort}.R35).}
    \textbf{A,} The values of the extracted sensitivity calibration parameter
    $\alpha$. It has different value for each pixel because the calibration data
    that takes part in the extraction equation has different values for each
    pixel.
    \textbf{B,} Per-row calibration data for $\alpha$.
    \textbf{C,} Per-column calibration data for $\alpha$.
    \textbf{D,} Per-pixel calibration data remainder for $\alpha$.
    \textbf{E,} The mathematical extraction equation for calibration parameter $\alpha$.
    \textbf{F}, \textbf{G}, and \textbf{H,} Representation uncertainty of
    already-extracted calibration parameters $\alpha_{\mathrm{CP0}}$,
    $\alpha_{\mathrm{CP1}}$, and $\mathrm{TGC}$.
    \textbf{J,} Representation uncertainty of just-extracted sensitivity
    calibration parameter for the center pixel ($a_{524})$.
    \label{fig:CalibrationExtraction}}
\end{minipage}
\end{figure*}

\begin{table*}[ht]
    \centering
    
    \caption{Performance effects of taking temperature-pixel uncertainty
    into account for edge detection with the Canny filter.}
    \begin{tabular}{rl|c|ccc}
        \toprule
        && \textbf{Conventional} & \multicolumn{2}{c}{\textbf{Uncertainty-Aware}} \\
        \textbf{Scene} & \textbf{Metric} &  & p80 & p90 & p99 \\
        \midrule
        \rowcolor{a} \textbf{Calibration Source 60 \textdegree C}  & Accuracy    & 0.7083 & 1.0000  & 1.0000  & 1.0000  \\
        \rowcolor{b}                                               & Precision   & - & -  & -  & -  \\
        \rowcolor{a}                                               & Sensitivity & - & -  & -  & -  \\
        \midrule
        \rowcolor{a} \textbf{Ferrite 50 \textdegree C}              & Accuracy    & 0.9779 & 0.9935  & 0.9987  & 1.0000  \\
        \rowcolor{b}                                                & Precision   & 0.7536 & 0.9123  & 0.9811  & 1.0000  \\
        \rowcolor{a}                                                & Sensitivity & 1.0000   & 1.0000    & 1.0000    & 1.0000    \\
        \midrule
        \rowcolor{a} \textbf{Heater Plate Corners}                  & Accuracy    & 0.9596 & 0.9701  & 0.9596  & 0.9401  \\
        \rowcolor{b}                                                & Precision   & 0.7732 & 1.0000  & 1.0000  & 1.0000  \\
        \rowcolor{a}                                                & Sensitivity & 0.8929   & 0.7262    & 0.6310    & 0.4524    \\
        \midrule
        \rowcolor{a} \textbf{Heater Plate Diagonal}                 & Accuracy    & 0.9544 & 0.9531  & 0.9518  & 0.9310  \\
        \rowcolor{b}                                                & Precision   & 0.7414 & 0.8395  & 0.8857  & 0.8519  \\
        \rowcolor{a}                                                & Sensitivity & 0.9451   & 0.7473    & 0.6813    & 0.5055    \\
        \midrule
        \rowcolor{a} \textbf{Heater Plate Curve}                    & Accuracy    & 0.9414 & 0.9440  & 0.9271  & 0.8919  \\
        \rowcolor{b}                                                & Precision   & 0.7623 & 1.0000  & 1.0000  & 1.0000  \\
        \rowcolor{a}                                                & Sensitivity & 0.8532   & 0.6055    & 0.4862    & 0.2385    \\
        \midrule
        \rowcolor{a} \textbf{Metal Hob}                             & Accuracy    & 0.9362 & 0.9857  & 0.9935  & 0.9948  \\
        \rowcolor{b}                                                & Precision   & 0.3553 & 0.7353  & 0.8929  & 0.9600  \\
        \rowcolor{a}                                                & Sensitivity & 1.0000 & 0.9259  & 0.9259  & 0.8889    \\
        \midrule
        \rowcolor{a} \textbf{Face}                                  & Accuracy    & 0.9049 & 0.9622  & 0.9531  & 0.9518  \\
        \rowcolor{b}                                                & Precision   & 0.4857 & 0.7941  & 0.7797  & 0.8810  \\
        \rowcolor{a}                                                & Sensitivity & 0.9855   & 0.7826    & 0.6667    & 0.5362    \\
        \midrule
        \rowcolor{a} \textbf{Thumb Up}                              & Accuracy    & 0.8984 & 0.9740  & 0.9753  & 0.9688  \\
        \rowcolor{b}                                                & Precision   & 0.4552 & 0.9423  & 0.9796  & 0.9773  \\
        \rowcolor{a}                                                & Sensitivity & 0.9242   & 0.7424    & 0.7273    & 0.6515    \\
        \midrule
        \rowcolor{a} \textbf{Mac Mini M4}                           & Accuracy    & 0.9427 & 0.9700  & 0.9700  & 0.9609  \\
        \rowcolor{b}                                                & Precision   & 0.6451 & 0.8701  & 0.9130  & 1.0000  \\
        \rowcolor{a}                                                & Sensitivity & 1.0000   & 0.8375    & 0.7875    & 0.6250    \\
        \midrule
        \rowcolor{a} \textbf{USB Hub}                               & Accuracy      & 0.9154 & 0.9831 & 0.9805 & 0.9688 \\
        \rowcolor{b}                                                & Precision     & 0.4836 & 0.9138 & 0.9259 & 0.9512 \\
        \rowcolor{a}                                                & Sensitivity   & 0.9672 & 0.8689 & 0.8197 & 0.6393 \\
        \midrule
        \rowcolor{a} \textbf{(Totals:)}                             & Accuracy & 0.9139 & 0.9736 & 0.9710 & 0.9608 \\
        \rowcolor{b}                                                & Precision & 0.4910 & 0.8935 & 0.9262 & 0.9543 \\
        \rowcolor{a}                                                & Sensitivity & 0.9405 & 0.7746 & 0.7074 & 0.5556 \\
        \bottomrule\\[-6ex]
        \multicolumn{6}{p{10cm}}{\scriptsize The subject is against a non-uniform emissive background in scenes USB Hub, Mac Mini M4, Thumb Up, Face, and Metal Hob.}
    \end{tabular}
    \label{TableEdgeDetectionMetricsFull}
    
\end{table*}

\begin{table*}
    \centering
    \caption{Uncertainty statistics for different calibration data storage
    scenarios. Scenarios that need more memory result to less epistemic
    uncertainty in the sensor output, with the IEEE-754 32-bit scenario leading
    to 99.99\% smaller error statistics across all datasets. While the IEEE-754
    16-bit requires more memory than the TwiceBits scenario, it yields less
    stable benefit compared to the nearest-integer rounding baseline.
    \label{TableUncertaintyStatisticsByScenarioPctChange}}
    \begin{tabular}{l|rrrr}
    \toprule
    \multicolumn{1}{r}{(Scenario:)}& FP16 & FP32 & TwiceBits   & TwoMoreBits \\
    \textbf{Metric}           & \multicolumn{4}{c}{\textbf{\%Change}}   \\
    \midrule
    \rowcolor{a}MinMAE     & -76.9172 & -99.9980 & -93.8324 & -75.0021 \\
    \rowcolor{a}MeanMAE    & -79.2225 & -99.9982 & -93.8263 & -75.0030 \\
    \rowcolor{a}MaxMAE     & -84.0131 & -99.9986 & -93.8185 & -75.0074 \\
    \rowcolor{b}MinMaxAE   & -79.4713 & -99.9980 & -94.1962 & -75.2442 \\
    \rowcolor{b}MeanMaxAE  & -81.9085 & -99.9983 & -94.2508 & -75.3493 \\
    \rowcolor{b}MaxMaxAE   & -86.2310 & -99.9987 & -94.3053 & -75.6145 \\
    \rowcolor{a}MinMRE     & -77.1726 & -99.9980 & -93.8320 & -75.0035 \\
    \rowcolor{a}MeanMRE    & -79.2099 & -99.9982 & -93.8268 & -75.0050 \\
    \rowcolor{a}MaxMRE     & -83.9950 & -99.9986 & -93.8211 & -75.0106 \\
    \rowcolor{b}MinMaxRE   & -79.7551 & -99.9981 & -94.1884 & -75.2225 \\
    \rowcolor{b}MeanMaxRE  & -81.9814 & -99.9983 & -94.2321 & -75.2751 \\
    \rowcolor{b}MaxMaxRE   & -86.2583 & -99.9987 & -94.2479 & -75.3987 \\
    \rowcolor{a}MinStd     & -82.9334 & -99.9979 & -93.8439 & -75.0042 \\
    \rowcolor{a}MeanStd    & -85.0140 & -99.9982 & -93.8369 & -75.0061 \\
    \rowcolor{a}MaxStd     & -88.5936 & -99.9986 & -93.8294 & -75.0144 \\
    \rowcolor{b}MinCI95    & -82.7544 & -99.9980 & -93.8570 & -75.0077 \\
    \rowcolor{b}MeanCI95   & -84.8464 & -99.9982 & -93.8491 & -75.0078 \\
    \rowcolor{b}MaxCI95    & -88.4642 & -99.9986 & -93.8415 & -75.0184 \\
    \bottomrule
    \end{tabular}
\end{table*}

\end{document}